%% file: main_text.tex
\documentclass[reprint,,amssymb,twocolumn,superscriptaddress,aps,prx,10pt]{revtex4-2}

\usepackage{latexsym,amsmath,amssymb,bm,euscript}
\usepackage{float}
\usepackage{amsmath}
\usepackage{amssymb}
\usepackage{amsthm}

\usepackage{amsfonts} 
\usepackage{nicematrix}

\usepackage{tabularx}
\newcolumntype{C}{>{\centering\arraybackslash}X}
\newcolumntype{R}{>{\raggedleft\arraybackslash}X}
\newcolumntype{L}{>{\raggedright\arraybackslash}X}

\usepackage{physics} 
\usepackage{siunitx}
\DeclareSIUnit{\atom}{atom}
\usepackage{bbold}
\usepackage[version=4]{mhchem}
\usepackage{graphicx}
\usepackage[svgnames]{xcolor} 
\usepackage{hyperref}
\hypersetup{
    colorlinks = true,
    linkcolor  = DodgerBlue,
    citecolor  = CornflowerBlue,
    urlcolor   = DarkSlateBlue,
    }

\usepackage{cleveref} 
\crefname{appendix}{Appendix}{Appendices}
\crefname{equation}{Eq.}{Eqs.}
\crefname{figure}{Fig.}{Figs.}
\crefname{table}{Table}{Tables}
\crefname{section}{Section}{Sections}
\crefname{enumi}{case}{cases}
\creflabelformat{appendix}{[#2#1#3]}

\usepackage{silence}
\newcommand{\nc}{\newcommand}

\nc{\webirvsp}{\href{https://github.com/zjwang11/irvsp}{\texttt{IRVSP} }}
\nc{\webirtb }{\href{https://github.com/zjwang11/irvsp}{\texttt{IR2TB} }}
\nc{\webBRdecomp}{\href{http://tm.iphy.ac.cn/UnconvMat.html}{\texttt{BRdecomp} }}
\nc{\webposabr}{\href{https://github.com/zjwang11/UnconvMat/blob/master/src_pos2aBR.tar.gz}{\texttt{POS2ABR} }}

\nc{\red}[1]{{\color{OrangeRed}{#1}}}
\nc{\redit}[1]{{\color{OrangeRed}{\textit{#1}}}}
\nc{\RED}[1]{{\red{\uppercase{#1}}}}
\nc{\green}[1]{{\color{Teal}{#1}}}
\nc{\blue}[1]{{\color{RoyalBlue}{#1}}}

\nc{\djz}[1]{{\color{Crimson}{\;\text{JD:}\;#1}}}

\nc{\addref}{\red{\emph{(add reference here)}} }

\nc{\prlsection}[1]{\textcolor{blue}{\textit{#1}.---}}

\nc{\eq}{=&\;}
\nc{\eqv}{\equiv&\;}

\nc{\cre}[2]{\hat{#1}_{#2}^{\dagger}}
\nc{\ann}[2]{\hat{#1}_{#2}}
\nc{\ua}{\uparrow}
\nc{\da}{\downarrow}

\newcommand{\be}[0]{\begin{equation}}
\newcommand{\ee}[0]{\end{equation}}

\def\ba#1\ea{\begin{align}#1\end{align}}

\def\kk{{\vb{k}}}
\def\qq{{\vb{q}}}

\def\GG{{\vb{G}}}

\def\RR{{\vb{R}}}

\newcommand{\boxedeq}[1]{
\begin{center}
\fbox{
\begin{minipage}{\linewidth}
#1
\end{minipage}
}
\end{center}
}

\makeatletter
\let\oldcref\cref

\newcommand{\AddCrefMap}[2]{%
  \expandafter\def\csname crefmap@#1\endcsname{#2}%
}

\renewcommand{\cref}[1]{%
  \expandafter\ifx\csname crefmap@#1\endcsname\relax
    \oldcref{#1}%
  \else
    \csname crefmap@#1\endcsname
  \fi
}
\makeatother
\AddCrefMap{app:THF_first}{Appendix~II~of the Supplementary Material (SM)~\cite{SM}}
\AddCrefMap{app:THF}{Appendix~II~of the SM~\cite{SM}}

\AddCrefMap{app:hyb_exp}{Appendix~III~of the SM~\cite{SM}}

\AddCrefMap{app:ana_der}{Appendix~IV~of the SM~\cite{SM}}
\AddCrefMap{app:num_res}{Appendix~V~of the SM~\cite{SM}}
\AddCrefMap{app:single_part_nu_0}{Appendix~VI~of the SM~\cite{SM}}
\AddCrefMap{app:J_exp}{Appendix~VII~of the SM~\cite{SM}}

\begin{document}


\title{Twisted Bilayer Graphene Lifetimes At Integer Fillings: An Analytic Result
}

\author{Haoyu~Hu$^{\dagger}$}
\affiliation{Department of Physics, Princeton University, Princeton, New Jersey 08544, USA}
\thanks{These authors contributed equally to this work.}

\author{Yuelin Shao$^{\dagger}$}
\affiliation{Donostia International Physics Center, P. Manuel de Lardizabal 4, 20018 Donostia-San Sebastian, Spain}
\thanks{These authors contributed equally to this work.}

\author{Lorenzo~Crippa}
\affiliation{Institut f\"ur Theoretische Physik und Astrophysik and W\"urzburg-Dresden Cluster of Excellence ctd.qmat, Universit\"at W\"urzburg, 97074 W\"urzburg, Germany}

\author{Dumitru~C\u{a}lug\u{a}ru}
\affiliation{Department of Physics, Princeton University, Princeton, New Jersey 08544, USA}
\affiliation{Rudolf Peierls Centre for Theoretical Physics, University of Oxford, Oxford OX1 3PU, United Kingdom}

\author{Giorgio~Sangiovanni}
\affiliation{Institut f\"ur Theoretische Physik und Astrophysik and W\"urzburg-Dresden Cluster of Excellence ctd.qmat, Universit\"at W\"urzburg, 97074 W\"urzburg, Germany}

\author{Tim~Wehling}
\affiliation{I. Institute of Theoretical Physics, University of Hamburg, Notkestrasse 9, 22607 Hamburg, Germany}
\affiliation{The Hamburg Centre for Ultrafast Imaging, 22761 Hamburg, Germany}

\author{Leonid I. Glazman}
\affiliation{Department of Physics and Yale Quantum Institute, Yale University, New Haven, Connecticut 06520, USA}

\author{B.~Andrei~Bernevig}
\email{bernevig@princeton.edu}
\affiliation{Department of Physics, Princeton University, Princeton, New Jersey 08544, USA}
\affiliation{Donostia International Physics Center, P. Manuel de Lardizabal 4, 20018 Donostia-San Sebastian, Spain}
\affiliation{IKERBASQUE, Basque Foundation for Science, Bilbao, Spain}

\begin{abstract}

Twisted bilayer graphene near integer fillings hosts correlated single-particle excitations whose dispersion and linewidth are increasingly accessible experimentally.
We study these excitations using the topological heavy-fermion model, which captures both strong correlations and band topology of twisted bilayer graphene.
In the decoupled limit, where both the single-particle $fc$ hybridization and the Hund’s coupling between $f$ and $c$ electrons are absent, the model admits exact solutions in which free Dirac fermions coexist with interacting $f$ electrons that form {zero-width} Hubbard bands.
By treating the $fc$ hybridization and Hund’s coupling perturbatively around this solvable limit, we obtain analytical results for the single-particle self-energy. From the resulting self-energy, we derive explicit expressions for \emph{both} the dispersion renormalization and scattering rates of both Hubbard-band excitations and low-energy Dirac modes, thereby establishing an analytical framework for understanding correlated excitations in twisted bilayer graphene. We analyze the scattering of the two kinds ($\Gamma_3$ and $\Gamma_{1,2}$) of  Dirac electrons, and find that it has different mechanisms. We briefly investigate the effect of strain.  We compare these analytical expressions with DMFT results on the same model.

\end{abstract}
\maketitle

\prlsection{Introduction}
Twisted bilayer graphene (TBG) near the magic angle~\cite{bistritzer_moire_2011} exhibits correlated phenomena~\cite{CAO18,KER19,XIE19,SHA19,JIA19,CHO19,POL19,YAN19,LU19,STE20,SAI20,SER20,CHE20b,WON20,CHO20,NUC20,CHO21,SAI21,LIU21c,PAR21c,WU21a,CAO21,DAS21,TSC21,PIE21,STE21,CHO21a,XIE21d,DAS22,NUC23,YU23c}, unconventional superconductivity~\cite{CAO18a,YAN19,LU19,STE20,SAI20,DE21a,OH21,TIA23,DI22a,CAL22d}, and exotic quantum phases~\cite{TOM19,CAO20,ZON20,LIS21,BEN21,LIA21c,ROZ21,SAI21a,LU21,HES21,DIE23,HUB22,GHA22,JAO22,PAU22,GRO22,ZHO23a}. 
Extensive theoretical efforts have been devoted to unraveling the rich phenomenology of magic-angle TBG. These studies span the development of models\cite{LOP07,SUA10,bistritzer_moire_2011,UCH14,WIJ15,DAI16,JAI16,NAM17,EFI18,KAN18,ZOU18,PO19,LIU19a,TAR19,MOR19,LI19,FAN19,KHA19,CAR19a,CAR19,RAD19,KWA20,CAR20,TRI20,WIL20,PAR20,CAR20a,FU20,HUA20a,CAL20,WU21d,REN21,HEJ21,CAL21,BER21,BER21a,WAN21a,RAM21,SHI21,LEI21,CAO21a,SHE21,WU21c,KOS18,LED21,GUE22,DAV22,CLA22,LIN22,SAM22,KAN23b,VAF23,SHI23}, investigations of correlated insulating and metallic phases \cite{OCH18,THO18,XU18b,KOS18,PO18a,VEN18,YUA18,DOD18,PAD18,KEN18,RAD18,LIU19,HUA19,WU19,CLA19,KAN19,SEO19,DA19,ANG19,XIE20b,BUL20b,CHA20,REP20,CEA20,ZHA20,KAN20a,BUL20a,CHI20b,SOE20,CHR20,EUG20,WU20,VAF20,XIE21,KAN21,LIU21,DA21,LIU21a,THO21,KWA21a,LIA21,ZHA21,PAR21a,VAF21,KWA21b,CHE21,POT21,XIE21b,XIE21a,LED21,CHA21,KWA21,HOF22,WAG22,CHR22,BRI22,CAL22d,HON22,ZHA23a,BLA22,XIE23a,KWA23,YU23a,WAN23c,FER20,DAT23,RAI23a}, and analyses of unconventional superconductivity \cite{PhysRevB.110.045133,GUO18,YUA18,XU18,DOD18,PO18a,LIU18a,VEN18,ISO18,PEL18,KEN18,WU18,GUI18,GON19,HUA19,ROY19,WU19a,WU19,YOU19,CLA19,LIA19,HU19a,JUL20,XIE20,CHI20a,LOP20,KON20,CHR20,WAN21,KHA21,LEW21,FER21,QIN21,PHO21,CHO21d,LAK21,CHO21c,LI22c,FIS22,YU22,SCA22,CHR22,CHA22,KWA22,WAG23a,GON23,WAG23,WAN24}. 
Beyond studies of ground states and ordered phases, recent quantum twisting microscope experiments also provide direct access to both the dispersion and the linewidth of excitations~\cite{xiao2025interactingenergybandsmagic,Birkbeck2025}.
This raises a central question: what microscopic processes control the lifetime of single-particle excitations in TBG, and can they be understood analytically?

This problem is challenging because TBG combines two ingredients whose interplay is inherently difficult to treat: strong correlations and nontrivial band topology. The extreme flatness of the moiré bands makes electron-electron interactions the dominant energy scale, while the bands themselves retain topological structure. TBG therefore provides a setting in which topology and strong correlations are inseparably intertwined. 
A natural framework for describing this physics is the topological heavy-fermion (THF) model~\cite{song_magic-angle_2022,calugaru_tbg_2023,shi_heavy-fermion_2022}. In this picture, localized $f$ electrons subject to strong Hubbard interaction coexist with dispersive $c$ electrons forming Dirac nodes. The nontrivial band topology arises from their hybridization. 
Within the THF framework, substantial theoretical efforts, encompassing both analytical and numerical approaches, have been devoted to understanding correlation effects in TBG \cite{hu_symmetric_2023,zhou_kondo_2024,rai_dynamical_2023,datta_heavy_2023,song_magic-angle_2022,shi_heavy-fermion_2022,chou_kondo_2023,lau_topological_2023,wang_molecular_2024,youn_hundness_2024,hu2025projectedsolvabletopologicalheavy, ledwith_nonlocal_2024}. 

A key advantage of the THF model is that it admits an exactly solvable decoupled limit when both the $fc$ hybridization and the Hund's coupling between $f$ and $c$ electrons vanish~\cite{PhysRevLett.131.026502,hu2025projectedsolvabletopologicalheavy}, provided that all density-density interactions other than the onsite Hubbard interaction of $f$ electron are treated at the mean-field level, as discussed in the next section.
In this limit, free Dirac $c$ fermions coexist with interacting $f$ electrons that form local moments and zero-width Hubbard bands.
This solvable point provides an analytical foothold for a controlled treatment of correlated excitations. It also makes it possible to controlably ask how the sharp excitations of the decoupled problem are broadened once the two sectors are weakly recoupled.

\begin{figure}
    \centering
    \includegraphics[width=\linewidth]{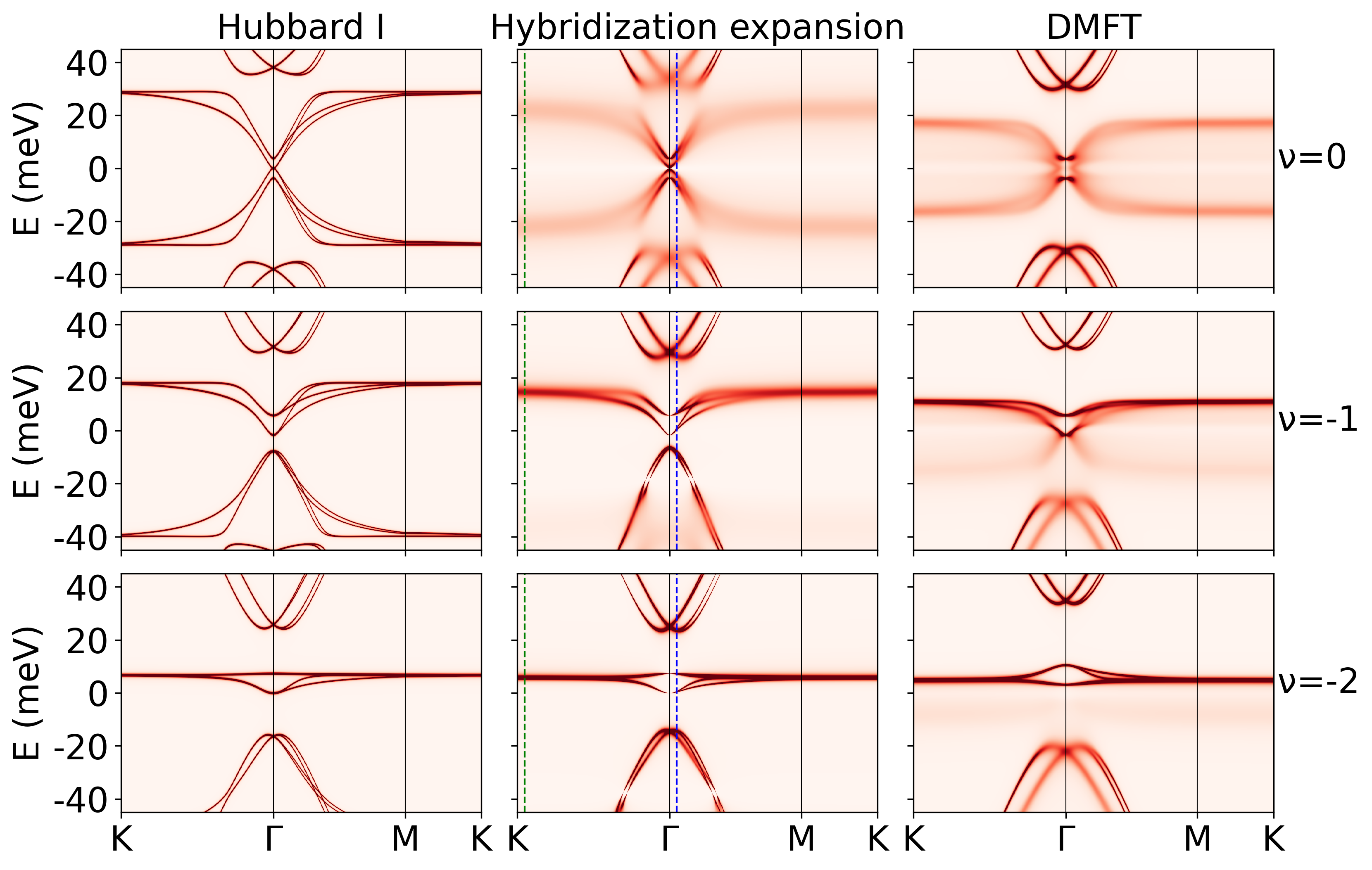}
    \caption{Single-particle excitation spectra at integer fillings $\nu=0,-1,-2$ (top to bottom), obtained within the Hubbard-I approximation (left), hybridization expansion (middle), and DMFT (right).
    To match the DMFT results, we evaluate the hybridization function \cref{eq:main:hyb_fun_M_ne_0} using a reduced effective hybridization parameter $\kappa^{eff}=\kappa/4$.
    }
    \label{fig:main:single_part_exct}
\end{figure}

In this letter, we develop such a controlled expansion in the zero-temperature limit.
Treating the $fc$ hybridization and Hund’s coupling perturbatively around the exactly solvable decoupled limit, we derive analytical expressions for the single-particle self-energy at integer fillings.
These results allow us to determine both the dispersion renormalization and the scattering rates of the Hubbard-band excitations and of the low-energy Dirac modes.
We show that $fc$ hybridization shifts the Hubbard bands toward the Fermi energy. This is consistent with the DMFT results \cite{crippa2025dynamicalcorrelationeffectstwisted, rai_dynamical_2023, datta_heavy_2023} and shows that the Hubbard-band separation is \emph{not} precisly $U$ but can be heavily renormalized. The $fc$ hybridization also generates a finite lifetime. 
For gapless modes near $\Gamma$, $fc$ hybridization renormalizes the Dirac dispersion and induces a scattering rate that is linear in $|\kk|$ in the flat-band limit~\cite{song_magic-angle_2022}, while producing a finite scattering rate away from the ideal flat-band limit. 
Additionally, we obtain the dispersion renormalization and damping rate of the low-energy Dirac node induced by Hund's coupling.
Our results (Fig[\ref{fig:main:single_part_exct}] and Table[\ref{tab:main:summary_cnp}]) provide a transparent analytical framework for single-particle excitation lifetimes in TBG and a direct complement to numerical approaches such as dynamical mean-field theory (DMFT), with which we compare.

\prlsection{Model} We employ the THF model for TBG~\cite{song_magic-angle_2022}, which captures the coexistence of strongly correlated localized $f$ electrons and dispersive Dirac $c$ electrons.
Unless noted otherwise, all results in this work are obtained at zero temperature.
The system is described by the following Hamiltonian (see \cref{app:THF_first})
\ba 
 H=H_0+H_{U_1}+H_{U_2}+H_V+H_{W}+H_J-\mu N
\ea 
Here, $H_0$ contains the $c$-electron kinetic energy and the $fc$ hybridization, $H_{U_1}$ and $H_{U_2}$ denote the on-site and nearest neighbor Hubbard interactions of the $f$ electron, $H_V$, and $H_W$ describe density-density interactions within the $c$ sector and between the $f$ and $c$ sectors, and $H_J$ is the Hund's coupling between $f$ and $c$ electrons.

We denote the $c$- and $f$-electron operators by $c^\dag_{\kk,a \eta s}$ and $f_{\RR,\alpha \eta s}$, respectively. Here $\kk$ and $\RR$ label the momentum and real-space position associated with the moir\'e unit cell, while $\alpha$, $\eta$, and $s$ denote orbital, valley, and spin indices, respectively. The chemical potential is denoted by $\mu$, and 
$N=\sum_{\kk,a \eta s}c^{\dagger}_{\kk,a \eta s}c_{\kk,a \eta s}+\sum_{\kk,\alpha \eta s}f^{\dagger}_{\kk,\alpha \eta s}f_{\kk,\alpha \eta s}$ 
represents the total electron number. 
We define the fillings relative to charge neutrality of the $f$ and $c$ electrons as $\nu_f = \langle \sum_{\kk,\alpha\eta s}:f^{\dagger}_{\kk,\alpha \eta s}f_{\kk,\alpha \eta s}:\rangle/N_M$ and $\nu_c =\langle \sum_{\kk,a\eta s}:c^{\dagger}_{\kk,a\eta s}c_{\kk,a\eta s}:\rangle/N_M$
where $::$ denotes normal ordering with respect to charge neutrality and $N_M$ is the total number of moiré unit cells. The total filling is $\nu=\nu_c+\nu_f$.

To capture excitation lifetimes, we retain the dominant local Hubbard interaction $H_{U_1}$ exactly and treat $H_{U_2}$, $H_{V}$, $H_{W}$ at the mean-field level, replacing them by $H_{U_2}^{MF}$, $H_{V}^{MF}$, and $H_{W}^{MF}$.
We similarly decompose the Hund’s coupling into a mean-field contribution and a residual interaction, $H_J=H_J^{MF}+H_J'$.
The explicit expressions for these mean-field Hamiltonians can be found in \cref{app:THF}.
The resulting effective Hamiltonian consists of the Dirac $c$ sector $H_c$, the atomic $f$ sector $H_f$, the $fc$ hybridization $H_{cf}$, and the residual Hund's coupling $H_J'$.

The $c$-electron Hamiltonian reads
\begin{align}
    H_c=&\sum_{\eta s} \sum_{aa'} \sum_{\kk} ( H^{(c,\eta)}_{aa'}(\kk) 
  ) c_{\kk,a\eta s}^\dagger c_{\kk,a'\eta s} \nonumber\\
   H^{(c,\eta)}(\kk) = &
  \begin{pmatrix}
    (\epsilon_{c,1}-\mu)\sigma_0 & v_\star (\eta k_x \sigma_0 + ik_y \sigma_z) \\
    v_\star (\eta k_x \sigma_0 - ik_y \sigma_z) & (\epsilon_{c,2}-\mu)\sigma_0+ M\sigma_x 
  \end{pmatrix}
  \label{eq:Hc}
\end{align}
where $v_\star$ is the Dirac velocity of the $c$ electrons,  $\epsilon_{c,1}$ and $\epsilon_{c,2}$ encode the mean-field shifts generated by $H_{V}$, $H_W$, $H_{J}$ (see \cref{app:THF}).
At charge neutrality, $\mu=\epsilon_{c,1}=\epsilon_{c,2}=0$, the $c$-electron sector hosts a linear Dirac node for $M=0$, whereas finite $M$ produces a quadratic node. As we show below, this distinction qualitatively changes the low-energy damping of the Dirac excitations.
The Hamiltonian of the interacting localized $f$ electron reads
\begin{align}
    H_f=&\sum_{\RR,\alpha\eta s}(\epsilon_f-\mu) f_{\RR ,\alpha \eta s}^\dag f_{\RR,\alpha \eta s}\label{eq:Hf}\\
  &+ {\frac{U_1}{2} \sum_{\RR,\alpha \eta s\ne \alpha'\eta's'}f_{\RR,\alpha\eta s}^\dag f_{\RR,\alpha\eta s} f_{\RR,\alpha'\eta's'}^\dag f_{\RR,\alpha'\eta's'}}   \nonumber
\end{align}
where $\epsilon_f$ characterizes the mean-field contributions from $H_{U_2},H_{W},H_{J}$ as well as the single-particle contribution obtained by expanding the normal ordering of $H_{U_1}$ (see \cref{app:THF}).
Since different sites are decoupled, $H_f$ reduces to an atomic problem and produces zero-width Hubbard bands at zero temperature.

While $H_c$ and $H_f$ alone yield sharp excitations, the $fc$ hybridization
\begin{align}
    H_{cf}=&\sum_{\eta s,a \alpha} \sum_{\kk}
  \bigg(  H_{a\alpha}^{(cf,\eta)}(\kk) 
    c_{\kk,a\eta s}^\dagger f_{\kk, \alpha \eta s} + h.c. 
  \bigg)\nonumber\\
  H^{(cf,\eta)}(\kk) =& \gamma 
  \begin{pmatrix}
    \sigma_0 + \tilde{v}_{\star}^\prime(\eta k_x \sigma_x + k_y \sigma_y) \\
    0_{2\times2}
  \end{pmatrix}
  \label{eq:Hcf}
\end{align}
provides the primary mechanism by which the sharp excitations of the decoupled limit acquire finite lifetimes. Here $\gamma$ sets the overall strength of the hybridization between the localized and itinerant sectors. We parameterize the momentum-dependent part by $\tilde{v}_\star^\prime$, scaled by $\gamma$, so that $\gamma=0$ cleanly turns off all hybridization terms.
The four $c$-electron orbitals transform as the irreducible representations $\Gamma_3\oplus\Gamma_1\oplus\Gamma_2$, whereas the two $f$ orbitals form the two-dimensional irreducible representation $\Gamma_3$. Consequently, near the $\Gamma$ point only the $\Gamma_3$ sector of the $c$ electrons hybridizes directly with the $f$ electrons.
In addition, the Hund's coupling
\begin{align}
    H_{J}=&-\frac{J}{2N_M}\sum_{\alpha\eta s}\sum_{\alpha'\eta's'}\sum_{\kk,\kk',\qq}[\eta\eta'+(-1)^{\alpha+\alpha'}]\label{eq:HJ}\\
    &:f^{\dagger}_{\kk+\qq,\alpha\eta s}f_{\kk,\alpha'\eta's'}: :c^{\dagger}_{\kk'-\qq,(\alpha'+2)\eta's'}c_{\kk',(\alpha+2)\eta s}: \nonumber
\end{align}
provides an additional mechanism that also broadens the single-particle linewidths. In contrast to $H_{cf}$, the Hund's coupling acts on the $\Gamma_1\oplus\Gamma_2$ sector of the $c$ electrons rather than the $\Gamma_3$ sector.

Throughout this manuscript, we use the parameter values reported in Ref.~\cite{song_magic-angle_2022}.
The dominant energy scale is the half-bandwidth of the $c$ electrons in the first moir\'e Brillouin zone, $v_\star\sqrt{\Omega_M/\pi} \approx 122\,\text{meV}$, where $\Omega_M$ is the area of the moir\'e Brillouin zone. The other relevant energy scales are the Hubbard interaction $U_1 = 57.95 \text{meV}$, the $fc$ hybridization strength $\gamma = -24.75\text{meV}$, the Hund's coupling $J = 16.38\text{meV}$, and $M=3.69$meV.
This hierarchy implies that local $f$-electron correlations are strong, while the hybridization and Hund's coupling act as the leading perturbations governing the linewidths. However, our results turn out to be identical to those obtained in the opposite limit \cite{scattering1,scattering2} suggesting- as mentioned in \cite{hu2025projectedsolvabletopologicalheavy}, that the THF results are not strongly dependent on the ratio $U/\gamma$, due to the light fermion linear dispersion. 

\begin{figure}
    \centering
    \includegraphics[width=1.0\linewidth]{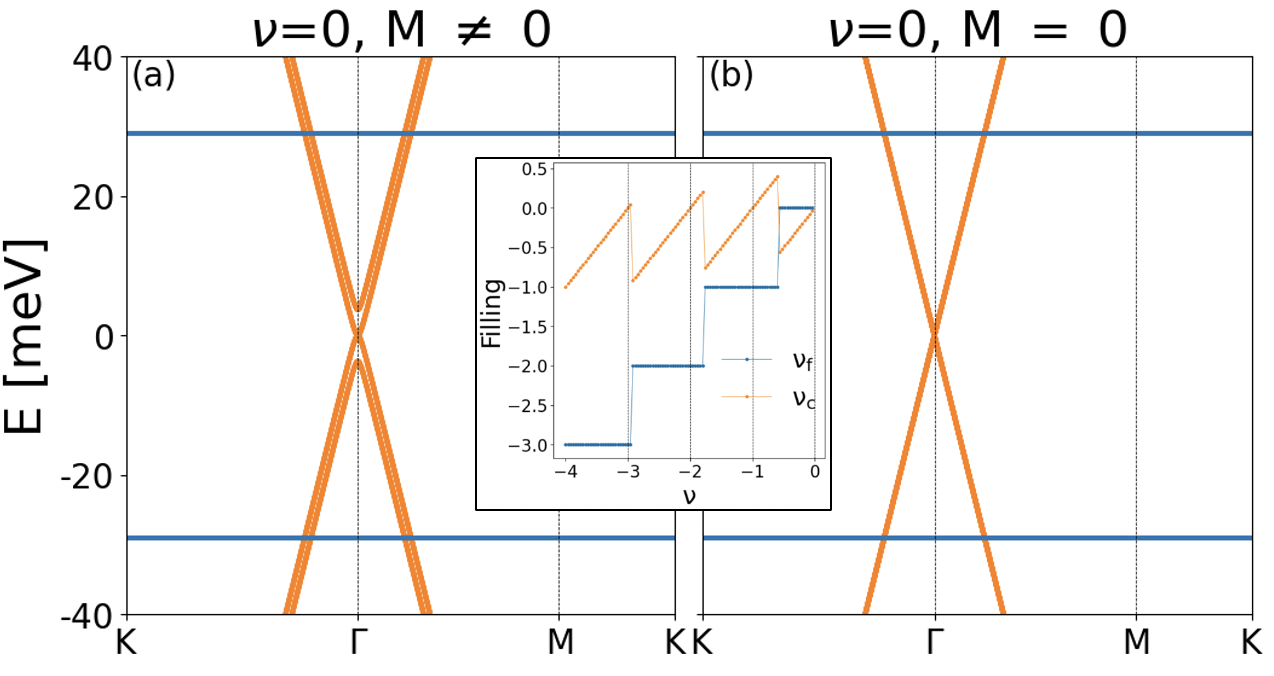}
    \caption{Exactly solvable decoupled limit of the THF model, $\gamma=J=0$.
(a,b) Single-particle excitation spectra at $\nu=0$ for $M=0$ (a) and $M\neq 0$ (b). The blue lines denote the Hubbard bands formed by the $f$ electrons, while the orange lines represent the gapless Dirac node of the $c$ electrons, exhibiting linear dispersion in (a) and quadratic dispersion in (b). The inset, reproduced from\cite{hu_kondo_2023}, shows $f$- and $c$-electron fillings $\nu_f$ and $\nu_c$ as functions of the total filling $\nu$ - in the presence of translational invariance.
 }
    \label{fig:zero_hyb}
\end{figure}

\prlsection{Decoupled limit}
We begin from the exactly solvable limit $\gamma=J=0$~\cite{PhysRevLett.131.026502,hu2025projectedsolvabletopologicalheavy}, which provides the sharp excitations that are broadened once hybridization and Hund's coupling are restored. In this limit, the $c$ and $f$ sectors decouple, and both the ground state and the single-particle spectrum are obtained analytically.
We focus on integer fillings $\nu=0,\pm1,\pm2$, for which the exact ground state places all charge in the localized sector: $\nu_f=\nu$ and $\nu_c=0$ [see the inset of Fig.~\ref{fig:zero_hyb}].
For simplicity we take $\epsilon_{c,1}=\epsilon_{c,2}$, which is exact at $\nu=0$. The $c$ sector then hosts a Dirac node pinned at the Fermi energy, with linear dispersion for $M=0$ and quadratic dispersion for $M\neq0$ [Fig.~\ref{fig:zero_hyb}(a,b)].
The localized $f$ electrons form free local moments~\cite{hu2025projectedsolvabletopologicalheavy}, producing atomic Hubbard bands. The corresponding charge-$\pm1$ excitations occur at energies $\Delta_{\pm1}$ (\cref{app:ana_der}) and satisfy $\Delta_{+1}+\Delta_{-1}=U_1$ [Fig.~\ref{fig:zero_hyb}(a,b)]. The exact $f$-electron self-energy is local and diagonal in orbital, valley, and spin:
\ba 
\label{eq:main:self_energy_zero_hyb}
&\Sigma_{f,\alpha \eta s,\alpha'\eta's'}(i\omega,\kk)\bigg|_{\text{decoupled limit}}
= \delta_{\alpha \eta s, \alpha'\eta's'}\Sigma_{f,loc}(i\omega)\nonumber\\
&
\Sigma_{f,loc}(i\omega)= 
\frac{\bigg(1- \frac{4\nu_f^2}{N_f^2}\bigg)U_1^2}{4 
\bigg[ i\omega - \frac{\nu_f}{N_f}U_1-\frac{\Delta_{+1}-\Delta_{-1}}{2} \bigg] 
}
\ea 
where $N_f=8$ is the number of $f$ flavors.

\prlsection{Hybridization expansion}~
We now perturb away from the decoupled limit by introducing a finite $fc$ hybridization while keeping the Hund's coupling fixed at $J=0$.
The resulting self-energy can be obtained analytically in perturbation theory in $\gamma$.
To order $\gamma^2$, the self-energy at integer fillings $\nu=0,\pm 1,\pm 2$ remains diagonal in orbital, valley, and spin space and is given by (see \cref{app:ana_der} for the derivation):
\ba 
\label{eq:main:self_energy}
&\Sigma_{f,\alpha \eta s,\alpha'\eta's'}(i\omega,\kk)
\approx  \delta_{\alpha \eta s, \alpha'\eta's'}\Sigma_{f,loc}(i\omega)\nonumber\\
&
\qquad\times\bigg[1
+ \frac{ (1+N_f)[\gamma^2\Delta(i\omega)] }{ i\omega - \frac{\nu_f}{N_f}U_1
-\frac{\Delta_{+1}-\Delta_{-1}}{2} 
}
 \bigg] +\mathcal{O}(\gamma^4)
\ea 
where we define the hybridization function as
\ba 
\label{eq:main:hyb_fun_def}
&\gamma^2\Delta(i\omega) = \frac{1}{ N_M}\\
&\times
\sum_{\kk} 
\bigg[ H^{(fc,\eta)}(\kk) \cdot [i\omega \mathbb{I} - H^{(c,\eta)}(\kk)]^{-1}\cdot H^{(cf,\eta)}(\kk)
\bigg]_{\alpha ,\alpha}\nonumber
\ea

 \begin{table*}[t]
    \centering
    \begin{tabular}{c|c|c}
         & $\kappa \ne 0, j=0$ (hybridization) & $\kappa =0, j\ne 0$ (Hund's coupling)\\
         \hline 
      Hubbard bands& $ U_1(N_f+1)\pi^2\kappa/4$ & --- \\ 
      Linear Dirac node near $\Gamma$ point $(M=0)$ & $\propto\kappa(N_f+1)|v_\star\kk|$ & $\propto j |v_\star\kk| $  \\ 
      Quadratic Dirac node near $\Gamma$ point $(M\ne 0)$  & $\propto \kappa (N_f+1) M$  & 
      $\propto j|v_{\star}\kk|^4/M^3$
    \end{tabular}
    \caption{Leading-order scattering rates at charge neutrality ($\nu=0$) for the Hubbard bands and low-energy dispersive bands near the $\Gamma$ point, shown separately for $fc$ hybridization and Hund's coupling.
    For Hund's coupling, we only consider the low-energy dispersive bands.
    }
    \label{tab:main:summary_cnp}
\end{table*}  

At integer fillings, the hybridization function \cref{eq:main:hyb_fun_def} can be evaluated analytically in the case of "chiral limit symmetry" with $v_\star^\prime =0 $~\cite{song_magic-angle_2022}.
In the case of $M=0$, the conduction electrons have a linear Dirac dispersion, and the hybridization function becomes
\ba 
\label{eq:main:hyb_fun_M_0}
\gamma^2 \Delta(i\omega) = -\kappa i\omega \pi \log\bigg( 
\frac{|v_\star\Lambda_c|^2 +\omega^2}{\omega^2 }
\bigg) 
\ea 
where the dimensionless parameter
\ba 
\kappa = \gamma^2/(v_\star^2\Omega_M)\approx 0.013 
\ea
characterizes the strength of the hybridization relative to the $c$-electron bandwidth, $\Lambda_c$ is the momentum cutoff of the $c$ electrons and is taken to be $\Lambda_c =\sqrt{ \Omega_M/\pi} $. This is the microscopic expression \cite{hu2025projectedsolvabletopologicalheavy} of the parameter $s$ used in \cite{ledwith_nonlocal_2024}.
With a finite $M$, the linear Dirac node becomes quadratic, resulting in a modified hybridization function,
\ba 
\label{eq:main:hyb_fun_M_ne_0}
 &\gamma^2\Delta_M(i\omega) = \gamma^2 \Delta(i\omega)  \nonumber\\
 &+ \frac{\kappa \pi M}{2}\log\bigg( \frac{i\omega+M}{i\omega-M}
 \bigg) 
 + \frac{\kappa \pi i\omega}{2}\log(\frac{M^2+\omega^2}{\omega^2})
\ea 
Detailed derivations of \cref{eq:main:hyb_fun_M_0,eq:main:hyb_fun_M_ne_0} are given in \cref{app:ana_der}.

{
It is worth noting that \cref{eq:main:hyb_fun_def} is different from the hybridization function used in DMFT calculations of the THF model. DMFT takes indeed the effect of the other lattice sites of the periodic Anderson model into account via auxiliary bath fermions that are self-consistently adjusted \cite{PhysRevLett.131.166501,PhysRevX.14.031045,CrippaStrain,MaxFragility}.}

\begin{figure}
    \centering
    \includegraphics[width=\linewidth]{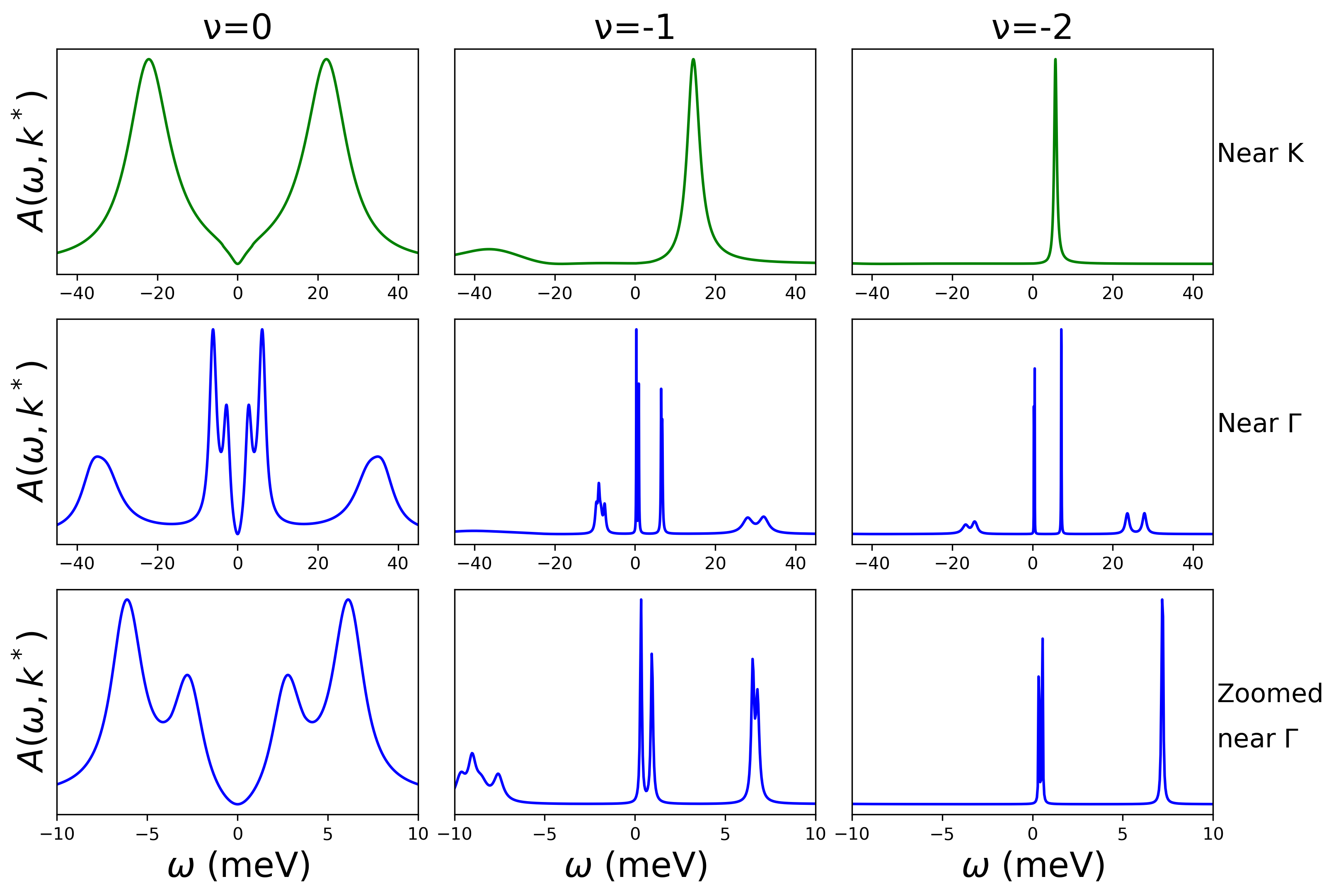}
    \caption{Line cuts of the hybridization-expansion spectra near $K$ (top) and $\Gamma$ (middle), taken along the green and blue dashed paths in \cref{fig:main:single_part_exct}. The bottom panel shows a zoom-in of the low-frequency spectra near the $\Gamma$ point. From left to right, the three columns correspond to $\nu=0$, $-1$, and $-2$, respectively.}
    \label{fig:main:spec_line_cut}
\end{figure}

\prlsection{Correlated single-particle spectra}
Using the self-energy and hybridization functions in Eqs.~\eqref{eq:main:self_energy} and \eqref{eq:main:hyb_fun_M_ne_0}, we obtain the single-particle spectra shown in Fig.~\ref{fig:main:single_part_exct} (detailed in \cref{app:num_res}). 
For comparison, we also show spectra from DMFT~\cite{PhysRevX.14.031045} and from the Hubbard-I approximation~\cite{hu2025projectedsolvabletopologicalheavy}, where the $f$-electron self-energy takes the atomic form $\Sigma_{f,\mathrm{loc}}(i\omega)$ in \cref{eq:main:self_energy_zero_hyb}. 
To align the hybridization-expansion results more closely with DMFT, we use the same parameters as in Ref.~\cite{PhysRevX.14.031045}, including finite $M$ and $v_\star'$, but evaluate the hybridization function \cref{eq:main:hyb_fun_M_ne_0} with a reduced effective hybridization parameter $\kappa^{eff}=\kappa/4$ (numerically realized by doubling the $c$-electron bandwidth, corresponding to taking $v_{\star}^{eff} = 2v_{\star}$ in the hybridization function).
\Cref{fig:main:spec_line_cut} shows line cuts of the hybridization-expansion spectra near $K$ and $\Gamma$, taken along the green and blue dashed paths in \cref{fig:main:single_part_exct}.
The results from using the original $\kappa$ are provided in \cref{app:num_res} for reference.

Within the Hubbard-I approximation, the $f$ electrons contribute atomic Hubbard bands through $\Sigma_{f,loc}(i\omega)$, while the bare lattice hybridization mixes these states with the dispersive $c$-electron bands. At this level, all excitations remain sharp and undamped. Including the leading self-energy correction from the hybridization expansion, Eq.~\eqref{eq:main:self_energy}, produces two effects: an additional renormalization of the dispersion and a finite linewidth. 
As shown in Fig.~\ref{fig:main:single_part_exct}, these leading-order results qualitatively reproduce the corresponding DMFT features across integer fillings, indicating that the expansion already captures the microscopic origin of the damping and band renormalization. The quantitative reproduction of DMFT requires a lower $\kappa$. The justification for introducing the effective $\kappa^{eff}$ is discussed in the next section.

\prlsection{Analytical results for the spectra}
We now derive the corresponding dispersion renormalization and scattering rates analytically. We focus on charge neutrality, $\nu=0$. 
Using Eqs.~\eqref{eq:main:self_energy}, \eqref{eq:main:hyb_fun_M_0}, and \eqref{eq:main:hyb_fun_M_ne_0}, we determine the excitation energies and scattering rates from the poles of the Green's function analytically. We focus on the Hubbard bands near the Brillouin-zone edge and the gapless modes near $\Gamma$ (see \cref{app:single_part_nu_0}).

Near the Brillouin-zone edge, the $f$ electrons form Hubbard bands at energies $\pm U_1/2$ within the Hubbard-I approximation. Including the self-energy correction induced by hybridization, the Hubbard-band energy is renormalized to
{\small
\ba 
E_{Hubbard} \approx \frac{U_1}{2} 
\bigg[ 
1 - \frac{\kappa}{2}
 \pi (N_f+1)  \log( \frac{4 \gamma^2}{\kappa \pi U_1^2} )  
+\mathcal{O}(\kappa^2)
\bigg] 
\label{eq:main:E_Hubbard}
\ea 
}
Thus, the Hubbard bands are shifted toward the Fermi energy. In addition, a finite scattering rate emerges,
\ba 
\frac{1}{\tau_{Hubbard}}
\approx 
\frac{U_1\pi^2 (N_f+1) }{4}(\kappa+\mathcal{O}(\kappa^2))
\ea 
which scales linearly with $U_1 \kappa$. 
Using the original value $\kappa \approx 0.013$, we obtain $E_{Hubbard} \approx 13.6~\mathrm{meV}$ and $\tau_{Hubbard}^{-1} \approx 16.7~\mathrm{meV}$, whereas the DMFT results of Ref.~\cite{PhysRevX.14.031045} give $E_{Hubbard}^{DMFT} \approx 16.4~\mathrm{meV}$ and $(\tau_{Hubbard}^{DMFT})^{-1} \approx 3.0~\mathrm{meV}$. The leading-order hybridization expansion therefore overestimates both the energy renormalization and the scattering rate of the Hubbard bands.
Indeed, the first-order correction in $\kappa$ in \cref{eq:main:E_Hubbard} is $\frac{\kappa}{2}\pi (N_f+1)\log\!\left(\frac{4\gamma^2}{\kappa \pi U_1^2}\right)\approx 0.53$ which is not small, indicating that higher-order corrections are likely important for a quantitatively accurate description of the Hubbard bands. Identical expressions and values occur for the $s$ parameter expansion in \cite{scattering1,scattering2, hu2025projectedsolvabletopologicalheavy, ledwith_nonlocal_2024}. To partially account for these higher-order effects, we use an effective hybridization parameter $\kappa^{eff}$ when generating the hybridization-expansion spectra in \cref{fig:main:single_part_exct}. We find that $\kappa^{eff}=\kappa/4$ gives $E_{Hubbard}^{eff} \approx 23.3~\mathrm{meV}$ and $(\tau_{Hubbard}^{eff})^{-1} \approx 4.2~\mathrm{meV}$, in reasonable agreement with the DMFT results.

We next consider the gapless excitations formed by the $c$ electrons near the $\Gamma$ point, analyzing the cases $M=0$ and $M\ne 0$ separately. In the $M=0, v_\star^\prime=0$ limit, the Hubbard-I approximation yields a linear Dirac dispersion, $
\pm \frac{U_1/2}{\sqrt{U_1^2/4+\gamma^2}}|v_\star\kk|
$, near the $\Gamma$ point. 
Including the self-energy correction from hybridization leads to a renormalization of the Dirac velocity,
{\small
\ba \label{HubbardBandSeparation}
E_{\kk,Dirac}
\approx &  \frac{U_1|v_\star \kk| }{2\sqrt{U_1^2/4+\gamma^2}}
\bigg[ 
1 -  \kappa 
\frac{2\pi (N_f+1)\gamma^2}{4\gamma^2 +U_1^2} \nonumber\\
&\qquad\times\log\bigg( \frac{1}{\kappa}\frac{\gamma^2(U_1^2+4\gamma^2)}{\pi U_1^2|v_\star\kk|^2}
\bigg) +\mathcal{O}(\kappa^2)
\bigg] 
\ea 
}
A finite scattering rate also appears at finite $\kk$,
\ba \label{scatteringrategamma3}
\frac{1}{\tau_{\kk,Dirac}}\approx
\bigg[ \frac{(N_f+1)U_1 \gamma^2 \pi^2}{4 (U_1^2/4+\gamma^2)^{3/2}}\kappa 
+\mathcal{O}(\kappa^2)
\bigg] 
|v_\star \kk|
\ea 
which vanishes linearly in $|\kk|$. Taking the numerical values of the parameters, we find $({\tau_{\kk,Dirac}})^{-1}\approx 0.18 
|v_\star \kk|$ for the original $\kappa$ and $0.045|v_\star \kk|$ for the effective $\kappa^{eff}$.

With finite $M$, the low-energy dispersion changes qualitatively: a quadratic Dirac node emerges near the $\Gamma$ point. Within the Hubbard-I approximation, the dispersion takes the form $\frac{U_1^2/4}{M(\gamma^2 +U_1^2/4)}|v_\star\kk|^2$. The self-energy correction at order $\kappa$ renormalizes it to
{\small 
\ba 
& E_{\kk,Dirac}^M\approx  
 \frac{U_1^2/4}{M(\gamma^2 +U_1^2/4)}|v_\star\kk|^2
\bigg[
1 -  \frac{  (N_f+1) \gamma^2 \pi \kappa}{\gamma^2+U_1^2/4}
\nonumber\\
&\qquad\times\bigg(
\log\bigg( \frac{1}{\kappa}\frac{\gamma^2(4\gamma^2 +U_1^2)}{\pi U_1^2|v_\star\kk|^2}\bigg)
- 1
\bigg) +\mathcal{O}(\kappa^2)
\bigg] 
\ea 
}
The corresponding scattering rate near the $\Gamma$ point is
\ba \label{scatteringrategamma3mass}
\frac{1}{\tau_{\kk,Dirac}^M} 
\approx  
\frac{(N_f+1)M\gamma^2 \pi^2}{2(\gamma^2 +U_1^2/4) } 
( \kappa +\mathcal{O}(\kappa^2)) 
\ea 
which is proportional to $M\kappa $. Taking the numerical values of the parameters, we find a relatively weak scattering rate $({\tau_{\kk,Dirac}^M})^{-1} \approx 0.9$meV for the original $\kappa$ and $0.225$meV for $\kappa^{eff}$. 
A summary of the scattering rates induced by $fc$ hybridization is provided in \cref{tab:main:summary_cnp}.

\prlsection{Effect of Hund's coupling}~
We next isolate the effect of Hund's coupling by setting $\gamma=0$ and expanding perturbatively in $J$ at charge neutrality, $\nu=0$. 
In the decoupled limit, the $c$ electrons remain gapless, while the half-filled $f$ sector hosts fluctuating local moments. Through $H_J$, these fluctuations generate a self-energy correction for the $\Gamma_1\oplus\Gamma_2$ $c$ electrons [\cref{eq:HJ}].
Introducing the dimensionless coupling
\ba
j=\frac{J^2}{|v_\star|^2\Omega_M}\approx 0.006,
\ea
we obtain the dispersion renormalization and scattering rates at order $j$ (see \cref{app:J_exp} for details).

For $M=0$, the Dirac dispersion is renormalized to
\ba 
E_{\kk,Dirac}^{j}\approx 
|v_\star \kk| \bigg[1 -\frac{31\pi j}{56}\log\bigg( \frac{|v_\star \Lambda_c|^2}{|v_\star \kk|^2 }\bigg) +\mathcal{O}(j^2)\bigg] 
\ea 
while the scattering rate becomes
\ba \label{scatteringrategamma12}
\frac{1}{\tau_{\kk,Dirac}^j}\approx 
\bigg[\frac{31}{56}\pi^2 j
+\mathcal{O}(j^2)\bigg]|v_\star \kk| 
\ea 
which therefore vanishes linearly in $|\kk|$. Numerically, this gives $(\tau_{\kk,Dirac}^j)^{-1}\approx 0.033|v_\star \kk| $, which is small compared with the hybridization-induced broadening.

For finite $M$, the low-energy mode becomes quadratic, with renormalized dispersion
\ba 
E_{\kk,Dirac}^{j,M}\approx 
\bigg[ 
1  +\frac{j\pi}{14}
+\mathcal{O}(j^2)
\bigg]
\frac{|v_\star\kk|^2}{M}
\ea 
while the scattering rate is further suppressed and scales as $j|v_\star\kk|^4$,
\ba 
\frac{1}{\tau_{\kk,Dirac}^{j,M}}\approx\bigg[\frac{29}{56}\pi^2 j+\mathcal{O}(j^2)\bigg]\frac{|v_{\star}\kk|^4}{M^3}. 
\ea 
which evaluates to $(\tau_{\kk,Dirac}^{j,M})^{-1}\approx 0.031|v_{\star}\kk|^4/M^3$.
This absence of an $O(j|v_{\star}\kk|^2)$ scattering rate is a symmetry consequence. Hund's coupling \cref{eq:HJ} acts on the $\Gamma_1\oplus\Gamma_2$ sector of the $c$ electrons, whereas the quadratic low-energy mode near $\Gamma$ is predominantly $\Gamma_3$. Hence the $\Gamma_3$ mode acquires lifetime by the hybridization expansion, while the $\Gamma_{1,2}$ aquire it through the Hund's coupling. Under finite strain, local-moment fluctuations in the $f$ sector are suppressed at charge neutrality, and no low-energy scattering rate is generated at order $j$ either (see \cref{app:J_exp}).
These results are summarized in \cref{tab:main:summary_cnp}.

\prlsection{Summary and Discussion}
We have addressed several experimentally important questions for the interacting bands of TBG: (1). At what energy are the Hubbard bands. We find Eq(\ref{eq:main:E_Hubbard}) a renormalization to lower values, consistent with DMFT, and which should be considered in future analysis of QTM experiments. (2) How sharp are the light $c$-electron  excitations around the $\Gamma$- point?  We find two mechanisms, for the two types of light electrons: the hybridization between $f$ and $\Gamma_3$-type $c$ electrons is responsible for their linear in $k$ lifetime (Eq(\ref{scatteringrategamma3})) in the flat band limit or their finite (Eq(\ref{scatteringrategamma3mass})) small constant scattering for finite bandwith. By contrast, the Hund's coupling $J$ is responsible for the linear scattering rate of the $\Gamma_{1,2}$  electrons (Eq(\ref{scatteringrategamma12})) in the flat band limit.

Our hybridization method valid for $\gamma < U$ gives almost identical expressions with  the projected THF method  $\gamma >U$ ~\cite{scattering1,scattering2}. As suggested in \cite{hu2025projectedsolvabletopologicalheavy}, the relevant ratio is between $U$ and the hybridization function at $\omega=U$. Since the hybridization function contains the DOS of a Dirac fermion and is proportional to $\omega$, the $U$ cancels out, and the THF model is valid in both limits. We can further ask: what value of $U$ would we obtain if we plugged in the experimental separation of the Hubbard bands ~\cite{xiao2025interactingenergybandsmagic} in Eq(\ref{HubbardBandSeparation}). The answer is $U=64$ meV, higher that the values used in all literature so far, and higher than the energy splitting in ~\cite{xiao2025interactingenergybandsmagic}. This urges caution in interpreting the energy splitting of the Hubbard bands in ~\cite{xiao2025interactingenergybandsmagic} directly as $U$. The Hubbard band spacing renormalizes from $U$ to lower values Eq[\ref{HubbardBandSeparation}]. Our work demonstrates that the solvable THF limit offers a systematic analytical route to correlated excitation dynamics in moiré systems, qualitatively complementing numerical approaches such as DMFT but still showing their need for quantitative agreement with experiment. We note that DMFT \cite{CrippaStrain, rai_dynamical_2023} spectral function agrees with ~\cite{xiao2025interactingenergybandsmagic} without tuning, and with parameters $U, \gamma$ that are computed from microscopics \cite{song_magic-angle_2022}. 


\begin{acknowledgments}
We thank R. Valenti, O. Vafek and E. Berg and Y. Vituri, S. Ilani, D. Efetov  for discussions. H. H. was supported by the Gordon and Betty Moore Foundation through Grant No.GBMF8685 towards the Princeton theory program, the Gordon and Betty Moore Foundation’s EPiQS Initiative (Grant No. GBMF11070), Global Collaborative Network Grant at Princeton University,the Simons Investigator Grant No. 404513,  NSF-MERSEC (Grant No. MERSEC DMR 2011750), Simons Collaboration on New Frontiers in Superconductivity (Grant no. SFI-MPS-NFS-00006741-01), Princeton Catalysis Initiative (PCI), and the Schmidt Foundation Fund at the Princeton University. Research at Yale was supported by NSF Grant No.\ DMR-2410182 and by the 	Air Force Office of Scientific Research under Award No.\ FA95502510287. B.A.B. was supported by Office of
Basic Energy Sciences, Material Sciences and Engineer-
ing Division, U.S. Department of Energy (DOE) under
Contracts No. DE-SC0016239.
Y.S. was supported by the European Research Council (ERC) under the European Union’s Horizon 2020 research and innovation program (Grant Agreement No. 101020833), and the Gipuzkoa Quantum 2025 grant (ref: 2025-QUAN-000009-01).
G.~S. acknowledges support from the Deutsche Forschungsgemeinschaft (DFG, German Science Foundation) through FOR 5249-449872909 ``QUAST'' as well as through the Germany's Excellence Strategy–EXC2147 ``ctd.qmat'' (project‐id 390858490) and is grateful to Max Fischer and Carl Lehmann for useful discussions.
\end{acknowledgments}

\bibliography{tbg}

\input{Appendix_input}

\end{document}

%% file: Appendix_input.tex
\onecolumngrid
\begin{center}
\textbf{Supplementary Material of\\
Twisted Bilayer Graphene Lifetimes At Integer Fillings: An Analytic Result}\\ 
\end{center}
\onecolumngrid

\renewcommand{\thefigure}{S\arabic{figure}}

\renewcommand{\thetable}{S\arabic{table}}

\renewcommand{\thesection}{\Roman{section}}

\renewcommand{\theequation}{S\arabic{equation}}

\tableofcontents
\appendix
\renewcommand{\thesection}{\Roman{section}}

\clearpage
\section{Summary}
In this note, we investigate self-energy corrections generated by the single-particle $fc$ hybridization and by the $fc$ interaction, namely the Hund's coupling term $H_J$.
Accordingly, two small parameters are introduced to perform systematic series expansions in each case:
\ba 
\label{eq:define_small_para}
\kappa = \frac{\gamma^2}{|v_\star|^2\Omega_M},\quad j = \frac{J^2}{|v_\star|^2\Omega_M }.
\ea 
Here, $\gamma$ denotes the $fc$ hybridization strength, $J$ is the Hund's coupling strength between $f$ and $c$ electrons, $v_\star$ is the velocity of the $c$ electrons, and $\Omega_M$ is the area of the moir\'e Brillouin zone.
Taking the parameters of Ref.~\cite{PhysRevLett.129.047601}, we obtain 
\ba 
\kappa =0.013,\quad j =0.006 
\ea

The results at charge neutrality ($\nu=0$) are summarized in \cref{tab:summary_cnp}.

\begin{table}[h!]
    \centering
    \begin{tabular}{c|c|c}
         & Hybridization expansion with $j=0$& Self-energy effect from $H_J$ with $\kappa=0$ \\
         \hline 
      Hubbard bands& $ U_1(N_f+1)\pi^2\kappa/4$ &  \\ 
      Gapless excitation (linear dispersion) near $\Gamma$ point $(M=0)$ & $\propto\kappa(N_f+1)|v_\star\kk|$ & $\propto j |v_\star\kk| $  \\ 
      Gapless excitation (quadratic dispersion) near $\Gamma$ point $(M\ne 0)$ & $\propto \kappa (N_f+1) M$  &
      $\propto j|v_{\star}\kk|^4/M^3$
    \end{tabular}
    \caption{Scattering rates at charge neutrality for the Hubbard bands and the low-energy dispersive bands. The latter exhibit quadratic dispersion when $M \neq 0$ and linear dispersion at $M = 0$. $M$ denotes the half bandwidth of the flat band. $U_1$ denotes the Hubbard interaction strength of $f$ electrons. 
    $N_f=8$ denotes the number of $f$ flavors. }
    \label{tab:summary_cnp}
\end{table}

{
\section{The topological heavy fermion model}\label{app:THF}
The grand-canonical topological heavy fermion (THF) model for twisted bilayer graphene (TBG) is written as
\begin{equation}
    H=H_0+H_{U_1}+H_{U_2}+H_V+H_{W}+H_J-\mu N
\end{equation}
where $N=\sum_{\kk,a\eta s}c^{\dagger}_{\kk,a\eta s}c_{\kk,a\eta s}+\sum_{\kk,\alpha\eta s}f^{\dagger}_{\kk,\alpha\eta s}f_{\kk,\alpha\eta s}$ is the total electron number operator.
The non-interacting part $H_0$ is written as
\begin{align}
    H_{0}=&\sum_{\eta s} \sum_{aa'} \sum_{\kk} ( H^{0,(c,\eta)}_{aa'}(\kk+\GG) 
  ) c_{\kk,a\eta s}^\dagger c_{\kk,a'\eta s}
  +  \sum_{\eta s \alpha s} \sum_{\kk}
  \bigg(  H_{a\alpha}^{0,(cf,\eta)}(\kk) 
    c_{\kk,a\eta s}^\dagger f_{\kk, \alpha \eta s} + h.c. 
  \bigg)\nonumber\\
  &+\sum_{\eta s}\sum_{\alpha\alpha'}\sum_{\kk}H^{0,(ff,\eta)}_{\alpha\alpha'}(\kk){f}^{\dagger}_{{\kk},\alpha\eta s}{f}_{\kk,\alpha'\eta s}
\end{align}
where the matrix elements are
\begin{align}
    & H^{0,(c,\eta)}(\kk) = 
  \begin{pmatrix}
    0_{2\times 2} & v_\star (\eta k_x \sigma_0 + ik_y \sigma_z) \\
    v_\star (\eta k_x \sigma_0 - ik_y \sigma_z) &  M\sigma_x 
  \end{pmatrix} \\
  &
  H^{0,(cf,\eta)}(\kk) = 
  \begin{pmatrix}
    \gamma \sigma_0 + v_\star' (\eta k_x \sigma_x + k_y \sigma_y) \\
    0_{2\times2}
  \end{pmatrix}\\
  &H^{0,(ff,\eta)}(\kk)\approx0_{2\times 2}
\end{align}
The interaction terms are
\begin{align}
    H_{U_1}=&\frac{U_1}{2}\sum_{\alpha\eta s}\sum_{\alpha' \eta' s'}\sum_{\RR}:f^{\dagger}_{\RR,\alpha\eta s}f_{\RR,\alpha\eta s}::f^{\dagger}_{\RR,\alpha'\eta's'}f_{\RR,\alpha'\eta's'}:,\\
    H_{U_2}=&\frac{U_1}{2}\sum_{\alpha\eta s}\sum_{\alpha' \eta' s'}\sum_{\langle\RR,\RR'\rangle}:f^{\dagger}_{\RR,\alpha\eta s}f_{\RR,\alpha\eta s}::f^{\dagger}_{\RR',\alpha'\eta's'}f_{\RR',\alpha'\eta's'}:,\\
    H_V=&\frac{1}{2\Omega_0N_M}\sum_{\eta s a}\sum_{\eta' s' a'}\sum_{\kk,\kk',\qq}V(\qq):c^{\dagger}_{\kk,a\eta s}c_{\kk+\qq,a\eta s}::c^{\dagger}_{\kk'+\qq,a'\eta's'}c_{\kk',a'\eta's'}:,\;(|\kk|,|\kk'|,|\kk+\qq|,|\kk'+\qq|<\Lambda_c),\\
    H_{W}=&\frac{1}{N_M}\sum_{\alpha\eta s}\sum_{a'\eta's'}\sum_{\kk,\kk',\qq}W_{a'}:f^{\dagger}_{\kk+\qq,\alpha\eta s}f_{\kk,\alpha\eta s}::c^{\dagger}_{\kk'-\qq,a'\eta's'}c_{\kk',a'\eta's'}:,\;(|\kk'|,|\kk'-\qq|<\Lambda_c),\\
    H_{J}=&-\frac{J}{2N_M}\sum_{\alpha\eta s}\sum_{\alpha'\eta's'}\sum_{\kk,\kk',\qq}[\eta\eta'+(-1)^{\alpha+\alpha'}]:f^{\dagger}_{\kk+\qq,\alpha\eta s}f_{\kk,\alpha'\eta's'}::c^{\dagger}_{\kk'-\qq,\alpha'+2\eta's'}c_{\kk',\alpha+2\eta s}:,\;(|\kk'|,|\kk'-\qq|<\Lambda_c).
\end{align}
where $N_M$ is the number of moir\'e unit cell, $\Omega_0$ is the unit cell size and 
\begin{equation}
    f^{\dagger}_{\RR,\alpha\eta s}=\frac{1}{\sqrt{N_M}}\sum_{\kk}f^{\dagger}_{\kk,\alpha\eta s}e^{-i\kk\cdot\RR}
    \label{eq:momentum-fourier}
\end{equation}
We also introduce the normal order notation, which is defined as
\begin{align}
    :c^{\dagger}_{\kk,a\eta s}c_{\kk',a'\eta's'}:=&c^{\dagger}_{\kk,a\eta s}c_{\kk',a'\eta's'}-\frac{1}{2}\delta_{\kk\kk'}\delta_{aa'}\delta_{\eta\eta'}\delta_{ss'}\\
    :f^{\dagger}_{\kk,\alpha\eta s}f_{\kk',\alpha'\eta's'}:=&f^{\dagger}_{\kk,\alpha\eta s}f_{\kk',\alpha'\eta's'}-\frac{1}{2}\delta_{\kk\kk'}\delta_{\alpha\alpha'}\delta_{\eta\eta'}\delta_{ss'}
\end{align}

We define the relative filling number of each orbital as 
\begin{align}
    \nu_{c,a}\equiv&\frac{1}{N_{M}}\sum_{\kk,\eta s}\bigg(\langle c^{\dagger}_{\kk,a\eta s}c_{\kk,a\eta s}\rangle-\frac{1}{2}\bigg)\\
    \nu_{f,\alpha}\equiv&\frac{1}{N_{M}}\sum_{\kk,\eta s}\bigg(\langle f^{\dagger}_{\kk,\alpha\eta s}f_{\kk,\alpha\eta s}\rangle-\frac{1}{2}\bigg)
    \label{eq:relative_filling}
\end{align}
In the symmetric phase with $C_{2z}T$ symmetry, the filling numbers satisfy the constraints: $\nu_{c,1}=\nu_{c,2}, \;\nu_{c,3}=\nu_{c,4}$ and $\nu_{f,1}=\nu_{f,2}$.
In addition, the total relative filling numbers for $c$ and $f$ orbitals are defined as
\begin{equation}
    \nu_{c}=\sum_{a=1}^{4}\nu_{c,a},\;\nu_f=\sum_{\alpha=1}^{2}\nu_{f,\alpha}.
    \label{eq:nuc_nuf}
\end{equation}
Treating $H_{U_2},H_{V},H_{W}$ and $H_{J}$ in the Hartree level, we obtain their contributions to the single particle Hamiltonian.
\begin{align}
    H_{U_2}^{\text{MF}}=&6U_2\nu_f\sum_{\RR,\alpha\eta s}f^{\dagger}_{\RR,\alpha\eta s}f_{\RR,\alpha\eta s}\\
    H_{V}^{\text{MF}}=&\frac{V(0)}{\Omega_0}\nu_c\sum_{\kk,a\eta s}c^{\dagger}_{\kk,a\eta s}c_{\kk,a\eta s}\\
    H_{W}^{\text{MF}}=&\bigg(\sum_{a'}W_{a'}\nu_{c,a'}\bigg)\sum_{\kk,\alpha\eta s}f^{\dagger}_{\kk,\alpha\eta s}f_{\kk,\alpha\eta s}+\nu_f\sum_{\kk,a\eta s}W_{a}c^{\dagger}_{\kk,a\eta s}c_{\kk,a\eta s}\\
    H_{J}^{\text{MF}}=&-\frac{J}{4}\sum_{\kk,\alpha\eta s}\nu_{c,\alpha+2}f^{\dagger}_{\kk,\alpha\eta s}f_{\kk,\alpha\eta s}-\frac{J}{4}\sum_{\kk,\alpha\eta s}\nu_{f,\alpha}c^{\dagger}_{\kk,\alpha+2\eta s}c_{\kk,\alpha+2\eta s}
\end{align}
For the $H_{U_1}$ term, it is convenient to expand the normal ordering and rewrite it as
\begin{align}
    H_{U_1}=&\frac{U_1}{2}\sum_{\alpha\eta s}\sum_{\alpha' \eta' s'}\sum_{\RR}(f^{\dagger}_{\RR,\alpha\eta s}f_{\RR,\alpha\eta s}-1/2)(f^{\dagger}_{\RR,\alpha'\eta's'}f_{\RR,\alpha'\eta's'}-1/2)\nonumber\\
    =&-\frac{N_fU_1}{2}\sum_{\alpha\eta s}\sum_{\RR}f^{\dagger}_{\RR,\alpha\eta s}f_{\RR,\alpha\eta s}+\frac{U_1}{2}\sum_{\alpha\eta s}\sum_{\alpha'\eta' s'}\sum_{\RR}f^{\dagger}_{\RR,\alpha\eta s}f_{\RR,\alpha\eta s}f^{\dagger}_{\RR,\alpha'\eta's'}f_{\RR,\alpha'\eta's'}\nonumber\\
    =&-\frac{(N_f-1)U_1}{2}\sum_{\alpha\eta s}\sum_{\RR}f^{\dagger}_{\RR,\alpha\eta s}f_{\RR,\alpha\eta s}+\frac{U_1}{2}\sum_{\alpha\eta s\ne \alpha'\eta' s'}\sum_{\RR}f^{\dagger}_{\RR,\alpha\eta s}f_{\RR,\alpha\eta s}f^{\dagger}_{\RR,\alpha'\eta's'}f_{\RR,\alpha'\eta's'}
\end{align}
where $N_f=8$ is the total number of $f$-electron flavors.
The effective Hamiltonian can then be written as
\begin{align}
    H'=&H_0+\epsilon_{c,1}\sum_{\kk,\eta s}\sum_{a=1,2}c^{\dagger}_{\kk,a\eta s}c_{\kk,a\eta s}+\epsilon_{c,2}\sum_{\kk,\eta s}\sum_{a=3,4}c^{\dagger}_{\kk,a\eta s}c_{\kk,a\eta s}+\epsilon_{f}\sum_{\kk,\alpha\eta s}f^{\dagger}_{\kk,\alpha\eta s}f_{\kk,\alpha\eta s}\nonumber\\
    &+\frac{U_1}{2}\sum_{\alpha\eta s\ne \alpha'\eta' s'}\sum_{\RR}f^{\dagger}_{\RR,\alpha\eta s}f_{\RR,\alpha\eta s}f^{\dagger}_{\RR,\alpha'\eta's'}f_{\RR,\alpha'\eta's'}-\mu N
\end{align}
where $\epsilon_{c,1}$, $\epsilon_{c,2}$, and $\epsilon_f$ are the Hartree energy shifts arising from $H_{U_2}$, $H_V$, $H_W$, $H_J$, and the reordering of $H_{U_1}$. These quantities are defined as
\begin{align}
    \epsilon_{c,1}=&\frac{V(0)}{\Omega_0}\nu_c+W_1\nu_f\\
    \epsilon_{c,2}=&\frac{V(0)}{\Omega_0}\nu_c+W_3\nu_f-\frac{J}{8}\nu_f\\
    \epsilon_{f}=&6U_2\nu_f +2W_1\nu_{c,1}+2W_3\nu_{c,3}-\frac{J}{4}\nu_{c,3}-\frac{(N_f-1)}{2}U_1
    \label{eq:hartree-energy}
\end{align}

}

\section{Hybridization expansion}
\label{app:hyb_exp}
In this section, we evaluate the self-energy using the hybridization expansion method. We start from the mean-field Hamiltonian introduced in the previous section:
\ba 
\label{eq:generic_Ham}
H' =& \sum_{\eta s} \sum_{aa'} \sum_{\kk} ( H^{(c,\eta)}_{a,a'}(\kk+\GG) 
  ) c_{\kk,a\eta s}^\dagger c_{\kk,a'\eta s}
  + \frac1{\sqrt{N_M}} \sum_{\eta s,a \alpha } \sum_{\RR} \sum_{\kk}
  \bigg( e^{- i\kk\cdot\RR } H_{a\alpha}^{(cf,\eta)}(\kk) 
    c_{\kk,a\eta s}^\dagger f_{\RR ,\alpha \eta s} + h.c. 
  \bigg) \nonumber\\ 
  &+ \sum_{\RR,\alpha\eta s}(\epsilon_f-\mu) f_{\RR ,\alpha \eta s}^\dag f_{\RR,\alpha \eta s}
  + {\frac{U_1}{2} \sum_{\RR,\alpha \eta s\ne \alpha'\eta's'}f_{\RR,\alpha\eta s}^\dag f_{\RR,\alpha\eta s} f_{\RR,\alpha'\eta's'}^\dag f_{\RR,\alpha'\eta's'}}
  \, .
\ea
where 
\begin{align} \label{eq:HcHcf}
 & H^{(c,\eta)}(\kk) = 
  \begin{pmatrix}
    \sigma_0(\epsilon_{c,1}-\mu) & v_\star (\eta k_x \sigma_0 + ik_y \sigma_z) \\
    v_\star (\eta k_x \sigma_0 - ik_y \sigma_z) & \sigma_0(\epsilon_{c,2}-\mu)+ M\sigma_x 
  \end{pmatrix} \nonumber\\
  &
  H^{(cf,\eta)}(\kk) = 
  \begin{pmatrix}
    \gamma \sigma_0 + v_\star' (\eta k_x \sigma_x + k_y \sigma_y) \\
    0_{2\times2}
  \end{pmatrix}
\end{align}
Here, $\epsilon_f$, $\epsilon_{c,1}$ and $\epsilon_{c,2}$ characterize the Hartree-Fock contributions from $H_{V}$, $H_W$, $H_{J}$ and $H_{U_2}$ as well as the single-particle contribution obtained by expanding the normal ordering of $H_{U_1}$\cite{hu2025projectedsolvabletopologicalheavy}.
Their specific definitions were given by \cref{eq:hartree-energy}.
$f_{\RR,\alpha\eta s}$ and $c_{\kk,a\eta s}$ denote the annihilation operators for $f$ and $c$ electrons, respectively.
We then integrate out the $c$ electrons, which leads to the following action involving only the $f$ electrons (see also Ref.~\cite{hu2025projectedsolvabletopologicalheavy})
\ba 
S = &\sum_{\kk,\alpha \eta s,\alpha'\eta's',i\omega}
f_{\kk,\alpha\eta s}^\dag(i\omega) 
\bigg[ 
(-i\omega-\mu +\epsilon_f) \delta_{\alpha \eta s,\alpha'\eta's'} 
+\gamma^2\Delta_{\kk}^{\alpha\eta s, \alpha'\eta's'}(i\omega)
\bigg] 
f_{\kk,\alpha'\eta'
s'}(i\omega) 
\nonumber\\
&
+ {\int_\tau \frac{U_1}{2}\sum_{\RR,\alpha\eta s \ne \alpha'\eta's'}
f_{\RR,\alpha\eta s}^\dag(\tau) f_{\RR,\alpha\eta s} (\tau)f_{\RR,\alpha'\eta's'}^\dag (\tau)f_{\RR,\alpha'\eta's'}(\tau)}
\ea 
The hybridization function is defined as 
\ba 
\label{eq:def_hyb_k}
\gamma^2 \Delta_\kk^{\alpha \eta s,\alpha'\eta's'}
(i\omega) 
= \delta_{\eta s, \eta's'}\sum_{n,a_1a_2}[H^{(fc,\eta)}(\kk)]_{\alpha a_1}
U^{\eta,c}_{\kk,a_1n} \frac{1}{i\omega - E^{\eta,c}_{\kk,n}}
U^{\eta, c,*}_{\kk,a_2n} [H^{(fc,\eta)}(\kk)]^*_{\alpha' a_2}
\ea 
where we have introduced the eigenvalues and eigenvectors of $H^{(c,\eta)}$
\ba 
\sum_{a'}H^{(c,\eta)}_{aa'} U_{\kk,a'n}^{\eta,c} = E_{\kk,n}^{\eta,c} U_{\kk,an}^{\eta,c}\, .
\ea 
Note that we have factored out an overall prefactor $\gamma^2$ in order to facilitate the perturbative expansion. Setting $\gamma^2=0$ therefore indicates that all $fc$ hybridization is turned off.

We first discuss symmetry constraints on the hybridization function. In the $f$-orbital subspace, the corresponding symmetry representations are
\begin{equation}
    D^{f}(C_{3z})=e^{i\frac{2\pi}{3}\sigma_z\tau_z},\quad D^f(C_{2x})=\sigma_x\tau_0,\quad D^{f}(C_{2z})=\sigma_x\tau_x
    \label{eq:sym_repres}
\end{equation}
where $\sigma$ and $\tau$ are Pauli matrices acting on the orbital and valley degrees of freedom, respectively.
As will be shown below, only the local hybridization function enters our calculation. It is defined as
\begin{equation}
    \Delta^{\alpha\eta s,\alpha'\eta' s'}(i\omega)\equiv\frac{1}{N_{M}}\sum_{\kk}\Delta^{\alpha\eta s,\alpha'\eta' s'}_{\kk}(i\omega)
    \label{eq:hyb_sym}
\end{equation}
Due to the spin $\mathrm{SU}(2)$ and valley $\mathrm{U}_{V}(1)$ symmetry, the hybridization function is diagonal in valley space and independent of spin (\cref{eq:def_hyb_k}).
Thus, we have 
\begin{equation}
    \Delta^{\alpha\eta s,\alpha'\eta' s'}(i\omega)=\delta_{\eta s,\eta' s'}\Delta_{\eta}^{\alpha,\alpha'}(i\omega)
\end{equation}
In orbital space, $\Delta_{\eta}^{\alpha,\alpha'}$ can be expanded in the Pauli-matrix basis as
$\Delta_{\eta}^{\alpha,\alpha'}=\sum_i \Delta_{\eta}^{i}\sigma_i^{\alpha,\alpha'}$.
$C_{3z}$ symmetry requires that $\Delta_{\eta}^{i=x,y}=0$ and $C_{2x}$ symmetry requires that $\Delta_{\eta}^{i=y,z}=0$. 
Thus, the only nonzero component is $\Delta_{\eta}^{i=0}$, which implies $\Delta_{\eta}^{\alpha,\alpha'}=\delta_{\alpha,\alpha'}\Delta_{\eta}$.
Using in addition the $C_{2z}$ symmetry, which swaps the two valleys, we find that $\Delta_{\eta=+}=\Delta_{\eta=-}=\Delta$.
In summary, in the presence of spin $\mathrm{SU}(2)$, valley $\mathrm{U}_{V}(1)$, $C_{3z}$, $C_{2x}$ and $C_{2z}$ symmetries, the local hybridization function takes the form
\begin{equation}
    \Delta^{\alpha\eta s,\alpha'\eta's'}(i\omega)=\delta_{\alpha\eta s,\alpha'\eta' s'}\Delta(i\omega)
    \label{eq:flavor_U8}
\end{equation}
which exhibits a flavor $\mathrm{U}(8)$ symmetry.

To simplify the notation, we use single indices $i$ and $j$ to collectively label the orbital $\alpha$, valley $\eta$, and spin $\sigma$ degrees of freedom of the $f$ electrons.
Then the action can be written as
\ba 
  S =& \sum_{\kk,ij,i\omega}f_{\kk,i}^\dag(i\omega)[(- i\omega-\mu {+ \epsilon_f}) \delta_{i,j}+ \gamma^2 \Delta^{i,j}_{\kk}(i\omega)] f_{\kk,j}(i\omega) {+\frac{U_1}{2} \int_\tau \sum_{i\ne j,\RR} f_{\RR,i}^\dag(\tau) f_{\RR,i} (\tau)f_{\RR,j}^\dag (\tau)f_{\RR,j}(\tau)}
  \label{eq:action-hyb}
\ea 
Since $H'$ (\cref{eq:generic_Ham}) is quadratic in the $c$ electron operators, integrating out the $c$ electron degree of freedom to obtain the action $S$ (\cref{eq:action-hyb}) is exact.

We define the atomic Hamiltonian as
\begin{align}
    H_{atom}=&(\epsilon_{f}-\mu)\sum_{\kk,i}f^{\dagger}_{\kk,i}f_{\kk,i}+\frac{U_1}{2}\sum_{i\ne j}\sum_{\RR}f^{\dagger}_{\RR,i}f_{\RR,i}f^{\dagger}_{\RR,j}f_{\RR,j}\nonumber\\
    =&(\epsilon_{f}-\mu)\sum_{\RR,i}f^{\dagger}_{\RR,i}f_{\RR,i}+\frac{U_1}{2}\sum_{i\ne j}\sum_{\RR}f^{\dagger}_{\RR,i}f_{\RR,i}f^{\dagger}_{\RR,j}f_{\RR,j}
    \label{eq:atomic_Hamiltonian}
\end{align}
where we have used the relation between $f^{\dagger}_{\kk,i}$ and $f^{\dagger}_{\RR,i}$ in \cref{eq:momentum-fourier}.
If we treat $H_{U_1}$ at the Hartree-Fock level, then the mean-field atomic Hamiltonian and action can be written as
\begin{align}
    H_{atom}^{\mathrm{MF}}=&\bigg[\epsilon_f-\mu+\frac{(N_f-1)n_f}{N_f}U_1\bigg]\sum_{\RR,i}f^{\dagger}_{\RR,i}f_{\RR,i}=\bigg[\epsilon_f-\mu+\frac{(N_f-1)n_f}{N_f}U_1\bigg]\sum_{\kk,i}f^{\dagger}_{\kk,i}f_{\kk,i}\nonumber\\
    S^{\text{MF}}=&\sum_{\kk,ij,i\omega}f^{\dagger}_{\kk,i}(i\omega)\bigg[\bigg(-i\omega-\mu+\epsilon_f+\frac{(N_f-1)n_f}{N_f}U_1\bigg)\delta_{i,j}+\gamma^2\Delta^{i,j}_{\kk}(i\omega)\bigg]f_{\kk,j}(i\omega)
    \label{eq:atom_MF}
\end{align}
where $n_f=\nu_f+N_f/2$ (see definition of $\nu_f$ in \cref{eq:nuc_nuf,eq:relative_filling}).
Thus, the interacting Green's function at the Hartree-Fock level is
\ba 
  &[G_{f}^{{\text{MF}}}(i\omega,\kk)]^{-1}_{ij}
  = [ i\omega+\mu -\epsilon_f - \frac{(N_f-1){n_f}U_1}{N_f} ]\delta_{i,j} -\gamma^2 \Delta_{\kk}^{i,j}(i\omega) 
  \label{eq:Green_MF}
\ea 
where $G^{\text{MF}}_f$ includes the Hartree-Fock contributions from $H_{U_2},H_V, H_{W}, H_{J}$ (see definition of $\epsilon_f$ in \cref{eq:hartree-energy}) and $H_{U_1}$ in the symmetric state.

We are particularly interested in the Green's function and the dynamical self-energy of the $f$ electrons, defined as
\begin{equation}
    G_{f,ij}(i\omega,\kk) = -\int_\tau \langle  f_{\kk,i}(\tau) f_{\kk,j}^\dag(0)\rangle e^{i\omega \tau } 
    \label{eq:def_Green_f}
\end{equation} 
\ba 
\Sigma_{f,ij}(i\omega,\kk) 
=[G^{{\text{MF}}}_{f}(i\omega,\kk)]^{-1}_{ij}- [G_{f}(i\omega,\kk)]^{-1}_{ij} 
\label{eq:dyson_eq}
\ea 
To achieve this, we solve the model defined by the action $S$ (\cref{eq:action-hyb}) using a hybridization expansion, treating $\gamma^2$ as a small parameter.
We first compute the Green's function $G_{f,ij}(i\omega,\kk)$ perturbatively to order $\gamma^2 \Delta_{\kk}^{i,j}$. We then use the relation between the Green's function and the self-energy at order $\gamma^2$ (from \cref{eq:dyson_eq}) to extract the self-energy to the same order.

Expanding in powers of $\gamma^2$, we write the Green's function and self-energy as
\ba 
&G_{f,ij}(i\omega,\kk) 
= G^{(0)}_{f,ij}(i\omega,\kk) 
+ G^{(1)}_{f,ij}(i\omega,\kk)  + \mathcal{O}(\gamma^4) \nonumber\\ 
&\Sigma_{f,ij}(i\omega,\kk)
= \Sigma^{(0)}_{f,ij}(i\omega,\kk) 
+ \Sigma^{(1)}_{f,ij}(i\omega,\kk)  + \mathcal{O}(\gamma^4) 
\label{eq:perturb_exp_Gf_Sigma_f}
\ea 
where $G_f^{(n)}$ denotes the interacting Green's function at order $\gamma^{2n}$, and $\Sigma_f^{(n)}$ denotes the self-energy at order $\gamma^{2n}$. 
Form the Dyson equation \cref{eq:dyson_eq}, we have 
\ba 
\Sigma_{f,ij}(i\omega,\kk) 
=&[G^{{\text{MF}}}_{f}(i\omega,\kk)]^{-1}_{ij}- [G_{f}(i\omega,\kk)]^{-1}_{ij} \nonumber\\ 
= &[G^{{\text{MF}}}_{f}(i\omega,\kk)]^{-1}_{ij}
- 
\bigg[ G_{f}^{(0)}(i\omega,\kk)
+ G_{f}^{(1)}(i\omega,\kk)\bigg]_{ij}^{-1} + \mathcal{O}(\gamma^4) \nonumber\\ 
=& 
[G^{{\text{MF}}}_{f}(i\omega,\kk)]^{-1}_{ij}
- 
\bigg[ G_{f}^{(0)}(i\omega,\kk)\bigg]^{-1}_{ij}
+\bigg[ [G_{f}^{(0)}(i\omega,\kk)]^{-1}
G_f^{(1)}(i\omega,\kk) [G_{f}^{(0)}(i\omega,\kk)]^{-1}\bigg]_{ij}+ \mathcal{O}(\gamma^4)
\label{eq:Sigma_f_expand}
\ea 
Combining \cref{eq:def_Green_f,eq:perturb_exp_Gf_Sigma_f,eq:Sigma_f_expand}, we find 
\ba 
&\Sigma_{f,ij}^{(0)}(i\omega,\kk) 
= [i\omega+\mu - \epsilon_f - \frac{(N_f-1)n_fU_1}{N_f} ]\delta_{i,j} 
-[ G_f^{(0)}(i\omega,\kk)]^{-1}_{ij} \nonumber\\ 
&
\Sigma_{f,ij}^{(1)}(i\omega,\kk)
= -\gamma^2 \Delta_{\kk}^{ij}(i\omega) 
+\bigg[ [G_{f}^{(0)}(i\omega,\kk)]^{-1}
G_f^{(1)}(i\omega,\kk) [G_{f}^{(0)}(i\omega,\kk)]^{-1}\bigg]_{ij}
\label{eq:Sigma_f_pert_Gf_pert}
\ea  
Accordingly, by evaluating the Green’s function to order $\gamma^2$, we can also obtain the self-energy up to the same order through \cref{eq:Sigma_f_pert_Gf_pert}. 

We note that, at zeroth order in $\gamma$, $G_f^{(0)}$ corresponds to the Green's function in the $\gamma=0$ limit, namely, the atomic Green's function. The associated self-energy $\Sigma_f^{(0)}$ then reduces to the Hubbard-I self-energy evaluated in Ref.~\cite{hu2025projectedsolvabletopologicalheavy}.

\subsection{Formulas for $G_f^{(1)}$ and $\Sigma_f^{(1)}$}
We now derive the $\mathcal{O}(\gamma^2)$ contribution to the self-energy, $\Sigma_{f,ij}^{(1)}$, by calculating $G_{f}^{(1)}(i\omega,\kk)$. 
To carry out the perturbative calculation, we first separate the action into two parts:
\begin{align}
\label{eq:action_sep}
    &S = S_{atom} +\gamma^2 S_{hyb} \nonumber\\ 
    &S_{atom} = \sum_{\RR,i}f_{\RR,i}^\dag(-i\omega-\mu  ) f_{\RR,i} + \frac{U_1}{2} \int_\tau \sum_\RR [n_{f,\RR}(\tau)- n_0]^2  \nonumber\\ 
    & S_{hyb} = \sum_{\kk,ij,\omega}f_{\kk,i}^\dag(i\omega) \Delta_{\kk}^{ij}(i\omega) f_{\kk,j}(i\omega) 
\end{align}
where ${n}_{f,\RR} = \sum_m f_{\RR,m}^\dag f_{\RR,m}$ and $n_0 = \frac{1}{2} -\frac{\epsilon_f}{U_1}$. 
{
In \cref{eq:action_sep} we have rewritten the atomic Hamiltonian defined in \cref{eq:atomic_Hamiltonian} as 
\begin{align}
    H_{atom}=&(\epsilon_{f}-\mu)\sum_{\RR,i}f^{\dagger}_{\RR,i}f_{\RR,i}+\frac{U_1}{2}\sum_{i\ne j}\sum_{\RR}f^{\dagger}_{\RR,i}f_{\RR,i}f^{\dagger}_{\RR,j}f_{\RR,j}\nonumber\\
    =&\bigg(\epsilon_f-\frac{U_1}{2}\bigg)\sum_{\RR,i}f^{\dagger}_{\RR,i}f_{\RR,i}+\frac{U_1}{2}\sum_{i}\sum_{j}\sum_{\RR}f^{\dagger}_{\RR,i}f_{\RR,i}f^{\dagger}_{\RR,j}f_{\RR,j}-\mu \sum_{\RR}n_{f,\RR}\nonumber\\
    =&\frac{U_1}{2}\sum_{\RR}\bigg[2\bigg(\frac{\epsilon_f}{U_1}-\frac{1}{2}\bigg){n}_{f,\RR}+{n}_{f,\RR}^2\bigg]-\mu \sum_{\RR}n_{f,\RR}\nonumber\\
    =&\frac{U_1}{2}\sum_{\RR}(n_{f,\RR}-n_0)^2-\mu\sum_{\RR}n_{f,\RR}+\text{const.}
    \label{eq:atomic_Hamiltonian_rewrite}
\end{align}
}

Then, up to order $\gamma^2$, the partition function is
\begin{align}
\label{eq:partition_function_expand}
Z = &
\int D[f,f^\dag ]{e^{-S}} \nonumber\\ 
\approx & 
\int D[f,f^\dag ]
(1-\gamma^2 S_{hyb})
e^{-S_{atom}} + \mathcal{O}(\gamma^4)\nonumber\\ 
\approx &Z_{\gamma=0}
-\gamma^2 Z_{\gamma=0}\sum_{ij,\kk,i\omega} \Delta^{ij}_\kk(i\omega)\langle f_{\kk,i}^\dag(i\omega) f_{\kk,j}(i\omega) \rangle_{\gamma=0}  +\mathcal{O}(\gamma^4)
\end{align}
where $Z_{\gamma=0}$ denotes the partition function in the $\gamma=0$ limit,
\begin{align}
    Z_{\gamma=0} = \int D[f,f^\dag]e^{-S_{atom} }
\end{align}
The expectation value of an operator $O$ in the $\gamma=0$ limit is defined as 
\begin{align}
    \langle O\rangle_{\gamma=0} = \frac{1}{Z_{\gamma=0}}
    \int D[f,f^\dag] Oe^{-S_{atom}}\label{eq:exp_path_integral}
\end{align}
We note that, in order to evaluate the partition function $Z$, we need to calculate the Green's function of the $f$ electrons in the $\gamma=0$ limit (\cref{eq:partition_function_expand}), namely, the atomic Green's function defined as 
\ba 
G_{f,ij}^{(0)}(i\omega,\kk) 
=-\langle f_{\kk,j}(i\omega) f_{\kk,i}^\dag(i\omega)  \rangle_{\gamma=0}=  -\int_{0}^\beta d \tau \langle T_{\tau}  f_{\kk,i}(\tau) f_{\kk,j}^\dag(0)\rangle_{\gamma=0} e^{i\omega \tau }
\ea 
Since $f$ electrons on different sites $\RR$ are decoupled in the atomic limit ($\gamma=0$), the atomic Green's function is independent of $\kk$. Therefore, we have 
\ba 
G_{f,ij}^{(0)}(\tau,\kk) 
= -\frac{1}{N_M}\sum_\RR \langle T_\tau f_{\RR,i}(\tau) f^\dag_{\RR,j}(0)\rangle_{\gamma=0}
\ea 
with $N_M$ the number of moir\'e unit cells. 
In the absence of strain, which is the case considered here, the atomic limit has an enlarged $\mathrm{U}(8)$ symmetry \cite{hu2025projectedsolvabletopologicalheavy}. This implies that the atomic Green's function is diagonal in the flavor basis:
\ba 
G_{f,ij}^{(0)}(\tau,\kk) \propto \delta_{i,j}
\ea 
Moreover, by translation symmetry, the expectation value $\langle T_{\tau} f_{\RR,i}(\tau) f^\dag_{\RR,j}(0)\rangle_{\gamma=0}$ is independent of $\RR$, and  we can therefore introduce the local Green's function as 
\begin{align}
    &G_{f,loc}(\tau-\tau') =- \langle T_\tau f_{\RR,i}(\tau) f_{\RR,i}^\dag(\tau')\rangle_{\gamma=0} 
\end{align}
such that 
\ba 
G_{f,ij}^{(0)}(\tau,\kk) = \delta_{i,j}G_{f,loc}(\tau) 
\label{eq:G_loc_tau}
\ea 
We define the Fourier transformation of the local Green's function as

\ba 
G_{f, loc}(i\omega) 
=  -\int_{0}^\beta d \tau   \langle T_\tau f_{\RR,i}(\tau) f_{\RR,i}^\dag(0)\rangle_{\gamma=0}    e^{i\omega \tau }
\label{eq:atomic_green_def}
\ea 
{
Since this is a fermionic system, $G_{f,\mathrm{loc}}(\tau)$ satisfies
\begin{equation}
    G_{f,loc}(\tau+\beta)=-G_{f,loc}(\tau)\implies G_{f,loc}(\tau)=\frac{1}{\beta}\sum_{i\omega}G_{f,loc}(i\omega)e^{-i\omega\tau}
\end{equation}
where $i\omega$ is the Matsubara frequency for Fermion.
The Fourier transformation could be calculated as
\begin{equation}
    G_{f,loc}(i\omega)=\int_0^{\beta}d\tau\;G_{f,loc}(\tau)e^{i\omega\tau}=\int_{-\beta}^{0}G_{f,loc}(\tau+\beta)e^{i\omega(\tau+\beta)}=\int_{-\beta}^{0}d\tau\;G_{f,loc}(\tau)e^{i\omega\tau}
\end{equation}
In the last equality, we used the fact that $e^{i\omega\beta}=-1$ for fermionic Matsubara frequencies.
The formula with integration over $[0,\beta]$ gives
\begin{align}
    G_{f,loc}(i\omega)=&-\int_0^{\beta}d\tau\;\langle T_{\tau}f_{\RR,i}(\tau>0)f^{\dagger}_{\RR,i}(0)\rangle_{\gamma=0}e^{i\omega\tau}=-\int_0^{\beta}d\tau\;\langle f_{\RR,i}(\tau)f^{\dagger}_{\RR,i}(0)\rangle_{\gamma=0}e^{i\omega\tau}=-\langle f_{\RR,i}(i\omega) f^{\dagger}_{\RR,i}(i\omega)\rangle_{\gamma=0}
\end{align}
while the formula with integration over $[-\beta,0]$ gives
\begin{align}
    G_{f,loc}(i\omega)=&-\int_{-\beta}^{0}d\tau\;\langle T_{\tau}f_{\RR,i}(\tau<0)f^{\dagger}_{\RR,i}(0)\rangle_{\gamma=0}e^{i\omega\tau}=\int_{-\beta}^{0}d\tau\;\langle f^{\dagger}_{\RR,i}(0)f_{\RR,i}(\tau)\rangle_{\gamma=0}e^{i\omega\tau}=\langle  f^{\dagger}_{\RR,i}(i\omega)f_{\RR,i}(i\omega)\rangle_{\gamma=0}
\end{align}
We have used the definition of $f(i\omega)$ as
\begin{equation}
    f(i\omega)\equiv\frac{1}{\sqrt{\beta}}\int_0^{\beta}d\tau\;f(\tau)e^{i\omega\tau}=\frac{1}{\sqrt{\beta}}\int_{-\beta}^{0}d\tau\;f(\tau+\beta)e^{i\omega(\tau+\beta)}=\frac{1}{\sqrt{\beta}}\int_{-\beta}^{0}d\tau \;f(\tau)e^{i\omega\tau}
\end{equation}
In summary, 
\begin{equation}
    G_{f,loc}(i\omega)=-\langle f_{\RR,i}(i\omega) f^{\dagger}_{\RR,i}(i\omega)\rangle_{\gamma=0}=\langle  f^{\dagger}_{\RR,i}(i\omega)f_{\RR,i}(i\omega)\rangle_{\gamma=0}\label{eq:G_loc_freq}
\end{equation}
Another way to understand this result is that in \cref{eq:partition_function_expand}, $f^{\dagger}_{\kk,i}(i\omega)$ and $f_{\kk,j}(i\omega)$ are not operators, but Grassmann numbers in the path integral language.
They therefore anticommute, which gives $\langle f^{\dagger}_{\kk,i}(i\omega)f_{\kk,j}(i\omega)\rangle_{\gamma=0}=-\langle f_{\kk,j}(i\omega) f^{\dagger}_{\kk,i}(i\omega)\rangle_{\gamma=0}$
}
Using \cref{eq:G_loc_tau,eq:G_loc_freq}, the partition function can be written as (\cref{eq:partition_function_expand}) 
\begin{align}
\label{eq:partiaion_expand}
    Z \approx Z_{\gamma=0} 
    \bigg[ 1-\gamma^2 \sum_{j,\kk,i\omega} \Delta^{jj}_{\kk}(i\omega) G_{f,loc}(i\omega)
    \bigg] + \mathcal{O}(\gamma^4)
\end{align}

In \cref{eq:Sigma_f_pert_Gf_pert}, we need the first-order correction in $\gamma^2$ to the $f$-electron Green's function, namely $G_f^{(1)}$.
This requires us to evaluate the Green's function in the presence of the $cf$ hybridization, defined as
\begin{equation}
    G_{f,ij}(\tau-\tau',\kk)\equiv-\frac{1}{Z}\int D[f,f^\dag] T_\tau f_{\kk,i}(\tau) f_{\kk,j}^\dag(\tau') e^{-S}
\end{equation}
We now evaluate the corresponding path integral using hybridization expansion:
\begin{align}
    &-\int D[f,f^\dag] T_\tau f_{\kk,i}(\tau) f_{\kk,j}^\dag(\tau') {e^{-S}}\nonumber\\ 
    \approx & 
    -\int D[f,f^\dag] T_\tau f_{\kk,i}(\tau) f_{\kk,j}^\dag(\tau') e^{-S_{atom}} \nonumber\\ 
    &
    +\gamma^2 \int D[f,f^\dag] T_\tau f_{\kk,i}(\tau) f_{\kk,j}^\dag(\tau') \sum_{\kk',i'j',i\omega'}\Delta_{\kk'}^{i'j'}(i\omega') f_{\kk',i'}^\dag(i\omega')f_{\kk',j'}(i\omega')e^{-S_{atom}} + \mathcal{O}(\gamma^4)
\end{align}
We notice that 
\begin{align}
   & \sum_{\kk'\omega',i'j'}\Delta_{\kk'}^{i'j'}(i\omega')f^\dag_{\kk',i'}(i\omega') f_{\kk',j'}(i\omega') \nonumber\\ 
    =& \frac{1}{\beta} \int_{\tau_1,\tau_2,\tau_3}\sum_{\kk',i'j',\omega'}\Delta_{\kk'}^{i'j'}(\tau_3)f^\dag_{\kk',i'}(\tau_1) f_{\kk',j'}(\tau_2)
    e^{i\omega'(\tau_2-\tau_1+\tau_3)}\nonumber\\ 
    =&\int_{\tau_1,\tau_2}\sum_{i'j'}\sum_{\kk'}
    \Delta_{\kk'}^{i'j'}(\tau_1-\tau_2)f_{\kk',i'}^\dag(\tau_1) f_{\kk',j'}(\tau_2) \nonumber\\
 =&\int_{\tau_1,\tau_2}\sum_{i'j'}\frac{1}{N_M}\sum_{\kk',\RR_3,\RR_4}
    \Delta_{\kk'}^{i'j'}(\tau_1-\tau_2)f_{\RR_3,i'}^\dag(\tau_1) f_{\RR_4,j'}(\tau_2) e^{-i\kk'\cdot(\RR_4-\RR_3)}
\end{align}
{where we used the abbreviation $\int_{\tau}=\int_0^{\beta}d\tau$ for simplicity.
In the following text, we will always use $\int_{\tau}$ for the integration over interval $[0,\beta]$. For other intervals, the integration will be written out explicitly.
}
We have also used the following Fourier transforms:
\ba 
\Delta_\kk^{ij}(\tau) = \frac{1}{\beta}\sum_{i\omega} \Delta_\kk^{ij}(i\omega) e^{-i\omega \tau },\quad f_{\kk,i}(i\omega) = \frac{1}{\sqrt{\beta}}\int_\tau f_{\kk,i}(\tau) e^{i\omega \tau }
\ea 
Then we find 
\ba 
 &-\int D[f,f^\dag] T_\tau f_{\kk,i}(\tau) f_{\kk,j}^\dag(\tau') {e^{-S}} \nonumber\\ 
    \approx & -\int D[f,f^\dag]T_\tau f_{\kk,i}(\tau) f_{\kk,j}^\dag(\tau') e^{-S_{atom}}\nonumber\\ 
    &
    +\gamma^2 \int D[f,f^\dag] T_\tau f_{\kk,i}(\tau) f_{\kk,j}^\dag(\tau') 
    \int_{\tau_1,\tau_2}\sum_{i'j'}\frac{1}{N_M}\sum_{\kk',\RR_3,\RR_4}
    \Delta_{\kk'}^{i'j'}(\tau_1-\tau_2)f_{\RR_3,i'}^\dag(\tau_1) f_{\RR_4,j'}(\tau_2) e^{-i\kk'\cdot(\RR_4-\RR_3)} e^{-S_{atom}}
    + \mathcal{O}(\gamma^4) \nonumber \\ 
=&
Z_{\gamma=0}
\bigg[\delta_{i,j}
G_{f,loc}(\tau-\tau') + \gamma^2 \int_{\tau_1,\tau_2}\frac{1}{N_M^2}\sum_{\RR_1,\RR_2,\RR_3,\RR_4}\sum_{i'j',\kk'}
 \nonumber\\ 
&
\Delta_{\kk'}^{i'j'}(\tau_1-\tau_2) e^{-i\kk\cdot(\RR_1-\RR_2)-i\kk'(\RR_4-\RR_3)}
\langle T_\tau f_{\RR_1,i}(\tau)f_{\RR_2,j}^\dag(\tau') f_{\RR_3,i'}^\dag(\tau_1)f_{\RR_4,j'}(\tau_2)\rangle_{\gamma=0}
\bigg] + \mathcal{O}(\gamma^4) 
\label{eq:ffdag_expand}
\ea

We now aim to evaluate 
\begin{align}
   \langle T_\tau f_{\RR_1,i}(\tau) f_{\RR_2,j}^\dag(\tau') f_{\RR_3,i'}^\dag(\tau_1) f_{\RR_4,j'}(\tau_2)\rangle_{\gamma=0}
\end{align}
Since the action at $\gamma=0$ describes an atomic Hamiltonian with decoupled sites, the non-zero contribution can arise only from the following situations, where the number of created and annihilated $f$ electrons is the same on each site:
\begin{itemize}
    \item 
    $\RR_1=\RR_2,\RR_3=\RR_4,\RR_1\ne \RR_3$. 
    \item $\RR_1=\RR_3,\RR_4=\RR_2,\RR_1\ne \RR_4$. 
    \item $\RR_1=\RR_2=\RR_3=\RR_4$. 
\end{itemize} 
Because this is an atomic problem, the $f$ electrons on different sites are decoupled. Therefore, the first two cases are easily evaluated:
\begin{itemize}
    \item $\RR_1=\RR_2,\RR_3=\RR_4,\RR_1\ne \RR_3$: 
    \ba 
  &  \langle T_\tau f_{\RR_1,i}(\tau) f_{\RR_1,j}^\dag(\tau') f_{\RR_3,i'}^\dag(\tau_1) f_{\RR_3,j'}(\tau_2)\rangle_{\gamma=0} =    \langle T_\tau f_{\RR_1,i}(\tau) f_{\RR_1,j}^\dag(\tau') \rangle_{\gamma=0}  \langle T_\tau f_{\RR_3,i'}^\dag(\tau_1) f_{\RR_3,j'}(\tau_2)\rangle_{\gamma=0}  \nonumber\\ 
    =& -\delta_{i,j}
    \delta_{i',j'}G_{f,loc}(\tau-\tau') 
    G_{f,loc}(\tau_2-\tau_1) 
    \ea
    
    \item $\RR_1=\RR_3,\RR_2=\RR_4,\RR_1\ne \RR_4$: 
    \ba 
    &\langle T_\tau f_{\RR_1,i}(\tau) f_{\RR_4,j}^\dag(\tau') f_{\RR_1,i'}^\dag(\tau_1) f_{\RR_4,j'}(\tau_2)\rangle_{\gamma=0} 
    = - \langle T_\tau f_{\RR_1,i}(\tau)f_{\RR_1,i'}^\dag(\tau_1)\rangle_{\gamma=0}\langle T_\tau  f_{\RR_4,j}^\dag(\tau')  f_{\RR_4,j'}(\tau_2)\rangle_{\gamma=0} \nonumber\\
    =& \delta_{i,i'}
    \delta_{j,j'}G_{f,loc}(\tau-\tau_1) 
    G_{f,loc}(\tau_2-\tau') 
    \ea
\end{itemize}
Using the results of the first two cases, we rewrite the four-point correlation function as
\begin{align}
   &\langle T_\tau f_{\RR_1,i}(\tau) f_{\RR_2,j}^\dag(\tau') f_{\RR_3,i'}^\dag(\tau_1) f_{\RR_4,j'}(\tau_2)\rangle_{\gamma=0}\nonumber\\
   =&\delta_{\RR_1,\RR_2}\delta_{\RR_3,\RR_4}(1-\delta_{\RR_1,\RR_3})\delta_{i,j}\delta_{i',j'}(-1)G_{f,loc}(\tau-\tau')G_{f,loc}(\tau_2-\tau_1)\nonumber\\
   &+{\delta_{\RR_1,\RR_3}\delta_{\RR_2,\RR_4}(1-\delta_{\RR_1,\RR_4})\delta_{i,i'}\delta_{j,j'}G_{f,loc}(\tau-\tau_1)
    G_{f,loc}(\tau_2-\tau')}\nonumber\\
    &+\delta_{\RR_1,\RR_2}\delta_{\RR_3,\RR_4}\delta_{\RR_1,\RR_3}\langle T_\tau f_{\RR,i}(\tau) f_{\RR,j}^\dag(\tau') f_{\RR,i'}^\dag(\tau_1) f_{\RR,j'}(\tau_2)\rangle_{\gamma=0}\nonumber\\
    =&\delta_{\RR_1,\RR_2}\delta_{\RR_3,\RR_4}\delta_{i,j}\delta_{i',j'}(-1)G_{f,loc}(\tau-\tau')G_{f,loc}(\tau_2-\tau_1)    +{\delta_{\RR_1,\RR_3}\delta_{\RR_2,\RR_4}\delta_{i,i'}\delta_{j,j'}G_{f,loc}(\tau-\tau_1)
    G_{f,loc}(\tau_2-\tau')} \nonumber\\
    &+\delta_{\RR_1,\RR_2}\delta_{\RR_3,\RR_4}\delta_{\RR_1,\RR_3}\tilde{F}^{iji'j'}(\tau,\tau',\tau_1,\tau_2)\label{eq:introduce_tildeF}
\end{align}
where $\tilde{F}^{iji'j'}$ is defined as
\ba 
\label{eq:def_tildeF}
\tilde{F}^{iji'j'}(\tau,\tau',\tau_1,\tau_2)
\equiv & 
\langle T_\tau f_{\RR,i}(\tau) f_{\RR,j}^\dag(\tau') f_{\RR,i'}^\dag(\tau_1) f_{\RR,j'}(\tau_2)\rangle_{\gamma=0} \nonumber\\
&
+ \delta_{i,j}\delta_{i'j'}
G_{f,loc}(\tau-\tau')G_{f,loc}(\tau_2-\tau_1) 
-{\delta_{i,i'}\delta_{j,j'}G_{f,loc}(\tau-\tau_1)
    G_{f,loc}(\tau_2-\tau')}
\ea 
From \cref{eq:introduce_tildeF}, we see that the four-point correlation function is decomposed into two parts: one involving products of two Green's functions, and the other encoded in $\tilde{F}^{iji'j'}$, which represents the remaining local contribution.
Combining \cref{eq:ffdag_expand,eq:introduce_tildeF}, we obtain
\begin{align}
    &-\int D[f,f^\dag] T_\tau f_{\kk,i}(\tau) f_{\kk,j}^\dag(\tau') {e^{-S}} \nonumber\\
      \approx & Z_{\gamma=0} 
    \bigg\{G_{f,loc}(\tau-\tau')\delta_{i,j}
    + 
    \int_{\tau_1,\tau_2}\sum_{\kk'i'} \gamma^2
    \Delta^{i'i'}_{\kk'}(\tau_1-\tau_2) 
    \bigg[ 
    - \delta_{i,j}G_{f,loc}(\tau-\tau')G_{f,loc}(\tau_2-\tau_1)
    \bigg] \nonumber\\ 
    &
+\int_{\tau_1,\tau_2}\gamma^2\Delta^{ij}_{\kk}(\tau_1-\tau_2)G_{f,loc}(\tau-\tau_1)G_{f,loc}(\tau_2-\tau')+ \int_{\tau_1,\tau_2}\sum_{i'j'}\gamma^2  \frac{1}{N_M}\sum_{\kk'}\Delta_{\kk'}^{i'j'}(\tau_1-\tau_2) \tilde{F}^{iji'j'}(\tau,\tau',\tau_1,\tau_2) 
    \bigg\} \label{Equation1}
\end{align}

From \cref{eq:flavor_U8},
we note that the local hybridization function, $\frac{1}{N_M}\sum_{\kk'}\Delta_{\kk'}^{ij}$, also respects the flavor U$(8)$ symmetry (also see Ref.~\cite{PhysRevX.14.031045}).
We can introduce the symmetric local hybridization function $\Delta(\tau)$ through
\ba 
\label{eq:def_hyb_loc}
\frac{1}{N_M}\sum_{\kk}\Delta_{\kk}^{ij}(\tau)
= \delta_{i,j}\Delta(\tau) 
\ea 
We therefore have 
\ba 
 &-\int D[f,f^\dag] T_\tau f_{\kk,i}(\tau) f_{\kk,j}^\dag(\tau') {e^{-S}}  \nonumber\\
    \approx & Z_{\gamma=0} 
    \bigg\{G_{f,loc}(\tau-\tau')\delta_{i,j}
    + 
    \int_{\tau_1,\tau_2}\sum_{\kk'i'} \gamma^2
    \Delta^{i'i'}_{\kk'}(\tau_1-\tau_2) 
    \bigg[ 
    - \delta_{i,j}G_{f,loc}(\tau-\tau')G_{f,loc}(\tau_2-\tau_1)
    \bigg] \nonumber\\ 
    &
+\int_{\tau_1,\tau_2}\gamma^2\Delta^{ij}_{\kk}(\tau_1-\tau_2)G_{f,loc}(\tau-\tau_1)G_{f,loc}(\tau_2-\tau')+ \int_{\tau_1,\tau_2}\sum_{i'}\gamma^2  \Delta(\tau_1-\tau_2) \tilde{F}^{iji'i'}(\tau,\tau',\tau_1,\tau_2) 
    \bigg\} \nonumber\\ 
=&Z_{\gamma=0} 
    \bigg\{G_{f,loc}(\tau-\tau')\delta_{i,j}
    + 
    \int_{\tau_1,\tau_2}\sum_{\kk'i'} \gamma^2
    \Delta^{i'i'}_{\kk'}(\tau_1-\tau_2) 
    \bigg[ 
    - \delta_{i,j}G_{f,loc}(\tau-\tau')G_{f,loc}(\tau_2-\tau_1)
    \bigg] \nonumber\\ 
    &
+\int_{\tau_1,\tau_2}\gamma^2\Delta^{ij}_{\kk}(\tau_1-\tau_2)G_{f,loc}(\tau-\tau_1)G_{f,loc}(\tau_2-\tau')+ \int_{\tau_1,\tau_2}\sum_{i'}\gamma^2  \Delta(\tau_1-\tau_2) \tilde{F}^{iii'i'}(\tau,\tau',\tau_1,\tau_2) \delta_{i,j}
    \bigg\}
\ea 
{where we have used the fact that $F^{iji'i'} \propto \delta_{i,j}$. The argument is as follows.
Since the atomic Hamiltonian \cref{eq:atomic_Hamiltonian_rewrite} has flavor U$(8)$ symmetry, the flavor pairs $(ij')$ and $(ji')$ must be the same in the correlation function $\tilde{F}^{iji'j'}$.
Therefore, it takes the form
\begin{equation}
    \tilde{F}^{iji'j'}\propto\delta_{i,j}\delta_{i',j'}f_1+\delta_{i,i'}\delta_{j',j}f_2
    \label{eq:correlation_U8}
\end{equation}
which implies
\begin{equation}
    \tilde{F}^{iji'i'}\propto \delta_{i,j}f_1+\delta_{i,i'}\delta_{i',j}f_2\propto \delta_{i,j}\label{eq:Fiji'i'}
\end{equation}
}

To simplify the notation, we define
\ba 
\label{eq:def_f_w}
&F^{ij}(\tau,\tau',\tau_1,\tau_2) =\tilde{F}^{iijj}(\tau,\tau',\tau_1,\tau_2) \nonumber\\ 
&    F^{ij}(\tau,\tau',\tau_1,\tau_2)=F^{ij}(\tau-\tau',0,\tau_1-\tau',\tau_2-\tau')= \frac{1}{\beta^2}\sum_{i\omega_1,i\omega_2,i\omega}F^{ij}(i\omega,i\omega_1,i\omega_2) e^{-i\omega (\tau-\tau')+i\omega_1(\tau_1-\tau')-i\omega_2(\tau_2-\tau') }
\ea 
More explicitly, using \cref{eq:def_tildeF}, $F^{ij}$ is given by
\ba 
\label{eq:def_f}
F^{ij}(\tau,\tau',\tau_1,\tau_2)=&\langle T_\tau f_{\RR,i}(\tau) f_{\RR,i}^\dag(\tau') f_{\RR,j}^\dag(\tau_1) f_{\RR,j}(\tau_2)\rangle_{\gamma=0} \nonumber\\
&
+ 
G_{f,loc}(\tau-\tau')G_{f,loc}(\tau_2-\tau_1) 
-{\delta_{i,j}G_{f,loc}(\tau-\tau_1)
    G_{f,loc}(\tau_2-\tau')}
\ea 
Expressing $\tilde{F}^{iijj}$ via $F^{ij}$, we find 
\ba 
 &-\int D[f,f^\dag] T_\tau f_{\kk,i}(\tau) f_{\kk,j}^\dag(\tau') {e^{-S}}  \nonumber\\ 
 \approx &Z_{\gamma=0} 
    \bigg\{G_{f,loc}(\tau-\tau')\delta_{i,j}
    + 
    \int_{\tau_1,\tau_2}\sum_{\kk'i'} \gamma^2
    \Delta^{i'i'}_{\kk'}(\tau_1-\tau_2) 
    \bigg[ 
    - \delta_{i,j}G_{f,loc}(\tau-\tau')G_{f,loc}(\tau_2-\tau_1)
    \bigg] \nonumber\\ 
    &
+\int_{\tau_1,\tau_2}\gamma^2\Delta^{ij}_{\kk}(\tau_1-\tau_2)G_{f,loc}(\tau-\tau_1)G_{f,loc}(\tau_2-\tau')+ \int_{\tau_1,\tau_2}\sum_{i'}\gamma^2  \Delta(\tau_1-\tau_2) F^{ii'}(\tau,\tau',\tau_1,\tau_2) \delta_{i,j}
    \bigg\}
    \ea 
This expression can also be transformed to Matsubara frequency space as
\begin{align}
\label{eq:fdag_f_matrubara_expand}
   &\frac{1}{\beta}\int_{\tau,\tau'}\bigg[-\int D[f,f^\dag] T_\tau f_{\kk,i}(\tau) f_{\kk,j}^\dag(\tau') {e^{-S}}  \bigg] e^{i\omega(\tau-\tau')}\nonumber\\ \nonumber\\ 
     \approx & 
   Z_{\gamma=0}
   \bigg[G_{f,loc}(i\omega)\delta_{i,j}\nonumber\\
   &- \delta_{i,j}\frac{1}{\beta^4}\sum_{i\omega',i\omega_1,i\omega_2}\gamma^2 \sum_{\kk',i'}\Delta_{\kk'}^{i'i'}(i\omega')G_{f,loc}(i\omega_1)G_{f,loc}(i\omega_2) \int_{\tau_1,\tau_2,\tau,\tau'} e^{-i\omega'(\tau_1-\tau_2) -i\omega_1(\tau-\tau') -i\omega_2(\tau_2-\tau_1)+i\omega(\tau-\tau')}  \nonumber\\ 
   &+\gamma^2 \frac{1}{\beta^4}
   \sum_{i\omega',i\omega_1,i\omega_2}\Delta_\kk^{ij}(i\omega')G_{f,loc}(i\omega_1)G_{f,loc}(i\omega_2)\int_{\tau_1,\tau_2,\tau,\tau'}e^{-i\omega'(\tau_1-\tau_2)-i\omega_1(\tau-\tau_1)-i\omega_2(\tau_2-\tau') + i\omega(\tau-\tau')} \nonumber\\ 
   &+\delta_{i,j}\frac{\gamma^2}{\beta^4}\sum_{i\omega',i\omega_1,i\omega_2,i\omega_3}\sum_{i'}\Delta(i\omega')F^{ii'}(i\omega_3,i\omega_1,i\omega_2)\int_{\tau_1,\tau_2,\tau,\tau'}
   e^{-i\omega'(\tau_1-\tau_2)-i\omega_3(\tau-\tau')+i\omega_1(\tau_1-\tau')-i\omega_2(\tau_2-\tau')+i\omega(\tau-\tau')}
   \bigg]
   \nonumber\\ 
   =& Z_{\gamma=0}
   \bigg[ G_{f,loc}(i\omega) \delta_{i,j}
   -\delta_{i,j}\gamma^2\sum_{\kk',i',i\omega'}\Delta_{\kk'}^{i'i'}(i\omega')G_{f,loc}(i\omega') G_{f,loc}(i\omega) \nonumber\\
   &+ \gamma^2 \Delta_\kk^{i,j}(i\omega)G_{f,loc}(i\omega) G_{f,loc}(i\omega) +\gamma^2\delta_{i,j}\sum_{i\omega',i'}\Delta(i\omega') F^{ii'}(i\omega, i\omega',i\omega')
   \bigg] 
\end{align}

Therefore, within hybridization expansion, the Green's function reads (combining \cref{eq:partiaion_expand,eq:fdag_f_matrubara_expand})
\begin{align}
    G_{f,ij}(i\omega,\kk) =& \frac{1}{Z}\int_\tau  - \langle f_{\kk,i}(\tau) f^\dag_{\kk,j}(0)\rangle  e^{i\omega \tau} \nonumber\\ 
    \approx&\bigg[ 1-\gamma^2 \sum_{j,\kk,i\omega} \Delta^{jj}_{\kk}(i\omega) G_{f,loc}(i\omega)
    \bigg]^{-1}\times\bigg[ G_{f,loc}(i\omega) \delta_{i,j}
   -\delta_{i,j}\gamma^2\sum_{\kk',i',i\omega'}\Delta_{\kk'}^{i'i'}(i\omega')G_{f,loc}(i\omega') G_{f,loc}(i\omega) \nonumber\\
   &+ \gamma^2 \Delta_\kk^{ij}(i\omega)G_{f,loc}(i\omega) G_{f,loc}(i\omega) +\gamma^2\delta_{i,j}\sum_{i\omega',i'}\Delta(i\omega') F^{ii'}(i\omega, i\omega',i\omega')
   \bigg]  \nonumber\\
    \approx &\delta_{i,j}G_{f,loc}(i\omega)  +  \gamma^2 \Delta_\kk^{ij}(i\omega) G_{f,loc}(i\omega)G_{f,loc}(i\omega) +  \gamma^2\delta_{i,j}\sum_{i\omega',i'}\Delta(i\omega') F^{ii'}(i\omega, i\omega',i\omega') + \mathcal{O}(\gamma^4)
\end{align}
By comparing the above equation with the first equation in \cref{eq:perturb_exp_Gf_Sigma_f}, we obtain
\ba 
&G^{(0)}_{f,ij}(i\omega,\kk) 
= \delta_{i,j}G_{f,loc}(i\omega) \nonumber\\ 
&G^{(1)}_{f,ij}(i\omega,\kk) 
=   \gamma^2 \Delta_\kk^{ij}(i\omega) G_{f,loc}(i\omega)G_{f,loc}(i\omega) +  \gamma^2\delta_{i,j}\sum_{i\omega',i'}\Delta(i\omega') F^{ii'}(i\omega, i\omega',i\omega') 
\ea 
In addition, from \cref{eq:Sigma_f_pert_Gf_pert}, the self-energy at zeroth order in $\gamma$ reads
\ba 
\Sigma_{f,ij}^{(0)}(i\omega,\kk) 
= \delta_{i,j}
\bigg[ 
i\omega+\mu -\epsilon_f - \frac{(N_f-1)n_fU_1}{N_f}
-1/G_{f,loc}(i\omega)
\bigg].  
\ea 
For convenience, we define the atomic self-energy as 
\begin{equation}
    \Sigma_{f,loc}(i\omega)=i\omega+\mu -\epsilon_f - \frac{(N_f-1)n_fU_1}{N_f}
-1/G_{f,loc}(i\omega)
\label{eq:Sigma0_def}
\end{equation}
The zeroth order self-energy is simply $\Sigma_{f,ij}^{(0)}(i\omega,\kk) =\delta_{i,j}\Sigma_{f,loc}(i\omega)$.
In addition, the self-energy at order $\gamma^2$ reads
\ba 
\label{eq:F_to_Sigma_1}
\Sigma_{f,ij}^{(1)}(i\omega)
=&-\gamma^2 \Delta_{\kk}^{ij}(i\omega) 
+ \gamma^2 
\Delta_{\kk}^{ij}(i\omega) 
+ \gamma^2 \delta_{i,j}\sum_{i\omega',i'}
\Delta(i\omega')F^{ii'}(i\omega,i\omega',i\omega') \frac{1}{[G_{f,loc}(i\omega)]^2} \nonumber\\
=& 
\gamma^2 \delta_{i,j}\sum_{i\omega',i'}
\Delta(i\omega')F^{ii'}(i\omega,i\omega',i\omega') \frac{1}{[G_{f,loc}(i\omega)]^2}
\ea 
We therefore conclude that, by calculating $F^{ii'}$ (\cref{eq:def_f}), which is related to the four-fermion correlation function in the atomic limit, we can extract the self-energy at order $\gamma^2$.

\section{Correlation functions in the atomic limit}
In this section, we evaluate the $f$ electron correlation function $F^{ij}(\tau,\tau',\tau_1,\tau_2)$ defined in \cref{eq:def_f} and the corresponding Fourier transformation $F^{ij}(i\omega,i\omega_1,i\omega_2)$ defined in \cref{eq:def_f_w} in the atomic limit $\gamma=0$.

\subsection{Atomic Green's function}
We first discuss the atomic Hamiltonian and the corresponding Green's function.
At the $\gamma=0$ limit, the atomic Hamiltonian of $f$ electrons (\cref{eq:action_sep,eq:atomic_Hamiltonian_rewrite}), with $H_{U_1},H_{V},H_{W}$ and $H_J$ included only through their Hartree-Fock contributions can be written as 
\ba 
\label{eq:H_atomic_def}
H_{atom} = \frac{U_1}{2} 
\sum_{\RR} \bigg(\hat{n}_{f,\RR} -n_0
\bigg) ^2 -\mu\sum_{\RR}\hat{n}_{f,\RR}
\ea 
where the $f$ electron density operator is defined as 
\ba 
\label{eq:density_op_f}
\hat{n}_{f,\RR}
= \sum_i f_{\RR,i}^\dag f_{\RR,i}
\ea 
{
Using \cref{eq:hartree-energy}, $n_0$ is given by
\begin{equation}
    n_0=\frac{1}{2}-\frac{\epsilon_f}{U_1}=\frac{N_f}{2}-\frac{1}{U_1}\bigg(6U_2\nu_f+2W_1\nu_{c,1}+2W_3\nu_{c,3}-\frac{J}{4}\nu_{c,3}\bigg)
\end{equation}
In general, $n_0$ should be determined self-consistently.
In the zero-hybridization limit where $\nu_f=\nu$, $\nu_{c,a}=0$ and assuming $U_2\ll U_1$, then obtain
\begin{equation}
    n_0\approx \frac{N_f}{2},\qquad (\nu_{c,a}\sim0,\;U_2\ll U_1).
\end{equation}

Since different sites are decoupled in \cref{eq:H_atomic_def}, each site can be solved independently.
We denote the single site atomic Hamiltonian by
\begin{equation}
    H_{atom,\RR}=\frac{U_1}{2}(\hat{n}_{f,\RR}-n_0)^2-\mu\hat{n}_{f,\RR}
    \label{eq:atomic_single_site}
\end{equation}
For an eigenstate with $f$-electron occupation $\tilde{n}\in\{0,\cdots,N_f\}$, the corresponding energy is
\begin{equation}
    E(\tilde{n})=\frac{U_1}{2}(\tilde{n}-n_0)^2-\mu \tilde{n}
\end{equation}
The atomic Green's function is then
\begin{equation}
    G_{f,loc}(i\omega)=\frac{1}{Z}\sum_{\tilde{n}=0}^{N_f-1}\binom{N_f-1}{\tilde{n}}\frac{e^{-\beta E(\tilde{n})}+e^{-\beta E(\tilde{n}+1)}}{i\omega+E(\tilde{n})-E(\tilde{n}+1)}=\frac{1}{N_f}\sum_{\tilde{n}=0}^{N_f-1}\frac{(N_f-\tilde{n})p_{\tilde{n}}+(\tilde{n}+1)p_{\tilde{n}+1}}{i\omega+E(\tilde{n})-E(\tilde{n}+1)}
    \label{eq:atomic_green_initial}
\end{equation}
where
\begin{equation}
    p_{\tilde{n}}=\frac{e^{-\beta E(\tilde{n})}}{Z}\binom{N_f}{\tilde{n}}
\end{equation}
}
At zero temperature, when the ground-state filling $n_f$ satisfies $n_f\in \{1,...,7\}$, we can define the charge $\pm 1$ excitation energies $\Delta_{\pm 1}$.
In the atomic limit, these are given by
\ba 
\label{eq:define_charge_gap}
\Delta_{\pm 1} = E(n_f\pm1)-E(n_f) = U _1\bigg[ 
\frac{1}{2} 
\pm \bigg(n_f-n_0\bigg) 
\bigg]\mp \mu = U_1\bigg[ 
\frac{1}{2} \pm\bigg( n_f - \frac{1}{2}+\frac{\epsilon_f}{U_1}\bigg) 
\bigg] \mp \mu
\ea 
where we have also used $n_0 = 1/2-\epsilon_f/U_1$ (see the definition of $n_0$ below \cref{eq:action_sep}). 
{
At zero temperature, the ground state dominates, so $p_{n_f}=1$ and $p_{\tilde{n}\ne n_f}\ll 1$
}
Thus, the atomic Green's function takes the form (also see Ref.~\cite{hu2025projectedsolvabletopologicalheavy})

\boxedeq{
\ba 
\label{eq:atomic_Green_fun}
    G_{f,loc}(i\omega)= \frac{1}{N_f}
    \bigg(\frac{n_f}{i\omega + \Delta_{-1}} +\frac{N_f-n_f}{i\omega-\Delta_{+1}}\bigg)
\ea 
}
The corresponding atomic (Hubbard-I) self-energy $\Sigma_{f,loc}(i\omega)$ (\cref{eq:Sigma0_def}) reads 
\boxedeq{
\ba  \Sigma_{f,loc}(i\omega)
=& \frac{n_f(N_f-n_f)U_1^2}{N_f^2 }
\bigg[ 
\frac{1}{i\omega 
- \frac{n_f\Delta_{+1} - (N_f-n_f)\Delta_{-1} }{N_f}
}
\bigg] \label{eq:atomic-self-energy}
\ea 
}
In the atomic limit, the zero-temperature partition function for a single site is 

\begin{equation}
    Z_{atom,\RR}=\binom{N_f}{n_f}{e^{-\beta E_{gs}}}\label{eq:partition_single_site}
\end{equation}
where 
\begin{equation}
    E_{gs}\equiv E(n_f)=\frac{U_1}{2}(n_f-n_0)^2-\mu n_f.
\end{equation}
The prefactor $\binom{N_f}{n_f}$ counts the number of ways to choose $n_f$ occupied flavors out of $N_f$.
Since different sites are decoupled, the partition function for the full lattice is simply
\begin{equation}
    Z_{\gamma=0}=(Z_{atom,\RR})^{N_M}
\end{equation}
where $N_M$ is the number of moir\'e cell.

We now examine the relation between the chemical potential $\mu$ and the ground-state occupation $n_f$ in the zero-temperature limit.
For $n_f$ to be the ground-state occupation, we must have
\begin{equation}
    \Delta_{\pm1}=U_1\bigg[\frac{1}{2}\pm (n_f-n_0)\bigg]\mp \mu>0
\end{equation}
which gives
\begin{equation}
    U_1\bigg[-\frac{1}{2}+(n_f-n_0)\bigg]<\mu<U_1\bigg[\frac{1}{2}+(n_f-n_0)\bigg]
\end{equation}
At exact integer filling $n_f$ near zero temperature, the chemical potential is fixed by requiring particle and hole excitations to be symmetric, which implies
\begin{equation}
   \lim_{\beta\to +\infty} [E(n_f-1)-E(n_f+1)]=\lim_{\beta\to +\infty}\frac{1}{\beta}\log\bigg[\frac{n_f(n_f+1)}{(N_f-n_f)(N_f-n_f+1)}\bigg]=0
\end{equation}
Thus, in the low-temperature limit we obtain
\begin{equation}
    \mu\approx U_1(n_f-n_0)
\end{equation}
Using this formula for chemical potential, we find that
\begin{equation}
    \Delta_{\pm1}\approx\frac{U_1}{2}
\end{equation}
and the atomic Green's function \cref{eq:atomic_Green_fun} becomes
\begin{equation}
    G_{f,loc}(i\omega)=\frac{1}{N_f}\bigg(\frac{n_f}{i\omega+U_1/2}+\frac{N_f-n_f}{i\omega-U_1/2}\bigg)
\end{equation}

\subsection{Evaluation of $F^{ij}$} 
We now evaluate $F^{ij}(i\omega,i\omega',i\omega')$. 
From \cref{eq:def_f_w,eq:def_f}, we have
\ba 
F^{ij}(i\omega,i\omega',i\omega') 
= &\frac{1}{\beta^2}\int_{\tau,\tau',\tau_1,\tau_2}
\langle T_\tau f_{\RR,i}(\tau) f_{\RR,i}^\dag(\tau') f_{\RR,j}^\dag(\tau_1)f_{\RR,j}(\tau_2)\rangle_{\gamma=0} e^{i\omega(\tau-\tau')-i\omega'(\tau_1-\tau')+i\omega'(\tau_2-\tau')} \nonumber\\ 
&+G_{f,loc}(i\omega)G_{f,loc}(i\omega')
-\delta_{i,j}
\delta_{\omega,\omega'} 
G_{f,loc}(i\omega)G_{f,loc}(i\omega)
\ea 
We first consider the four-point correlation function
\ba 
\langle T_\tau f_{\RR,i}(\tau) f_{\RR,i}^\dag(\tau') f_{\RR,j}^\dag(\tau_1) f_{\RR,j}(\tau_2)\rangle_{\gamma=0}
=& - 
\langle T_\tau 
f_{\RR,j}^\dag(\tau_1) f_{\RR,j}(\tau_2)f_{\RR,i}^\dag(\tau') f_{\RR,i}(\tau)  \rangle_{\gamma=0} \nonumber\\ 
=& - 
\langle T_\tau 
f_{\RR,j}^\dag(\tau_1-\tau) f_{\RR,j}(\tau_2-\tau)f_{\RR,i}^\dag(\tau'-\tau) f_{\RR,i}(0)  \rangle_{\gamma=0}
\ea 
where, in the last line, we have used the imaginary-time translational invariance, so that the correlator only depends on the imaginary time differences $\tau_1-\tau,\tau_2-\tau,\tau'-\tau$.
Then the expression for $F^{ij}$ becomes
\ba 
\label{eq:F_nm_from_ffff}
F^{ij}(i\omega,i\omega',i\omega') 
= & - 
\frac{1}{\beta}\int_{\tau_1,\tau_2,\tau_3} \langle T_\tau 
f_{\RR,j}^\dag(\tau_1) f_{\RR,j}(\tau_2)f_{\RR,i}^\dag(\tau_3) f_{\RR,i}(0)  \rangle_{\gamma=0} e^{-i\omega\tau_3 +i\omega'\tau_2-i\omega'\tau_1 } \nonumber\\ 
& +G_{f,loc}(i\omega)G_{f,loc}(i\omega')
-\delta_{i,j}
\delta_{\omega,\omega'} 
G_{f,loc}(i\omega)G_{f,loc}(i\omega)
\ea 
Therefore, we can obtain $F^{ij}$ via the following correlation function and the corresponding Fourier transform
\ba 
\label{eq:FT_of_four-point}
&\langle T_\tau 
f_{\RR,j}^\dag(\tau_1) f_{\RR,j}(\tau_2)f_{\RR,i}^\dag(\tau_3) f_{\RR,i}(0)  \rangle_{\gamma=0} \nonumber\\ 
\text{and   }\quad\quad &\frac{1}{\beta}\int_{\tau_1,\tau_2,\tau_3} \langle T_\tau 
f_{\RR,j}^\dag(\tau_1) f_{\RR,j}(\tau_2)f_{\RR,i}^\dag(\tau_3) f_{\RR,i}(0)  \rangle_{\gamma=0} e^{-i\omega\tau_3+i\omega'\tau_2-i\omega'\tau_1 }
\ea 

{
\subsubsection{Correlation function in the eigenstate basis}\label{app:path2Lehmann}
Previously, the expectation value $\langle \cdots\rangle_{\gamma=0}$ was defined in path-integral formalism in \cref{eq:exp_path_integral}.
{
For multiple operators, the corresponding expectation value is
\begin{equation}
    \langle O_N(\tau_N)\cdots O_2(\tau_2) O_{1}(\tau_1)\rangle_{\gamma=0}=\frac{1}{Z_{\gamma=0}}\int D[f^{\dagger}(\tau),f(\tau)]\;O_N(\tau_N)\cdots O_{2}(\tau_2)O_{1}(\tau_1)e^{-S_{atom}}\label{eq:multi_operator_path_integral}
\end{equation}
For an operator $\hat{O}=\hat{O}[\hat{f}^{\dagger},\hat{f}]$ constructed by the fermionic creation and annihilation operators, the corresponding Grassmann expression $O(\tau)$ is obtained by replacing the operators by the Grassmann numbers, i.e.,
\begin{equation}
    O(\tau)\equiv O[f^{\dagger}(\tau),f(\tau)]
\end{equation}
In general, it is not necessary to include the time-ordering operator explicitly in the path-integral formula, because the required minus signs are automatically encoded in the Grassmann algebra.
For example,
\begin{equation}
    T_{\tau}O_1(\tau_1)O_2(\tau_2)=\begin{cases}
        O_1(\tau_1)O_2(\tau_2),&\tau_1>\tau_2\\
        (-1)^{\xi_1\xi_2}O_2(\tau_2)O_1(\tau_1),&\tau_2>\tau_1
    \end{cases}
    \label{eq:time_order}
\end{equation}
where $\xi_i$ is the number of fermionic operators appearing in operator $\hat{O}_i$.
On the other hand, because $O_i(\tau_i)$ are Grassmann expressions, they already satisfy $O_1(\tau_1)O_{2}(\tau_2)=(-1)^{\xi_1\xi_2}O_2(\tau_2)O_1(\tau_1)$.
This means that in the path integral formula, we always have
\begin{equation}
    T_{\tau}O_1(\tau_1)O_{2}(\tau_2)=O_1(\tau_1)O_{2}(\tau_2)\label{eq:Grassmann_time_order}
\end{equation}
}

We now show that the path integral expression $\langle O_N(\tau_N)\cdots  O_{1}(\tau_1)\rangle_{\gamma=0}$ is equivalent to the operator expression {$\frac{1}{Z_{\gamma=0}}\mathrm{Tr}[{T_{\tau}}e^{-\beta H_{atom}}\hat{O}_N(\tau_N)\cdots \hat{O}_1(\tau_1)]$} which can be written as
{
\begin{align}
    &\frac{1}{Z_{\gamma=0}}\mathrm{Tr}[{T_{\tau}}e^{-\beta H_{atom}}\hat{O}_N(\tau_N)\cdots \hat{O}_{2}(\tau_2)\hat{O}_1(\tau_1)]\nonumber\\
    =&{
    \frac{1}{Z_{\gamma=0}}(-1)^{\xi_{\pi}}\mathrm{Tr}[e^{-\beta H_{atom}}\hat{O}_{\pi(N)}(\tau_{\pi(N)})\cdots \hat{O}_{\pi(2)}(\tau_{\pi(2)})\hat{O}_{\pi(1)}(\tau_{\pi(1)})]
    }\nonumber\\
    =&{
    \frac{1}{Z_{\gamma=0}}(-1)^{\xi_{\pi}}\mathrm{Tr}[U(\beta,\tau_{\pi(N)})\hat{O}_{\pi(N)}U(\tau_{\pi(N)},\tau_{\pi(N-1)})\cdots U(\tau_{\pi(3)},\tau_{\pi(2)})\hat{O}_{\pi(2)}U(\tau_{\pi(2)},\tau_{\pi(1)})\hat{O}_{\pi(1)}U(\tau_{\pi(1)},0)]
    }\label{eq:exp_trace}
\end{align}
In the expression above, $\pi$ is the permutation that orders the times $\tau_{1,\cdots N}$ and $\xi_{\pi}$ counts the number of swaps of fermionic operators required in the time-ordering process.
}
The partition function in operator language is defined as $Z_{\gamma=0}=\mathrm{Tr}(e^{-\beta H_{atom}})$ and the time evolution of operator $\hat{O}(\tau)$ is controlled by $H_{atom}$ as 
\begin{equation}
    \hat{O}(\tau)=e^{\tau H_{atom}}\hat{O}e^{-\tau H_{atom}}
\end{equation}
Besides, the time evolution operator is defined as 
\begin{equation}
    U(\tau,\tau')=U(\tau-\tau')=\exp[-(\tau-\tau')H_{atom}]
\end{equation}
By inserting the identity in terms of fermionic coherent states,
\begin{equation}
    1=\int df^{\dagger}df\;e^{-f^{\dagger}f}|{f}\rangle\langle{f^{\dagger}}|
\end{equation}
The partition function is 
\begin{align}
    Z_{\gamma=0}=&\lim_{M\to\infty}\int\prod_{j=0}^{M-1}df^{\dagger}(\tau^{j})df(\tau^j)\;\exp\bigg(-\sum_{j=0}^{M-1}f^{\dagger}(\tau^j)f(\tau^j)\bigg)\prod_{j=0}^{M-1}\langle f^{\dagger}(\tau^{j+1})|e^{-\frac{\beta}{M}H_{atom}}|f(\tau^j)\rangle\nonumber\\
    =&\lim_{M\to\infty}\int\prod_{j=0}^{M-1}df^{\dagger}(\tau^{j})df(\tau^j)\;\exp\bigg(-\sum_{j=0}^{M-1}f^{\dagger}(\tau^j)f(\tau^j)\bigg)\prod_{j=0}^{M-1}e^{f^{\dagger}(\tau^{j+1})f(\tau^j)-\frac{\beta}{M}H_{atom}[f^{\dagger}(\tau^{j+1}),f(\tau^{j})]}\nonumber\\
    =&\int D[f^{\dagger}(\tau),f(\tau)]\;\exp\bigg[-\int_{0}^{\beta}d\tau\;\big(f^{\dagger}(\tau)\partial_{\tau}f(\tau)+H_{atom}[f^{\dagger}(\tau),f(\tau)]\big)\bigg]\nonumber\\
    =&\int D[f^{\dagger}(\tau),f(\tau)]e^{-S_{atom}}\label{eq:partition_path_integral_new}
\end{align}

Similarly,
\begin{align}
    &{
   (-1)^{\xi_{\pi}}\mathrm{Tr}[U(\beta,\tau_{\pi(N)})\hat{O}_{\pi(N)}U(\tau_{\pi(N)},\tau_{\pi(N-1)})\cdots U(\tau_{\pi(3)},\tau_{\pi(2)})\hat{O}_{\pi(2)}U(\tau_{\pi(2)},\tau_{\pi(1)})\hat{O}_{\pi(1)}U(\tau_{\pi(1)},0)]
    }\nonumber\\
    =&{
    (-1)^{\xi_{\pi}}\int D[f^{\dagger}(\tau),f(\tau)]\;O_{\pi(N)}(\tau_{\pi(N)})\cdots O_{\pi(1)}(\tau_{\pi(1)})e^{-S_{atom}}
    }
    \nonumber\\
    =&\int D[f^{\dagger}(\tau),f(\tau)]\;O_N(\tau_N)\cdots O_1(\tau_1)e^{-S_{atom}}\label{eq:exp_path_integral_new}
\end{align}
{
In the last equality, we use the fact that the permutation $\pi$ generates the same sign whether one exchanges fermionic operators or Grassmann variables.
Combining \cref{eq:multi_operator_path_integral}, \cref{eq:exp_trace}, \cref{eq:partition_path_integral_new} and \cref{eq:exp_path_integral_new}, we find that
\begin{equation}
    \langle O_{N}(\tau_N)\cdots O_{1}(\tau_1)\rangle_{\gamma=0}=\frac{1}{Z_{\gamma=0}}\mathrm{Tr}[T_{\tau}e^{-\beta H_{atom}}\hat{O}_{N}(\tau_N)\cdots \hat{O}_{1}(\tau_1)]\label{eq:exp_equivalence}
\end{equation}
Here \cref{eq:partition_path_integral_new} and \cref{eq:exp_path_integral_new} recover the path integral definition of the expectation value at $\gamma=0$ given by \cref{eq:exp_path_integral}.

If all operators act on a single site $\RR$, we then have
\begin{align}
   \langle O_{N,\RR}(\tau_N)\cdots O_{1,\RR}(\tau_1)\rangle_{\gamma=0}=&\frac{1}{Z_{\gamma=0}}\mathrm{Tr}[T_{\tau}e^{-\beta H_{atom}}\hat{O}_{N,\RR}(\tau_N)\cdots \hat{O}_{1,\RR}(\tau_1)] \nonumber\\
   =&\frac{1}{Z_{\gamma=0}}\mathrm{Tr}[e^{-\beta\sum_{\RR'\ne \RR}H_{atom,\RR'}}]\mathrm{Tr}[T_{\tau}e^{-\beta H_{atom,\RR}}\hat{O}_{N,\RR}(\tau_N)\cdots \hat{O}_{1,\RR}(\tau_1)]\nonumber\\
   =&\frac{1}{Z_{atom,\RR}}\mathrm{Tr}[T_{\tau}e^{-\beta H_{atom,\RR}}\hat{O}_{N,\RR}(\tau_N)\cdots \hat{O}_{1,\RR}(\tau_1)]\label{eq:exp_equivalence_single_site}
\end{align}
Using the fact that $[H_{atom,\RR'\ne\RR},\hat{O}_{\RR}]=0$, we have
\begin{equation}
    \hat{O}_{\RR}(\tau)\equiv e^{\tau H_{atom}}\hat{O}_{\RR}e^{-\tau H_{atom}}=e^{\tau H_{atom,\RR}}\hat{O}_{\RR}e^{-\tau H_{atom,\RR}}
\end{equation}
Besides, we can also define the time evolution operator for a single site as 
\begin{equation}
    U_{\RR}(\tau,\tau')\equiv e^{-(\tau-\tau')H_{atom,\RR}}
\end{equation}
}

In the path-integral formalism, the trace in \cref{eq:exp_trace} is expanded using fermionic coherent states. For the atomic problem, however, the many-body eigenstates are easily obtained, so it is more convenient to work in the eigenstate basis.
Let $|s\rangle$ denote an eigenstate of the single-site atomic Hamiltonian with eigenvalue $E_s$, i.e.,
\begin{equation}
    H_{atom,\RR}|s\rangle =E_s |s\rangle
\end{equation}
For simplicity, we assume the times are already ordered as $\beta>\tau_N>\cdots>\tau_2>\tau_1>0$. Then the expectation value \cref{eq:exp_equivalence_single_site} can be written in the eigenstate basis as
\begin{align}
    &\langle O_{\RR}(\tau_N)\cdots O_{2,\RR}(\tau_2) O_{1,\RR}(\tau_1)\rangle_{\gamma=0}\\
    =&\frac{1}{Z_{atom,\RR}}\sum_{s_{1},s_{2},\cdots s_{N}}\langle s_N|U_{\RR}(\beta,\tau_N)O_{N,\RR}|s_{N-1}\rangle \cdots\langle s_2|U_{\RR}(\tau_3,\tau_2)O_{2,\RR}|s_1\rangle\langle s_1|U_{\RR}(\tau_2,\tau_1)O_{1,\RR} U_{\RR}(\tau_1,0)|s_N\rangle\nonumber\\
    =&\frac{1}{Z_{atom,\RR}}\sum_{s_1,s_2,\cdots,s_N}e^{-(\beta+\tau_1-\tau_{N}) E_{s_N}-\sum_{n=1}^{N-1}(\tau_{n+1}-\tau_{n})E_{s_n}}\langle s_N|O_{N,\RR}|s_{N-1}\rangle\cdots \langle s_2|O_{2,\RR}|s_1\rangle\langle s_1|O_{1,\RR}|s_N\rangle\label{eq:path_to_eigenstates}
\end{align}
{In general cases, we should make the operators time-ordered first and then expand over the eigenstate basis.}
\cref{eq:path_to_eigenstates} will be used below to evaluate the four-point correlation function.

}

{
\subsubsection{Summary of the results}
We summarize here the evaluation of $F^{ij}$ defined in \cref{eq:F_nm_from_ffff} as
\ba 
F^{ij}(i\omega,i\omega',i\omega') 
=  - I_{ij} +G_{f,loc}(i\omega)G_{f,loc}(i\omega')
-\delta_{i,j}
\delta_{\omega,\omega'} 
G_{f,loc}(i\omega)G_{f,loc}(i\omega)
\label{eq:F_summary}
\ea 
where $I_{ij}$ is the Fourier transformation of $\langle T_\tau 
f_{\RR,j}^\dag(\tau_1) f_{\RR,j}(\tau_2)f_{\RR,i}^\dag(\tau_3) f_{\RR,i}(0)  \rangle_{\gamma=0}$ defined by
\begin{equation}
    I_{ij}(\omega,\omega',\omega')=\frac{1}{\beta}\int_{\tau_1,\tau_2,\tau_3} \langle T_\tau 
f_{\RR,j}^\dag(\tau_1) f_{\RR,j}(\tau_2)f_{\RR,i}^\dag(\tau_3) f_{\RR,i}(0)  \rangle_{\gamma=0} e^{-i\omega\tau_3 +i\omega'\tau_2-i\omega'\tau_1 }
\label{eq:Iij_summary}
\end{equation}
and $G_{f,loc}$ is the atomic Green's function derived in \cref{eq:atomic_Green_fun} and given by
\boxedeq{
\ba 
    G_{f,loc}(i\omega)= \frac{1}{N_f}
    \bigg(\frac{n_f}{i\omega + \Delta_{-1}} +\frac{N_f-n_f}{i\omega-\Delta_{+1}}\bigg)
    \label{eq:G_atom_summary}
\ea 
}

For the case of $i=j$, we obtain (proof from \cref{eq:Fii_prove_start} to \cref{eq:ffff_ii_freq})
\boxedeq{
\begin{align}
       I_{ii}(\omega,\omega',\omega')
   =&  \frac{n_f}{N_f} \frac{1- \delta_{\omega\omega '}}{(i\omega'+\Delta_{-1})(i\omega+ \Delta_{-1})}   +  \frac{N_f- n_f}{N_f}  \frac{1- \delta_{\omega\omega '}}{(i\omega' -\Delta_{+1})(i\omega-\Delta_{+1})} 
   \label{eq:Iii_summary}
\end{align}
}
and $F^{ii}$ is (proof in \cref{eq:F_ii_proof})
\boxedeq{
\begin{align}
    F^{ii}(i\omega,i\omega',i\omega')
    =&{\frac{n_f(N_f-n_f)}{N_f^2}}
\frac{(\Delta_{-1}+\Delta_{+1})^2 }{(i\omega + \Delta_{-1})(i\omega -\Delta_{+1})(i\omega' - \Delta_{+1})(i\omega'+\Delta_{-1})}(-1+\delta_{\omega ,\omega'})
\label{eq:Fii_summary}
\end{align}
}

For the case of $i\ne j$, we write 
\begin{equation}
    I_{i\ne j}(\omega,\omega',\omega')=I_{i\ne j}^0(\omega,\omega',\omega')+I_{i\ne j}^1(\omega,\omega',\omega')
    \label{eq:I0ij_I1ij_def}
\end{equation}
Here, $I_{i\ne j}^1(\omega,\omega',\omega')$ is proportional to $1/\beta$ whereas $I_{i\ne j}^0(\omega,\omega',\omega')$ does not contain such a prefactor.
The explicit expressions are (proof from \cref{eq:sum_over_Fij_start} to \cref{eq:sum_Iij_beta}))
\boxedeq{
\begin{align}
 &I^{0}_{i\ne j}(\omega,\omega',\omega')\nonumber\\
 =&  \frac{n_f(n_f-1)}{N_f(N_f-1)} \frac{1}{(i\omega'+ \Delta_{-1})(i\omega + \Delta_{-1})}    +  \frac{(N_f - n_f)(N_f - n_f-1)}{N_f (N_f-1)} \frac{1}{ (i\omega'- \Delta_{+1})(i\omega- \Delta_{+1})}  \nonumber \\ 
 & +
  \frac{n_f(N_f- n_f)}{ N_f(N_f-1)} \frac{  2 i \omega' i\omega + (i \omega' + i \omega) (\Delta_{-1}   - \Delta_{+1} ) - 2 \Delta_{+1}  \Delta_{-1}   -(\Delta_{-1} + \Delta_{+1})^2 \delta_{\omega \omega' }  
  }{{(i \omega - \Delta_{+1} )(i\omega'+ \Delta_{-1}) ( i \omega'- \Delta_{+1})(i\omega + \Delta_{-1})}  } \label{eq:I0ij_summary}\\
  &I^{1}_{i\ne j}(\omega,\omega',\omega')\nonumber\\
  =& \frac{1 }{\beta} \bigg\{\frac{n_f(n_f-1)}{ N_f(N_f-1)} \frac{i\omega' + i\omega + 2\Delta_1  }{ (i\omega + \Delta_1)^2(i\omega' + \Delta_1)^2}     \frac{ 2\Delta_1 -\Delta_{2} } {i\omega+   i \omega'+ \Delta_{2} } +\frac{ (N_f - n_f)  (N_f -  n_f-1)}{N_f(N_f-1)} \frac{i\omega'+ i\omega- 2\Delta_1}{(i\omega- \Delta_1)^2(i\omega'- \Delta_1)^2 }  \frac{ 2 \Delta_1- \Delta_{2}}{ i \omega+ i\omega'- \Delta_{2} }  \nonumber \\
   & -
  \frac{n_f(N_f- n_f)}{ N_f(N_f-1)} \frac{4\Delta_1 }{(i\omega'- \Delta_1)^2(  i \omega'+ \Delta_1)^2( i\omega- \Delta_1)^2 (  i \omega + \Delta_1)^2} \bigg[     (i \omega)^2 (i \omega')^2   +( (i \omega)^2+ (i \omega')^2)  
   \Delta_1^2  - 3 \Delta_1^4  \bigg ] \bigg\}\nonumber\\\label{eq:I1ij_summary}
\end{align}
}
The corresponding expression for $F^{i\ne j}$ is (proof in \cref{eq:F_ij_proof})
\boxedeq{
\begin{align}
    &F^{i\ne j}(i\omega,i\omega',i\omega')=-I_{i\ne j}(\omega,\omega',\omega')+G_{f,loc}(i\omega)G_{f,loc}(i\omega')\nonumber\\
    =&\frac{n_f(N_f- n_f)}{ N_f^2(N_f-1)} \frac{  (\Delta_{+1}+  \Delta_{-1} )^2  (1+ N_f \delta_{\omega, \omega'})
  }{{(i \omega - \Delta_{+1} )(i\omega'+ \Delta_{-1}) ( i \omega'- \Delta_{+1})(i\omega + \Delta_{-1})}  }   - I^1_{i\ne j}(i\omega,i\omega',i\omega') + \mathcal{O}( e^{- \beta \frac{U_1}{2} })  
  \label{eq:Fij_summary}
\end{align}
}

Once $F^{ij}$ has been determined, the self-energy at order $\gamma^2$ is given by \cref{eq:F_to_Sigma_1}:
\ba 
\Sigma_{f,ij}^{(1)}(i\omega)= 
 \gamma^2 \delta_{i,j}\sum_{i\omega',i'}
\Delta(i\omega')F^{ii'}(i\omega,i\omega',i\omega') \frac{1}{[G_{f,loc}(i\omega)]^2} 
\label{eq:self1_summary}
\ea 
The final result is (\cref{eq:F_ij_final})
\boxedeq{
\ba 
\Sigma_{f,ij}^{(1)}(i\omega)= 
 \gamma^2 \delta_{i,j} \Delta(i\omega) {\frac{n_f(N_f-n_f)}{N_f^2}}
\frac{(\Delta_{-1}+\Delta_{+1})^2 (N_f+1)   }{\left[i \omega + \Delta_{-1} - \frac{n_f}{N_f}(\Delta_{+1}+ \Delta_{-1})\right]^2}  - \gamma^2 \delta_{i,j} 
 \sum_{i\omega'}
\Delta(i\omega')L^1(i\omega,i\omega') 
\label{eq:Sigma1_summary}
\ea 
}
where $L^1(i\omega,i\omega')$ is defined in \cref{eq:L1_def}
\begin{align}
    & L^1(i\omega,i\omega') = \frac{ (N_f-1)  I^1_{i\ne j}(i\omega,i\omega',i\omega') (i \omega + \Delta_{-1})^2(i \omega - \Delta_{+1})^2 }{\left[ i \omega + \Delta_{-1} - \frac{n_f}{N_f}(\Delta_{+1}+ \Delta_{-1})  \right]^2}
\end{align}
which accounts for the contribution of $I_{i\ne j}^{1}(\omega,\omega',\omega')$ to the self-energy with the prefactor $1/\beta$.

\begin{table}[]
    \centering
    \begin{tabular}{c|c|c|c}
    \hline
        Quantity & definition & final expression & proofs \\\hline
        $G_{f,loc}$ & \cref{eq:atomic_green_def} & \cref{eq:G_atom_summary} &\cref{eq:atomic_green_initial} to \cref{eq:atomic_Green_fun}
        
        \\\hline
        $I_{ii}$ & \cref{eq:Iij_summary} & \cref{eq:Iii_summary} & \cref{eq:Fii_prove_start} to \cref{eq:ffff_ii_freq}), also needs \cref{eq:ii_integral_sum}
        \\\hline
        $F^{ii}$ & \cref{eq:F_summary} & \cref{eq:Fii_summary} & \cref{eq:F_ii_proof}\\\hline
        $I_{i\ne j}$ & \cref{eq:Iij_summary},\cref{eq:I0ij_I1ij_def} & \cref{eq:I0ij_summary},\cref{eq:I1ij_summary}
        &\cref{eq:Fij_no1,eq:Fij_no2,eq:Fij_no3,eq:Fij_no4,eq:Fij_no5,eq:Fij_no6}
        \\
        & &(original in \cref{eq:sum_Iij0},\cref{eq:sum_Iij_beta}) 
        &(original from \cref{eq:sum_over_Fij_start} to \cref{eq:sum_Iij_beta})\\\hline
        $F_{i\ne j}$ & \cref{eq:F_summary} & \cref{eq:Fij_summary} & \cref{eq:F_ij_proof} \\\hline
        $\Sigma^{(1)}_{f,ij}$ & \cref{eq:self1_summary} & \cref{eq:Sigma1_summary} &
        \cref{eq:Sigma1_proof_start} to \cref{eq:Sigma1_final}
        \\\hline
    \end{tabular}
    \caption{Summary of physical quantities, including their definitions, final expressions and the corresponding derivations.}
    \label{tab:quantity_summary}
\end{table}

}

\subsubsection{A useful integral identity}
Before evaluating \(F^{ij}(i\omega,i\omega',i\omega')\), we introduce the following integral, which will be used repeatedly below. It is defined by
\begin{equation}
   I(\omega_1,\omega_2,\omega_3,E_1,E_2,E_3)\equiv  \frac{1}{\beta}\int_0^\beta d\tau_3 \int_{\tau_3}^\beta d\tau_2 \int_{\tau_2}^\beta d\tau_1\; e^{-\tau_3 (E_{3}+i\omega_3)+\tau_2 (E_2+i\omega_2)-\tau_1 (E_{1}+i\omega_1)}
\end{equation}
A straightforward evaluation gives
\begin{align}
  &I(\omega_1,\omega_2,\omega_3,E_1,E_2,E_3)\nonumber\\
  =&\frac{1}{\beta}\int_0^\beta d\tau_3 \int_{\tau_3}^\beta d\tau_2 \;e^{-\tau_3(E_3+i\omega_3)+\tau_2(E_2+i\omega_2)}\frac{e^{-\tau_2(E_1+i\omega_1)}-e^{-\beta(E_1+i\omega_1)}}{E_1+i\omega_1}\nonumber\\
  =&\frac{1}{\beta(E_1+i\omega_1)}\int_0^\beta d\tau_3\;e^{-\tau_3(E_3+i\omega_3)} \bigg[\frac{e^{\beta(E_2-E_1+i\omega_2-i\omega_1)}-e^{\tau_3(E_2-E_1+i\omega_2-i\omega_1)}}{E_2-E_1+i\omega_2-i\omega_1}(1-\delta_{E_1E_2}\delta_{\omega_1\omega_2})+(\beta-\tau_3)\delta_{E_1E_2}\delta_{\omega_1\omega_2}\nonumber\\
  &+e^{-\beta E_1}\frac{e^{\beta(E_2+i\omega_2)}-e^{\tau_3(E_2+i\omega_2)}}{E_2+i\omega_2}  \bigg]\nonumber\\
  =&\frac{1}{\beta(E_1+i\omega_1)}\bigg\{\frac{1-\delta_{E_1E_2}\delta_{\omega_1\omega_2}}{E_2-E_1+i\omega_2-i\omega_1}\bigg[e^{\beta(E_2-E_1)}\frac{1-e^{-\beta(E_3+i\omega_3)}}{E_3+i\omega_3}-\frac{1-e^{-\beta(E_3-E_2+E_1+i\omega_3-i\omega_2+i\omega_1)}}{E_3-E_2+E_1+i\omega_3-i\omega_2+i\omega_1}  \bigg]  \nonumber\\
  &+\delta_{E_1E_2}\delta_{\omega_1\omega_2}\frac{-1+e^{-\beta(E_3+i\omega_3)}+\beta(E_3+i\omega_3)}{(E_3+i\omega_3)^2}  \nonumber\\
  &+\frac{e^{-\beta E_1}}{E_2+i\omega_2}\bigg[-e^{\beta E_2}\frac{1-e^{-\beta(E_3+i\omega_3)}}{E_3+i\omega_3}-\frac{1-e^{-\beta(E_3-E_2+i\omega_3-i\omega_2)}}{E_3-E_2+i\omega_3-i\omega_2}(1-\delta_{E_2E_3}\delta_{\omega_2\omega_3})   -\beta\delta_{E_2E_3}\delta_{\omega_2\omega_3}\bigg]  \bigg\}\nonumber\\
  =&  \frac{1}{\beta(E_1+i\omega_1)}\bigg\{\frac{1-\delta_{E_1E_2}\delta_{\omega_1\omega_2}}{E_2-E_1+i\omega_2-i\omega_1}\bigg[e^{\beta(E_2-E_1)}\frac{1+e^{-\beta E_3}}{E_3+i\omega_3}-\frac{1+e^{-\beta(E_3-E_2+E_1)}}{E_3-E_2+E_1+i\omega_3-i\omega_2+i\omega_1}  \bigg]  \nonumber\\
  &+\delta_{E_1E_2}\delta_{\omega_1\omega_2}\frac{-1-e^{-\beta E_3}+\beta(E_3+i\omega_3)}{(E_3+i\omega_3)^2}  \nonumber\\
  &+\frac{e^{-\beta E_1}}{E_2+i\omega_2}\bigg[-e^{\beta E_2}\frac{1+e^{-\beta E_3}}{E_3+i\omega_3}-\frac{1-e^{-\beta(E_3-E_2)}}{E_3-E_2+i\omega_3-i\omega_2}(1-\delta_{E_2E_3}\delta_{\omega_2\omega_3})   -\beta\delta_{E_2E_3}\delta_{\omega_2\omega_3}\bigg]  \bigg\}   \nonumber
\end{align}

After some algebra, this expression can be rewritten as
\begin{align}
   & I(\omega_1,\omega_2,\omega_3,E_1,E_2,E_3)\nonumber\\
   =&\frac{1}{\beta(E_1+i\omega_1)}\bigg\{\frac{1-\delta_{E_1E_2}\delta_{\omega_1\omega_2}}{E_2-E_1+i\omega_2-i\omega_1}\bigg[\frac{e^{-\beta(E_1-E_2)}(1+e^{-\beta E_3})}{{E_3}+i\omega_3}-\frac{1+e^{-\beta(E_3-E_2+E_1)}}{E_3-E_2+E_1+i\omega_3-i\omega_2+i\omega_1}  \bigg]\nonumber\\
   & -\delta_{E_1E_2}\delta_{\omega_1\omega_2}\frac{1+e^{-\beta E_3}}{(E_3+i\omega_3)^2}-\frac{e^{-\beta(E_1-E_2)}(1+e^{-\beta E_3})}{(E_2+i\omega_2)(E_3+i\omega_3)}-\frac{e^{-\beta E_1}[1-e^{-\beta(E_3-E_2)}]}{(E_2+i\omega_2)(E_3-E_2+i\omega_3-i\omega_2)}(1-\delta_{E_2E_3}\delta_{\omega_2\omega_3})   \bigg\}\nonumber\\
   &+ \frac{\delta_{E_1E_2}\delta_{\omega_1\omega_2}}{(E_1+i\omega_1)(E_3+i\omega_3)} -   
\frac{e^{-\beta E_1} \delta_{E_2E_3}\delta_{\omega_2\omega_3} }{(E_1+i\omega_1)(E_2+i\omega_2)}\label{eq:integral_full_expression}
\end{align}

\subsubsection{Evaluation of $\langle T_\tau 
f_{\RR,i}^\dag(\tau_1) f_{\RR,i}(\tau_2)f_{\RR,i}^\dag(\tau_3) f_{\RR,i}(0)  \rangle_{\gamma=0}$ for $i=j$}
We now evaluate \cref{eq:FT_of_four-point} for the case of $i=j$ 
\ba 
\langle T_\tau 
f_{\RR,i}^\dag(\tau_1) f_{\RR,i}(\tau_2)f_{\RR,i}^\dag(\tau_3) f_{\RR,i}(0)  \rangle_{\gamma=0}
\ea 
where we take $\tau_1,\tau_2,\tau_3 \in (0,\beta]$. 
A nonzero contribution arises only when, after time ordering, the two creation operators are not adjacent. Equivalently, the two annihilation operators must also not be adjacent.
Therefore, nonzero contributions arise only in the following two time-ordering sectors:
\begin{equation}
    \tau_1>\tau_2>\tau_3,\;\tau_3>\tau_2>\tau_1
\end{equation}
\begin{itemize}
    \item For $\tau_1>\tau_2>\tau_3$, using \cref{eq:path_to_eigenstates} we have
 \begin{align}
    &\langle f_{\RR,i}^\dag(\tau_1) f_{\RR,i}(\tau_2) f_{\RR,i}^\dag(\tau_3)f_{\RR,i}(0)\rangle\nonumber\\
    =& \frac{1}{Z_{atom,\RR}}
    \sum_{n,m}  \langle n|f_{\RR,i}^\dag |m\rangle \langle m| f_{\RR,i}|n\rangle \langle n | f_{\RR,i}^\dag|m\rangle \langle m| f_{\RR,i}|n\rangle e^{-(\beta-\tau_1)E_n - (\tau_1-\tau_2)E_{m} -(\tau_2-\tau_3)E_n - \tau_3 E_m} \nonumber\\ 
    =& \frac{1}{Z_{atom,\RR}}
    \sum_{n,m}  \langle n|f_{\RR,i}^\dag |m\rangle \langle m| f_{\RR,i}|n\rangle \langle n| f_{\RR,i}^\dag|m\rangle \langle m| f_{\RR,i}|n\rangle e^{-\beta E_n + \tau_1(E_n-E_m) +\tau_2(E_m-E_n) + \tau_3(E_n-E_m)}
\end{align}
Here, $|n\rangle$ denotes the eigenstate of the atomic Hamiltonian $H_{atom,\RR}$ (\cref{eq:atomic_single_site}) with energy $E_n$. 
Moreover, since the flavor occupation operator commutes with the atomic Hamiltonian, $[f^{\dagger}_{\RR,i}f_{\RR,i},H_{atom,\RR}]=0$, we chose the energy eigenstates $\ket{n}$ to be simultaneous eigenstates of the flavor number operators $f^{\dagger}_{\RR,i}f_{\RR,i}$.
We next evaluate the product of matrix elements $\langle n|f_{\RR,i}^\dag |m\rangle \langle {m}| f_{\RR,i}|n\rangle \langle n| f_{\RR,i}^\dag|m\rangle \langle m| f_{\RR,i}|n\rangle $.
This product is nonzero only when $\ket{m}=e^{i\phi}f_{\RR,i}\ket{n}$.
Since the overall phase is irrelevant, we choose $\phi=0$ for simplicity.
Then the matrix element becomes
\begin{equation}
    \langle n|f_{\RR,i}^\dag |m\rangle \langle {m}| f_{\RR,i}|n\rangle \langle n| f_{\RR,i}^\dag|m\rangle \langle m| f_{\RR,i}|n\rangle =(\bra{n}f^{\dagger}_{\RR,i}f_{\RR,i}\ket{n})^4=\bra{n}f^{\dagger}_{\RR,i}f_{\RR,i}\ket{n}
\end{equation}
After Fourier transforming (\cref{eq:FT_of_four-point}), we obtain 
\begin{align}
    & \frac{1}{Z_{atom,\RR}}
    \sum_{n,m}\langle n|f_{\RR,i}^\dag |m\rangle \langle m| f_{\RR,i}|n\rangle \langle n| f_{\RR,i}^\dag|m\rangle \langle m| f_{\RR,i}|n\rangle \nonumber\\
    &\times\frac{1}{\beta}\int_0^{\beta}d\tau_3 \int_{\tau_3}^{\beta} d\tau_2\int_{\tau_2}^{\beta} d\tau_1\;e^{-\beta E_n + \tau_1(E_n-E_m) +\tau_2(E_m-E_n) + \tau_3(E_n-E_m)} e^{-i\omega \tau_3 + i\omega'\tau_2 - i\omega'\tau_1}\nonumber\\
    =&\frac{1}{Z_{atom,\RR}}
    \sum_{n,m}\langle n|f_{\RR,i}^\dag |m\rangle \langle m| f_{\RR,i}|n\rangle \langle n| f_{\RR,i}^\dag|m\rangle \langle m| f_{\RR,i}|n\rangle e^{-\beta E_n}I(\omega',\omega',\omega,E_m-E_n,E_m-E_n,E_m-E_n)\nonumber\\
    =&\sum_{\substack{n\\\ket{m}=f_{\RR,i}\ket{n}}}\bra{n}f^{\dagger}_{\RR,i}f_{\RR,i}\ket{n}\frac{e^{-\beta E_n}}{Z_{atom,\RR}}I(\omega',\omega',\omega,E_m-E_n,E_m-E_n,E_m-E_n)
    \label{eq:ii_1}
\end{align}

\item For $\tau_3>\tau_2>\tau_1$, we similarly obtain 
\begin{align}
    &\langle T_\tau f_{\RR,i}^\dag(\tau_1) f_{\RR,i}(\tau_2) f_{\RR,i}^\dag(\tau_3)f_{\RR,i}(0)\rangle  
\nonumber\\ 
=&-\langle   f_{\RR,i}^\dag(\tau_3)  f_{\RR,i}(\tau_2) f_{\RR,i}^\dag(\tau_1)f_{\RR,i}(0)\rangle   \nonumber\\
=& -
 \frac{1}{Z_{atom,\RR}}
    \sum_{n,m}  \langle n|f_{\RR,i}^\dag |m \rangle \langle m| f_{\RR,i}|n  \rangle \langle n |f_{\RR,i}^\dag |m\rangle \langle m| f_{\RR,i}|n\rangle e^{-(\beta-\tau_3)E_n - (\tau_3-\tau_2)E_{m} -(\tau_2-\tau_1)E_n - \tau_1 E_m} 
\end{align}
Again taking $\ket{m}=f_{\RR,i}\ket{n}$, the product of matrix elements becomes
\begin{align}
    \langle n|f_{\RR,i}^\dag |m \rangle \langle m| f_{\RR,i}|n  \rangle \langle n |f_{\RR,i}^\dag |m\rangle \langle m| f_{\RR,i}|n\rangle=(\bra{n}f^{\dagger}_{\RR,i}f_{\RR,i}\ket{n})^4=\bra{n}f^{\dagger}_{\RR,i}f_{\RR,i}\ket{n}
\end{align}
Thus, the Fourier transformation is
\begin{align}
  & -
 \frac{1}{Z_{atom,\RR}}
    \sum_{n,m}  \langle n|f_{\RR,i}^\dag |m \rangle \langle m| f_{\RR,i}|n  \rangle \langle n |f_{\RR,i}^\dag |m\rangle \langle m| f_{\RR,i}|n\rangle \nonumber\\
    &\times\frac{1}{\beta}\int_0^{\beta}d\tau_1\int_{\tau_1}^{\beta}d\tau_2\int_{\tau_2}^{\beta}d\tau_3\; e^{-\beta E_n-\tau_1(E_m-E_n)+\tau_2(E_m-E_n)-\tau_3(E_m-E_n)-i\omega'\tau_1+i\omega'\tau_2-i\omega\tau_3 }\nonumber\\
    =& -
    \frac{1}{Z_{atom,\RR}}\sum_{n,m}  \langle n|f_{\RR,i}^\dag |m \rangle \langle m| f_{\RR,i}|n  \rangle \langle n |f_{\RR,i}^\dag |m\rangle \langle m| f_{\RR,i}|n\rangle e^{-\beta E_n} I(\omega,\omega',\omega',E_m-E_n,E_m-E_n,E_m-E_n)\nonumber\\
    =&-\sum_{\substack{n\\\ket{m}=f_{\RR,i}\ket{n}}}\bra{n}f^{\dagger}_{\RR,i}f_{\RR,i}\ket{n}\frac{e^{-\beta E_n}}{Z_{atom,\RR}}I(\omega,\omega',\omega',E_m-E_n,E_m-E_n,E_m-E_n)
    \label{eq:ii_2}
\end{align}
\end{itemize}

Summing \cref{eq:ii_1,eq:ii_2} and using \cref{eq:ii_integral_sum}, we obtain
\begin{align}
    &\frac{1}{\beta}\int_0^\beta d\tau_3 \int_{\tau_3}^\beta d\tau_2 \int_{\tau_2}^\beta d\tau_1 
\langle f_{\RR,i}^\dag(\tau_1) f_{\RR,i}(\tau_2) f_{\RR,i}^\dag(\tau_3)f_{\RR,i}(0)\rangle e^{-i\omega \tau_3 + i\omega'\tau_2 - i\omega'\tau_1}\nonumber\\ 
=&\sum_{\substack{n\\
\ket{m}=f_{\RR,i}\ket{n}
}}
\bra{n}f^{\dagger}_{\RR,i}f_{\RR,i}\ket{n}\frac{e^{-\beta E_n}}{Z_{atom,\RR}}[I(\omega',\omega',\omega,E,E,E)-I(\omega,\omega',\omega',E,E,E)]\bigg|_{E=E_m-E_n}\nonumber\\
=&\sum_{\substack{n\\
\ket{m}=f_{\RR,i}\ket{n}
}}
\bra{n}f^{\dagger}_{\RR,i}f_{\RR,i}\ket{n}\frac{e^{-\beta E_n}}{Z_{atom,\RR}}\frac{(1- \delta_{\omega\omega '}) [1+ e^{-\beta (E_m- E_n)} ]}{(E_m- E_n+i\omega')(E_m - E_n+i\omega)}
\label{eq:Fii_prove_start}
\end{align}
In the zero-temperature limit, and using \cref{eq:partition_single_site}, the previous equation becomes
\begin{align}
    &\lim_{\beta\to+\infty}\sum_{\substack{n\\
\ket{m}=f_{\RR,i}\ket{n}
}}
\bra{n}f^{\dagger}_{\RR,i}f_{\RR,i}\ket{n}\frac{e^{-\beta E_n}}{Z_{atom,\RR}}\frac{(1- \delta_{\omega\omega '}) [1+ e^{-\beta (E_m- E_n)} ]}{(E_m- E_n+i\omega')(E_m - E_n+i\omega)}\nonumber\\
=&\binom{N_f}{n_f}^{-1}\lim_{\beta\to+\infty}\sum_{\substack{n\\
\ket{m}=f_{\RR,i}\ket{n}
}}
\bra{n}f^{\dagger}_{\RR,i}f_{\RR,i}\ket{n}e^{-\beta(E_{n}-E_{gs})}\frac{(1- \delta_{\omega\omega '}) [1+ e^{-\beta (E_m- E_n)} ]}{(E_m- E_n+i\omega')(E_m - E_n+i\omega)}\nonumber\\
 =&   \binom{N_f}{n_f}^{-1}\lim_{\beta\to+\infty}\sum_{\substack{n\\\ket{m}=f_{\RR,i}\ket{n}}}\bra{n}f^{\dagger}_{\RR,i}f_{\RR,i}\ket{n} \frac{(1- \delta_{\omega\omega '}) [e^{-\beta (E_n - E_{gs})}+ e^{-\beta (E_m- E_{gs})} ]}{(E_m- E_n+i\omega')(E_m - E_n+i\omega)}
    \nonumber \\ 
    =& {\binom{N_f}{n_f}}^{-1}\sum_{\substack{n \in GS \\\ket{m}=f_{\RR,i}\ket{n}}}\bra{n}f^{\dagger}_{\RR,i}f_{\RR,i}\ket{n} \frac{(1- \delta_{\omega\omega '}) }{(E_m- E_n+i\omega')(E_m - E_n+i\omega)}  \nonumber \\ 
    & + {\binom{N_f}{n_f}}^{-1}\sum_{\substack{n  \\\ket{m}=f_{\RR,i}\ket{n}, \;\; m\in GS}}\bra{n}f^{\dagger}_{\RR,i}f_{\RR,i}\ket{n} \frac{(1- \delta_{\omega\omega '}) }{(E_m- E_n+i\omega')(E_m - E_n+i\omega)}  \nonumber \\ 
    =& {\binom{N_f}{n_f}}^{-1}\sum_{\substack{n \in GS \\\ket{m}=f_{\RR,i}\ket{n}}}\bra{n}f^{\dagger}_{\RR,i}f_{\RR,i}\ket{n} \frac{(1- \delta_{\omega\omega '}) }{(\Delta_{-1}+i\omega')(\Delta_{-1}+i\omega)}  \nonumber \\ 
    & +  {\binom{N_f}{n_f}}^{-1}\sum_{\substack{n  \\\ket{m}=f_{\RR,i}\ket{n}, \;\; m\in GS}}\bra{n}f^{\dagger}_{\RR,i}f_{\RR,i}\ket{n} \frac{(1- \delta_{\omega\omega '}) }{(-\Delta_{+1}+i\omega')(-\Delta_{+1}+i\omega)}  
\end{align}
The first sum, $ \sum_{\substack{n \in GS \\\ket{m}=f_{\RR,i}\ket{n}}}$, runs over all ground states $\ket{n}$ in which flavor $i$ is occupied; the number of such ground states is $\binom{N_f-1}{n_f-1}$. 
The second sum, $\sum_{\substack{n  \\\ket{m}=f_{\RR,i}\ket{n}, \;\; m\in GS}}$, runs over all ground states $\ket{m}$ in which flavor $i$ is unoccupied; the number of such states is $\binom{N_f-1}{n_f}$. 
Putting everything together, we find that in the zero-temperature limit the Fourier transform of the four-point correlation function for $i=j$ is
\boxedeq{
\begin{align}
       I_{ii}(\omega,\omega',\omega')\equiv &\frac{1}{\beta}\int_0^\beta d\tau_3 \int_{\tau_3}^\beta d\tau_2 \int_{\tau_2}^\beta d\tau_1 
\langle f_{\RR,i}^\dag(\tau_1) f_{\RR,i}(\tau_2) f_{\RR,i}^\dag(\tau_3)f_{\RR,i}(0)\rangle e^{-i\omega \tau_3 + i\omega'\tau_2 - i\omega'\tau_1}\nonumber\\ 
=&\sum_{\substack{n\\
\ket{m}=f_{\RR,i}\ket{n}
}}
\bra{n}f^{\dagger}_{\RR,i}f_{\RR,i}\ket{n}\frac{e^{-\beta E_n}}{Z_{atom,\RR}}[I(\omega',\omega',\omega,E,E,E)-I(\omega,\omega',\omega',E,E,E)]\bigg|_{E=E_m-E_n}\nonumber\\
   =&  \frac{n_f}{N_f} \frac{1- \delta_{\omega\omega '}}{(i\omega'+\Delta_{-1})(i\omega+ \Delta_{-1})}   +  \frac{N_f- n_f}{N_f}  \frac{1- \delta_{\omega\omega '}}{(i\omega' -\Delta_{+1})(i\omega-\Delta_{+1})} 
   \label{eq:ffff_ii_freq}
\end{align}
}

Finally, combining \cref{eq:atomic_Green_fun,eq:F_nm_from_ffff,eq:ffff_ii_freq}, we obtain 
\ba 
\label{eq:F_ii_proof}
&F^{ii}(i\omega,i\omega',i\omega')  \nonumber\\ 
= &-\frac{1}{\beta}\int_{\tau_1,\tau_2,\tau_3} \langle T_\tau 
f_{\RR,i}^\dag(\tau_1) f_{\RR,i}(\tau_2)f_{\RR,i}^\dag(\tau_3) f_{\RR,i}(0)  \rangle_{\gamma=0} e^{-i\omega\tau_3+i\omega'\tau_2-i\omega'\tau_1 }   +G_{f,loc}(i\omega)G_{f, loc}(i\omega') -{\delta_{i\omega,i\omega'}G_{f,loc}(i\omega)G_{f,loc}(i\omega')  }\nonumber\\ 
=&(-1+\delta_{\omega,\omega'})\bigg[\frac{n_f}{N_f}\frac{1}{(i\omega+\Delta_{-1})(i\omega'+\Delta_{-1})}+\frac{N_f-n_f}{N_f}\frac{1}{(i\omega-\Delta_{+1})(i\omega'-\Delta_{+1})}\nonumber\\
&-\frac{1}{N_f^2}\bigg(\frac{n_f}{i\omega+\Delta_{-1}}+\frac{N_f-n_f}{i\omega-\Delta_{+1}}\bigg)\bigg(\frac{n_f}{i\omega'+\Delta_{-1}}+\frac{N_f-n_f}{i\omega'-\Delta_{+1}}\bigg)\bigg]\nonumber\\
=&\frac{1}{N_f^2}\frac{(-1+\delta_{\omega,\omega'})}{(i\omega+\Delta_{-1})(i\omega-\Delta_{+1})(i\omega'+\Delta_{+1})(i\omega'-\Delta_{-1})}\bigg\{n_fN_f(i\omega-\Delta_{+1})(i\omega'-\Delta_{+1})+(N_f-n_f)N_f(i\omega+\Delta_{-1})(i\omega'+\Delta_{-1})\nonumber\\
&-[n_f(i\omega-\Delta_{+1})+(N_f-n_f)(i\omega+\Delta_{-1})][n_f(i\omega'-\Delta_{+1})+(N_f-n_f)(i\omega'+\Delta_{-1})]\bigg\}\nonumber\\
=&\frac{1}{N_f^2}\frac{(-1+\delta_{\omega,\omega'})}{(i\omega+\Delta_{-1})(i\omega-\Delta_{+1})(i\omega'+\Delta_{+1})(i\omega'-\Delta_{-1})}\bigg[n_f(N_f-n_f)(i\omega-\Delta_{+1})(i\omega'-\Delta_{+1})\nonumber\\
&+(N_f-n_f)n_f(i\omega+\Delta_{-1})(i\omega'+\Delta_{-1})-n_f(N_f-n_f)(i\omega-\Delta_{+1})(i\omega'+\Delta_{-1})-(N_f-n_f)n_f(i\omega+\Delta_{-1})(i\omega'-\Delta_{+1})\bigg]\nonumber\\
=&\frac{n_f(N_f-n_f)}{N_f^2}\frac{(-1+\delta_{\omega,\omega'})}{(i\omega+\Delta_{-1})(i\omega-\Delta_{+1})(i\omega'+\Delta_{-1})(i\omega'-\Delta_{+1})}\nonumber\\
&\times\bigg[  (i\omega-\Delta_{+1})(i\omega'-\Delta_{+1})+(i\omega+\Delta_{-1})(i\omega'+\Delta_{-1})-(i\omega-\Delta_{+1})(i\omega'+\Delta_{-1})-(i\omega+\Delta_{-1})(i\omega'-\Delta_{+1})\bigg]\nonumber\\
=& 
{\frac{n_f(N_f-n_f)}{N_f^2}}
\frac{(\Delta_{-1}+\Delta_{+1})^2 }{(i\omega + \Delta_{-1})(i\omega -\Delta_{+1})(i\omega' - \Delta_{+1})(i\omega'+\Delta_{-1})}(-1+\delta_{\omega ,\omega'})
\ea 
In summary, we obtain
\boxedeq{
\begin{align}
    F^{ii}(i\omega,i\omega',i\omega')=&-I_{ii}(\omega,\omega',\omega')+G_{f,loc}(i\omega)G_{f, loc}(i\omega') -{\delta_{i\omega,i\omega'}G_{f,loc}(i\omega)G_{f,loc}(i\omega')  }\nonumber\\
    =&{\frac{n_f(N_f-n_f)}{N_f^2}}
\frac{(\Delta_{-1}+\Delta_{+1})^2 }{(i\omega + \Delta_{-1})(i\omega -\Delta_{+1})(i\omega' - \Delta_{+1})(i\omega'+\Delta_{-1})}(-1+\delta_{\omega ,\omega'})
\label{eq:F_ii_final}
\end{align}
}

\subsubsection{Evaluation of $\langle T_\tau 
f_{\RR,j}^\dag(\tau_1) f_{\RR,j}(\tau_2)f_{\RR,i}^\dag(\tau_3) f_{\RR,i}(0)  \rangle_{\gamma=0}$ for $i\ne j$}

We next consider the case $i\ne j$:
\ba 
\langle T_\tau 
f_{\RR,j}^\dag(\tau_1) f_{\RR,j}(\tau_2)f_{\RR,i}^\dag(\tau_3) f_{\RR,i}(0)  \rangle_{\gamma=0}
\ea 

There are six possible time-ordering sectors:
\begin{align}
    &\tau_1>\tau_2 > \tau_3 \nonumber\\ 
    &\tau_1>\tau_3>\tau_2 \nonumber\\ 
    &\tau_2 > \tau_1 > \tau_3 \nonumber\\ 
    & \tau_2 > \tau_3 > \tau_1 \nonumber\\
    &\tau_3>\tau_1>\tau_2 \nonumber\\ 
    &\tau_3>\tau_2>\tau_1 
\end{align} 

\begin{itemize}
    \item 
For $\tau_1 > \tau_2 > \tau_3$, we find 
\begin{align}
    &\langle f_{\RR,j}^\dag(\tau_1) f_{\RR,j}(\tau_2) f_{\RR,i}^\dag(\tau_3)f_{\RR,i}(0)\rangle\nonumber\\
    =& \frac{1}{Z_{atom,\RR}}
    \sum_{n,s,m}  \langle n|f_{\RR,j}^\dag |m\rangle \langle m| f_{\RR,j}|n\rangle \langle n | f_{\RR,i}^\dag|s\rangle \langle s| f_{\RR,i}|n\rangle e^{-(\beta-\tau_1)E_n - (\tau_1-\tau_2)E_{m} -(\tau_2-\tau_3)E_n - \tau_3 E_s} \nonumber\\ 
    =& \frac{1}{Z_{atom,\RR}}
    \sum_{n,s,m}  \langle n|f_{\RR,j}^\dag |m\rangle \langle m| f_{\RR,j}|n\rangle \langle n| f_{\RR,i}^\dag|s\rangle \langle s| f_{\RR,i}|n\rangle e^{-\beta E_n + \tau_1(E_n-E_m) +\tau_2(E_m-E_n) + \tau_3(E_n-E_s)} 
\end{align}
Denote $\ket{s}=f_{\RR,i}\ket{n}$, $\ket{m}=f_{\RR,j}\ket{n}$, the matrix element is evaluated as 
\begin{equation}
    \langle n|f_{\RR,j}^\dag |m\rangle \langle m| f_{\RR,j}|n\rangle \langle n| f_{\RR,i}^\dag|s\rangle \langle s| f_{\RR,i}|n\rangle=(\bra{n}f^{\dagger}_{\RR,j}f_{\RR,j}\ket{n})^2(\bra{n}f^{\dagger}_{\RR,i}f_{\RR,i}\ket{n})^2=\bra{n}f^{\dagger}_{\RR,j}f_{\RR,j}\ket{n}\bra{n}f^{\dagger}_{\RR,i}f_{\RR,i}\ket{n}
\end{equation}
After Fourier transforming, we have 
\begin{align}
  & \frac{1}{Z_{atom,\RR}}
    \sum_{n,s,m}\langle n|f_{\RR,j}^\dag |m\rangle \langle m| f_{\RR,j}|n\rangle \langle n| f_{\RR,i}^\dag|s\rangle \langle s| f_{\RR,i}|n\rangle \nonumber\\
    &\times\frac{1}{\beta}\int_0^{\beta}d\tau_3 \int_{\tau_3}^{\beta} d\tau_2\int_{\tau_2}^{\beta} d\tau_1\;e^{-\beta E_n + \tau_1(E_n-E_m) +\tau_2(E_m-E_n) + \tau_3(E_n-E_s)} e^{-i\omega \tau_3 + i\omega'\tau_2 - i\omega'\tau_1}\nonumber\\
    =&\frac{1}{Z_{atom,\RR}}
    \sum_{n,s,m}\langle n|f_{\RR,j}^\dag |m\rangle \langle m| f_{\RR,j}|n\rangle \langle n| f_{\RR,i}^\dag|s\rangle \langle s| f_{\RR,i}|n\rangle e^{-\beta E_n}I(\omega',\omega',\omega,E_m-E_n,E_m-E_n,E_s-E_n)\nonumber\\
    =&\sum_{\substack{n\\\ket{s}=f_{\RR,i}\ket{n},\ket{m}=f_{\RR,j}\ket{n} }}\bra{n}f^{\dagger}_{\RR,j}f_{\RR,j}\ket{n}\bra{n}f^{\dagger}_{\RR,i}f_{\RR,i}\ket{n}\frac{e^{-\beta E_n}}{Z_{atom,\RR}}I(\omega',\omega',\omega,E_m-E_n,E_m-E_n,E_s-E_n) \nonumber \\ 
    =&
    {
    \sum_{\substack{n\\\ket{s}=f_{\RR,i}\ket{n},\ket{m}=f_{\RR,j}\ket{n} }}\bra{n}f^{\dagger}_{\RR,j}f_{\RR,j}\ket{n}\bra{n}f^{\dagger}_{\RR,i}f_{\RR,i}\ket{n}\frac{e^{-\beta E_n}}{Z_{atom,\RR}}I(\omega',\omega',\omega,E_m-E_n,E_m-E_n,E_m-E_n)
    }
\end{align} 
In the last line we used $E_s = E_m$ as the two states contain the same number of $f$ electrons.
{
Using \cref{eq:integral_full_expression_same_energy}, the integral evaluates to
\begin{align}
    & I(\omega',\omega',\omega,E= E_m-E_n,E_m-E_n,E_m-E_n)  \nonumber \\ 
   =& \frac{1}{(E+i\omega')(E+i\omega)}
   \bigg[-\frac{1+e^{-\beta E}}{\beta} 
   \bigg(
   \frac{1}{E+i\omega}+ \frac{1}{E+i\omega'}
   \bigg)  
   + 1 - e^{-\beta E} \delta_{\omega \omega' } 
   \bigg]
\end{align}
Taking the zero-temperature limit $\beta\to\infty$, we obtain
\begin{align}
   & e^{-\beta(E_n- E_{gs})}I(\omega',\omega',\omega,E= E_m-E_n,E_m-E_n,E_m-E_n)\bigg|_{\beta \rightarrow \infty}  \nonumber \\ 
   =& \frac{1}{(E+i\omega')(E+i\omega)}
   \bigg[-\frac{\delta_{n,gs}+\delta_{m,gs}}{\beta} \bigg(
   \frac{1}{E+i\omega}+ \frac{1}{E+i\omega'}
   \bigg)  + \delta_{n,gs} -\delta_{m,gs} \delta_{\omega \omega' } 
   \bigg] 
\end{align}
and the Fourier transform becomes
\begin{align}
& {\binom{N_f}{n_f}}^{-1} \sum_{\substack{n\\\ket{s}=f_{\RR,i}\ket{n},\ket{m}=f_{\RR,j}\ket{n} }}\bra{n}f^{\dagger}_{\RR,j}f_{\RR,j}\ket{n}\bra{n}f^{\dagger}_{\RR,i}f_{\RR,i}\ket{n} e^{- \beta (E_n - E_{gs}) } I(\omega',\omega',\omega,E_m-E_n,E_m-E_n,E_m-E_n)\bigg|_{\beta\to \infty} \nonumber \\ 
 =& {\binom{N_f}{n_f}}^{-1} 
 \bigg\{ \sum_{n\in GS}
 \frac{\bra{n}f^{\dagger}_{\RR,j}f_{\RR,j}\ket{n}\bra{n}f^{\dagger}_{\RR,i}f_{\RR,i}\ket{n}}{(\Delta_{-1}+i\omega')(\Delta_{-1}+i\omega)}
 \bigg[-\frac{1}{\beta} 
 \bigg(
 \frac{1}{\Delta_{-1}+i\omega}+ \frac{1}{\Delta_{-1}+i\omega'}
 \bigg)  + 1 
 \bigg]  \nonumber \\ 
 &+ \sum_{\substack{n\\m\in GS}}\frac{\bra{n}f^{\dagger}_{\RR,j}f_{\RR,j}\ket{n}\bra{n}f^{\dagger}_{\RR,i}f_{\RR,i}\ket{n}}{(i\omega'- \Delta_{+1})(i\omega- \Delta_{+1})}
 \bigg[
 -\frac{1}{\beta} \bigg(\frac{1}{i\omega- \Delta_{+1}}+ \frac{1}{i\omega'- \Delta_{+1}}\bigg)   - \delta_{\omega \omega' } 
 \bigg]
 \bigg\} 
 \nonumber \\ 
 =&  \frac{n_f(n_f-1)}{N_f(N_f-1) (\Delta_{-1}+i\omega')(\Delta_{-1}+i\omega)}
 \bigg[-\frac{1}{\beta} \bigg(\frac{1}{\Delta_{-1}+i\omega}+ \frac{1}{\Delta_{-1}+i\omega'}\bigg)  + 1 
 \bigg]  \nonumber \\ 
 &+ \frac{n_f(N_f- n_f)}{N_f(N_f-1) (i\omega'- \Delta_{+1})(i\omega- \Delta_{+1})}
 \bigg[-\frac{1}{\beta} 
 \bigg(\frac{1}{i\omega- \Delta_{+1}}+ \frac{1}{i\omega'- \Delta_{+1}}
 \bigg)   - \delta_{\omega \omega' } \bigg]
 \label{eq:Fij_no1}
\end{align}
In previous equation, $\sum_{n\in GS}$ counts the ground states in which both flavors $i$ and $j$ are occupied, while $\sum_{\substack{n\\m\in GS}}$ counts the corresponding charge $+1$ excitation states. Their multiplicities are $\binom{N_f-2}{n_f-2}$ and $\binom{N_f-2}{n_f-1}$, respectively.

}

\item 
For $\tau_1>\tau_3>\tau_2$, we find 

\ba 
 &\langle T_\tau f_{\RR,j}^\dag(\tau_1) f_{\RR,j}(\tau_2) f_{\RR,i}^\dag(\tau_3)f_{\RR,i}(0)\rangle \nonumber\\
 =&- \langle  f_{\RR,j}^\dag(\tau_1)f_{\RR,i}^\dag(\tau_3) f_{\RR,j}(\tau_2) f_{\RR,i}(0)\rangle 
 \nonumber\\ 
 =& (-1)\frac{1}{Z_{atom,\RR}}
    \sum_{n,s,m,u}  \langle n|f_{\RR,j}^\dag |m \rangle \langle m| f_{\RR,i}^\dag|u \rangle \langle u | f_{\RR,j}|s\rangle \langle s| f_{\RR,i}|n\rangle e^{-(\beta-\tau_1)E_n - (\tau_1-\tau_3)E_{m} -(\tau_3-\tau_2)E_u - \tau_2 E_s}
\ea 
Denote $\ket{s}=f_{\RR,i}\ket{n}$, $\ket{u}=f_{\RR,j}\ket{s}=f_{\RR,j}f_{\RR,i}\ket{n}$ and $\ket{m}=f_{\RR,j}\ket{n}$.
For $i\ne j$, the matrix element is evaluated as
\begin{align}
    &\langle n|f_{\RR,j}^\dag |m \rangle \langle m| f_{\RR,i}^\dag|u \rangle \langle u | f_{\RR,j}|s\rangle \langle s| f_{\RR,i}|n\rangle\nonumber\\
    =&\bra{n}f^{\dagger}_{\RR,j}f_{\RR,j}\ket{n}\bra{n}f^{\dagger}_{\RR,j}f^{\dagger}_{\RR,i}f_{\RR,j}f_{\RR,i}\ket{n}\bra{n}f^{\dagger}_{\RR,i}f^{\dagger}_{\RR,j}f_{\RR,j}f_{\RR,i}\ket{n}\bra{n}f^{\dagger}_{\RR,i}f_{\RR,i}\ket{n}\nonumber\\
    =&-\bra{n}f^{\dagger}_{\RR,j}f_{\RR,j}\ket{n}(\bra{n}f^{\dagger}_{\RR,j}f_{\RR,j}f^{\dagger}_{\RR,i}f_{\RR,i}\ket{n})^2\bra{n}f^{\dagger}_{\RR,i}f_{\RR,i}\ket{n}\nonumber\\
    =&-(\bra{n}f^{\dagger}_{\RR,j}f_{\RR,j}\ket{n}\bra{n}f^{\dagger}_{\RR,i}f_{\RR,i}\ket{n})^3\nonumber\\
    =&-\bra{n}f^{\dagger}_{\RR,j}f_{\RR,j}\ket{n}\bra{n}f^{\dagger}_{\RR,i}f_{\RR,i}\ket{n}
\end{align}
where we have used the fact $\ket{n}$ is also the eigenstate of the flavor number operators $f^{\dagger}_{\RR,i}f_{\RR,i}$ for all $i$, so that $\bra{n}f^{\dagger}_{\RR,j}f_{\RR,j}f^{\dagger}_{\RR,i}f_{\RR,i}\ket{n}=\bra{n}f^{\dagger}_{\RR,j}f_{\RR,j}\ket{n}\bra{n}f^{\dagger}_{\RR,i}f_{\RR,i}\ket{n}$.
The Fourier transform gives 
\begin{align}
  &(-1)\frac{1}{Z_{atom,\RR}}
    \sum_{n,s,m,u}  \langle n|f_{\RR,j}^\dag |m \rangle \langle m| f_{\RR,i}^\dag|u \rangle \langle u | f_{\RR,j}|s\rangle \langle s| f_{\RR,i}|n\rangle \nonumber\\
    &\times \frac{1}{\beta}\int_{0}^{\beta}d\tau_2\int_{\tau_2}^{\beta}d \tau_3\int_{\tau_3}^{\beta} d\tau_1 \; e^{-\beta E_n-\tau_2(E_s-E_u)+\tau_3(E_m-E_u)-\tau_1(E_m-E_n)+i\omega'\tau_2-i\omega\tau_3-i\omega'\tau_1} \nonumber\\
    =&(-1)\frac{1}{Z_{atom,\RR}}
    \sum_{n,s,m,u}  \langle n|f_{\RR,j}^\dag |m \rangle \langle m| f_{\RR,i}^\dag|u \rangle \langle u | f_{\RR,j}|s\rangle \langle s| f_{\RR,i}|n\rangle e^{-\beta E_n}I(\omega',-\omega,-\omega',E_m-E_n,E_m-E_u,E_s-E_u)\nonumber\\
    =&\sum_{\substack{n\\\ket{s}=f_{\RR,i}\ket{n},\ket{u}=f_{\RR,j}f_{\RR,i}\ket{n},\ket{m}=f_{\RR,j}\ket{n}}}\bra{n}f^{\dagger}_{\RR,j}f_{\RR,j}\ket{n}\bra{n}f^{\dagger}_{\RR,i}f_{\RR,i}\ket{n}
    \frac{e^{-\beta E_n}}{Z_{atom,\RR}}I(\omega',-\omega,-\omega',E_m-E_n,E_m-E_u,E_s-E_u)  \nonumber\\
    =&
    {\sum_{\substack{n\\\ket{s}=f_{\RR,i}\ket{n},\ket{u}=f_{\RR,j}f_{\RR,i}\ket{n},\ket{m}=f_{\RR,j}\ket{n}}}\bra{n}f^{\dagger}_{\RR,j}f_{\RR,j}\ket{n}\bra{n}f^{\dagger}_{\RR,i}f_{\RR,i}\ket{n}
    \frac{e^{-\beta E_n}}{Z_{atom,\RR}}I(\omega',-\omega,-\omega',E_m-E_n,E_m-E_u,E_m-E_u)}
\end{align}
In the last line, we used $E_s= E_m $ as the two states contain the same number of $f$ electrons.
{
Using \cref{eq:integral_full_expression}, the integral evaluates to 
\begin{align}
   & I(\omega',-\omega,-\omega',E= E_m-E_n,E'= E_m-E_u,E_m-E_u) \nonumber\\
   =&\frac{1}{\beta(E+i\omega')}\bigg\{\frac{1-\delta_{EE'}\delta_{\omega',- \omega}}{E'-E- i \omega-i\omega'}\bigg[\frac{e^{-\beta(E-E')}(1+e^{-\beta E'})}{E'- i \omega'}-\frac{1+e^{-\beta E}}{E+ i \omega}  \bigg] -\frac{1+e^{-\beta E'}}{(E'- i \omega') } \bigg[ \frac{\delta_{EE'}\delta_{\omega', - \omega} }{(E'- i \omega')}+\frac{e^{-\beta(E-E')}}{(E'- i \omega)} \bigg] \bigg\}\nonumber\\
   &+ \frac{\delta_{EE'}\delta_{\omega', - \omega} - e^{-\beta E} \delta_{ \omega \omega'} }{(E+i\omega')(E'- i \omega')}  
\end{align}
Taking the zero-temperature limit, we obtain
\begin{align}
   & e^{-\beta(E_n- E_{gs})} I(\omega',-\omega,-\omega',E= E_m-E_n,E'= E_m-E_u,E_m-E_u) \bigg|_{\beta \rightarrow \infty}\nonumber\\
   =&\frac{1}{\beta(E+i\omega')}\bigg\{\frac{1-\delta_{E_n E_u }\delta_{\omega',- \omega}}{E'-E- i \omega-i\omega'}\bigg[\frac{\delta_{u, gs}+\delta_{m,gs} }{E'- i \omega'}-\frac{\delta_{n, gs} +\delta_{m, gs}}{E+ i \omega}  \bigg] -\nonumber\\ & - \frac{\delta_{n, gs} +\delta_{m, gs} }{(E'- i \omega') } \frac{\delta_{E_nE_u }\delta_{\omega', - \omega} }{(E'- i \omega')}
  -   \frac{\delta_{u, gs} + \delta_{m, gs} }{(E'- i \omega')(E'- i \omega)}  \bigg\}+ \frac{\delta_{E_n E_u }\delta_{\omega', - \omega} \delta_{n,gs} - \delta_{m, gs} \delta_{ \omega \omega'} }{(E+i\omega')(E'- i \omega')}  
\nonumber \\ 
\end{align}
and the Fourier transform becomes
\begin{align}
 &  {\binom{N_f}{n_f}}^{-1}  \sum_{\substack{n\\\ket{s}=f_{\RR,i}\ket{n},\ket{u}=f_{\RR,j}f_{\RR,i}\ket{n},\ket{m}=f_{\RR,j}\ket{n}}}\bra{n}f^{\dagger}_{\RR,j}f_{\RR,j}\ket{n}\bra{n}f^{\dagger}_{\RR,i}f_{\RR,i}\ket{n}\nonumber\\
 &\qquad\qquad\qquad\qquad\times
 e^{- \beta (E_n - E_{gs}) } I(\omega',-\omega,-\omega',E_m-E_n,E_m-E_u,E_m-E_u)\bigg|_{\beta\to\infty} \nonumber\\
   =& {\binom{N_f}{n_f}}^{-1}  \sum_{\substack{n\\\ket{s}=f_{\RR,i}\ket{n},\ket{u}=f_{\RR,j}f_{\RR,i}\ket{n},\ket{m}=f_{\RR,j}\ket{n}}}\bra{n}f^{\dagger}_{\RR,j}f_{\RR,j}\ket{n}\bra{n}f^{\dagger}_{\RR,i}f_{\RR,i}\ket{n} \nonumber \\ 
   &\bigg[
   \frac{1}{\beta(i\omega'- \Delta_{+1})} \frac{1-\delta_{\Delta_{+1} \Delta_{-1} }\delta_{\omega',- \omega}}{\Delta_{+1}-\Delta_{-1}- i \omega-i\omega'}
   \bigg(
   \frac{\delta_{m,gs} }{- \Delta_{-1} - i \omega'}-\frac{\delta_{m, gs}}{-\Delta_{+1}+ i \omega}  \bigg)
   - \frac{1}{\beta(-\Delta_{+1}+i\omega')} \frac{\delta_{m, gs} \delta_{\Delta_{+1}\Delta_{-1} }\delta_{\omega', - \omega} }{(\Delta_{-1}+ i \omega')^2}
    \nonumber \\ 
   & -  \frac{1}{\beta(i\omega'-\Delta_{+1})} \frac{ \delta_{m, gs} }{(\Delta_{-1}+ i \omega')( \Delta_{-1}+ i \omega)}  +\frac{  \delta_{m, gs} \delta_{ \omega \omega'} }{(i\omega'-\Delta_{+1})(\Delta_{-1}+ i \omega')}  \nonumber\\ 
   &+  \frac{1}{\beta(\Delta_{+1} - \Delta_{+2} +i\omega')} \frac{1}{\Delta_{+2} - i \omega-i\omega'}\frac{\delta_{u, gs}}{\Delta_{+1} - i \omega'}  -  \frac{1}{\beta(\Delta_{+1} - \Delta_{+2} +i\omega')} \frac{\delta_{u, gs}  }{(\Delta_{+1}- i \omega')(\Delta_{+1} - i \omega)}   \nonumber\\ 
   &+ \frac{1}{\beta(\Delta_{-1}+i\omega')} \frac{1}{ \Delta_{-2} + i \omega+i\omega'}\frac{\delta_{n, gs} }{\Delta_{-1} + i \omega}
   \bigg]
 \nonumber \\      
 =& \frac{(N_f- n_f)n_f}{ N_f(N_f-1)}\frac{1}{(i\omega'- \Delta_{+1})(\Delta_{-1} + i \omega')}\bigg[ \frac{1}{ \beta ( i\omega- \Delta_{+1}) } -  \frac{ 1}{\beta(  i \omega + \Delta_{-1})}  +  \delta_{ \omega \omega'} \bigg]  \nonumber\\ 
 &+  \frac{(N_f- n_f)(N_f - n_f -1) }{N_f(N_f-1)} \frac{1}{\beta(\Delta_{+1} - \Delta_{+2} +i\omega')(\Delta_{+1}- i \omega')} \bigg[ \frac{1}{\Delta_{+2} - i \omega-i\omega'} -   \frac{1 }{(\Delta_{+1} - i \omega)}   \bigg] \nonumber\\ 
 &+ \frac{n_f(n_f-1)}{N_f(N_f-1)} \frac{1}{\beta(\Delta_{-1}+i\omega')} \frac{1}{ \Delta_{-2} + i \omega+i\omega'}\frac{1 }{\Delta_{-1} + i \omega}  
 \nonumber \\ 
 =&
  \frac{(N_f- n_f)n_f}{ N_f(N_f-1)}\frac{1}{(i\omega'- \Delta_{+1})(\Delta_{-1} + i \omega')}\bigg[ \frac{1}{ \beta ( i\omega- \Delta_{+1}) } -  \frac{ 1}{\beta(  i \omega + \Delta_{-1})}  +  \delta_{ \omega \omega'} \bigg]  \nonumber\\ 
  &+  \frac{(N_f- n_f)(N_f - n_f -1) }{N_f(N_f-1)} \frac{1}{\beta(\Delta_{+1}- i \omega') (\Delta_{+2} - i \omega-i\omega') (\Delta_{+1} - i \omega)}     \nonumber\\ 
  &+ \frac{n_f(n_f-1)}{N_f(N_f-1)} \frac{1}{\beta(\Delta_{-1}+i\omega')} \frac{1}{ \Delta_{-2} + i \omega+i\omega'}\frac{1 }{\Delta_{-1} + i \omega}  
  \label{eq:Fij_no2}
\end{align}
where $\Delta_{\pm2}=E(n_f\pm2)-E(n_f)$ are the energies of charge $\pm 2$ excitations respectively.
Besides, we also use that $\sum_{\substack{n\\u\in GS}}$ counts the charge $+2$ excitation states in which both $i,j$ flavors are occupied. And the multiplicity is $\binom{N_f-2}{n_f}$

}

\item 
For $\tau_2>\tau_1>\tau_3$, we have 
\ba 
 &\langle T_\tau f_{\RR,j}^\dag(\tau_1) f_{\RR,j}(\tau_2) f_{\RR,i}^\dag(\tau_3)f_{\RR,i}(0)\rangle \nonumber\\
 =&- \langle f_{\RR,j}(\tau_2)   f_{\RR,j}^\dag(\tau_1)f_{\RR,i}^\dag(\tau_3) f_{\RR,i}(0)\rangle  \nonumber\\ 
 =& 
  (-1)\frac{1}{Z_{atom,\RR}}
    \sum_{n,s,m}  \langle n|f_{\RR,j} |m \rangle \langle m| f_{\RR,j}^\dag|n  \rangle \langle n |f_{\RR,i}^\dag |s\rangle \langle s| f_{\RR,i}|n\rangle e^{-(\beta-\tau_2)E_n - (\tau_2-\tau_1)E_{m} -(\tau_1-\tau_3)E_n - \tau_3 E_s} 
\ea 
Denote $\ket{s}=f_{\RR,i}\ket{n}$ and $\ket{m}=f^{\dagger}_{\RR,j}\ket{n}$, then the matrix element is evaluated as
\begin{align}
    \langle n|f_{\RR,j} |m \rangle \langle m| f_{\RR,j}^\dag|n  \rangle \langle n |f_{\RR,i}^\dag |s\rangle \langle s| f_{\RR,i}|n\rangle=(\bra{n}f_{\RR,j}f^{\dagger}_{\RR,j}\ket{n})^2(\bra{n}f^{\dagger}_{\RR,i}f_{\RR,i}\ket{n})^2=
    \bra{n}f_{\RR,j}f^{\dagger}_{\RR,j}\ket{n}\bra{n}f^{\dagger}_{\RR,i}f_{\RR,i}\ket{n}
\end{align}
And the Fourier transform is
\begin{align}
  &(-1)\frac{1}{Z_{atom,\RR}}
  \sum_{n,s,m}  \langle n|f_{\RR,j} |m \rangle \langle m| f_{\RR,j}^\dag|n  \rangle \langle n |f_{\RR,i}^\dag |s\rangle \langle s| f_{\RR,i}|n\rangle\nonumber\\
  &\times\frac{1}{\beta}\int_0^{\beta}d\tau_3\int_{\tau_3}^{\beta}d\tau_1\int_{\tau_1}^{\beta}d\tau_2\; e^{-\beta E_n -\tau_3(E_s-E_n)+\tau_1(E_m-E_n)-\tau_2(E_m-E_n)-i\omega\tau_3+i\omega'\tau_2-i\omega'\tau_1 }\nonumber\\
  =&(-1)\frac{1}{Z_{atom,\RR}}
  \sum_{n,s,m}  \langle n|f_{\RR,j} |m \rangle \langle m| f_{\RR,j}^\dag|n  \rangle \langle n |f_{\RR,i}^\dag |s\rangle \langle s| f_{\RR,i}|n\rangle e^{-\beta E_n}I(-\omega',-\omega',\omega,E_m-E_n,E_m-E_n,E_s-E_n)\nonumber\\
  =&-\sum_{\substack{n\\
  \ket{s}=f_{\RR,i}\ket{n},\ket{m}=f^{\dagger}_{\RR,j}\ket{n}
  }}
  \bra{n}f_{\RR,j}f^{\dagger}_{\RR,j}\ket{n}\bra{n}f^{\dagger}_{\RR,i}f_{\RR,i}\ket{n}\frac{e^{-\beta E_n}}{Z_{atom,\RR}}
  I(-\omega',-\omega',\omega,E_m-E_n,E_m-E_n,E_s-E_n)\nonumber\\
  =& 
  {
  -\sum_{\substack{n\\
  \ket{s}=f_{\RR,i}\ket{n},\ket{m}=f^{\dagger}_{\RR,j}\ket{n}
  }}
  (1-\bra{n}f^{\dagger}_{\RR,j} f_{\RR,j} \ket{n}) \bra{n}f^{\dagger}_{\RR,i}f_{\RR,i}\ket{n}\frac{e^{-\beta E_n}}{Z_{atom,\RR}}
  I(-\omega',-\omega',\omega,E_m-E_n,E_m-E_n,E_s-E_n) 
  }
\end{align}
{
Using \cref{eq:integral_full_expression}, the integral is
\begin{align}
&  I(-\omega',-\omega',\omega,E= E_m-E_n,E_m-E_n, E'= E_s-E_n) \nonumber\\
   =&\frac{1}{\beta(E- i \omega')}\bigg\{-\frac{1+e^{-\beta E'}}{(E'+i\omega)}\bigg[ \frac{1}{(E'+i\omega)}  + \frac{1}{(E-  i \omega')}\bigg] -\frac{e^{-\beta E}-e^{-\beta E'}}{(E-  i \omega')(E'-E+i\omega+   i \omega')}(1-\delta_{EE'}\delta_{-   \omega'\omega})   \bigg\}\nonumber\\
   &+ \frac{1 - e^{-\beta E} \delta_{EE'}\delta_{-   \omega'\omega}  }{(E- i \omega')(E'+i\omega)}  \nonumber\\   
\end{align}
Taking zero-temperature limit, we obtain
\begin{align}
& e^{-\beta(E_n- E_{gs})} I(-\omega',-\omega',\omega,E= E_m-E_n,E_m-E_n, E'= E_s-E_n)\bigg|_{\beta \rightarrow \infty} \nonumber\\
   =&\frac{1}{\beta(E- i \omega')}
   \bigg\{-\frac{\delta_{n,gs} +\delta_{s,gs}}{(E'+i\omega)}\bigg[ \frac{1}{(E'+i\omega)}  + \frac{1}{(E-  i \omega')}\bigg] -\frac{\delta_{m, gs}-\delta_{s, gs}}{(E-  i \omega')(E'-E+i\omega+   i \omega')}(1-\delta_{E_sE_m }\delta_{-   \omega'\omega})   
   \bigg\}\nonumber\\
   &+ \frac{\delta_{n, gs}- \delta_{m,gs} \delta_{E_s,E_m }\delta_{-   \omega'\omega}  }{(E- i \omega')(E'+i\omega)}  
\end{align}
and the Fourier transform becomes
\begin{align}
    &   -{\binom{N_f}{n_f}}^{-1}  \sum_{\substack{n\\
  \ket{s}=f_{\RR,i}\ket{n},\ket{m}=f^{\dagger}_{\RR,j}\ket{n}
  }}
  (1-\bra{n}f^{\dagger}_{\RR,j} f_{\RR,j} \ket{n}) \bra{n}f^{\dagger}_{\RR,i}f_{\RR,i}\ket{n}\nonumber\\
  &\qquad\qquad\qquad\qquad \times e^{- \beta (E_n - E_{gs}) }
  I(-\omega',-\omega',\omega,E_m-E_n,E_m-E_n,E_s-E_n)\bigg|_{\beta\to\infty}  \nonumber \\ 
    =& {\binom{N_f}{n_f}}^{-1}  \sum_{\substack{n\\
  \ket{s}=f_{\RR,i}\ket{n},\ket{m}=f^{\dagger}_{\RR,j}\ket{n}
  }}
  (1-\bra{n}f^{\dagger}_{\RR,j} f_{\RR,j} \ket{n}) \bra{n}f^{\dagger}_{\RR,i}f_{\RR,i}\ket{n}  \nonumber\\
   &\bigg\{ 
   \frac{1}{\beta(E- i \omega')}\frac{\delta_{n,gs} }{(E'+i\omega)}\bigg[ \frac{1}{(E'+i\omega)}  + \frac{1}{(E-  i \omega')}\bigg]   - \frac{\delta_{n, gs} }{(E- i \omega')(E'+i\omega)}    \nonumber \\ 
   &+
\frac{1}{\beta(E- i \omega')}\frac{\delta_{s,gs}}{(E'+i\omega)}\bigg[ \frac{1}{(E'+i\omega)}  + \frac{1}{(E-  i \omega')}\bigg] +\frac{1}{\beta(E- i \omega')} \frac{-\delta_{s, gs}}{(E-  i \omega')(E'-E+i\omega+   i \omega')}(1-\delta_{E_sE_m }\delta_{-   \omega'\omega})    \nonumber \\ 
&+ \frac{1}{\beta(E- i \omega')} \frac{\delta_{m, gs}}{(E-  i \omega')(E'-E+i\omega+   i \omega')}(1-\delta_{E_sE_m }\delta_{-   \omega'\omega})  + \frac{\delta_{m,gs} \delta_{E_s,E_m }\delta_{-   \omega'\omega}  }{(E- i \omega')(E'+i\omega)}  \bigg\}   \nonumber \\ 
=&\frac{n_f(N_f- n_f)}{ N_f(N_f -1)}   \frac{1}{(\Delta_{+1}- i \omega')(\Delta_{-1}  +i\omega)} \bigg [
\frac{1}{\beta} \bigg( \frac{1}{\Delta_{-1}+i\omega}  + \frac{1}{\Delta_{+1} -  i \omega'}\bigg)   - 1 
\bigg]    \nonumber \\ 
&+
\frac{(N_f - n_f)(N_f- n_f-1)}{ N_f (N_f-1)} \frac{1}{\beta(\Delta_{+2}- \Delta_{+1} - i \omega')} 
\bigg[ \frac{1}{ i\omega - \Delta_{+1}}\bigg(    \frac{1}{\Delta_{+2} - \Delta_{+1} -  i \omega'} - \frac{1}{ \Delta_{+1} -i\omega}\bigg) \nonumber\\
&+ \frac{1}{(\Delta_{+2} - \Delta_{+1} - i \omega')} \frac{1}{(\Delta_{+2} -i\omega-   i \omega')} 
\bigg]    \nonumber \\ 
&+ \frac{n_f(n_f-1)}{ N_f (N_f-1)} \frac{1}{\beta(- \Delta_{-1} - i \omega')^2} \frac{1} {(\Delta_{-2} +i\omega+   i \omega')}    
\label{eq:Fij_no3}
\end{align}
In this case, $\sum_{n\in GS}$, $\sum_{\substack{n\\ s\in GS}}$, $\sum_{\substack{n\\ m\in GS}}$ count the ground state, charge $+1$ excitation and charge $-1$ excitation states $\ket{n}$ that contain the $i$-the flavor but not contain the $j$-th flavor. The multiplicities of them are $\binom{N_f-2}{n_f-1}$, $\binom{N_f-2}{n_f}$ and $\binom{N_f-2}{n_f-2}$, respectively.

}

 \item 
For $\tau_2>\tau_3 >\tau_1$, we find 
\ba 
&\langle T_\tau f_{\RR,j}^\dag(\tau_1) f_{\RR,j}(\tau_2) f_{\RR,i}^\dag(\tau_3)f_{\RR,i}(0)\rangle \nonumber\\
 =& \langle f_{\RR,j}(\tau_2)   f_{\RR,i}^\dag(\tau_3)
 f_{\RR,j}^\dag(\tau_1) f_{\RR,i}(0)\rangle   \nonumber
 \\ 
 =& 
 \frac{1}{Z_{atom,\RR}}
    \sum_{n,s,m,u}  \langle n|f_{\RR,j} |m \rangle \langle m| f_{\RR,i}^\dag|u  \rangle \langle u |f_{\RR,j}^\dag |s\rangle \langle s| f_{\RR,i}|n\rangle e^{-(\beta-\tau_2)E_n - (\tau_2-\tau_3)E_{m} -(\tau_3-\tau_1)E_u - \tau_1 E_s} 
\ea 
Denote $\ket{s}=f_{\RR,i}\ket{n}$, $\ket{u}=f^{\dagger}_{\RR,j}\ket{s}=f^{\dagger}_{\RR,j}f_{\RR,i}\ket{n}$ and $\ket{m}=f^{\dagger}_{\RR,j}\ket{n}$, then the matrix element is evaluated as
\begin{align}
    &\langle n|f_{\RR,j} |m \rangle \langle m| f_{\RR,i}^\dag|u  \rangle \langle u |f_{\RR,j}^\dag |s\rangle \langle s| f_{\RR,i}|n\rangle\nonumber\\
    =&\bra{n}f_{\RR,j}f^{\dagger}_{\RR,j}\ket{n}\bra{n}f_{\RR,j}f^{\dagger}_{\RR,i}f^{\dagger}_{\RR,j}f_{\RR,i}\ket{n}\bra{n}f^{\dagger}_{\RR,i}f_{\RR,j}f^{\dagger}_{\RR,j}f_{\RR,i}\ket{n}\bra{n}f^{\dagger}_{\RR,i}f_{\RR,i}\ket{n}\nonumber\\
    =&-\bra{n}f_{\RR,j}f^{\dagger}_{\RR,j}\ket{n}(\bra{n}f_{\RR,j}f^{\dagger}_{\RR,j}f^{\dagger}_{\RR,i}f_{\RR,i}\ket{n})^2\bra{n}f^{\dagger}_{\RR,i}f_{\RR,i}\ket{n}\nonumber\\
    =&-\bra{n}f_{\RR,j}f^{\dagger}_{\RR,j}\ket{n}\bra{n}f^{\dagger}_{\RR,i}f_{\RR,i}\ket{n}
\end{align}
And the Fourier transform is
\begin{align}
  &\frac{1}{Z_{atom,\RR}}
    \sum_{n,s,m,u}  \langle n|f_{\RR,j} |m \rangle \langle m| f_{\RR,i}^\dag|u  \rangle \langle u |f_{\RR,j}^\dag |s\rangle \langle s| f_{\RR,i}|n\rangle\nonumber\\
    &\times \frac{1}{\beta}\int_{0}^{\beta}d\tau_1\int_{\tau_1}^{\beta}d\tau_3\int_{\tau_3}^{\beta}d\tau_2\;e^{-\beta E_n -\tau_1(E_s-E_u)+\tau_3(E_m-E_u)-\tau_2(E_m-E_n)-i\omega'\tau_1-i\omega\tau_3+i\omega'\tau_2} \nonumber\\
    =&\frac{1}{Z_{atom,\RR}}
    \sum_{n,s,m,u}  \langle n|f_{\RR,j} |m \rangle \langle m| f_{\RR,i}^\dag|u  \rangle \langle u |f_{\RR,j}^\dag |s\rangle \langle s| f_{\RR,i}|n\rangle e^{-\beta E_n}I(-\omega',-\omega,\omega',E_m-E_n,E_m-E_u,E_s-E_u)\nonumber\\
    =&-\sum_{\substack{n\\
    \ket{s}=f_{\RR,i}\ket{n},\ket{u}=f^{\dagger}_{\RR,j}f_{\RR,i}\ket{n},\ket{m}=f^{\dagger}_{\RR,j}\ket{n}
    }}
    \bra{n}f_{\RR,j}f^{\dagger}_{\RR,j}\ket{n}\bra{n}f^{\dagger}_{\RR,i}f_{\RR,i}\ket{n}
    \frac{e^{-\beta E_n}}{Z_{atom,\RR}}I(-\omega',-\omega,\omega',E_m-E_n,E_m-E_u,E_s-E_u)
    \nonumber\\
    =&
    {
    -\sum_{\substack{n\\
    \ket{s}=f_{\RR,i}\ket{n},\ket{u}=f^{\dagger}_{\RR,j}f_{\RR,i}\ket{n},\ket{m}=f^{\dagger}_{\RR,j}\ket{n}
    }}
    (1- \bra{n}f^{\dagger}_{\RR,j} f_{\RR,j} \ket{n})\bra{n}f^{\dagger}_{\RR,i}f_{\RR,i}\ket{n}
    \frac{e^{-\beta E_n}}{Z_{atom,\RR}}I(-\omega',-\omega,\omega',E_m-E_n,E_m-E_n,E_s-E_n)
    }
\end{align}
{
Using \cref{eq:integral_full_expression}, the integral is 
\begin{align}
   &   I(-\omega',-\omega,\omega',E=E_m-E_n,E_m-E_n,E'=E_s-E_n)  \nonumber\\
   =&\frac{1}{\beta(E-i \omega')}\bigg\{- \frac{(1+e^{-\beta E'} )}{ (E'+i \omega')  } \bigg[ \frac{(1-\delta_{\omega' \omega}) }{(E'+i \omega)}  +\frac{\delta_{\omega' \omega}}{(E'+i \omega')}+\frac{1}{(E-i \omega)}\bigg] -\frac{e^{-\beta E}-e^{-\beta E'}}{(E-i \omega)(E'-E+i \omega'+i \omega)}(1-\delta_{EE'}\delta_{- \omega \omega'})   \bigg\}\nonumber\\
   &+ \frac{\delta_{\omega' \omega}- e^{-\beta E} \delta_{EE'}\delta_{- \omega \omega'}}{(E-i \omega')(E'+i \omega')}  \nonumber \\ 
\end{align}
Taking zero-temperature limit, we obtain
\begin{align}
   & e^{-\beta(E_n- E_{gs})}  I(-\omega',-\omega,\omega',E=E_m-E_n,E_m-E_n,E'=E_s-E_n)|_{\beta \rightarrow \infty}  \nonumber\\
   =&\frac{1}{\beta(E-i \omega')}\bigg\{- \frac{(\delta_{n, gs}+\delta_{s, gs} )}{ (E'+i \omega')  } \bigg[ \frac{1 }{E'+i \omega}  +\frac{1}{E-i \omega}\bigg] -\frac{\delta_{m,gs}-\delta_{s, gs}}{(E-i \omega)(E'-E+i \omega'+i \omega)}(1-\delta_{E_mE_s}\delta_{- \omega \omega'})   \bigg\}\nonumber\\
   &+ \frac{\delta_{n,gs} \delta_{\omega' \omega}- \delta_{m,gs} \delta_{E_mE_s}\delta_{- \omega \omega'}}{(E-i \omega')(E'+i \omega')}  \nonumber \\ 
\end{align}
and the Fourier transform becomes
\begin{align}
&      - {\binom{N_f}{n_f}}^{-1} \sum_{\substack{n\\
    \ket{s}=f_{\RR,i}\ket{n},\ket{u}=f^{\dagger}_{\RR,j}f_{\RR,i}\ket{n},\ket{m}=f^{\dagger}_{\RR,j}\ket{n}
    }}
    (1- \bra{n}f^{\dagger}_{\RR,j} f_{\RR,j} \ket{n})\bra{n}f^{\dagger}_{\RR,i}f_{\RR,i}\ket{n}\nonumber\\
    &\qquad\qquad\qquad\qquad\times
    e^{- \beta (E_n - E_{gs}) }I(-\omega',-\omega,\omega',E_m-E_n,E_m-E_n,E_s-E_n) \nonumber\\
   =&\frac{n_f(N_f- n_f)}{N_f (N_f-1)} \frac{1}{({\Delta_{+1}}-i \omega')({\Delta_{-1}}+i \omega')}\bigg[ \frac{1}{ \beta  } \bigg( \frac{1 }{{\Delta_{-1}}+i \omega}  +\frac{1}{{\Delta_{+1}}-i \omega}\bigg)    -  \delta_{\omega' \omega} \bigg]   \nonumber \\ 
   & + \frac{(N_f- n_f)(N_f- n_f -1)}{ N_f(N_f-1)} \frac{1}{\beta(\Delta_{+2} - \Delta_{+1} -i \omega')}\bigg[ \frac{1 }{  i \omega'- \Delta_{+1}  } \bigg( \frac{1 }{- \Delta_{+1} +i \omega}  +\frac{1}{\Delta_{+2} - \Delta_{+1} -i \omega}\bigg) \nonumber\\
   &+\frac{1}{(\Delta_{+2}- \Delta_{+1} -i \omega)(\Delta_{+2} -i \omega'- i \omega)}   \bigg]\nonumber \\ 
   &  +\frac{n_f(n_f-1)}{ N_f(N_f-1)} \frac{1}{\beta(- \Delta_{-1} -i \omega')} \frac{1}{(- \Delta_{-1} -i \omega)(\Delta_{-2} +i \omega'+i \omega)}
   \label{eq:Fij_no4}
   \end{align} 

}

\item For $\tau_3>\tau_1>\tau_2$, we find 
\ba 
&\langle T_\tau f_{\RR,j}^\dag(\tau_1) f_{\RR,j}(\tau_2) f_{\RR,i}^\dag(\tau_3)f_{\RR,i}(0)\rangle  \nonumber\\
=&\langle   f_{\RR,i}^\dag(\tau_3) f_{\RR,j}^\dag(\tau_1) f_{\RR,j}(\tau_2) f_{\RR,i}(0)\rangle   \nonumber\\
=& 
 \frac{1}{Z_{atom,\RR}}
    \sum_{n,s,u}  \langle n|f_{\RR,i}^\dag |s \rangle \langle s| f_{\RR,j}^\dag|u  \rangle \langle u |f_{\RR,j} |s\rangle \langle s| f_{\RR,i}|n\rangle e^{-(\beta-\tau_3)E_n - (\tau_3-\tau_1)E_{s} -(\tau_1-\tau_2)E_u - \tau_2 E_s} 
\ea 
Denote $\ket{s}=f_{\RR,i}\ket{n}$ and $\ket{u}=f_{\RR,j}\ket{s}=f_{\RR,j}f_{\RR,i}\ket{n}$, then the matrix element is 
\begin{align}
    &\langle n|f_{\RR,i}^\dag |s \rangle \langle s| f_{\RR,j}^\dag|u  \rangle \langle u |f_{\RR,j} |s\rangle \langle s| f_{\RR,i}|n\rangle=(\bra{n}f^{\dagger}_{\RR,i}f_{\RR,i}\ket{n})^2(\bra{n}f^{\dagger}_{\RR,i}f^{\dagger}_{\RR,j}f_{\RR,j}f_{\RR,i}\ket{n})^2=\bra{n}f^{\dagger}_{\RR,i}f_{\RR,i}\ket{n}\bra{n}f^{\dagger}_{\RR,j}f_{\RR,j}\ket{n}
\end{align}
And the Fourier transform is
\begin{align}
  &\frac{1}{Z_{atom,\RR}}
    \sum_{n,s,u}  \langle n|f_{\RR,i}^\dag |s \rangle \langle s| f_{\RR,j}^\dag|u  \rangle \langle u |f_{\RR,j} |s\rangle \langle s| f_{\RR,i}|n\rangle \nonumber\\
    &\times\frac{1}{\beta}\int_{0}^{\beta}d\tau_2\int_{\tau_2}^{\beta}d\tau_1\int_{\tau_1}^{\beta}d\tau_3\; e^{-\beta E_n-\tau_2(E_s-E_u)+\tau_1(E_s-E_u)-\tau_3(E_s-E_n)+i\omega'\tau_2-i\omega'\tau_1-i\omega\tau_3}\nonumber\\
    =&\frac{1}{Z_{atom,\RR}}
    \sum_{n,s,u}  \langle n|f_{\RR,i}^\dag |s \rangle \langle s| f_{\RR,j}^\dag|u  \rangle \langle u |f_{\RR,j} |s\rangle \langle s| f_{\RR,i}|n\rangle e^{-\beta E_n}I(\omega,-\omega',-\omega',E_s-E_n,E_s-E_u,E_s-E_u)\nonumber\\
    =&\sum_{\substack{n\\
    \ket{s}=f_{\RR,i}\ket{n},\ket{u}=f_{\RR,j}f_{\RR,i}\ket{n}
    }}
    \bra{n}f^{\dagger}_{\RR,i}f_{\RR,i}\ket{n}\bra{n}f^{\dagger}_{\RR,j}f_{\RR,j}\ket{n}
    \frac{e^{-\beta E_n}}{Z_{atom,\RR}}I(\omega,-\omega',-\omega',E_s-E_n,E_s-E_u,E_s-E_u) \nonumber \\     
    =&{\sum_{\substack{n\\
    \ket{s}=f_{\RR,i}\ket{n},\ket{u}=f_{\RR,j}f_{\RR,i}\ket{n}= f_{\RR,j} \ket{s}
    }}
    \bra{n}f^{\dagger}_{\RR,i}f_{\RR,i}\ket{n}\bra{n}f^{\dagger}_{\RR,j}f_{\RR,j}\ket{n}
    \frac{e^{-\beta E_n}}{Z_{atom,\RR}}I(\omega,-\omega',-\omega',E_s-E_n,E_s-E_u,E_s-E_u) }
\end{align}
{
Using \cref{eq:integral_full_expression}, the integral is 
\begin{align}
   &  I(\omega,-\omega',-\omega',E= E_s-E_n,E'=E_s-E_u,E'=E_s-E_u)  \nonumber\\
   =&\frac{1}{\beta(E+i \omega)}\bigg\{\frac{1-\delta_{EE'}\delta_{\omega,- \omega'}}{E'-E- i \omega'-i \omega}\bigg[\frac{e^{-\beta(E-E')}(1+e^{-\beta E'})}{E'- i \omega'}-\frac{1+e^{-\beta E}}{E+i \omega}  \bigg] -\frac{1+e^{-\beta E'}}{(E'- i \omega')^2} (\delta_{EE'}\delta_{\omega,- \omega'}    +e^{-\beta(E-E')} )   \bigg\}\nonumber\\
   &+ \frac{\delta_{EE'}\delta_{\omega,- \omega'} - e^{-\beta E}}{(E+i \omega)(E'- i \omega')}
\end{align}
Taking zero-temperature limit, we obtain
\begin{align}
   &  e^{-\beta(E_n- E_{gs})} I(\omega,-\omega',-\omega',E= E_s-E_n,E'=E_s-E_u,E'=E_s-E_u)|_{\beta \rightarrow \infty}  \nonumber\\
   =&\frac{1}{\beta(E+i \omega)}\bigg\{\frac{1-\delta_{E_nE_u }\delta_{\omega,- \omega'}}{E'-E- i \omega'-i \omega}\bigg[\frac{\delta_{u, gs}+\delta_{s, gs}}{E'- i \omega'}-\frac{\delta_{n, gs} +\delta_{s, gs}}{E+i \omega}  \bigg] -\nonumber \\ & - \frac{\delta_{n, gs} +\delta_{s, gs} }{(E'- i \omega')^2} \delta_{E_nE_u}\delta_{\omega,- \omega'}    -  \frac{\delta_{u, gs} +\delta_{s, gs} }{(E'- i \omega')^2}   \bigg\}+ \frac{\delta_{E_n E_u}\delta_{\omega,- \omega'}\delta_{n, gs}  - \delta_{s, gs} }{(E+i \omega)(E'- i \omega')} 
\end{align}
and the Fourier transform becomes
\begin{align}
       &  {\binom{N_f}{n_f}}^{-1} \sum_{\substack{n\\
    \ket{s}=f_{\RR,i}\ket{n},\ket{u}=f_{\RR,j}f_{\RR,i}\ket{n}= f_{\RR,j} \ket{s}
    }}
    \bra{n}f^{\dagger}_{\RR,i}f_{\RR,i}\ket{n}\bra{n}f^{\dagger}_{\RR,j}f_{\RR,j}\ket{n}\nonumber\\
    &\qquad\qquad\qquad\qquad\times
     e^{- \beta (E_n - E_{gs}) }I(\omega,-\omega',-\omega',E_s-E_n,E_s-E_u,E_s-E_u)  \nonumber \\ =&\frac{(N_f- n_f)(N_f - n_f -1) }{N_f (N_f -1)} \frac{1}{\beta(\Delta_{+1} - \Delta_{+2} +i \omega)(\Delta_{+1} - i \omega') }\bigg( \frac{1}{\Delta_{+2} - i \omega'-i \omega}   -  \frac{1 }{\Delta_{+1} - i \omega')}   \bigg )  \nonumber \\ 
     & + \frac{n_f(N_f - n_f)}{ N_f(N_f-1)}
     \bigg\{ \frac{1}{\beta(- \Delta_{+1} +i \omega)}\bigg[\frac{1-\delta_{\Delta_{+1}\Delta_{-1} } \delta_{\omega,- \omega'}}{(\Delta_{-1} + i \omega')(- \Delta_{+1} +i \omega)  } -\frac{1}{(\Delta_{-1}+ i \omega')^2} (1+ \delta_{\Delta_{+1} \Delta_{-1} }\delta_{\omega,- \omega'} )   \bigg]\nonumber\\
     &+  \frac{  1 }{(- \Delta_{+1} +i \omega)( \Delta_{-1} + i \omega')}  
     \bigg\}  \nonumber \\ 
     & + \frac{n_f(n_f-1)}{N_f(N_f-1)} \frac{1}{\beta(\Delta_{-1}+i \omega)}\bigg\{\frac{1}{\Delta_{-2}+ i \omega'+i \omega}\frac{1 }{E+i \omega}  \bigg\} \nonumber \\ 
    =&\frac{(N_f- n_f)(N_f - n_f -1) }{N_f (N_f -1)} \frac{1}{\beta(\Delta_{+1} - i \omega') (\Delta_{+2} - i \omega'-i \omega)(\Delta_{+1} - i \omega') } + \nonumber \\ 
     & + \frac{n_f(N_f - n_f)}{ N_f(N_f-1)} \frac{1}{(- \Delta_{+1} +i \omega)(\Delta_{-1}+ i \omega')}\bigg[\frac{1}{\beta} \bigg( \frac{1}{- \Delta_{+1} +i \omega  } -\frac{1}{\Delta_{-1}+ i \omega'} \bigg)   + 1\bigg]   \nonumber \\ 
     & + \frac{n_f(n_f-1)}{N_f(N_f-1)} \frac{1}{\beta(\Delta_{-1}+i \omega)^2}\frac{1}{\Delta_{-2}+ i \omega'+i \omega}
     \label{eq:Fij_no5}
   \end{align}

}

\item For $\tau_3>\tau_2>\tau_1$, we have 
\begin{align}
    &\langle T_\tau f_{\RR,j}^\dag(\tau_1) f_{\RR,j}(\tau_2) f_{\RR,i}^\dag(\tau_3)f_{\RR,i}(0)\rangle \nonumber\\ 
=&-\langle   f_{\RR,i}^\dag(\tau_3)  f_{\RR,j}(\tau_2) f_{\RR,j}^\dag(\tau_1)f_{\RR,i}(0)\rangle   \nonumber\\
=& -
 \frac{1}{Z_{atom,\RR}}
    \sum_{n,s,u}  \langle n|f_{\RR,i}^\dag |s \rangle \langle s| f_{\RR,j}|u  \rangle \langle u |f_{\RR,j}^\dag |s\rangle \langle s| f_{\RR,i}|n\rangle e^{-(\beta-\tau_3)E_n - (\tau_3-\tau_2)E_{s} -(\tau_2-\tau_1)E_u - \tau_1 E_s} 
\end{align}
Denote $\ket{s}=f_{\RR,i}\ket{n}$ and $\ket{u}=f^{\dagger}_{\RR,j}\ket{s}=f^{\dagger}_{\RR,j}f_{\RR,i}\ket{n}$, then the matrix element is
\begin{align}
    &\langle n|f_{\RR,i}^\dag |s \rangle \langle s| f_{\RR,j}|u  \rangle \langle u |f_{\RR,j}^\dag |s\rangle \langle s| f_{\RR,i}|n\rangle=(\bra{n}f^{\dagger}_{\RR,i}f_{\RR,i}\ket{n}\bra{n}f^{\dagger}_{\RR,i}f_{\RR,j}f^{\dagger}_{\RR,j}f_{\RR,i}\ket{n})^2=\bra{n}f^{\dagger}_{\RR,i}f_{\RR,i}\ket{n}\bra{n}f_{\RR,j}f^{\dagger}_{\RR,j}\ket{n}
\end{align}
And the Fourier transform is
\begin{align}
  & -
 \frac{1}{Z_{atom,\RR}}
    \sum_{n,s,u}  \langle n|f_{\RR,i}^\dag |s \rangle \langle s| f_{\RR,j}|u  \rangle \langle u |f_{\RR,j}^\dag |s\rangle \langle s| f_{\RR,i}|n\rangle \nonumber\\
    &\times\frac{1}{\beta}\int_0^{\beta}d\tau_1\int_{\tau_1}^{\beta}d\tau_2\int_{\tau_2}^{\beta}d\tau_3\; e^{-\beta E_n-\tau_1(E_s-E_u)+\tau_2(E_s-E_u)-\tau_3(E_s-E_n)-i\omega'\tau_1+i\omega'\tau_2-i\omega\tau_3 }\nonumber\\
    =& -
    \frac{1}{Z_{atom,\RR}}\sum_{n,s,u}  \langle n|f_{\RR,i}^\dag |s \rangle \langle s| f_{\RR,j}|u  \rangle \langle u |f_{\RR,j}^\dag |s\rangle \langle s| f_{\RR,i}|n\rangle e^{-\beta E_n} I(\omega,\omega',\omega',E_s-E_n,E_s-E_u,E_s-E_u)\nonumber\\
    =&-\sum_{\substack{n\\
    \ket{s}=f_{\RR,i}\ket{n},\ket{u}=f^{\dagger}_{\RR,j}f_{\RR,i}\ket{n}
    }}
    \bra{n}f^{\dagger}_{\RR,i}f_{\RR,i}\ket{n}\bra{n}f_{\RR,j}f^{\dagger}_{\RR,j}\ket{n}
    \frac{e^{-\beta E_n}}{Z_{atom,\RR}}I(\omega,\omega',\omega',E_s-E_n,E_s-E_u,E_s-E_u) \nonumber\\
    =&{ -\sum_{\substack{n\\
    \ket{s}=f_{\RR,i}\ket{n},\ket{u}=f^{\dagger}_{\RR,j}f_{\RR,i}\ket{n}
    }}
    \bra{n}f^{\dagger}_{\RR,i}f_{\RR,i}\ket{n} (1- \bra{n}f^{\dagger}_{\RR,j}f_{\RR,j}\ket{n})
    \frac{e^{-\beta E_n}}{Z_{atom,\RR}}I(\omega,\omega',\omega',E_s-E_n,E_s-E_n,E_s-E_n) }
\end{align}
{
Using \cref{eq:integral_full_expression_same_energy}, the integral is
\begin{align}
   & I(\omega,\omega',\omega',E=E_s-E_n,E_s-E_n,E_s-E_n)  \nonumber\\
   =&\frac{1}{{(E+i\omega)(E+i\omega')} }\bigg[  -\frac{1+e^{-\beta E}}{\beta}\bigg\{\frac{1-\delta_{\omega\omega'}}{(E+i\omega)}  +\frac{1+\delta_{\omega\omega'}}{(E+i\omega')}\bigg\}
   + \delta_{\omega\omega'}- e^{-\beta E} \bigg]
\end{align}
Taking zero-temperature limit, we obtain
\begin{align}
   &  e^{-\beta(E_n- E_{gs})}I(\omega,\omega',\omega',E=E_s-E_n,E_s-E_n,E_s-E_n)|_{\beta \rightarrow \infty}  \nonumber\\
   =&\frac{1}{{(E+i\omega)(E+i\omega')} }\bigg[  -\frac{\delta_{n, gs} +\delta_{s,gs}}{\beta}\bigg\{\frac{1-\delta_{\omega\omega'}}{(E+i\omega)}  +\frac{1+\delta_{\omega\omega'}}{(E+i\omega')}\bigg\}
   + \delta_{\omega\omega'}\delta_{n,gs} - \delta_{s,gs} \bigg]
\end{align}
And the Fourier transform becomes
\begin{align}
    &    - {\binom{N_f}{n_f}}^{-1} \sum_{\substack{n\\
    \ket{s}=f_{\RR,i}\ket{n},\ket{u}=f^{\dagger}_{\RR,j}f_{\RR,i}\ket{n}
    }}
    \bra{n}f^{\dagger}_{\RR,i}f_{\RR,i}\ket{n} (1- \bra{n}f^{\dagger}_{\RR,j}f_{\RR,j}\ket{n})\nonumber\\
    &\qquad\qquad\qquad\qquad\times
   e^{- \beta (E_n - E_{gs}) } I(\omega,\omega',\omega',E_s-E_n,E_s-E_n,E_s-E_n) )\nonumber \\ 
   =&\frac{n_f(N_f- n_f)}{ N_f(N_f-1)} \frac{1}{{(\Delta_{-1} +i\omega)(\Delta_{-1}+i\omega')} }\bigg[  \frac{1 }{\beta}\bigg\{\frac{1-\delta_{\omega\omega'}}{(\Delta_{-1} +i\omega)}  +\frac{1+\delta_{\omega\omega'}}{(\Delta_{-1} +i\omega')}\bigg\}
   - \delta_{\omega\omega'}\bigg]\nonumber \\ 
   &+ \frac{(N_f - n_f)(N_f - n_f-1}{N_f (N_f-1)} \frac{1}{{(- \Delta_{+1} +i\omega)(-\Delta_{+1} +i\omega')} }\bigg[  \frac{1}{\beta}\bigg\{\frac{1-\delta_{\omega\omega'}}{(- \Delta_{+1}+i\omega)}  +\frac{1+\delta_{\omega\omega'}}{(-\Delta_{+1}+i\omega')}\bigg\}
  + 1 \bigg] \nonumber \\ 
   =&\frac{n_f(N_f- n_f)}{ N_f(N_f-1)} \frac{1}{{(\Delta_{-1} +i\omega)(\Delta_{-1}+i\omega')} }\bigg[  \frac{1 }{\beta}\bigg(\frac{1}{\Delta_{-1} +i\omega}  +\frac{1}{\Delta_{-1} +i\omega'}\bigg)
   - \delta_{\omega\omega'}\bigg]+\nonumber \\ 
   &+ \frac{(N_f - n_f)(N_f - n_f-1)}{N_f (N_f-1)} \frac{1}{{(- \Delta_{+1} +i\omega)(-\Delta_{+1} +i\omega')} }\bigg[  \frac{1}{\beta}\bigg(\frac{1}{- \Delta_{+1}+i\omega}  +\frac{1}{-\Delta_{+1}+i\omega'}\bigg)
  + 1 \bigg]
  \label{eq:Fij_no6}
\end{align}

}

\end{itemize}

{
Sum \cref{eq:Fij_no1,eq:Fij_no2,eq:Fij_no3,eq:Fij_no4,eq:Fij_no5,eq:Fij_no6} gives the Fourier transform of the four point correlation function
\begin{equation}
    \frac{1}{\beta}\int_{\tau_1,\tau_2,\tau_3} \langle T_\tau 
f_{\RR,j}^\dag(\tau_1) f_{\RR,j}(\tau_2)f_{\RR,i}^\dag(\tau_3) f_{\RR,i}(0)  \rangle_{\gamma=0} e^{-i\omega\tau_3+i\omega'\tau_2-i\omega'\tau_1 }=I^{0}_{i\ne j}(\omega,\omega',\omega')+I_{i\ne j}^1(\omega,\omega',\omega')
\end{equation}
which has two types of contribution.
The first kind $I^0$ does not have any $1/\beta$ prefactor, which reads:
\boxedeq{
\begin{align}
 I^{0}_{i\ne j}(\omega,\omega',\omega')=&  \frac{n_f(n_f-1)}{N_f(N_f-1)} \frac{1}{(i\omega'+ \Delta_{-1})(i\omega + \Delta_{-1})}    +  \frac{(N_f - n_f)(N_f - n_f-1)}{N_f (N_f-1)} \frac{1}{ (i\omega'- \Delta_{+1})(i\omega- \Delta_{+1})}  \nonumber \\ 
 & +
  \frac{n_f(N_f- n_f)}{ N_f(N_f-1)} \frac{  2 i \omega' i\omega + (i \omega' + i \omega) (\Delta_{-1}   - \Delta_{+1} ) - 2 \Delta_{+1}  \Delta_{-1}   -(\Delta_{-1} + \Delta_{+1})^2 \delta_{\omega \omega' }  
  }{{(i \omega - \Delta_{+1} )(i\omega'+ \Delta_{-1}) ( i \omega'- \Delta_{+1})(i\omega + \Delta_{-1})}  } 
\end{align}
}
Notice that the formula is invariant under $n_f \rightarrow N_f - n_f $ together with $\Delta_{+1}  \leftrightarrow - \Delta_{-1}$, which is a statement of the particle-hole symmetry in this problem. 
The second contribution is proportional to $1/\beta$ and is given by
\boxedeq{
\begin{align}
   &I^{1}_{i\ne j}(\omega,\omega',\omega')\nonumber\\
   =& \frac{1 }{\beta} \bigg\{\frac{n_f(n_f-1)}{ N_f(N_f-1)} \frac{i\omega' + i\omega + 2\Delta_1  }{ (i\omega + \Delta_1)^2(i\omega' + \Delta_1)^2}     \frac{ 2\Delta_1 -\Delta_{2} } {i\omega+   i \omega'+ \Delta_{2} } +\frac{ (N_f - n_f)  (N_f -  n_f-1)}{N_f(N_f-1)} \frac{i\omega'+ i\omega- 2\Delta_1}{(i\omega- \Delta_1)^2(i\omega'- \Delta_1)^2 }  \frac{ 2 \Delta_1- \Delta_{2}}{ i \omega+ i\omega'- \Delta_{2} }  \nonumber \\
   & -
  \frac{n_f(N_f- n_f)}{ N_f(N_f-1)} \frac{4\Delta_1 }{(i\omega'- \Delta_1)^2(  i \omega'+ \Delta_1)^2( i\omega- \Delta_1)^2 (  i \omega + \Delta_1)^2} \bigg[     (i \omega)^2 (i \omega')^2   +( (i \omega)^2+ (i \omega')^2)  
   \Delta_1^2  - 3 \Delta_1^4  \bigg ] \bigg\}\nonumber\\
\end{align}
}
Here, we have used $\Delta_{-1}=\Delta_{+1}=\Delta_{1}$ and $\Delta_{-2}=\Delta_{+2}=\Delta_{2}$ to simplify the results.
The details of the summation are given from \cref{eq:sum_over_Fij_start} to \cref{eq:sum_Iij_beta}.

}

{
Using \cref{eq:F_nm_from_ffff}, we then find that $F^{i\ne j}(i\omega,i\omega',i\omega')$ is

\begin{align}
    & F^{i\ne j}(i\omega,i\omega',i\omega') = - I^0_{i\ne j}(i\omega,i\omega',i\omega') + G_{f, loc}(i \omega) G_{f, loc}(i \omega' )- I^1_{i\ne j}(i\omega,i\omega',i\omega') + \mathcal{O}( e^{- \beta \frac{U_1}{2} }) \nonumber \\ 
=& -\frac{n_f(n_f-1)}{N_f(N_f-1)} \frac{1}{(i\omega'+ \Delta_{-1})(i\omega + \Delta_{-1})}    -  \frac{(N_f - n_f)(N_f - n_f-1)}{N_f (N_f-1)} \frac{1}{ (i\omega'- \Delta_{+1})(i\omega- \Delta_{+1})}   \nonumber \\ 
 & -
  \frac{n_f(N_f- n_f)}{ N_f(N_f-1)} \frac{ ( 2 i \omega' i\omega + (i \omega' + i \omega) (\Delta_{-1}   - \Delta_{+1} ) - 2 \Delta_{+1}  \Delta_{-1}   -(\Delta_{-1} + \Delta_{+1})^2 \delta_{\omega \omega' }  
  }{{(i \omega - \Delta_{+1} )(i\omega'+ \Delta_{-1}) ( i \omega'- \Delta_{+1})(i\omega + \Delta_{-1})}  }   \nonumber \\ 
& + \frac{1}{N_f^2}
    \bigg(\frac{n_f}{i\omega + \Delta_{-1}} +\frac{N_f-n_f}{i\omega-\Delta_{+1}}\bigg) 
    \bigg(\frac{n_f}{i\omega' + \Delta_{-1}} +\frac{N_f-n_f}{i\omega'-\Delta_{+1}}\bigg)  - I^1_{i\ne j}(i\omega,i\omega',i\omega') + \mathcal{O}( e^{- \beta \frac{U_1}{2} })   \nonumber \\ 
=&  \bigg[\frac{n_f^2}{N_f^2} -\frac{n_f(n_f-1)}{N_f(N_f-1)} \bigg]\frac{1}{(i\omega'+ \Delta_{-1})(i\omega + \Delta_{-1})}  +\bigg[\frac{(N_f-n_f)^2}{N_f^2}  -  \frac{(N_f - n_f)(N_f - n_f-1)}{N_f (N_f-1)}\bigg] \frac{1}{ (i\omega'- \Delta_{+1})(i\omega- \Delta_{+1})}  \nonumber \\ 
 & -
  \frac{n_f(N_f- n_f)}{ N_f(N_f-1)} \frac{  2 i \omega' i\omega + (i \omega' + i \omega) (\Delta_{-1}   - \Delta_{+1} ) - 2 \Delta_{+1}  \Delta_{-1}   -(\Delta_{-1} + \Delta_{+1})^2 \delta_{\omega \omega' }  
  }{{(i \omega - \Delta_{+1} )(i\omega'+ \Delta_{-1}) ( i \omega'- \Delta_{+1})(i\omega + \Delta_{-1})}  }  \nonumber \\ 
& + \frac{(N_f-n_f)n_f}{N_f^2}
   \frac{2 i\omega i\omega'+(i\omega+ i \omega')  (\Delta_{-1}-\Delta_{+1} )    -2 \Delta_{+1} \Delta_{-1}   }{(i\omega + \Delta_{-1})(i\omega'-\Delta_{+1})(i\omega-\Delta_{+1})(i\omega' + \Delta_{-1}) }   - I^1_{i\ne j}(i\omega,i\omega',i\omega') + \mathcal{O}( e^{- \beta \frac{U_1}{2} })   \nonumber \\ 
=& \frac{n_f(N_f-n_f)}{N_f^2(N_f-1)} \bigg[  \frac{1}{(i\omega'+ \Delta_{-1})(i\omega + \Delta_{-1})}  +\frac{1}{ (i\omega'- \Delta_{+1})(i\omega- \Delta_{+1})}\bigg]  \nonumber \\ 
 & - 
  \frac{n_f(N_f- n_f)}{ N_f^2(N_f-1)} \frac{ 2 i \omega' i\omega + (i \omega' + i \omega) (\Delta_{-1}   - \Delta_{+1} ) - 2 \Delta_{+1}  \Delta_{-1}    
  }{{(i \omega - \Delta_{+1} )(i\omega'+ \Delta_{-1}) ( i \omega'- \Delta_{+1})(i\omega + \Delta_{-1})}  }  \nonumber \\ 
&   +
  \frac{n_f(N_f- n_f)}{ N_f(N_f-1)} \frac{ (\Delta_{-1} + \Delta_{+1})^2 \delta_{\omega \omega' }  
  }{{(i \omega - \Delta_{+1} )(i\omega'+ \Delta_{-1}) ( i \omega'- \Delta_{+1})(i\omega + \Delta_{-1})}  }    - I^1_{i\ne j}(i\omega,i\omega',i\omega') + \mathcal{O}( e^{- \beta \frac{U_1}{2} })     \nonumber \\ 
 =& 
  \frac{n_f(N_f- n_f)}{ N_f^2(N_f-1)} \frac{ (i\omega'+ \Delta_{-1})(i\omega + \Delta_{-1})(i\omega'- \Delta_{+1})(i\omega- \Delta_{+1}) -(2 i \omega' i\omega + (i \omega' + i \omega) (\Delta_{-1}   - \Delta_{+1} ) - 2 \Delta_{+1}  \Delta_{-1}   ) 
  }{{(i \omega - \Delta_{+1} )(i\omega'+ \Delta_{-1}) ( i \omega'- \Delta_{+1})(i\omega + \Delta_{-1})}  }   \nonumber \\ 
&   
  \frac{n_f(N_f- n_f)}{ N_f(N_f-1)} \frac{ (\Delta_{-1} + \Delta_{+1})^2 \delta_{\omega \omega' }  
  }{{(i \omega - \Delta_{+1} )(i\omega'+ \Delta_{-1}) ( i \omega'- \Delta_{+1})(i\omega + \Delta_{-1})}  }    - I^1_{i\ne j}(i\omega,i\omega',i\omega') + \mathcal{O}( e^{- \beta \frac{U_1}{2} }) 
    \nonumber \\ 
=&  
  \frac{n_f(N_f- n_f)}{ N_f^2(N_f-1)} \frac{  (\Delta_{+1}+  \Delta_{-1} )^2  (1+ N_f \delta_{\omega, \omega'})
  }{{(i \omega - \Delta_{+1} )(i\omega'+ \Delta_{-1}) ( i \omega'- \Delta_{+1})(i\omega + \Delta_{-1})}  }   - I^1_{i\ne j}(i\omega,i\omega',i\omega') + \mathcal{O}( e^{- \beta \frac{U_1}{2} })  
  \label{eq:F_ij_proof}
\end{align}
By contrast, ghe equation for $F^{ii}$ is already correct to order $\mathcal{O}( e^{- \beta \frac{U_1}{2} })$ and does not contain a $1/\beta$ contribution. 

In summary, in the zero-temperature limit, we obtain
\boxedeq{
\begin{align}
    &F^{i\ne j}(i\omega,i\omega',i\omega')=-I_{i\ne j}(\omega,\omega',\omega')+G_{f,loc}(i\omega)G_{f,loc}(i\omega')\nonumber\\
    =&\frac{n_f(N_f- n_f)}{ N_f^2(N_f-1)} \frac{  (\Delta_{+1}+  \Delta_{-1} )^2  (1+ N_f \delta_{\omega, \omega'})
  }{{(i \omega - \Delta_{+1} )(i\omega'+ \Delta_{-1}) ( i \omega'- \Delta_{+1})(i\omega + \Delta_{-1})}  }   - I^1_{i\ne j}(i\omega,i\omega',i\omega') + \mathcal{O}( e^{- \beta \frac{U_1}{2} })  
  \label{eq:F_ij_final}
\end{align}
}

\subsubsection{Additional calculational details}
In this subsection, we provide the intermediate steps needed for the evaluation of $F^{ij}(i\omega,i\omega',i\omega')$.

In the calculation of $F^{ii}(i\omega,i\omega',i\omega')$, we need the special case of \cref{eq:integral_full_expression} in which all three energies are equal:
\begin{align}
   & I(\omega_1,\omega_2,\omega_3,E,E,E)\nonumber\\
   =&\frac{1}{\beta(E+i\omega_1)}\bigg\{\frac{1-\delta_{\omega_1\omega_2}}{i\omega_2-i\omega_1}\bigg[\frac{1+e^{-\beta E}}{E+i\omega_3}-\frac{1+e^{-\beta E}}{E+i\omega_3-i\omega_2+i\omega_1}  \bigg]
-\delta_{\omega_1\omega_2}\frac{1+e^{-\beta E}}{(E+i\omega_3)^2}-\frac{1+e^{-\beta E}}{(E+i\omega_2)(E+i\omega_3)}   \bigg\}\nonumber\\
   &+ \frac{\delta_{\omega_1\omega_2}}{(E+i\omega_1)(E+i\omega_3)} - 
\frac{e^{-\beta E} \delta_{\omega_2\omega_3} }{(E+i\omega_1)(E+i\omega_2)}
\label{eq:integral_full_expression_same_energy}
\end{align}
Hence,
\begin{align}
   & I(\omega',\omega',\omega,E,E,E)
   =-\frac{1+e^{-\beta E}}{\beta(E+i\omega')(E+i\omega)}\bigg\{
\frac{1}{(E+i\omega)}+\frac{1}{(E+i\omega')}   \bigg\} + \frac{1}{(E+i\omega')(E+i\omega)} - 
\frac{e^{-\beta E} \delta_{\omega \omega'} }{(E+i\omega')^2}\label{eq:integral_full_expressionwithparticularenergies} \nonumber \\ & I(\omega,\omega',\omega',E,E,E)
   =\frac{1+e^{-\beta E}}{\beta(E+i\omega)}\bigg\{\frac{1-\delta_{\omega \omega'}}{i\omega' -i\omega}\bigg[\frac{1}{E+i\omega' }-\frac{1}{E+i\omega}  \bigg]
-\frac{1+\delta_{\omega \omega'}}{(E+i\omega')^2}\bigg\}
   + \frac{\delta_{\omega\omega '}- e^{-\beta E} }{(E+i\omega)(E+i\omega')} 
\end{align}
Therefore,
\begin{align}
   & I(\omega',\omega',\omega,E,E,E)-  I(\omega,\omega',\omega',E,E,E)
   \nonumber \\ 
    =& -\frac{1+e^{-\beta E}}{\beta(E+i\omega')(E+i\omega)}\bigg\{
\frac{1}{(E+i\omega)}+\frac{1}{(E+i\omega')}   \bigg\} + \frac{1}{(E+i\omega')(E+i\omega)} - 
\frac{e^{-\beta E} \delta_{\omega \omega'} }{(E+i\omega')^2}\nonumber \\ 
&
   -\frac{1+e^{-\beta E}}{\beta(E+i\omega)}\bigg\{\frac{1-\delta_{\omega \omega'}}{i\omega' -i\omega}\bigg[\frac{1}{E+i\omega' }-\frac{1}{E+i\omega}  \bigg]
-\frac{1+\delta_{\omega \omega'}}{(E+i\omega')^2}\bigg\}
   - \frac{\delta_{\omega\omega '}- e^{-\beta E} }{(E+i\omega)(E+i\omega')}       \nonumber \\
 =& -\frac{1+e^{-\beta E}}{\beta(E+i\omega')(E+i\omega)}\bigg\{
\frac{1}{(E+i\omega)}+\frac{1}{(E+i\omega')}   \bigg\} + \frac{1- \delta_{\omega\omega '}+ e^{-\beta E} }{(E+i\omega')(E+i\omega)} - 
\frac{e^{-\beta E} \delta_{\omega \omega'} }{(E+i\omega')^2}\nonumber \\ 
&
   -\frac{1+e^{-\beta E}}{\beta(E+i\omega)}\bigg\{\frac{1-\delta_{\omega \omega'}}{i\omega' -i\omega}\bigg[\frac{1}{E+i\omega' }-\frac{1}{E+i\omega}  \bigg]
-\frac{1+\delta_{\omega \omega'}}{(E+i\omega')^2}\bigg\}  \nonumber \\ 
=& -\frac{1+e^{-\beta E}}{\beta(E+i\omega)}\bigg\{ \frac{1}{(E+i\omega')}\bigg(
\frac{1}{(E+i\omega)}+\frac{1}{(E+i\omega')} \bigg)  - \frac{1+\delta_{\omega \omega'}}{(E+i\omega')^2} + \frac{1-\delta_{\omega \omega'}}{i\omega' -i\omega}\bigg[\frac{1}{E+i\omega' }-\frac{1}{E+i\omega}  \bigg]\bigg\}  \nonumber\\ 
&+  \frac{1- \delta_{\omega\omega '}+ e^{-\beta E} }{(E+i\omega')(E+i\omega)} - 
\frac{e^{-\beta E} \delta_{\omega \omega'} }{(E+i\omega')^2}\nonumber \\ 
=&  -\frac{1+e^{-\beta E}}{\beta(E+i\omega)}\bigg\{ \frac{1}{(E+i\omega')}\bigg(
\frac{1}{(E+i\omega)}- \frac{\delta_{\omega \omega'}}{(E+i\omega')} \bigg)   + \frac{1-\delta_{\omega \omega'}}{i\omega' -i\omega}\bigg[\frac{1}{E+i\omega' }-\frac{1}{E+i\omega}  \bigg]\bigg\}  \nonumber\\ 
&+  \frac{1- \delta_{\omega\omega '}+ e^{-\beta E} }{(E+i\omega')(E+i\omega)} - 
\frac{e^{-\beta E} \delta_{\omega \omega'} }{(E+i\omega')^2} \nonumber \\ 
=&  -\frac{1+e^{-\beta E}}{\beta(E+i\omega)}\bigg\{ \frac{1}{(E+i\omega')}\bigg(
\frac{1}{(E+i\omega)}- \frac{\delta_{\omega \omega'}}{(E+i\omega')} \bigg)   -\frac{1-\delta_{\omega \omega'}}{(E+i\omega')(E+i\omega) }\bigg\}  \nonumber\\ 
&+  \frac{1+ e^{-\beta E} }{(E+i\omega')(E+i\omega)}- \frac{ \delta_{\omega\omega '}}{(E+i\omega')(E+i\omega)} - 
\frac{e^{-\beta E} \delta_{\omega \omega'} }{(E+i\omega')^2}  \nonumber \\ 
=&  -\frac{1+e^{-\beta E}}{\beta(E+i\omega)}\bigg\{ \frac{1}{(E+i\omega')}
\frac{1}{(E+i\omega)}- \frac{1}{(E+i\omega')} \frac{\delta_{\omega \omega'}}{(E+i\omega')}    -\frac{1-\delta_{\omega \omega'}}{(E+i\omega')(E+i\omega) } \bigg\}  \nonumber\\ 
&+  \frac{1+ e^{-\beta E} }{(E+i\omega')(E+i\omega)}- \frac{ \delta_{\omega\omega '}(1+ e^{-\beta E})}{(E+i\omega')(E+i\omega)} \nonumber \\ 
=&  \frac{(1- \delta_{\omega\omega '}) (1+ e^{-\beta E} )}{(E+i\omega')(E+i\omega)}\label{eq:ii_integral_sum}
\end{align}

In the calculation of $F^{i\neq j}(i\omega,i\omega',i\omega')$, we need to evaluate 
{
\begin{align}
   &  {\binom{N_f}{n_f}}^{-1}  \nonumber \bigg[ \sum_{\substack{n\\\ket{s}=f_{\RR,i}\ket{n}\\\ket{m}=f_{\RR,j}\ket{n} }}\bra{n}f^{\dagger}_{\RR,j}f_{\RR,j}\ket{n}\bra{n}f^{\dagger}_{\RR,i}f_{\RR,i}\ket{n} e^{- \beta (E_n - E_{gs}) } I(\omega',\omega',\omega,E_m-E_n,E_m-E_n,E_m-E_n)  \nonumber \\ 
   & + \sum_{\substack{n\\\ket{s}=f_{\RR,i}\ket{n}\\\ket{u}=f_{\RR,j}f_{\RR,i}\ket{n}\\\ket{m}=f_{\RR,j}\ket{n}}}\bra{n}f^{\dagger}_{\RR,j}f_{\RR,j}\ket{n}\bra{n}f^{\dagger}_{\RR,i}f_{\RR,i}\ket{n}
 e^{- \beta (E_n - E_{gs}) } I(\omega',-\omega,-\omega',E_m-E_n,E_m-E_u,E_m-E_u)  \nonumber \\ 
 &   -\sum_{\substack{n\\
  \ket{s}=f_{\RR,i}\ket{n}\\\ket{m}=f^{\dagger}_{\RR,j}\ket{n}
  }}
  (1-\bra{n}f^{\dagger}_{\RR,j} f_{\RR,j} \ket{n}) \bra{n}f^{\dagger}_{\RR,i}f_{\RR,i}\ket{n}  e^{- \beta (E_n - E_{gs}) }
  I(-\omega',-\omega',\omega,E_m-E_n,E_m-E_n,E_s-E_n)   \nonumber \\ 
  & -\sum_{\substack{n\\
    \ket{s}=f_{\RR,i}\ket{n}\\\ket{u}=f^{\dagger}_{\RR,j}f_{\RR,i}\ket{n}\\\ket{m}=f^{\dagger}_{\RR,j}\ket{n}
    }}
    (1- \bra{n}f^{\dagger}_{\RR,j} f_{\RR,j} \ket{n})\bra{n}f^{\dagger}_{\RR,i}f_{\RR,i}\ket{n}
    e^{- \beta (E_n - E_{gs}) }I(-\omega',-\omega,\omega',E_m-E_n,E_m-E_n,E_s-E_n)  \nonumber \\ 
    &+ \sum_{\substack{n\\
    \ket{s}=f_{\RR,i}\ket{n}\\\ket{u}=f_{\RR,j}f_{\RR,i}\ket{n}= f_{\RR,j} \ket{s}
    }}
    \bra{n}f^{\dagger}_{\RR,i}f_{\RR,i}\ket{n}\bra{n}f^{\dagger}_{\RR,j}f_{\RR,j}\ket{n}
     e^{- \beta (E_n - E_{gs}) }I(\omega,-\omega',-\omega',E_s-E_n,E_s-E_u,E_s-E_u)  \nonumber \\ &  -\sum_{\substack{n\\
    \ket{s}=f_{\RR,i}\ket{n}\\\ket{u}=f^{\dagger}_{\RR,j}f_{\RR,i}\ket{n}
    }}
    \bra{n}f^{\dagger}_{\RR,i}f_{\RR,i}\ket{n} (1- \bra{n}f^{\dagger}_{\RR,j}f_{\RR,j}\ket{n})
   e^{- \beta (E_n - E_{gs}) } I(\omega,\omega',\omega',E_s-E_n,E_s-E_n,E_s-E_n) \bigg]
\end{align} 
}
where the explicit expressions for the six terms are given in \cref{eq:Fij_no1,eq:Fij_no2,eq:Fij_no3,eq:Fij_no4,eq:Fij_no5,eq:Fij_no6}.
In these expressions, $\Delta_{\pm1}=E(n_f\pm1)-E(n_f)$ denote the energies of charge $\pm 1$ excitations and $\Delta_{\pm2}=E(n_f\pm2)-E(n_f)$ denote the energies of charge $\pm 2$ excitations.
In the zero temperature limit at exact integer filling, they satisfy $\Delta_{+1}\approx \Delta _{-1} = U_1/2$, $\Delta_{+2}\approx \Delta _{-2} = 2U_1$.

We now sum these contributions:
\begin{align}
&   \frac{1}{(i\omega'+ \Delta_{-1})(i\omega + \Delta_{-1})}\bigg\{ \frac{1 }{\beta} \frac{n_f(N_f- 2 n_f+ 1)}{ N_f(N_f-1)} (\frac{1}{ i\omega + \Delta_{-1}}  +\frac{1}{i\omega' + \Delta_{-1}})  +  \frac{n_f(n_f-1)}{N_f(N_f-1)}   -  \frac{n_f(N_f- n_f)}{ N_f(N_f-1)} \delta_{\omega\omega'} \bigg\}  \nonumber \\  
 &+ \frac{1}{ (i\omega'- \Delta_{+1})(i\omega- \Delta_{+1})}\bigg\{ \frac{1}{\beta} \frac{ (N_f - n_f)(N_f - 2 n_f-1)}{N_f(N_f-1)} (\frac{1}{i\omega- \Delta_{+1}}+ \frac{1}{i\omega'- \Delta_{+1}})   - \frac{n_f(N_f- n_f)}{N_f(N_f-1)} \delta_{\omega \omega' } \nonumber\\
 &\qquad\qquad+  \frac{(N_f - n_f)(N_f - n_f-1)}{N_f (N_f-1)} \bigg\}  \nonumber \\ 
 &+ 
  \frac{(N_f- n_f)n_f}{ N_f(N_f-1)}\frac{1}{(i\omega'- \Delta_{+1})(  i \omega'+ \Delta_{-1})}\bigg[ \frac{1}{ \beta ( i\omega- \Delta_{+1}) } -  \frac{ 1}{\beta(  i \omega + \Delta_{-1})}  +  \delta_{ \omega \omega'} \bigg]  \nonumber\\ 
  &+  \frac{(N_f- n_f)(N_f - n_f -1) }{N_f(N_f-1)} \frac{1}{\beta(\Delta_{+2} - i \omega-i\omega')} 
  \bigg[ \frac{1}{(\Delta_{+1} - i \omega')^2 } +   \frac{1}{(\Delta_{+1}- i \omega')  (\Delta_{+1} - i \omega)}     \bigg] \nonumber\\ 
   &+ \frac{n_f(N_f- n_f)}{ N_f(N_f -1)}   \frac{1}{(\Delta_{+1}- i \omega')(\Delta_{-1}  +i\omega)} \bigg [\frac{1}{\beta} \bigg( \frac{1}{\Delta_{-1}+i\omega}  + \frac{1}{\Delta_{+1} -  i \omega'}\bigg)   - 1 \bigg]    \nonumber \\ 
&+
\frac{(N_f - n_f)(N_f- n_f-1)}{ N_f (N_f-1)} \frac{1}{\beta(\Delta_{+2}- \Delta_{+1} - i \omega')} \bigg[ \frac{1}{ i\omega - \Delta_{+1}} \frac{1}{  i\omega- \Delta_{+1} }\nonumber\\
&\qquad\qquad+ \frac{1}{\Delta_{+2} - \Delta_{+1} - i \omega'} \left( \frac{1}{\Delta_{+2} -i\omega-   i \omega'}  +  \frac{1}{ i\omega - \Delta_{+1}} \right) \bigg]    \nonumber \\ 
&+ \frac{n_f(n_f-1)}{ N_f (N_f-1)} \frac{1} {\beta(\Delta_{-2} +i\omega+   i \omega')} \left( \frac{1}{   i \omega'+ \Delta_{-1}}       +  \frac{1}{i \omega+ \Delta_{-1}}  \right)^2  \nonumber \\ 
&+ \frac{n_f(N_f- n_f)}{N_f (N_f-1)} \frac{1}{({\Delta_{+1}}-i \omega')({\Delta_{-1}}+i \omega')}\bigg[ \frac{1}{ \beta  } \bigg( \frac{1 }{{\Delta_{-1}}+i \omega}  +\frac{1}{{\Delta_{+1}}-i \omega}\bigg)    -  \delta_{\omega' \omega} \bigg]   \nonumber \\ 
   & + \frac{(N_f- n_f)(N_f- n_f -1)}{ N_f(N_f-1)} \frac{1}{\beta(\Delta_{+2} - \Delta_{+1} -i \omega')}\bigg[ \frac{1 }{  i \omega'- \Delta_{+1}  }  \frac{1 }{- \Delta_{+1} +i \omega} \nonumber\\
   &\qquad\qquad+\frac{1}{\Delta_{+2}- \Delta_{+1} -i \omega}\left( \frac{1}{\Delta_{+2} -i \omega'- i \omega}    +   \frac{1 }{  i \omega'- \Delta_{+1}  }    \right) \bigg]  \nonumber \\
   & + \frac{n_f(N_f - n_f)}{ N_f(N_f-1)} \frac{1}{(- \Delta_{+1} +i \omega)(\Delta_{-1}+ i \omega')}\bigg[\frac{1}{\beta} \bigg( \frac{1}{- \Delta_{+1} +i \omega  } -\frac{1}{\Delta_{-1}+ i \omega'} \bigg)   + 1\bigg]   \nonumber \\ & 
   \label{eq:sum_over_Fij_start}
\end{align}

After further algebra, this becomes
\begin{align}
&   \frac{1}{(i\omega'+ \Delta_{-1})(i\omega + \Delta_{-1})}\bigg[ \frac{1 }{\beta} \frac{n_f(N_f- 2 n_f+ 1)}{ N_f(N_f-1)} \bigg(\frac{1}{ i\omega + \Delta_{-1}}  +\frac{1}{i\omega' + \Delta_{-1}}\bigg)  +  \frac{n_f(n_f-1)}{N_f(N_f-1)}   -  \frac{n_f(N_f- n_f)}{ N_f(N_f-1)} \delta_{\omega\omega'} \bigg]  \nonumber \\  
 &+ \frac{1}{ (i\omega'- \Delta_{+1})(i\omega- \Delta_{+1})}\bigg[ \frac{1}{\beta} \frac{ (N_f - n_f)(N_f - 2 n_f-1)}{N_f(N_f-1)} \bigg(\frac{1}{i\omega- \Delta_{+1}}+ \frac{1}{i\omega'- \Delta_{+1}}\bigg)   - \frac{n_f(N_f- n_f)}{N_f(N_f-1)} \delta_{\omega \omega' } \nonumber\\
 &\qquad\qquad+  \frac{(N_f - n_f)(N_f - n_f-1)}{N_f (N_f-1)} \bigg] \nonumber \\ 
 &+ 
  \frac{(N_f- n_f)n_f}{ N_f(N_f-1)}\frac{1}{(i\omega'- \Delta_{+1})(  i \omega'+ \Delta_{-1})}\bigg[ \frac{1}{ \beta ( i\omega- \Delta_{+1}) } -  \frac{ 1}{\beta(  i \omega + \Delta_{-1})}  +  \delta_{ \omega \omega'} \bigg]  \nonumber\\ 
  &+  \frac{(N_f- n_f)(N_f - n_f -1) }{N_f(N_f-1)} \frac{1}{\beta(\Delta_{+2} - i \omega-i\omega')} \bigg[ \frac{1}{(\Delta_{+1} - i \omega')^2 } +   \frac{1}{(\Delta_{+1}- i \omega')  (\Delta_{+1} - i \omega)}     \bigg] \nonumber\\ 
   &+ \frac{n_f(N_f- n_f)}{ N_f(N_f -1)}   \frac{1}{(\Delta_{+1}- i \omega')(\Delta_{-1}  +i\omega)} \bigg [\frac{1}{\beta} \bigg( \frac{1}{\Delta_{-1}+i\omega}  + \frac{1}{\Delta_{+1} -  i \omega'}\bigg)   - 1 \bigg]   \nonumber \\ 
&+
\frac{(N_f - n_f)(N_f- n_f-1)}{ N_f (N_f-1)} \frac{1}{\beta(\Delta_{+2}- \Delta_{+1} - i \omega')}  \frac{1}{ i\omega - \Delta_{+1}} \bigg(  \frac{1}{  i\omega- \Delta_{+1} }+  \frac{1}{\Delta_{+2} -i\omega-   i \omega'}    \bigg)    \nonumber \\ 
&+ \frac{n_f(n_f-1)}{ N_f (N_f-1)} \frac{1} {\beta(\Delta_{-2} +i\omega+   i \omega')} \left( \frac{1}{   i \omega'+ \Delta_{-1}}       +  \frac{1}{i \omega+ \Delta_{-1}}  \right)^2  \nonumber \\ 
&+ \frac{n_f(N_f- n_f)}{N_f (N_f-1)} \frac{1}{({\Delta_{+1}}-i \omega')({\Delta_{-1}}+i \omega')}\bigg[ \frac{1}{ \beta  } \bigg( \frac{1 }{{\Delta_{-1}}+i \omega}  +\frac{1}{{\Delta_{+1}}-i \omega}\bigg)    -  \delta_{\omega' \omega} \bigg]   \nonumber \\ 
   & + \frac{(N_f- n_f)(N_f- n_f -1)}{ N_f(N_f-1)} \frac{1}{\beta(\Delta_{+2} - \Delta_{+1} -i \omega')}     \frac{1 }{  i \omega'- \Delta_{+1}  }  \bigg(   \frac{1 }{- \Delta_{+1} +i \omega} + \frac{1}{\Delta_{+2} -i \omega'- i \omega}   \bigg)   \nonumber \\ 
   & + \frac{n_f(N_f - n_f)}{ N_f(N_f-1)} \frac{1}{(- \Delta_{+1} +i \omega)(\Delta_{-1}+ i \omega')}\bigg[\frac{1}{\beta} \bigg( \frac{1}{- \Delta_{+1} +i \omega  } -\frac{1}{\Delta_{-1}+ i \omega'} \bigg)   + 1\bigg]   \nonumber \\ 
   =&   \frac{1}{(i\omega'+ \Delta_{-1})(i\omega + \Delta_{-1})}\bigg[ \frac{1 }{\beta} \frac{n_f(N_f- 2 n_f+ 1)}{ N_f(N_f-1)} \bigg(\frac{1}{ i\omega + \Delta_{-1}}  +\frac{1}{i\omega' + \Delta_{-1}}\bigg)  +  \frac{n_f(n_f-1)}{N_f(N_f-1)}   -  \frac{n_f(N_f- n_f)}{ N_f(N_f-1)} \delta_{\omega\omega'} \bigg] \nonumber \\  
 &+ \frac{1}{ (i\omega'- \Delta_{+1})(i\omega- \Delta_{+1})}\bigg[ \frac{1}{\beta} \frac{ (N_f - n_f)(N_f - 2 n_f-1)}{N_f(N_f-1)} \bigg(\frac{1}{i\omega- \Delta_{+1}}+ \frac{1}{i\omega'- \Delta_{+1}}\bigg)   - \frac{n_f(N_f- n_f)}{N_f(N_f-1)} \delta_{\omega \omega' } \nonumber\\
 &\qquad\qquad+  \frac{(N_f - n_f)(N_f - n_f-1)}{N_f (N_f-1)} \bigg]  \nonumber \\ 
 &+ 
  \frac{(N_f- n_f)n_f}{ N_f(N_f-1)}\frac{1}{(i\omega'- \Delta_{+1})(  i \omega'+ \Delta_{-1})}\bigg[ \frac{1}{ \beta ( i\omega- \Delta_{+1}) } -  \frac{ 1}{\beta(  i \omega + \Delta_{-1})}  +  \delta_{ \omega \omega'} \bigg]  \nonumber\\ 
  &+  \frac{(N_f- n_f)(N_f - n_f -1) }{N_f(N_f-1)} \frac{1}{\beta(\Delta_{+2} - i \omega-i\omega')} \left( \frac{1}{(\Delta_{+1} - i \omega') } + \frac{1}{(\Delta_{+1} - i \omega) } \right)^2 \nonumber\\ 
   &+ \frac{n_f(N_f- n_f)}{ N_f(N_f -1)}   \frac{1}{(\Delta_{+1}- i \omega')(\Delta_{-1}  +i\omega)} \bigg [\frac{1}{\beta} \bigg( \frac{1}{\Delta_{-1}+i\omega}  + \frac{1}{\Delta_{+1} -  i \omega'}\bigg)   - 1 \bigg]    \nonumber \\   
&+ \frac{n_f(n_f-1)}{ N_f (N_f-1)} \frac{1} {\beta(\Delta_{-2} +i\omega+   i \omega')} \left( \frac{1}{   i \omega'+ \Delta_{-1}}       +  \frac{1}{i \omega+ \Delta_{-1}}  \right)^2  \nonumber \\ 
&+ \frac{n_f(N_f- n_f)}{N_f (N_f-1)} \frac{1}{({\Delta_{+1}}-i \omega')({\Delta_{-1}}+i \omega')}\bigg[ \frac{1}{ \beta  } \bigg( \frac{1 }{{\Delta_{-1}}+i \omega}  +\frac{1}{{\Delta_{+1}}-i \omega}\bigg)    -  \delta_{\omega' \omega} \bigg]      \nonumber \\ 
& + \frac{n_f(N_f - n_f)}{ N_f(N_f-1)} \frac{1}{(- \Delta_{+1} +i \omega)(\Delta_{-1}+ i \omega')}\bigg[\frac{1}{\beta} \bigg( \frac{1}{i \omega - \Delta_{+1}  } -\frac{1}{ i \omega'+ \Delta_{-1}} \bigg)   + 1\bigg] 
\end{align}

After further simplification,
\begin{align}
     & \frac{1}{(i\omega'+ \Delta_{-1})(i\omega + \Delta_{-1})}\bigg[ \frac{1 }{\beta} \frac{n_f(N_f- 2 n_f+ 1)}{ N_f(N_f-1)} \bigg(\frac{1}{ i\omega + \Delta_{-1}}  +\frac{1}{i\omega' + \Delta_{-1}}\bigg)  +  \frac{n_f(n_f-1)}{N_f(N_f-1)}   -  \frac{n_f(N_f- n_f)}{ N_f(N_f-1)} \delta_{\omega\omega'} \bigg] \nonumber \\  
 &+ \frac{1}{ (i\omega'- \Delta_{+1})(i\omega- \Delta_{+1})}\bigg[ \frac{1}{\beta} \frac{ (N_f - n_f)(N_f - 2 n_f-1)}{N_f(N_f-1)} \bigg(\frac{1}{i\omega- \Delta_{+1}}+ \frac{1}{i\omega'- \Delta_{+1}}\bigg)   - \frac{n_f(N_f- n_f)}{N_f(N_f-1)} \delta_{\omega \omega' } \nonumber\\
 &\qquad\qquad+  \frac{(N_f - n_f)(N_f - n_f-1)}{N_f (N_f-1)} \bigg]  \nonumber \\ 
 &+
  \frac{n_f(N_f- n_f)}{ N_f(N_f-1)}\frac{1}{(i\omega'- \Delta_{+1})(  i \omega'+ \Delta_{-1})}\bigg[ \frac{1}{ \beta ( i\omega- \Delta_{+1}) } -  \frac{ 1}{\beta(  i \omega + \Delta_{-1})}  +  \delta_{ \omega \omega'} \bigg]  
    \nonumber\\ 
   &+ \frac{n_f(N_f- n_f)}{ N_f(N_f -1)}   \frac{-1}{( i \omega'- \Delta_{+1})( i\omega + \Delta_{-1} )} \bigg [\frac{1}{\beta} \bigg( \frac{1}{\Delta_{-1}+i\omega}  + \frac{1}{\Delta_{+1} -  i \omega'}\bigg)   - 1 \bigg]    \nonumber \\ 
&+ \frac{n_f(N_f- n_f)}{N_f (N_f-1)} \frac{-1}{(i \omega'- \Delta_{+1})(i \omega'+ \Delta_{-1})}\bigg[ \frac{1}{ \beta  } \bigg( \frac{1 }{i \omega{+} \Delta_{-1}}  -\frac{1}{i \omega- \Delta_{+1}}\bigg)    -  \delta_{\omega' \omega} \bigg]      \nonumber \\ 
& + \frac{n_f(N_f - n_f)}{ N_f(N_f-1)} \frac{1}{(i \omega - \Delta_{+1} )( i \omega'+\Delta_{-1})}\bigg[\frac{1}{\beta} \bigg( \frac{1}{i \omega - \Delta_{+1}  } -\frac{1}{ i \omega'+ \Delta_{-1}} \bigg)   + 1\bigg]  \nonumber\\ 
  &+  \frac{(N_f- n_f)(N_f - n_f -1) }{N_f(N_f-1)} \frac{1}{\beta(\Delta_{+2} - i \omega-i\omega')} \left( \frac{1}{ i \omega'- \Delta_{+1}  } + \frac{1}{  i \omega- \Delta_{+1} } \right)^2     \nonumber \\ 
  &+ \frac{n_f(n_f-1)}{ N_f (N_f-1)} \frac{1} {\beta(\Delta_{-2} +i\omega+   i \omega')} \left( \frac{1}{   i \omega'+ \Delta_{-1}}       +  \frac{1}{i \omega+ \Delta_{-1}}  \right)^2   \nonumber \\ 
 =& \frac{1}{(i\omega'+ \Delta_{-1})(i\omega + \Delta_{-1})}\bigg[ \frac{1 }{\beta} \frac{n_f(N_f- 2 n_f+ 1)}{ N_f(N_f-1)} \bigg(\frac{1}{ i\omega + \Delta_{-1}}  +\frac{1}{i\omega' + \Delta_{-1}}\bigg)  +  \frac{n_f(n_f-1)}{N_f(N_f-1)}   -  \frac{n_f(N_f- n_f)}{ N_f(N_f-1)} \delta_{\omega\omega'} \bigg] \nonumber \\  
 &+ \frac{1}{ (i\omega'- \Delta_{+1})(i\omega- \Delta_{+1})}\bigg[ \frac{1}{\beta} \frac{ (N_f - n_f)(N_f - 2 n_f-1)}{N_f(N_f-1)} \bigg(\frac{1}{i\omega- \Delta_{+1}}+ \frac{1}{i\omega'- \Delta_{+1}}\bigg)   - \frac{n_f(N_f- n_f)}{N_f(N_f-1)} \delta_{\omega \omega' } \nonumber\\
 &\qquad\qquad+  \frac{(N_f - n_f)(N_f - n_f-1)}{N_f (N_f-1)} \bigg]  \nonumber \\ 
 & +
  \frac{n_f(N_f- n_f)}{ N_f(N_f-1)}\frac{1}{(i\omega'- \Delta_{+1})(  i \omega'+ \Delta_{-1})}\bigg[ \frac{1}{\beta} \bigg( \frac{2}{  i\omega- \Delta_{+1} } -  \frac{ 1}{  i \omega + \Delta_{-1}}  -  \frac{1 }{i \omega + \Delta_{-1}}   \bigg) +  2 \delta_{ \omega \omega'}      \bigg]   
    \nonumber\\ 
   &+ \frac{n_f(N_f- n_f)}{ N_f(N_f -1)} \bigg\{  \frac{-1}{( i \omega'- \Delta_{+1})( i\omega + \Delta_{-1} )} \bigg [\frac{1}{\beta}  \frac{i \omega'  -  i\omega- \Delta_{+1}- \Delta_{-1}}{(i\omega+ \Delta_{-1})( i \omega'- \Delta_{+1} )}   - 1 \bigg]   \nonumber\\
   &\qquad\qquad+ \frac{1}{(i \omega - \Delta_{+1} )( i \omega'+\Delta_{-1})}\bigg[\frac{1}{\beta}  \frac{i \omega'- i \omega + \Delta_{-1} + \Delta_{+1} }{(i \omega - \Delta_{+1}) ( i \omega'+ \Delta_{-1})}    + 1\bigg]  \bigg \}  \nonumber\\ 
  &+  \frac{(N_f- n_f)(N_f - n_f -1) }{N_f(N_f-1)} \frac{1}{\beta(\Delta_{+2} - i \omega-i\omega')} \left( \frac{1}{ i \omega'- \Delta_{+1}  } + \frac{1}{  i \omega- \Delta_{+1} } \right)^2    \nonumber \\ 
  &+ \frac{n_f(n_f-1)}{ N_f (N_f-1)} \frac{1} {\beta(\Delta_{-2} +i\omega+   i \omega')} \left( \frac{1}{   i \omega'+ \Delta_{-1}}       +  \frac{1}{i \omega+ \Delta_{-1}}  \right)^2 
\end{align}

We simplify this further:
\begin{align}
   &  \frac{1}{(i\omega'+ \Delta_{-1})(i\omega + \Delta_{-1})}\bigg[ \frac{1 }{\beta} \frac{n_f(N_f- 2 n_f+ 1)}{ N_f(N_f-1)} \bigg(\frac{1}{ i\omega + \Delta_{-1}}  +\frac{1}{i\omega' + \Delta_{-1}}\bigg)  +  \frac{n_f(n_f-1)}{N_f(N_f-1)}   -  \frac{n_f(N_f- n_f)}{ N_f(N_f-1)} \delta_{\omega\omega'} \bigg] \nonumber \\  
 &+ \frac{1}{ (i\omega'- \Delta_{+1})(i\omega- \Delta_{+1})}\bigg[ \frac{1}{\beta} \frac{ (N_f - n_f)(N_f - 2 n_f-1)}{N_f(N_f-1)} \bigg(\frac{1}{i\omega- \Delta_{+1}}+ \frac{1}{i\omega'- \Delta_{+1}}\bigg)   - \frac{n_f(N_f- n_f)}{N_f(N_f-1)} \delta_{\omega \omega' } \nonumber\\
 &\qquad\qquad+  \frac{(N_f - n_f)(N_f - n_f-1)}{N_f (N_f-1)} \bigg]  \nonumber \\ 
 & + 
  \frac{n_f(N_f- n_f)}{ N_f(N_f-1)} \bigg\{ \frac{1}{(i\omega'- \Delta_{+1})(  i \omega'+ \Delta_{-1})}\bigg[ \frac{1}{\beta} \bigg( \frac{2}{  i\omega- \Delta_{+1} } -  \frac{ 1}{ i \omega + \Delta_{-1}}  -  \frac{1 }{i \omega + \Delta_{-1}}   \bigg) +  2 \delta_{ \omega \omega'}      \bigg]   
    \nonumber\\ 
   & - \frac{1}{\beta}  \frac{i \omega'  -  i\omega- \Delta_{+1}- \Delta_{-1}}{(i\omega+ \Delta_{-1})^2( i \omega'- \Delta_{+1} )^2}    +\frac{1}{\beta}  \frac{i \omega'- i \omega + \Delta_{-1} + \Delta_{+1} }{(i \omega - \Delta_{+1})^2 ( i \omega'+ \Delta_{-1})^2}      +  \frac{1}{( i \omega'- \Delta_{+1})( i\omega + \Delta_{-1} )} 
   +  \frac{1}{(i \omega - \Delta_{+1} )( i \omega'+\Delta_{-1})}   \bigg \}  \nonumber\\ 
  &+  \frac{(N_f- n_f)(N_f - n_f -1) }{N_f(N_f-1)} \frac{1}{\beta(\Delta_{+2} - i \omega-i\omega')} \left( \frac{1}{ i \omega'- \Delta_{+1}  } + \frac{1}{  i \omega- \Delta_{+1} } \right)^2    + \nonumber \\ &+ \frac{n_f(n_f-1)}{ N_f (N_f-1)} \frac{1} {\beta(\Delta_{-2} +i\omega+   i \omega')} \left( \frac{1}{   i \omega'+ \Delta_{-1}}       +  \frac{1}{i \omega+ \Delta_{-1}}  \right)^2 
\end{align}

We now separate the terms proportional to $1/\beta$ from the $\beta$-independent terms.
First, the constant terms are
\begin{align}
   &  \frac{1}{(i\omega'+ \Delta_{-1})(i\omega + \Delta_{-1})}\bigg[ \frac{n_f(n_f-1)}{N_f(N_f-1)}   -  \frac{n_f(N_f- n_f)}{ N_f(N_f-1)} \delta_{\omega\omega'} \bigg] \nonumber\\ 
   & +  \frac{1}{ (i\omega'- \Delta_{+1})(i\omega- \Delta_{+1})}\bigg[   - \frac{n_f(N_f- n_f)}{N_f(N_f-1)} \delta_{\omega \omega' } +  \frac{(N_f - n_f)(N_f - n_f-1)}{N_f (N_f-1)} \bigg] + \nonumber \\ 
   & + 
  \frac{n_f(N_f- n_f)}{ N_f(N_f-1)} \bigg[ \frac{1}{(i\omega'- \Delta_{+1})(  i \omega'+ \Delta_{-1})} 2 \delta_{ \omega \omega'}         +  \frac{1}{( i \omega'- \Delta_{+1})( i\omega + \Delta_{-1} )} 
   +  \frac{1}{(i \omega - \Delta_{+1} )( i \omega'+\Delta_{-1})}   \bigg ] \nonumber \\ 
=&  \frac{n_f(n_f-1)}{N_f(N_f-1)} \frac{1}{(i\omega'+ \Delta_{-1})(i\omega + \Delta_{-1})}    +  \frac{(N_f - n_f)(N_f - n_f-1)}{N_f (N_f-1)} \frac{1}{ (i\omega'- \Delta_{+1})(i\omega- \Delta_{+1})}  \nonumber \\ 
& + 
  \frac{n_f(N_f- n_f)}{ N_f(N_f-1)} \bigg[ \left( \frac{2}{(i\omega'- \Delta_{+1})(  i \omega'+ \Delta_{-1})}   -  \frac{1}{(i\omega'+ \Delta_{-1})(i\omega + \Delta_{-1})}     -  \frac{1}{ (i\omega'- \Delta_{+1})(i\omega- \Delta_{+1})}   \right) \delta_{\omega \omega' }     \nonumber\\ 
  &   +  \frac{1}{( i \omega'- \Delta_{+1})( i\omega + \Delta_{-1} )} 
  + \frac{1}{(i \omega - \Delta_{+1} )( i \omega'+\Delta_{-1})}   \bigg ]   \nonumber \\
=& \frac{n_f(n_f-1)}{N_f(N_f-1)} \frac{1}{(i\omega'+ \Delta_{-1})(i\omega + \Delta_{-1})}    +  \frac{(N_f - n_f)(N_f - n_f-1)}{N_f (N_f-1)} \frac{1}{ (i\omega'- \Delta_{+1})(i\omega- \Delta_{+1})}  \nonumber \\ 
& + 
  \frac{n_f(N_f- n_f)}{ N_f(N_f-1)} \frac{1}{{(i \omega - \Delta_{+1} )(i\omega'+ \Delta_{-1}) ( i \omega'- \Delta_{+1})(i\omega + \Delta_{-1})}  }  \nonumber \\ 
  & \times \bigg[     2 i \omega' i\omega + (i \omega' + i \omega) (\Delta_{-1}   - \Delta_{+1} ) - 2 \Delta_{+1}  \Delta_{-1}   -(\Delta_{-1} + \Delta_{+1})^2 \delta_{\omega \omega' }  
  \bigg ]
  \label{eq:sum_Iij0}
\end{align}
Notice that this expression is invariant under $n_f \rightarrow N_f - n_f $ together with $\Delta_{+1}  \leftrightarrow - \Delta_{-1}$, which is a statement of the particle-hole symmetry in this problem.

For the terms proportional to $1/\beta$, we obtain:
\begin{align}
   & \frac{1 }{\beta} \frac{n_f(N_f- 2 n_f+ 1)}{ N_f(N_f-1)} \frac{i\omega' + i\omega + 2\Delta_{-1}  }{ (i\omega + \Delta_{-1})(i\omega' + \Delta_{-1})(i\omega'+ \Delta_{-1})(i\omega + \Delta_{-1})}   \nonumber \\  
 &+ \frac{1}{\beta} \frac{ (N_f - n_f)(N_f - 2 n_f-1)}{N_f(N_f-1)} \frac{i\omega'+ i\omega- 2\Delta_{+1}}{(i\omega- \Delta_{+1})(i\omega'- \Delta_{+1}) (i\omega'- \Delta_{+1})(i\omega- \Delta_{+1})}   \nonumber \\ 
 & + 
  \frac{n_f(N_f- n_f)}{ N_f(N_f-1)} \bigg[ \frac{2}{(i\omega'- \Delta_{+1})(  i \omega'+ \Delta_{-1})} \frac{1}{\beta} \bigg( \frac{1}{  i\omega- \Delta_{+1} } -  \frac{ 1}{ i \omega + \Delta_{-1}}    \bigg)    
    \nonumber\\  
& \qquad\qquad- \frac{1}{\beta}  \frac{i \omega'  -  i\omega- \Delta_{+1}- \Delta_{-1}}{(i\omega+ \Delta_{-1})^2( i \omega'- \Delta_{+1} )^2}    +\frac{1}{\beta}  \frac{i \omega'- i \omega + \Delta_{-1} + \Delta_{+1} }{(i \omega - \Delta_{+1})^2 ( i \omega'+ \Delta_{-1})^2}     \bigg ]  \nonumber\\   
&+  \frac{(N_f- n_f)(N_f - n_f -1) }{N_f(N_f-1)} \frac{1}{\beta(\Delta_{+2} - i \omega-i\omega')} \left( \frac{1}{ i \omega'- \Delta_{+1}  } + \frac{1}{  i \omega- \Delta_{+1} } \right)^2    + \nonumber \\ &+ \frac{n_f(n_f-1)}{ N_f (N_f-1)} \frac{1} {\beta(\Delta_{-2} +i\omega+   i \omega')} \left( \frac{1}{   i \omega'+ \Delta_{-1}}       +  \frac{1}{i \omega+ \Delta_{-1}}  \right)^2 
\end{align}
which are further simplified as
\begin{align}
   & \frac{1 }{\beta} \bigg\{\frac{n_f(N_f- 2 n_f+ 1)}{ N_f(N_f-1)} \frac{i\omega' + i\omega + 2\Delta_{-1}  }{ (i\omega + \Delta_{-1})^2(i\omega' + \Delta_{-1})^2} + \frac{ (N_f - n_f)(N_f - 2 n_f-1)}{N_f(N_f-1)} \frac{i\omega'+ i\omega- 2\Delta_{+1}}{(i\omega- \Delta_{+1})^2(i\omega'- \Delta_{+1})^2 }  \nonumber \\ 
   & + 
  \frac{n_f(N_f- n_f)}{ N_f(N_f-1)} \bigg[ \frac{2(\Delta_{-1}+ \Delta_{+1}) }{(i\omega'- \Delta_{+1})(  i \omega'+ \Delta_{-1})( i\omega- \Delta_{+1}) (  i \omega + \Delta_{-1})}     -   \frac{i \omega'  -  i\omega- \Delta_{+1}- \Delta_{-1}}{(i\omega+ \Delta_{-1})^2( i \omega'- \Delta_{+1} )^2}   \nonumber\\
  &\qquad\qquad+ \frac{i \omega'- i \omega + \Delta_{-1} + \Delta_{+1} }{(i \omega - \Delta_{+1})^2 ( i \omega'+ \Delta_{-1})^2}     \bigg ]  \nonumber\\   
  &+  \frac{(N_f- n_f)(N_f - n_f -1) }{N_f(N_f-1)} \frac{i \omega+ i \omega'-2 \Delta_{+1}}{(\Delta_{+2} - i \omega-i\omega')} \frac{ i \omega+ i \omega'-2 \Delta_{+1} }{ (i \omega'- \Delta_{+1} )^2(  i \omega- \Delta_{+1} )^2}    \nonumber\\
  &+  \frac{n_f(n_f-1)}{ N_f (N_f-1)} \frac{i \omega+  i \omega'+ 2\Delta_{-1} } {(i\omega+   i \omega'+ \Delta_{-2} )}  \frac{i \omega+  i \omega'+ 2\Delta_{-1}  }{  ( i \omega'+ \Delta_{-1})^2(i \omega+ \Delta_{-1})^2 }   \bigg\}\nonumber\\
  = & \frac{1 }{\beta} \bigg\{\frac{n_f}{ N_f(N_f-1)} \frac{i\omega' + i\omega + 2\Delta_{-1}  }{ (i\omega + \Delta_{-1})^2(i\omega' + \Delta_{-1})^2} \left[ (N_f- 2 n_f+ 1) +   (n_f-1) \frac{i \omega+  i \omega'+ 2\Delta_{-1} } {i\omega+   i \omega'+ \Delta_{-2} }  \right]
   \nonumber \\
   &+\frac{ (N_f - n_f)}{N_f(N_f-1)} \frac{i\omega'+ i\omega- 2\Delta_{+1}}{(i\omega- \Delta_{+1})^2(i\omega'- \Delta_{+1})^2 } \left[ (N_f - 2 n_f-1) +   (N_f - n_f -1)  \frac{i \omega+ i \omega'-2 \Delta_{+1}}{\Delta_{+2} - i \omega-i\omega'} \right]     \nonumber \\
   & + 
  \frac{n_f(N_f- n_f)}{ N_f(N_f-1)} \bigg[ \frac{2(\Delta_{-1}+ \Delta_{+1}) }{(i\omega'- \Delta_{+1})(  i \omega'+ \Delta_{-1})( i\omega- \Delta_{+1}) (  i \omega + \Delta_{-1})}     -   \frac{i \omega'  -  i\omega- \Delta_{+1}- \Delta_{-1}}{(i\omega+ \Delta_{-1})^2( i \omega'- \Delta_{+1} )^2}    \nonumber\\
  &\qquad\qquad+ \frac{i \omega'- i \omega + \Delta_{-1} + \Delta_{+1} }{(i \omega - \Delta_{+1})^2 ( i \omega'+ \Delta_{-1})^2}     \bigg ]   \bigg\}\nonumber\\
   =& \frac{1 }{\beta} \bigg\{\frac{n_f}{ N_f(N_f-1)} \frac{i\omega' + i\omega + 2\Delta_{-1}  }{ (i\omega + \Delta_{-1})^2(i\omega' + \Delta_{-1})^2} \left[ (N_f- 2 n_f+ 1) +   (n_f-1) \frac{i \omega+  i \omega'+ 2\Delta_{-1} } {i\omega+   i \omega'+ \Delta_{-2} }  \right]
   \nonumber \\ 
   &+\frac{ (N_f - n_f)}{N_f(N_f-1)} \frac{i\omega'+ i\omega- 2\Delta_{+1}}{(i\omega- \Delta_{+1})^2(i\omega'- \Delta_{+1})^2 } \left[ (N_f - 2 n_f-1) +   (N_f - n_f -1)  \frac{i \omega+ i \omega'-2 \Delta_{+1}}{\Delta_{+2} - i \omega-i\omega'} \right]    \nonumber \\ 
   & + 
  \frac{n_f(N_f- n_f)}{ N_f(N_f-1)} \frac{1}{(i\omega'- \Delta_{+1})^2(  i \omega'+ \Delta_{-1})^2( i\omega- \Delta_{+1})^2 (  i \omega + \Delta_{-1})^2}  \nonumber \\ 
  &\times \bigg[ 2(\Delta_{-1}+ \Delta_{+1}) (i\omega'- \Delta_{+1})(  i \omega'+ \Delta_{-1})( i\omega- \Delta_{+1}) (  i \omega + \Delta_{-1}) \nonumber\\
  &\qquad\qquad-(i \omega'  -  i\omega- \Delta_{+1}- \Delta_{-1})(  i \omega'+ \Delta_{-1})^2( i\omega- \Delta_{+1})^2   \nonumber \\ 
  & \qquad\qquad+ (i \omega'- i \omega + \Delta_{-1} + \Delta_{+1}) (i\omega + \Delta_{-1})^2 (i\omega'- \Delta_{+1})^2        \bigg ]   \bigg\}
\end{align}

\begin{align}
   & \frac{1 }{\beta} \bigg\{\frac{n_f}{ N_f(N_f-1)} \frac{i\omega' + i\omega + 2\Delta_{-1}  }{ (i\omega + \Delta_{-1})^2(i\omega' + \Delta_{-1})^2} \left[ (N_f- 2 n_f+ 1) +   (n_f-1) \frac{i \omega+  i \omega'+ 2\Delta_{-1} } {i\omega+   i \omega'+ \Delta_{-2} }  \right] 
   \nonumber \\ 
   &+\frac{ (N_f - n_f)}{N_f(N_f-1)} \frac{i\omega'+ i\omega- 2\Delta_{+1}}{(i\omega- \Delta_{+1})^2(i\omega'- \Delta_{+1})^2 } \left[ (N_f - 2 n_f-1) +   (N_f - n_f -1)  \frac{i \omega+ i \omega'-2 \Delta_{+1}}{\Delta_{+2} - i \omega-i\omega'} \right]    \nonumber \\ 
   & + 
  \frac{n_f(N_f- n_f)}{ N_f(N_f-1)} \frac{\Delta_{-1} + \Delta_{+1} }{(i\omega'- \Delta_{+1})^2(  i \omega'+ \Delta_{-1})^2( i\omega- \Delta_{+1})^2 (  i \omega + \Delta_{-1})^2}  \nonumber \\ &
  \times \bigg[  \bigg(  2 i\omega  i\omega' +(\Delta_{-1}- \Delta_{+1}) (i \omega + i \omega')- 2 \Delta_{-1}  \Delta_{+1} \bigg) \bigg((i\omega)^2 +(i \omega')^2    +(\Delta_{-1} - \Delta_{+1} )(i\omega + i\omega' ) -
   2 \Delta_{-1} \Delta_{+1} \bigg)  \bigg ]   \bigg\}
\end{align}

\begin{align}
   & \frac{1 }{\beta} \bigg\{\frac{n_f(n_f-1)}{ N_f(N_f-1)} \frac{i\omega' + i\omega + 2\Delta_{-1}  }{ (i\omega + \Delta_{-1})^2(i\omega' + \Delta_{-1})^2} \left[  -1 +    \frac{i \omega+  i \omega'+ 2\Delta_{-1} } {i\omega+   i \omega'+ \Delta_{-2} }  \right]
   \nonumber \\ 
   &+\frac{ (N_f - n_f)  (N_f -  n_f-1)}{N_f(N_f-1)} \frac{i\omega'+ i\omega- 2\Delta_{+1}}{(i\omega- \Delta_{+1})^2(i\omega'- \Delta_{+1})^2 } \left[  1 +     \frac{i \omega+ i \omega'-2 \Delta_{+1}}{\Delta_{+2} - i \omega-i\omega'} \right]    \nonumber \\ 
   & + 
  \frac{n_f(N_f- n_f)}{ N_f(N_f-1)} \frac{\Delta_{-1} + \Delta_{+1} }{(i\omega'- \Delta_{+1})^2(  i \omega'+ \Delta_{-1})^2( i\omega- \Delta_{+1})^2 (  i \omega + \Delta_{-1})^2}  \nonumber \\ 
  &\times \bigg[  \bigg(  2 i\omega  i\omega' +(\Delta_{-1}- \Delta_{+1}) (i \omega + i \omega')- 2 \Delta_{-1}  \Delta_{+1} \bigg) \bigg((i\omega)^2 +(i \omega')^2    +(\Delta_{-1} - \Delta_{+1} )(i\omega + i\omega' ) -
   2 \Delta_{-1} \Delta_{+1} \bigg)  \bigg ] \nonumber \\  
   &+ \frac{n_f(N_f-n_f)}{ N_f(N_f-1)} \frac{i\omega' + i\omega + 2\Delta_{-1}  }{ (i\omega + \Delta_{-1})^2(i\omega' + \Delta_{-1})^2} - \frac{ n_f (N_f - n_f)}{N_f(N_f-1)} \frac{i\omega'+ i\omega- 2\Delta_{+1}}{(i\omega- \Delta_{+1})^2(i\omega'- \Delta_{+1})^2 }  \bigg\} 
\end{align}

\begin{align}
   & \frac{1 }{\beta} \bigg\{\frac{n_f(n_f-1)}{ N_f(N_f-1)} \frac{i\omega' + i\omega + 2\Delta_{-1}  }{ (i\omega + \Delta_{-1})^2(i\omega' + \Delta_{-1})^2}     \frac{ 2\Delta_{-1} -\Delta_{-2} } {i\omega+   i \omega'+ \Delta_{-2} } \nonumber\\
   &+\frac{ (N_f - n_f)  (N_f -  n_f-1)}{N_f(N_f-1)} \frac{i\omega'+ i\omega- 2\Delta_{+1}}{(i\omega- \Delta_{+1})^2(i\omega'- \Delta_{+1})^2 }  \frac{\Delta_{+2} -2 \Delta_{+1}}{\Delta_{+2} - i \omega-i\omega'}   \nonumber \\ 
   & + 
  \frac{n_f(N_f- n_f)}{ N_f(N_f-1)} \frac{1 }{(i\omega'- \Delta_{+1})^2(  i \omega'+ \Delta_{-1})^2( i\omega- \Delta_{+1})^2 (  i \omega + \Delta_{-1})^2}   \bigg[ (\Delta_{-1} + \Delta_{+1})\nonumber\\
  &\times\bigg(  2 i\omega  i\omega' +(\Delta_{-1}- \Delta_{+1}) (i \omega + i \omega')- 2 \Delta_{-1}  \Delta_{+1} \bigg) \bigg((i\omega)^2 +(i \omega')^2    +(\Delta_{-1} - \Delta_{+1} )(i\omega + i\omega' ) -
   2 \Delta_{-1} \Delta_{+1} \bigg)    \nonumber \\  
   &+  (i\omega' + i\omega + 2\Delta_{-1})(i\omega- \Delta_{+1})^2(i\omega'- \Delta_{+1})^2   - (i\omega'+ i\omega- 2\Delta_{+1}) (i\omega + \Delta_{-1})^2(i\omega' + \Delta_{-1})^2\bigg ] \bigg\}
\end{align}

\begin{align}
   & \frac{1 }{\beta} \bigg\{\frac{n_f(n_f-1)}{ N_f(N_f-1)} \frac{i\omega' + i\omega + 2\Delta_{-1}  }{ (i\omega + \Delta_{-1})^2(i\omega' + \Delta_{-1})^2}     \frac{ 2\Delta_{-1} -\Delta_{-2} } {i\omega+   i \omega'+ \Delta_{-2} } \nonumber\\
   &+\frac{ (N_f - n_f)  (N_f -  n_f-1)}{N_f(N_f-1)} \frac{i\omega'+ i\omega- 2\Delta_{+1}}{(i\omega- \Delta_{+1})^2(i\omega'- \Delta_{+1})^2 }  \frac{\Delta_{+2} -2 \Delta_{+1}}{\Delta_{+2} - i \omega-i\omega'}   \nonumber \\ 
   & + 
  \frac{n_f(N_f- n_f)}{ N_f(N_f-1)} \frac{\Delta_{-1} + \Delta_{+1} }{(i\omega'- \Delta_{+1})^2(  i \omega'+ \Delta_{-1})^2( i\omega- \Delta_{+1})^2 (  i \omega + \Delta_{-1})^2}  \nonumber \\ 
  &\times \bigg[   - 2 (i \omega)^2 (i \omega')^2 +  
 2 (i \omega) (i \omega') (\Delta_{+1} - \Delta_{-1}) (i \omega + i \omega') - (i \omega + i \omega')^2 ( 
   \Delta_{+1}^2 + \Delta_{-1}^2) \nonumber\\
   &\qquad-  (i \omega + i \omega') (\Delta_{-1} - \Delta_{+1}) (\Delta_{-1}^2 + \Delta_{+1}^2)  + 
 4 (i \omega) (i \omega') \Delta_{-1}  \Delta_{+1} + 2 \Delta_{-1} \Delta_{+1} (\Delta_{-1}^2 + \Delta_{-1} \Delta_{+1} + \Delta_{+1}^2)  \bigg ] \bigg\}
\end{align}

Finally, using $\Delta_{+1} = \Delta_{-1}= \Delta_1$ and $\Delta_{+2} = \Delta_{-2} = \Delta_2$ we simplify the above expression to
\begin{align}
   & \frac{1 }{\beta} \bigg\{\frac{n_f(n_f-1)}{ N_f(N_f-1)} \frac{i\omega' + i\omega + 2\Delta_1  }{ (i\omega + \Delta_1)^2(i\omega' + \Delta_1)^2}     \frac{ 2\Delta_1 -\Delta_{2} } {i\omega+   i \omega'+ \Delta_{2} } +\frac{ (N_f - n_f)  (N_f -  n_f-1)}{N_f(N_f-1)} \frac{i\omega'+ i\omega- 2\Delta_1}{(i\omega- \Delta_1)^2(i\omega'- \Delta_1)^2 }  \frac{ 2 \Delta_1- \Delta_{2}}{ i \omega+ i\omega'- \Delta_{2} }   \nonumber \\
   & -
  \frac{n_f(N_f- n_f)}{ N_f(N_f-1)} \frac{4\Delta_1 }{(i\omega'- \Delta_1)^2(  i \omega'+ \Delta_1)^2( i\omega- \Delta_1)^2 (  i \omega + \Delta_1)^2} \bigg[     (i \omega)^2 (i \omega')^2   +( (i \omega)^2+ (i \omega')^2)  
   \Delta_1^2  - 3 \Delta_1^4  \bigg ] \bigg\}
   \label{eq:sum_Iij_beta}
\end{align}

}

\subsection{Self-energy from the hybridization expansion}

We are now in a position to evaluate the self-energy within the hybridization expansion. 
At order $\gamma^2$, the self-energy is related to $F^{ij}(i\omega,i\omega',i\omega')$ through \cref{eq:F_to_Sigma_1} 
\ba 
\Sigma_{f,ij}^{(1)}(i\omega)= 
 \gamma^2 \delta_{i,j}\sum_{i\omega',i'}
\Delta(i\omega')F^{ii'}(i\omega,i\omega',i\omega') \frac{1}{[G_{f,loc}(i\omega)]^2} 
\label{eq:Sigma1_proof_start}
\ea 
We first perform the sum over $i'$.
Using the explicit expressions of $F^{ii}(i\omega,i\omega',i\omega')$ and $F^{i\neq j}(i\omega,i\omega',i\omega')$ obtained in \cref{eq:F_ii_final,eq:F_ij_final}, 
we find
\begin{align}
     \sum_{i'}F^{ii'}(i\omega,i\omega',i\omega') =& F^{ii}(i\omega,i\omega',i\omega')+ \sum_{i'\ne i }F^{ii'}(i\omega,i\omega',i\omega') \nonumber \\  
     =& {\frac{n_f(N_f-n_f)}{N_f^2}}
\frac{(\Delta_{-1}+\Delta_{+1})^2 }{(i\omega + \Delta_{-1})(i\omega -\Delta_{+1})(i\omega' - \Delta_{+1})(i\omega'+\Delta_{-1})}(-1+\delta_{\omega ,\omega'})  \nonumber \\ 
&+   \frac{n_f(N_f- n_f)}{ N_f^2} \frac{  (\Delta_{+1}+  \Delta_{-1} )^2  (1+ N_f \delta_{\omega, \omega'})
  }{{(i \omega - \Delta_{+1} )(i\omega'+ \Delta_{-1}) ( i \omega'- \Delta_{+1})(i\omega + \Delta_{-1})}  }   - (N_f-1)  I^1_{i\ne j}(i\omega,i\omega',i\omega')  \nonumber \\ 
  =& {\frac{n_f(N_f-n_f)}{N_f^2}}
\frac{(\Delta_{-1}+\Delta_{+1})^2 }{(i\omega + \Delta_{-1})^2(i\omega -\Delta_{+1})^2} (N_f+1)\delta_{\omega ,\omega'}    - (N_f-1)  I^1_{i\neq j}(i\omega,i\omega',i\omega').
\end{align}
The local Green's function $G_{f,loc}(i\omega)$ is given by (\cref{eq:atomic_Green_fun})
\begin{align}
    &G_{f,loc}(i\omega) = \frac{i \omega + \Delta_{-1} - \frac{n_f}{N_f}(\Delta_{+1}+ \Delta_{-1})  }{ (i \omega + \Delta_{-1})(i \omega - \Delta_{+1}) }   
\end{align}

{f
Therefore, 
\begin{align}
   &   \sum_{i'}F^{ii'}(i\omega,i\omega',i\omega') \frac{1}{[G_{f,loc}(i\omega)]^2} \nonumber \\ 
= &  {\frac{n_f(N_f-n_f)}{N_f^2}}
\frac{(\Delta_{-1}+\Delta_{+1})^2 (N_f+1)\delta_{\omega ,\omega'}   }{\left[i \omega + \Delta_{-1} - \frac{n_f}{N_f}(\Delta_{+1}+ \Delta_{-1})\right]^2}   -  \frac{ (N_f-1)  I^1_{i\ne j}(i\omega,i\omega',i\omega') (i \omega + \Delta_{-1})^2(i \omega - \Delta_{+1})^2 }{\left[ i \omega + \Delta_{-1} - \frac{n_f}{N_f}(\Delta_{+1}+ \Delta_{-1})  \right]^2}
 \end{align} 
We now define another notation for the $1/\beta$ term
\begin{align}
    & L^1(i\omega,i\omega') = \frac{ (N_f-1)  I^1_{i\ne j}(i\omega,i\omega',i\omega') (i \omega + \Delta_{-1})^2(i \omega - \Delta_{+1})^2 }{\left[ i \omega + \Delta_{-1} - \frac{n_f}{N_f}(\Delta_{+1}+ \Delta_{-1})  \right]^2}
    \label{eq:L1_def}
\end{align}
where the numerator is
\begin{align}
   &(N_f-1) I^{1}_{i\ne j}(\omega,\omega',\omega')  (i \omega + \Delta_{-1})^2(i \omega - \Delta_{+1})^2  \nonumber\\
   =& \frac{1 }{\beta} \bigg\{\frac{n_f(n_f-1)}{ N_f} \frac{(i \omega - \Delta_{1})^2 (i\omega' + i\omega + 2\Delta_1 ) }{ (i\omega' + \Delta_1)^2}     \frac{ 2\Delta_1 -\Delta_{2} } {i\omega+   i \omega'+ \Delta_{2} } \nonumber \\ 
   &+ \frac{ (N_f - n_f)  (N_f -  n_f-1)}{N_f} \frac{ (i \omega + \Delta_{1})^2(i\omega'+ i\omega- 2\Delta_1)}{(i\omega'- \Delta_1)^2 }  \frac{ 2 \Delta_1- \Delta_{2}}{ i \omega+ i\omega'- \Delta_{2} }  \nonumber \\
   & -
  \frac{n_f(N_f- n_f)}{ N_f} \frac{4\Delta_1 }{(i\omega'- \Delta_1)^2(  i \omega'+ \Delta_1)^2} \bigg[     (i \omega)^2 (i \omega')^2   +( (i \omega)^2+ (i \omega')^2)  
   \Delta_1^2  - 3 \Delta_1^4  \bigg ] \bigg\}\nonumber\\
\end{align} 
To proceed, we now substitute $\Delta_2= 4 \Delta_1$

\begin{align}
   &(N_f-1) I^{1}_{i\ne j}(\omega,\omega',\omega')  (i \omega + \Delta_{-1})^2(i \omega - \Delta_{+1})^2  \nonumber\\
   =& -\frac{1 }{\beta} \frac{2\Delta_1 }{ (i\omega'- \Delta_1)^2(  i \omega'+ \Delta_1)^2}  \bigg\{\frac{n_f(n_f-1)}{ N_f}  (i\omega'- \Delta_1)^2 (i \omega - \Delta_{1})^2     \frac{ (i\omega' + i\omega + 2\Delta_1 )  } {i\omega+   i \omega'+ 4 \Delta_1 }  \nonumber \\
   &+  \frac{ (N_f - n_f)  (N_f -  n_f-1)}{N_f}  (  i \omega'+ \Delta_1)^2 (i \omega + \Delta_{1})^2 \frac{ (i\omega'+ i\omega- 2\Delta_1)}{ i \omega+ i\omega'- 4\Delta_{1} } \nonumber \\
   & +
2  \frac{n_f(N_f- n_f)}{ N_f} \bigg[     (i \omega)^2 (i \omega')^2   +( (i \omega)^2+ (i \omega')^2)  
   \Delta_1^2  - 3 \Delta_1^4  \bigg ] \bigg\}\nonumber\\
\end{align}

After further simplification, we obtain

\begin{align}
   &(N_f-1) I^{1}_{i\ne j}(\omega,\omega',\omega')  (i \omega + \Delta_{-1})^2(i \omega - \Delta_{+1})^2  \nonumber\\
   =& -\frac{1}{N_f} \frac{1 }{\beta} \frac{2\Delta_1 }{ (i\omega'- \Delta_1)^2(  i \omega'+ \Delta_1)^2(i\omega+   i \omega'+ 4 \Delta_1 ) (i\omega+   i \omega'- 4 \Delta_1 )}  \nonumber 
   \\ & \times \bigg\{n_f(n_f-1) (i\omega'- \Delta_1)^2 (i \omega - \Delta_{1})^2      (i\omega' + i\omega + 2\Delta_1 ) (i\omega+   i \omega'- 4 \Delta_1  )  \nonumber \\ 
   &\qquad+ (N_f - n_f)  (N_f -  n_f-1)  (  i \omega'+ \Delta_1)^2 (i \omega + \Delta_{1})^2 (i\omega'+ i\omega- 2\Delta_1)(i\omega+   i \omega'+ 4 \Delta_1 )  \nonumber \\
   & \qquad+
2  n_f(N_f- n_f) \bigg[     (i \omega)^2 (i \omega')^2   +( (i \omega)^2+ (i \omega')^2)  
   \Delta_1^2  - 3 \Delta_1^4  \bigg ](i\omega+   i \omega'- 4 \Delta_1 )(i\omega+   i \omega'+ 4 \Delta_1 )  \bigg\}
\end{align}
Using the results above, we obtain the self-energy to order $\gamma^2$
\ba 
\Sigma_{f,ij}^{(1)}(i\omega)= 
 \gamma^2 \delta_{i,j} \Delta(i\omega) {\frac{n_f(N_f-n_f)}{N_f^2}}
\frac{(\Delta_{-1}+\Delta_{+1})^2 (N_f+1)   }{\left[i \omega + \Delta_{-1} - \frac{n_f}{N_f}(\Delta_{+1}+ \Delta_{-1})\right]^2}  - \gamma^2 \delta_{i,j} 
 \sum_{i\omega'}
\Delta(i\omega')L^1(i\omega,i\omega') 
\label{eq:Sigma1_final}
\ea

}

Notice that the local self-energy (\cref{eq:atomic-self-energy}) is
\ba  \Sigma_{f,loc}(i\omega)
=& \frac{n_f(N_f-n_f)U_1^2}{N_f^2 }
\bigg[ 
\frac{1}{i\omega 
- \frac{n_f\Delta_{+1} - (N_f-n_f)\Delta_{-1} }{N_f}
}
\bigg]
\label{eq:Sigma_Hubbard_I}
\ea 
Thus, if we ignore the $1/\beta$ contribution to the self-energy, the total self-energy up to the order of $\gamma^2$ behaves as 
\ba 
\label{eq:Sigma_hyb_exp}
\Sigma_{f,ij}(i\omega,\kk) 
\approx & \delta_{i,j}\Sigma_{f,loc}(i\omega)
+\Sigma_{f,ij}^{(1)}(i\omega)  \nonumber\\ 
=  & \delta_{i,j} \frac{(N_f-n_f)n_fU_1^2}{N_f^2}\frac{1}{
i\omega - \frac{n_f\Delta_{+1} - (N_f-n_f)\Delta_{-1}}{N_f}
}
\bigg[1
+ {\frac{\gamma^2N_f}{U_1^2 }}\Delta(i\omega) \frac{{(\Delta_{-1}+\Delta_{+1})^2} + {\frac{(\Delta_{+1}+\Delta_{-1})^2}{N_f}}
}{ 
i\omega - \frac{n_f\Delta_{+1} - (N_f-n_f)\Delta_{-1}}{N_f}
}
\bigg] \nonumber\\
=&\delta_{i,j} \Sigma_{f,loc}(i\omega)
\bigg[1
+ {\frac{\gamma^2N_f}{U_1^2 }}\Delta(i\omega) \frac{{(\Delta_{-1}+\Delta_{+1})^2} + {\frac{(\Delta_{+1}+\Delta_{-1})^2}{N_f}}
}{ 
i\omega - \frac{n_f\Delta_{+1} - (N_f-n_f)\Delta_{-1}}{N_f}
}
\bigg] \nonumber 
\\ =& { \delta_{i,j} \Sigma_{f,loc}(i\omega)
\bigg[1
+ {\frac{\gamma^2(N_f+1) }{U_1^2 }}\Delta(i\omega) \frac{(\Delta_{+1}+\Delta_{-1})^2}{ 
i\omega - \frac{n_f\Delta_{+1} - (N_f-n_f)\Delta_{-1}}{N_f}
} \bigg] }
\ea 

\subsection{Hybridization function}
\label{app:hybridization_func}
The self-energy derived in the previous subsection depends explicitly on the hybridization function $\Delta(i\omega)$. In this subsection, we derive an analytic expression for $\Delta(i\omega)$.

In \cref{eq:HcHcf}, $H^{(c,\eta)}$ and $H^{(cf,\eta)}$ are defined as
\begin{align}
    H^{(c,\eta)}(\kk)=&\begin{pmatrix}
        \sigma_0(\epsilon_{c,1}-\mu) & v_{\star}(\eta k_x\sigma_0+ik_y \sigma_z)\\
        v_{\star}(\eta k_x\sigma_0-ik_y\sigma_z) & \sigma_0(\epsilon_{c,2}-\mu)+M\sigma_x
    \end{pmatrix}\nonumber\\
    H^{(cf,\eta)}(\kk) =& 
  \begin{pmatrix}
    \gamma \sigma_0 + v_\star' (\eta k_x \sigma_x + k_y \sigma_y) \\
    0_{2\times2}
  \end{pmatrix}
\end{align}
where (\cref{eq:hartree-energy})
\begin{align}
    \epsilon_{c,1}=&\frac{V(0)}{\Omega_0}\nu_c+W_1\nu_f\\
    \epsilon_{c,2}=&\frac{V(0)}{\Omega_0}\nu_c+W_3\nu_f-\frac{J}{8}\nu_f
\end{align}
Based on the analytic expression of the hybridization function \cref{eq:def_hyb_k}, we analyze the analytic behaviors of the self-energy. 
To obtain a closed analytic form, we make the following approximations:
1) We set $W_1=W_3$ and ignore the HF contribution of $H_J$ to the $c$ electrons, which yields $\epsilon_{c,1}=\epsilon_{c,2}$ (Note that the HF contribution of $H_J$ to the $f$ electrons is still kept).
2) We focus on the integer fillings $\nu=0,\pm 1, \pm 2$ where, in the zero-hybridization limit, we have $\nu_f =n_f-N_f/2= \nu$ and $\nu_c=0$. 
This will fix the chemical potential to $\mu=\epsilon_{c,1}=\epsilon_{c,2}$.
3) We will take the "chiral limit U($4$) symmetry " of the THF model with $v_{\star}'=0$ (note, however, that the parameters used in the final comparison are chosen \emph{away} from the chiral limit, to match the BM model \cite{PhysRevLett.129.047601}).

Under the approximations, the mean-field Hamiltonian for $c$ electrons reduces to 
\ba 
H_c = \sum_{\kk,\eta s}
\begin{bmatrix}
    c_{\kk,1 \eta s}^\dag 
    & c_{\kk,2\eta s}^\dag
    & c_{\kk,3\eta s}^\dag
    & c_{\kk,4\eta s}^\dag
\end{bmatrix}
\begin{bmatrix}
    & & v_\star(\eta k_x +ik_y) \\
    & & & v_\star(\eta k_x -ik_y)\\
    v_\star(\eta k_x-ik_y) & & & M\\
    & v_\star(\eta k_x+ik_y) & M
\end{bmatrix}
\begin{bmatrix}
    c_{\kk,1 \eta s}\\
    c_{\kk,2\eta s}\\
    c_{\kk,3\eta s}\\ 
    c_{\kk,4\eta s}
\end{bmatrix}
\ea 
The corresponding eigenvalues and eigenvectors are 
\ba 
&E^{\eta,c}_{\kk,n=1,2,3,4} = 
\frac{-M-\sqrt{4|v_\star\kk|^2+M^2}}{2}
,
\frac{M-\sqrt{4|v_\star\kk|^2+M^2}}{2}
,
\frac{-M+\sqrt{4|v_\star\kk|^2+M^2}}{2}
,
\frac{M+\sqrt{4|v_\star\kk|^2+M^2}}{2} \nonumber\\
&U^{\eta,c}_{\kk,an=1}=A_{1,\kk}\begin{bmatrix}
    \frac{-M+\sqrt{4|v_\star\kk|^2+M^2}}{2}v_\star (\eta k_x +ik_y)
    &
    \frac{M-\sqrt{4|v_\star\kk|^2+M^2}}{2}v_\star (\eta k_x -ik_y)
    &-|v_\star \kk|^2 
    & |v_\star \kk|^2 
\end{bmatrix}_a\nonumber\\
&U^{\eta,c}_{\kk,an=2}=A_{2,\kk}\begin{bmatrix}
    \frac{-M-\sqrt{4|v_\star\kk|^2+M^2}}{2}v_\star (\eta k_x +ik_y)
    &
    \frac{-M-\sqrt{4|v_\star\kk|^2+M^2}}{2}v_\star (\eta k_x -ik_y)
    &|v_\star \kk|^2 
    & |v_\star \kk|^2 
\end{bmatrix}_a\nonumber\\
&U^{\eta,c}_{\kk,an=3}=A_{2,\kk}\begin{bmatrix}
    \frac{-M-\sqrt{4|v_\star\kk|^2+M^2}}{2}v_\star (\eta k_x +ik_y)
    &
    \frac{M+\sqrt{4|v_\star\kk|^2+M^2}}{2}v_\star (\eta k_x -ik_y)
    &- |v_\star \kk|^2 
    & |v_\star \kk|^2 
\end{bmatrix}_a\nonumber\\
&U^{\eta,c}_{\kk,an=4}=A_{1,\kk}\begin{bmatrix}
    \frac{-M+\sqrt{4|v_\star\kk|^2+M^2}}{2}v_\star (\eta k_x +ik_y)
    &
    \frac{-M+\sqrt{4|v_\star\kk|^2+M^2}}{2}v_\star (\eta k_x -ik_y)
    &|v_\star \kk|^2 
    & |v_\star \kk|^2 
\end{bmatrix}_a
\label{eq:Hc_eigs_evcs}
\ea 
where the normalization factors are
\ba 
A_{1,\kk}= \frac{1}{|v_\star\kk| \sqrt{M^2+4|v_\star\kk|^2-M\sqrt{M^2+4|v_\star\kk|^2}}} \nonumber\\ 
A_{2,\kk}= \frac{1}{|v_\star\kk| \sqrt{M^2+4|v_\star\kk|^2+M\sqrt{M^2+4|v_\star\kk|^2}}} 
\label{eq:Hc_evcs_normalize}
\ea 
To emphasize the dependence on the parameter $M$, we denote the corresponding hybridization function by $\Delta_M(i\omega)$.
Using \cref{eq:def_hyb_k,eq:def_hyb_loc}m], the hybridization function reads 
\ba
\gamma^2\Delta_M(i\omega) 
=&\frac{1}{N_M}\sum_{\kk}\sum_{n,a_1a_2}[H^{(fc,\eta)}(\kk)]_{\alpha a_1}
U^{\eta,c}_{\kk,a_1n} \frac{1}{i\omega - E^{\eta,c}_{\kk,n}}
U^{\eta, c,*}_{\kk,a_2n} [H^{(fc,\eta)}(\kk)]^*_{\alpha a_2}\bigg|_{\alpha=1\;\mathrm{or}\;2}\nonumber\\
=&\gamma^2\frac{1}{N_{M}}\sum_{\kk}\sum_{n}\frac{|U^{\eta,c}_{\kk,\alpha n}|^2}{i\omega-E^{\eta,c}_{\kk,n}}\bigg|_{\alpha=1\;\mathrm{or}\;2}  \nonumber\\
= &\gamma^2 
\frac{1}{N_M}\sum_{\kk}
\bigg[ 
|A_{1,\kk}|^2
\bigg(\frac{-M+\sqrt{4|v_\star\kk|^2+M^2}}{2} \bigg)^2|v_\star\kk|^2 
\bigg( 
\frac{1}{i\omega - E^{\eta,c}_{\kk,n=1}}
+\frac{1}{i\omega -E^{\eta,c}_{\kk,n=4}}\bigg)
\nonumber\\
&
+|A_{2,\kk}|^2
\bigg(\frac{M+\sqrt{4|v_\star\kk|^2+M^2}}{2} \bigg)^2|v_\star\kk|^2 
\bigg( 
\frac{1}{i\omega - E^{\eta,c}_{\kk,n=2}}
+\frac{1}{i\omega -E^{\eta,c}_{\kk,n=3}}\bigg)
\bigg] \nonumber\\
=&\gamma^2\frac{1}{N_{M}}\sum_{\kk}\bigg[\frac{1}{4}\frac{(\sqrt{4|v_\star\kk|^2+M^2}-M)^2}{\sqrt{4|v_\star\kk|^2+M^2}(\sqrt{4|v_\star\kk|^2+M^2}-M)}\bigg( 
\frac{1}{i\omega - E^{\eta,c}_{\kk,n=1}}
+\frac{1}{i\omega -E^{\eta,c}_{\kk,n=4}}\bigg)\nonumber\\
&+\frac{1}{4}\frac{(\sqrt{4|v_\star\kk|^2+M^2}+M)^2}{\sqrt{4|v_\star\kk|^2+M^2}(\sqrt{4|v_\star\kk|^2+M^2}+M)}\bigg( 
\frac{1}{i\omega - E^{\eta,c}_{\kk,n=2}}
+\frac{1}{i\omega -E^{\eta,c}_{\kk,n=3}}\bigg)
\bigg]\nonumber\\
=&\gamma^2\frac{1}{N_M}\sum_{\kk}\bigg[
\frac{1}{4}\bigg(1-\frac{M}{\sqrt{M^2+4|v_\star\kk|^2}}\bigg)\bigg( 
\frac{1}{i\omega - E^{\eta,c}_{\kk,n=1}}
+\frac{1}{i\omega -E^{\eta,c}_{\kk,n=4}}\bigg)\nonumber\\
&+\frac{1}{4}\bigg(1+\frac{M}{\sqrt{M^2+4|v_\star\kk|^2}}\bigg)\bigg( 
\frac{1}{i\omega - E^{\eta,c}_{\kk,n=2}}
+\frac{1}{i\omega -E^{\eta,c}_{\kk,n=3}}\bigg)
\bigg]
\ea 
where the restriction $|_{\alpha=1\;\mathrm{or}\;2}$ indicates that only the $\alpha=1,2$ components couple directly to the $f$ orbitals.
Note that the $c$-electron spectrum is particle-hole symmetric at integer fillings of the entire system in the absence of $fc$ hybridization $\gamma$, i.e. $E^{\eta,c}_{\kk,n=1}=-E^{\eta,c}_{\kk,n=4}$ and $E^{\eta,c}_{\kk,n=2}=-E^{\eta,c}_{\kk,n=3}$. At these fillings, the $c$ electron is half filled, and the heavy $f$ electrons control the charge number.  
Using $\mathrm{Im}[\frac{1}{\omega-i0^+-E}] = \pi \delta(\omega-E)$, we introduce the following spectral density associated with the hybridization function 
\ba 
\label{eq:hyb_fun_spec_fun_def}
\rho_c^{hyb}(\epsilon) \equiv 
&\frac{1}{\pi}\mathrm{Im}\bigg[\Delta_M(i\omega\rightarrow \epsilon-i0^+)\bigg] 
\nonumber\\ 
=&\frac{\gamma^2}{\Omega_M} \int_{|\kk|<\Lambda_c}
\bigg[ 
\frac{1}{4}\bigg(1-\frac{M}{\sqrt{M^2+4|v_\star\kk|^2}}\bigg) [\delta(\epsilon- E^{\eta,c}_{\kk,n=1})+\delta(\epsilon -E^{\eta,c}_{\kk,n=4})] \nonumber\\
&+ 
\frac{1}{4}\bigg(1+\frac{M}{\sqrt{M^2+4|v_\star\kk|^2}}\bigg) [\delta(\epsilon- E^{\eta,c}_{\kk,n=2})+\delta(\epsilon -E^{\eta,c}_{\kk,n=3})]
\bigg]d^2k \nonumber\\
=&\frac{2\pi\gamma^2}{\Omega_M}
\int_0^{\Lambda_c}
\frac{1}{4} \bigg[
\bigg(1- \frac{M}{\sqrt{M^2+4|v_\star k|^2}}\bigg)
\delta( |\epsilon| - \frac{M+\sqrt{4|v_\star k|^2+M^2}}{2})\nonumber\\
&+
\bigg(1+ \frac{M}{\sqrt{M^2+4|v_\star k|^2}}\bigg)
\delta( |\epsilon| - \frac{-M+\sqrt{4|v_\star k|^2+M^2}}{2})
\bigg] kdk 
\ea 
To evaluate the first delta-function contribution, we introduce the variable $u=\frac{M+\sqrt{4|v_\star k|^2+M^2}}{2}\in[M,\frac{M+\sqrt{4|v_\star \Lambda_c|^2+M^2}}{2}]$, then $\sqrt{4|v_\star k|^2+M^2}=2u-M$, $4|v_{\star}|^2k^2=(2u-M)^2-M^2$ and $8|v_{\star}|^2 kdk=4(2u-M)du$.
There the integral is evaluated as
\begin{align}
    &\int_{0}^{\Lambda_c}\bigg(1- \frac{M}{\sqrt{M^2+4|v_\star k|^2}}\bigg)
\delta( |\epsilon| - \frac{M+\sqrt{4|v_\star k|^2+M^2}}{2}) kdk\nonumber\\
=&\int_{M}^{\frac{M+\sqrt{4|v_{\star}\Lambda_c|^2+M^2}}{2}}(1-\frac{M}{2u-M})\delta(|\epsilon|-u)\frac{(2u-M)du}{2|v_{\star}|^2}\nonumber\\
=&\frac{1}{|v_{\star}|^2}\int_{M}^{\frac{M+\sqrt{4|v_{\star}\Lambda_c|^2+M^2}}{2}}\delta(|\epsilon|-u)(u-M)du\nonumber\\
=&\frac{1}{|v_{\star}|^2}(|\epsilon|-M)\theta(|\epsilon|-M)\theta(\frac{M+\sqrt{4|v_{\star}\Lambda_c|^2+M^2}}{2}-|\epsilon|)
\label{eq:delta_func_integration}
\end{align}
Similarly, for the term involving $\delta( |\epsilon| - \frac{-M+\sqrt{4|v_\star k|^2+M^2}}{2})$, one simply replaces $M$ by $-M$ in \cref{eq:delta_func_integration}.
Substituting \cref{eq:delta_func_integration} and the $M\to -M$ counterpart into \cref{eq:hyb_fun_spec_fun_def}, we obtain
\begin{align}
    \rho_c^{hyb}(\epsilon)
    =& \frac{\pi\gamma^2}{2|v_\star|^2\Omega_M}
\bigg[ (
|\epsilon|-M)\theta(|\epsilon|-M)\theta( 
\frac{M+\sqrt{4|v_\star\Lambda_c|^2+M^2}}{2}
-|\epsilon|)
+ (|\epsilon|+M)
\theta(\frac{\sqrt{4|v_\star\Lambda_c|^2+M^2}-M}{2}-|\epsilon|)
\bigg] \nonumber\\
\approx & 
 \frac{\pi\gamma^2}{2|v_\star|^2\Omega_M}
\bigg[ (
|\epsilon|-M)\theta(|\epsilon|-M)
+ (|\epsilon|+M)
\bigg]\theta( |v_\star\Lambda_c|
-|\epsilon|) \nonumber\\
=&\frac{\pi\gamma^2}{2|v_\star|^2\Omega_M}
\bigg[ (
|\epsilon|-M)(1-\theta(M-|\epsilon|))
+ (|\epsilon|+M)
\bigg]\theta( |v_\star\Lambda_c|
-|\epsilon|) \nonumber\\
=& \frac{\pi \gamma^2}{|v_\star|^2\Omega_M}
|\epsilon| \theta(|v_\star\Lambda_c|-|\epsilon|)
+ (M-|\epsilon|)\frac{\pi \gamma^2}{2|v_\star|^2\Omega_M} \theta(M-|\epsilon|)
\label{eq:hyb_fun_spec_fun_form}
\end{align}
In the last step, we have used the fact that $|v_\star\Lambda_c|\gg M$. 
Thus, $\theta( 
\frac{\sqrt{4|v_\star\Lambda_c|^2+M^2}\pm M}{2}
-|\epsilon|)\approx \theta(|v_\star\Lambda_c|-|\epsilon|)$ and $\theta(|v_\star\Lambda_c|-|\epsilon|)\theta(M-|\epsilon|)=\theta(M-|\epsilon|)$.

The hybridization function can then be reconstructed from its spectral density through (note that $\mathrm{Im}[\frac{1}{\omega-i0^+-E}] = \pi \delta(\omega-E)$)
\ba 
\gamma^2 \Delta_M(i\omega) =&
\int_{\epsilon}\rho^{hyb}_c(\epsilon) \frac{d\epsilon}{i\omega-\epsilon}
\label{eq:Delta_rho_relation}
\ea 

using \cref{eq:hyb_fun_spec_fun_form}, we obtain
\begin{align}
\label{eq:self_energy_in_f_M}
   \gamma^2 \Delta_M(i\omega) =&\int \bigg[\frac{\pi \gamma^2}{|v_\star|^2\Omega_M}
|\epsilon| \theta(|v_\star\Lambda_c|-|\epsilon|)
+ (M-|\epsilon|)\frac{\pi \gamma^2}{2|v_\star|^2\Omega_M} \theta(M-|\epsilon|)\bigg]\frac{d\epsilon}{i\omega-\epsilon}\nonumber\\
=&\frac{\pi \gamma^2}{|v_\star|^2\Omega_M}\int_{-|v_{\star}\Lambda_c|}^{|v_{\star}\Lambda_c|}\frac{|\epsilon|d\epsilon}{i\omega-\epsilon}
+\frac{\pi \gamma^2}{2|v_\star|^2\Omega_M}\int_{-M}^{M}\frac{(M-|\epsilon|)d\epsilon}{i\omega-\epsilon}\nonumber\\
=&\frac{\pi \gamma^2}{|v_\star|^2\Omega_M}\int_{0}^{|v_{\star}\Lambda_c|}d\epsilon\bigg(\frac{\epsilon}{i\omega-\epsilon}+\frac{\epsilon}{i\omega+\epsilon}\bigg)
+\frac{\pi \gamma^2}{2|v_\star|^2\Omega_M}\int_{0}^{M}d\epsilon\bigg(\frac{M-\epsilon}{i\omega-\epsilon}+\frac{M-\epsilon}{i\omega+\epsilon}\bigg)\nonumber\\
=&\frac{\pi \gamma^2}{|v_\star|^2\Omega_M}[-i\omega\log(\epsilon-i\omega)-i\omega\log(\epsilon+i\omega)]\bigg|_{\epsilon=0}^{|v_{\star}\Lambda_c|}\nonumber\\
&+\frac{\pi \gamma^2}{2|v_\star|^2\Omega_M}[(i\omega-M)\log(\epsilon-i\omega)+(i\omega+M)\log(\epsilon+i\omega)]\bigg|_{\epsilon=0}^{M}\nonumber\\
=&-i\omega \frac{\pi\gamma^2}{|v_\star^2|\Omega_M}
\log\bigg( \frac{|v_\star\Lambda_c|^2 +\omega^2}{\omega^2} 
\bigg) +
{\color{black}\frac{\pi \gamma^2}{2|v_\star|^2\Omega_M} M \log\bigg( 
\frac{i\omega +M }{i\omega-M}\bigg)}+
{i\omega \frac{\pi\gamma^2}{2|v_\star^2|\Omega_M}
\log\bigg( \frac{M^2 +\omega^2}{\omega^2} 
\bigg)}
\end{align}
For later convenience, we define 
\ba 
\label{eq:def_fM}
f_M(i\omega)
 = \log\bigg( \frac{|v_\star\Lambda_c|^2+\omega^2}{\omega^2}\bigg)
 - {\color{black}
 {\frac{M}{2i\omega }\log\bigg( \frac{i\omega+M}{i\omega-M}
 \bigg)}}
 {-\frac{1}{2}\log\bigg(\frac{M^2+\omega^2}{\omega^2}\bigg)}
\ea 
 such that 
\ba 
\gamma^2\Delta_M(i\omega)
= -i\omega \kappa \pi f_M(i\omega) 
\ea 
where we introduced the dimensionless parameter
\begin{equation}
    \kappa=\frac{\gamma^2}{|v_\star|^2\Omega_M}
\end{equation}
to measure the hybridization strength.

After analytic continuation $i\omega\to \omega+i0^+$, the real-frequency expression reads 
\ba 
\gamma^2 \Delta_M(\omega+i0^+) = -\omega \kappa \pi f_M(\omega+i0^+)
\ea 
with 
{
\begin{align}
\label{eq:fM_real_freq}
    f_{M}(i\omega\to \omega+i0^+)=&\log\bigg[\frac{|v_{\star}\Lambda_c|^2-(\omega+i0^+)^2}{-(\omega+i0^+)^2}\bigg]-\frac{M}{2(\omega+i0^+)}\log\bigg[\frac{(\omega+i0^+)+M}{(\omega+i0^+)-M}\bigg]-\frac{1}{2}\log\bigg[\frac{M^2-(\omega+i0^+)^2}{-(\omega+i0^+)^2}\bigg]\nonumber\\
    =&\log\bigg[\frac{\omega^2-|v_{\star}\Lambda_c|^2}{\omega^2}+i\mathrm{sgn}(\omega)0^+\bigg]-\frac{M}{2\omega}\log\bigg(\frac{\omega+M}{\omega-M}-i0^+\bigg)-\frac{1}{2}\log\bigg[\frac{\omega^2-M^2}{\omega^2}+i\mathrm{sgn}(\omega)0^+\bigg]\nonumber\\
    =&\log\bigg(\bigg|\frac{|v_{\star}\Lambda_c|^2-\omega^2}{\omega^2}\bigg|\bigg)-\frac{M}{2\omega}\log\bigg(\bigg|\frac{\omega+M}{\omega-M}\bigg|\bigg)-\frac{1}{2}\log\bigg(\bigg|\frac{M^2-\omega^2}{\omega^2}\bigg|\bigg)
    \nonumber\\
    &+i\pi\mathrm{sgn}(\omega)\theta(|v_{\star}\Lambda_c|-|\omega|)+i\frac{\pi}{2}\frac{M}{\omega}\theta(M-|\omega|)-i\frac{\pi}{2}\mathrm{sgn}(\omega)\theta(M-|\omega|)\nonumber\\
    =&\log\bigg(\bigg|\frac{|v_{\star}\Lambda_c|^2-\omega^2}{\omega^2}\bigg|\bigg)-\frac{M}{2\omega}\log\bigg(\bigg|\frac{\omega+M}{\omega-M}\bigg|\bigg)-\frac{1}{2}\log\bigg(\bigg|\frac{M^2-\omega^2}{\omega^2}\bigg|\bigg)
    \nonumber\\
    &+i\pi\mathrm{sgn}(\omega)\theta(|v_{\star}\Lambda_c|-|\omega|)+i\frac{\pi}{2}\frac{M-|\omega|}{\omega}\theta(M-|\omega|)
\end{align}
}
To evaluate the logarithms unambiguously, we choose the branch cut of $\log(z)$ to be $z\in(-\infty,0]$.
Accordingly, for $x\in\mathbb{R}$, we have $\mathrm{Im}[\log(x\pm i0^+)]=\pm\pi\theta(-x)$.

In the chiral-flat limit $M=0$, the hybridization function reduces to 
\ba 
\label{eq:self_energy_in_f}
\gamma^2\Delta(i\omega) = \gamma^2 \Delta_M(i\omega)\bigg|_{M=0} =  -i\omega \pi \kappa f(i\omega) 
\ea 
with 
\ba 
f(i\omega) =f_M(i\omega)\bigg|_{M=0}= \log\bigg(\frac{|v_\star \Lambda_c|^2 +\omega^2}{\omega^2 } \bigg) 
\ea 
The corresponding real-frequency expression is
\ba 
\label{eq:f_fun_real_frequency}
f(\omega+i0^+) =  \log 
\bigg( \bigg| 
\frac{|v_\star \Lambda_c|^2 -\omega^2}{\omega^2 }\bigg| 
\bigg) 
+ i\pi \text{sgn}(\omega) \theta(|v_\star\Lambda_c|-|\omega|)
\ea 

\subsubsection{Symmetries of the hybridization function}
As shown in \cref{eq:sym_repres} to \cref{eq:flavor_U8}, the flavor U(8)-symmetric local hybridization function $\Delta(i\omega)$ keeps the spin SU(2), valley U$_{V}(1)$, $C_{3z}$, $C_{2x}$ and $C_{2z}$ symmetries.
Here we show that the hybridization function $\Delta_{M}(i\omega)$ derived in \cref{eq:self_energy_in_f_M} also preserves the particle-hole symmetry and time-reversal symmetry.

The local self-energy \cref{eq:atomic-self-energy} respects the particle-hole symmetry $P:\Delta_{+1}\leftrightarrow\Delta_{-1},n_f\to (N_f-n_f)$ as
\begin{align}
    P \Sigma_{f,loc}(i\omega)P^{-1}=&\frac{n_f(N_f-n_f)U_1^2}{N_f^2}\frac{1}{i\omega-\frac{(N_f-n_f)\Delta_{-1}-n_f\Delta_{+1}}{N_f}}\nonumber\\
    =&-\frac{n_f(N_f-n_f)U_1^2}{N_f^2}\frac{1}{-i\omega-\frac{n_f\Delta_{+1}-(N_f-n_f)\Delta_{-1}}{N_f}}\nonumber\\
    =&-\Sigma_{f,loc}(-i\omega)
\end{align}
In the flavor space of $f$ orbitals, the representation of particle-hole symmetry is $D^f(P)=i\sigma_z\tau_z$ ($\sigma$ for orbital and $\tau$ for valley).
This representation is diagonal in flavor space.
Therefore, the self-energy to order $\gamma^2$ \cref{eq:Sigma_hyb_exp} will transform under particle-hole symmetry as
\begin{align}
    P\Sigma_{f,ij}(i\omega,\kk)P^{-1}=&P\delta_{i,j}\Sigma_{f,loc}(i\omega)\bigg[1
+ {\frac{\gamma^2(N_f+1) }{U_1^2 }}\Delta_M(i\omega) \frac{(\Delta_{+1}+\Delta_{-1})^2}{ 
i\omega - \frac{n_f\Delta_{+1} - (N_f-n_f)\Delta_{-1}}{N_f}
} \bigg]P^{-1}\nonumber\\
=&\delta_{i,j}[-\Sigma_{f,loc}(-i\omega)]\bigg[1
+ {\frac{\gamma^2(N_f+1) }{U_1^2 }}\Delta_M(i\omega) \frac{(\Delta_{+1}+\Delta_{-1})^2}{ 
i\omega - \frac{(N_f-n_f)\Delta_{-1} - n_f\Delta_{+1}}{N_f}
} \bigg]\nonumber\\
=&\delta_{i,j}[-\Sigma_{f,loc}(-i\omega)]\bigg[1
- {\frac{\gamma^2(N_f+1) }{U_1^2 }}\Delta_M(i\omega) \frac{(\Delta_{+1}+\Delta_{-1})^2}{ 
-i\omega - \frac{n_f\Delta_{+1}-(N_f-n_f)\Delta_{-1}}{N_f}
} \bigg]
\end{align}
Since $\Sigma_{f,ij}(i\omega,\kk)$ must also satisfy the particle-hole symmetry $P\Sigma_{f,ij}(i\omega,\kk)P^{-1}=-\Sigma_{f,ij}(-i\omega,-\kk)$, this puts a constraint on the hybridization function:
\begin{equation}
    \Delta_M(-i\omega)=-\Delta_M(i\omega)
\end{equation}
This constraint is indeed satisfied by the explicit expression in \cref{eq:self_energy_in_f_M}:
\begin{align}
    \gamma^2\Delta_M(-i\omega)=&i\omega \frac{\pi\gamma^2}{|v_\star^2|\Omega_M}
\log\bigg( \frac{|v_\star\Lambda_c|^2 +\omega^2}{\omega^2} 
\bigg) +\frac{\pi \gamma^2}{2|v_\star|^2\Omega_M} M \log\bigg( 
\frac{-i\omega +M }{-i\omega-M}\bigg)-{i\omega \frac{\pi\gamma^2}{2|v_\star^2|\Omega_M}
\log\bigg( \frac{M^2 +\omega^2}{\omega^2} 
\bigg)}\nonumber\\
=&i\omega \frac{\pi\gamma^2}{|v_\star^2|\Omega_M}
\log\bigg( \frac{|v_\star\Lambda_c|^2 +\omega^2}{\omega^2} 
\bigg) -\frac{\pi \gamma^2}{2|v_\star|^2\Omega_M} M \log\bigg( 
\frac{i\omega +M }{i\omega-M}\bigg)-{i\omega \frac{\pi\gamma^2}{2|v_\star^2|\Omega_M}
\log\bigg( \frac{M^2 +\omega^2}{\omega^2} 
\bigg)}\nonumber\\
=&-\gamma^2\Delta_{M}(i\omega)
\end{align}

Similarly, we have 
\begin{align}
    \gamma^2[\Delta_M(i\omega)]^*=&i\omega \frac{\pi\gamma^2}{|v_\star^2|\Omega_M}
\log\bigg( \frac{|v_\star\Lambda_c|^2 +\omega^2}{\omega^2} 
\bigg) +\frac{\pi \gamma^2}{2|v_\star|^2\Omega_M} M \log\bigg( 
\frac{-i\omega +M }{-i\omega-M}\bigg)-{i\omega \frac{\pi\gamma^2}{2|v_\star^2|\Omega_M}
\log\bigg( \frac{M^2 +\omega^2}{\omega^2} 
\bigg)}\nonumber\\
=&\gamma^2\Delta_M(-i\omega)
\end{align}
Therefore, the self-energy also respects time-reversal symmetry:
\begin{align}
    T\Sigma_{f,ij}(i\omega,\kk)T^{-1}=&[\Sigma_{f,ij}(i\omega,-\kk)]^*\nonumber\\
    =&\delta_{i,j}[\Sigma_{f,loc}(i\omega)]^*\bigg[1
+ {\frac{\gamma^2(N_f+1) }{U_1^2 }}(\Delta_M(i\omega))^* \frac{(\Delta_{+1}+\Delta_{-1})^2}{ 
-i\omega - \frac{n_f\Delta_{+1} - (N_f-n_f)\Delta_{-1}}{N_f}
} \bigg]\nonumber\\
=&\Sigma_{f,ij}(-i\omega,-\kk)
\end{align}

\section{Numerical results}
\label{app:num_res}
\begin{figure}
    \centering
    \includegraphics[width=0.8\linewidth]{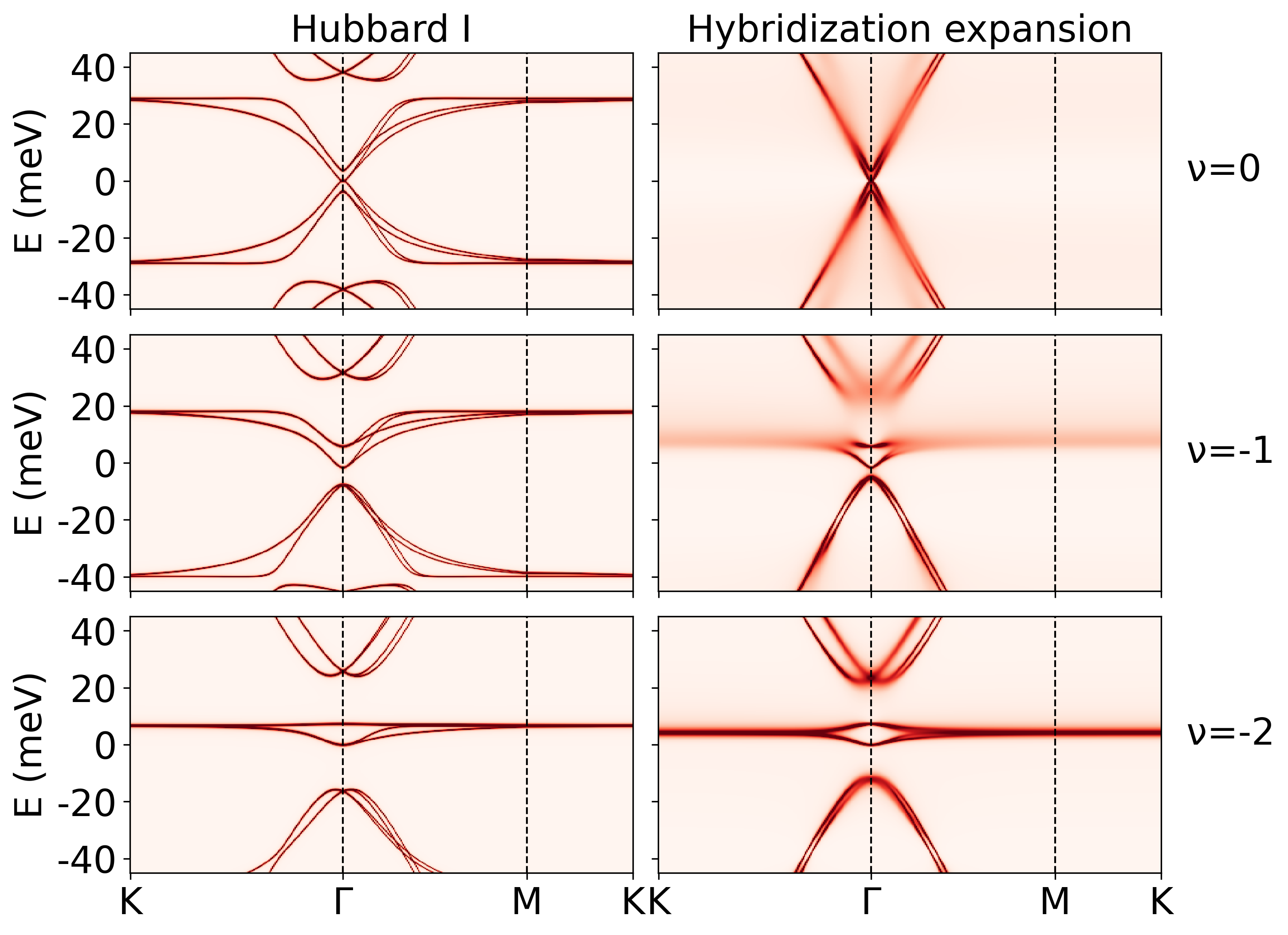}
    \caption{Spectral functions at filling $\nu=0,-1,-2$ obtained from Hubbard-I approximation \cref{eq:Sigma_Hubbard_I} and hybridization expansions \cref{eq:Sigma_hyb_exp}. We take the same parameters as in Ref.~\cite{PhysRevX.14.031045}. In generating the Hubbard-I spectral functions, we introduce an artificial damping $\delta = 0.1\,\mathrm{meV}$ to both $f$ and $c$ electrons, which amounts to replacing each Dirac delta function, $\delta(x)$, in the spectral function with a Lorentzian broadening $\frac{\delta/\pi}{x^2+\delta^2}$. Note that in the hybridization expansion (or the infinite $s\equiv \kappa$-sum series methods \cite{ledwith_nonlocal_2024, scattering1, scattering2}, if the original $\kappa \equiv s$ parameter is kept, the Hubbard bands are broadened too much. A reduction of  $\kappa \equiv s$  by a factor of 4  is necessary to match DMFT, which in its turn matches QTM. This might be due to the fact that $(N_f+1)\pi \kappa^2 \sim 1$ and hence higher orders are needed for these expansions to converge.}
    \label{fig:spec_hyb_exp}
\end{figure}

using the self-energy derived in \cref{eq:Sigma_hyb_exp}, we can also obtain the Green's function and the corresponding spectral functions numerically. 
In \cref{fig:spec_hyb_exp}, we illustrate the spectral function with the Hubbard-I approximation (\cref{eq:Sigma_Hubbard_I}) and the result from the hybridization expansion (\cref{eq:Sigma_hyb_exp}) at integer filling $\nu=0,-1,-2$. 
In \cref{fig:spec_hyb_exp}, we have used the full model with finite $M$ and $v_\star^\prime$.

Several general remarks are in order. First, the self-energy at order $\gamma^2$
renormalizes the Hubbard-band energies, shifting them toward the Fermi level.
Second, the self-energy introduces finite damping of the Hubbard bands. 
These trends are qualitatively consistent with the DMFT results reported in
Ref.~\cite{PhysRevX.14.031045}. 
In addition, both Hubbard bands at $\nu=0$, the lower Hubbard bands at $\nu=-1$ and $\nu=-2$ are strongly damped
and are barely seen in the figure, again in qualitative agreement with the DMFT results
of Ref.~\cite{PhysRevX.14.031045}.

\section{Self-energy and spectral function at charge neutrality in the chiral limit ($v_\star^\prime=0$)}
\label{app:chiral_limit}

In this section, we analytically investigate the properties of the spectral
function and the self-energy at charge neutrality ($\nu=0$) in the chiral limit
with $v_\star' = 0$. We consider two cases separately, corresponding to $M=0$
and $M \neq 0$. We first focus on the case $M=0$.

At charge neutrality, we have $n_f=N_f/2$ with $N_f=8$. The charge $\pm 1$ excitation energy reads (\cref{eq:define_charge_gap})
\ba 
\Delta_{\pm 1} = \frac{U_1}{2}
\ea 
The self-energy now behaves as (from \cref{eq:Sigma_hyb_exp})
\ba 
\Sigma_{f,ij}(i\omega,\kk) \approx \delta_{i,j}\Sigma^{CNP}(i\omega) 
\ea 
where 
\ba 
\label{eq:self_energy_CNP_hyb}
\Sigma_f^{CNP}(i\omega) 
=\frac{ U_1^2}{4i\omega}
\bigg[ 
1 + {\frac{ \gamma^2}{U_1^2} \Delta(i\omega) \frac{U_1^2(N_f+1)}{i\omega}}
\bigg] 
\ea 
We first consider the chiral-flat limit with $v_\star^\prime =0 , M=0$. The hybridization function is given in \cref{eq:self_energy_in_f}. Then 
\ba 
\Sigma^{CNP}_f(i\omega) 
= 
{ \frac{U_1^2}{4i\omega}
\bigg[ 
1-  (N_f+1)\pi\kappa f(i\omega) 
\bigg]} 
\label{eq:self-cnp}
\ea

We rewrite $\Sigma_f^{CNP}(i\omega)$ by introducing $Z_a(i\omega)$ 
\ba 
\label{eq:sigma_f_CNP}
&\Sigma_f^{CNP}(i\omega) 
=\frac{ U_1^2 Z_a(i\omega)}{4i\omega}  \nonumber\\ 
&Z_a(i\omega) = 
1 - (N_f+1)\pi \kappa f(i\omega) = 
1 -  (N_f+1)\frac{\pi\gamma^2}{|v_\star|^2\Omega_M}
\log\bigg(\frac{|v_\star \Lambda_c|^2 +\omega^2}{\omega^2 }\bigg) 
\ea 
The Hubbard-I self-energy is recovered in the limit of $\gamma\to 0$ and $Z_a(i\omega)=1$. 
While in the following, we will keep the entire $Z_{a}(i\omega)$ including the contribution to the order of $\gamma^2$.
This term will have imaginary part in the real frequency $i\omega\to \omega+i0^{+}$, which will introduce a finite lifetime to the $c$-electron.

We note that, with only the Hubbard-I self-energy, one can introduce an auxiliary fermion to rewrite the Hubbard-I Green's function, as shown in \cite{hu2025projectedsolvabletopologicalheavy,calugaruObtainingSpectralFunction2025}.
To get an analytic expression of the poles of the Green's function, we now take a similar approach.
This approach is equivalent to solve poles of Green's function directly.
However, by using the auxiliary fermion approach, it is more easy to compare with Hubbard I results.
Here, we introduce an auxiliary fermion $a_{\kk,\alpha\eta s}$ for each flavor of the $f$ electron at each site. We define the Green's function as
\ba 
 & [G^{aux,\alpha \eta s}(\tau,\kk)]_{ij} = -\langle T_\tau [\psi_{\kk,\alpha\eta s}]_i(\tau) [\psi_{\kk,\alpha\eta s}^\dag ]_j (0) \rangle  \nonumber\\ 
    &\psi_{\kk,\alpha\eta s}= \begin{bmatrix}
        f_{\kk,\alpha\eta s} & c_{\kk,\alpha\eta s}& c_{\kk,(\alpha+2) \eta s} & a_{\kk,\alpha \eta s}
    \end{bmatrix}
\ea 
which takes the form of 
\ba 
\label{eq:inv_G_aux_M=0}
&[G^{aux,\alpha \eta s}(i\omega,\kk)]^{-1}  \nonumber\\ 
= &\begin{bmatrix}
   i\omega & -\gamma & 0 &- \frac{U_1}{2} \\
    -\gamma & i\omega & -v_\star(\eta k_x +i(-1)^{\alpha+1}k_y) & 0 \\
    0 & -v_\star(\eta k_x-i(-1)^{\alpha+1}k_y) & i\omega &0 \\
    -\frac{U_1}{2} & 0 & 0& Z^{-1}_a(i\omega) i\omega 
\end{bmatrix}
\ea 
Note that the above equation is the definition of the Green’s function of the system with the auxiliary degrees of freedom.
We can observe that, by letting $Z_a(i\omega)=1$, we recover the effective non-interacting Green's function of the auxiliary system under Hubbard-I approximation. 

By taking the inverse of the matrix in \cref{eq:inv_G_aux_M=0}, we obtain the following Green's function for the $f$ electron. 
\ba 
\label{eq:G_aux_hyb_exp}
&[G^{aux,\alpha \eta s}(i\omega,\kk)]_{11} 
= \frac{1}{ i\omega -\gamma^2 \Delta_{\kk}^{\alpha \eta s,\alpha \eta s}(i\omega)
-\frac{U_1^2 Z_a(i\omega) }{4i\omega} 
}\nonumber\\
=&
 \frac{1}{i\omega -\gamma^2 \Delta_{\kk}^{\alpha \eta s,\alpha \eta s}(i\omega)
-\frac{U_1^2Z_a(i\omega)}{4 i\omega} 
} \nonumber\\
 =& \frac{1}{i\omega -\gamma^2 \Delta_{\kk}^{\alpha \eta s,\alpha \eta s}(i\omega )
-\Sigma^{CNP}(i\omega) 
}
\ea 
which is exactly the same interacting Green's function of $f$ electron that we have derived (see \cref{eq:def_Green_f,eq:dyson_eq,eq:sigma_f_CNP}. 
We emphasize that the introduction of $G^{{aux},\alpha \eta s}(i\omega,\kk)$ (\cref{eq:inv_G_aux_M=0}) is merely another way to present the interacting Green's function of the system. 
From \cref{eq:inv_G_aux_M=0}, we also observe that, at order $\gamma^2$, the self-energy generates an effective quasiparticle weight renormalization $Z_a(i\omega)$ for both the auxiliary fermion.

We now consider the trace of Green's function for the physical degrees of freedom 
\ba 
\mathcal{G}^{\alpha\eta s}(i\omega,\kk) 
= \text{Tr}[G^{aux,\alpha \eta s}(i\omega,\kk)P_{phys}] 
\ea 
where the projection matrix only keeps the physical degrees of freedom $P_{phys} = \text{diag}\{1,1,1,0\}$. 
The spectral function of the system is then defined as 
\ba 
\rho^{\alpha \eta s}(\omega,\kk) 
= -\frac{1}{\pi}\mathrm{Im}
\bigg[ 
\mathcal{G}^{\alpha \eta s}(i\omega\rightarrow \omega+i0^+,\kk) 
\bigg] 
\ea

We now investigate the Green's function. The Green's function takes the form of 
\ba 
\label{eq:inverse_inverse_Green}
G^{aux,\alpha \eta s}(i\omega,\kk) 
= \text{adj}([G^{aux,\alpha \eta s}(i\omega,\kk)]^{-1})/\text{det}([G^{aux,\alpha \eta s}(i\omega,\kk)]^{-1})
\ea 
 where $\text{adj}$ denotes the adjugate of the matrix. 
The determinant of the inverse of Green's function matrix (\cref{eq:G_aux_hyb_exp}) reads
\ba 
\label{eq:det_decompose}
\text{det}( [G^{aux,\alpha\eta s}(i\omega,\kk))]^{-1} ]= \frac{1}{Z_a(i\omega)}  
\bigg[ 
(i\omega)^4 -(i\omega)^2 
\bigg(|v_\star\kk|^2 +(\gamma^2 +\frac{U_1^2Z_a(i\omega)}{4})\bigg)
+ \frac{ U_1^2Z_a(i\omega)}{4}|v_\star\kk|^2 
\bigg] 
\ea 
Again, in the $Z_a(i\omega)=1$ limit, we recover the Hubbard-I limit, and the determinant simply reads 
\ba 
\label{eq:det_decompose_Hubbard_I}
\text{det}( [G^{aux,\alpha\eta s}(i\omega,\kk))]^{-1} ]=&
\bigg[ 
i\omega - 
\sqrt{ \frac{
|v_\star\kk|^2 +\gamma^2 + \frac{U_1^2}{4}
- \sqrt{
\bigg( |v_\star\kk|^2 +\gamma^2 + \frac{U_1^2}{4}
\bigg)^2 -U_1^2|v_\star\kk|^2 
}  }{2}
}
\bigg] \nonumber\\
&
\bigg[ 
i\omega + 
\sqrt{ \frac{
|v_\star\kk|^2 +\gamma^2 + \frac{U_1^2}{4}
- \sqrt{
\bigg( |v_\star\kk|^2 +\gamma^2 + \frac{U_1^2}{4}
\bigg)^2 -U_1^2|v_\star\kk|^2 
}  }{2}
}
\bigg] \nonumber\\
&
\bigg[ 
i\omega - 
\sqrt{ \frac{
|v_\star\kk|^2 +\gamma^2 + \frac{U_1^2}{4}
+ \sqrt{
\bigg( |v_\star\kk|^2 +\gamma^2 + \frac{U_1^2}{4}
\bigg)^2 -U_1^2|v_\star\kk|^2 
}  }{2}
}
\bigg] \nonumber\\
&
\bigg[ 
i\omega + 
\sqrt{ \frac{
|v_\star\kk|^2 +\gamma^2 + \frac{U_1^2}{4}
+ \sqrt{
\bigg( |v_\star\kk|^2 +\gamma^2 + \frac{U_1^2}{4}
\bigg)^2 -U_1^2|v_\star\kk|^2 
}  }{2}
}
\bigg] 
\ea 
which can be understood as $\prod_n (i\omega - E_{\kk,n})$ where $E_{\kk,n}$ describes the dispersion of the Hubbard-I bands. 
After introducing $Z_a(i\omega)$, $U_1^2$ is effectively renormalized to $U_1^2 Z_a(i\omega)$ (from \cref{eq:det_decompose}). We therefore replace $U_1^2$ by $U_1^2 Z_a(i\omega)$ in \cref{eq:det_decompose_Hubbard_I}, while also introducing an additional prefactor $1/Z_a(i\omega)$, as shown in \cref{eq:det_decompose}. This leads to 
\ba 
\label{eq:det_decompose_complext_pole}
\text{det}( [G^{aux,\alpha\eta s}(i\omega,\kk))]^{-1} ]=
\frac{1}{Z_a(i\omega) } \prod_{n=1}^4
\bigg[ 
i\omega - \mathcal{E}_{i\omega,\kk,n} 
\bigg] 
\ea 
where 
\ba 
&\mathcal{E}_{i\omega,\kk,n=1}
= \sqrt{ \frac{ \gamma^2 + Z_a(i\omega)U_1^2/4+|v_\star\kk|^2 
-\sqrt{ 
[\gamma^2 +Z_a(i\omega)U_1^2/4+|v_\star\kk|^2]^2-Z_a(i\omega)U_1^2|v_\star\kk|^2
}
}{2}} \nonumber\\ 
& \mathcal{E}_{i\omega,\kk,n=2}
= \sqrt{ \frac{ \gamma^2 + Z_a(i\omega)U_1^2/4+|v_\star\kk|^2 
+\sqrt{ 
[\gamma^2 +Z_a(i\omega)U_1^2/4+|v_\star\kk|^2]^2-Z_a(i\omega)U_1^2|v_\star\kk|^2
}
}{2}} \nonumber\\
&\mathcal{E}_{i\omega,\kk,n=3} = -\mathcal{E}_{i\omega,\kk,n=1} 
\nonumber\\
&\mathcal{E}_{i\omega,\kk,n=4} =-\mathcal{E}_{i\omega,\kk,n=2}
\ea 
Then the decomposition of the determinant (\cref{eq:det_decompose_complext_pole}) allows us to rewrite the trace of Green's function as (also using \cref{eq:inverse_inverse_Green})
\ba 
\mathcal{G}^{\alpha\eta s}(i\omega,\kk) &=\text{Tr}[G^{aux,\alpha \eta s}(i\omega,\kk)P_{phys}]= 
\frac{1}{\text{det}([G^{aux,\alpha \eta s}(i\omega,\kk)]^{-1})}
\text{Tr}
\bigg[\text{adj}([G^{aux,\alpha \eta s}(i\omega,\kk)]^{-1})P_{phys}
\bigg] 
\nonumber\\
&
= 
\sum_{n=1,...,4} \frac{\mathcal{A}_{i\omega,\kk}(-1)^{n+1}}{i\omega -\mathcal{E}_{i\omega,\kk,n}}
\ea 
where 
\ba 
& \mathcal{A}_{i\omega,\kk} 
= \frac{2 
\bigg( 2[\gamma^2 +|v_\star\kk|^2 -3(i\omega)^2] +U_1^2Z_a(i\omega) 
\bigg) }{Z_a(i\omega)\sqrt{ [4\gamma^2 +4|v_\star\kk|^2 +U_1^2Z_a(i\omega)]^2 -16U_1^2|v_\star\kk|^2 Z_a(i\omega) }}
\ea 
Such a decomposition characterizes the dispersion/spectral functions of the interacting system, as can be seen by revisiting the Hubbard--I approximation. Within the Hubbard-I approximation, we have $Z_a(i\omega)=1$ and we recover the Hubbard-I green's function discussed in Ref.~\cite{hu2025projectedsolvabletopologicalheavy} taking the form of 
\ba 
&\mathcal{G}^{\alpha \eta s}(i\omega,\kk) 
 \overset{\mathrm{Hubbard\, I}}{=} \sum_{n=1,...,4} \frac{A_{i\omega,\kk} (-1)^{n+1}}{i\omega - E_{\kk,n}} \nonumber\\
&
E_{\kk,n=1} = 
\sqrt{ 
\frac{\gamma^2 +U_1^2/4 + |v_\star\kk|^2 
-\sqrt{ (\gamma^2 +U_1^2/4+|v_\star \kk|^2)^2 - U_1^2|v_\star\kk|^2 }
}{2}
} \nonumber\\ 
&E_{\kk,n=2} = 
\sqrt{ 
\frac{\gamma^2 +U_1^2/4 + |v_\star\kk|^2 
+\sqrt{ (\gamma^2 +U_1^2/4+|v_\star \kk|^2)^2 - U_1^2|v_\star\kk|^2 }
}{2}
} \nonumber\\ 
&E_{\kk,n=3} = -E_{\kk,n=1}\nonumber\\ 
&E_{\kk,n=4}=-E_{\kk,n=2} \nonumber\\
& A_{i\omega,\kk} =\frac{2 
\bigg( 2[\gamma^2 +|v_\star\kk|^2 -3(i\omega)^2] +U_1^2 
\bigg) }{\sqrt{ [4\gamma^2 +4|v_\star\kk|^2 +U_1^2]^2 -16U_1^2|v_\star\kk|^2  }}
\ea 
The Green's function within the Hubbard-I approximation has four poles (in the real frequency) located at $E_{\kk,n=1/2/3/4}$ describing the corresponding excitation for each $\alpha \eta s$ sector as discussed in Ref.~\cite{hu2025projectedsolvabletopologicalheavy}. 

We next consider the correction introduced by $ Z_a(i\omega)$. By treating $\kappa$ as a small parameter, we find 
\ba 
&\mathcal{E}_{i\omega,\kk,n=1} 
\approx E_{\kk,n=1} 
\bigg[ 
1 -(N_f+1)\pi\kappa f(i\omega)  
\frac{U_1^2 }{16E_{\kk,n=1}^2}
\bigg( 1 - \frac{(\gamma^2 +U_1^2/4-|v_\star\kk|^2)}{\sqrt{ (\gamma^2+U_1^2/4+|v_\star\kk|^2)^2 -U_1^2|v_\star\kk|^2  }}
\bigg) 
\bigg] \nonumber\\ 
&\mathcal{E}_{i\omega,\kk,n=2} 
\approx E_{\kk,n=2} 
\bigg[ 
1 -(N_f+1)\pi\kappa f(i\omega)  
\frac{U_1^2 }{16E_{\kk,n=2}^2}
\bigg( 1 + \frac{(\gamma^2 +U_1^2/4-|v_\star\kk|^2)}{\sqrt{ (\gamma^2+U_1^2/4+|v_\star\kk|^2)^2 -U_1^2|v_\star\kk|^2  }}
\bigg) 
\bigg]  \label{eq:hyb_poles}
\ea 

We now work in the real-frequency to obtain the spectral function. In the real frequency, we notice that $f(\omega+i0^+)$ (see \cref{eq:f_fun_real_frequency})
\ba 
f(\omega+i0^+) =  \log 
\bigg( \bigg| 
\frac{|v_\star \Lambda_c|^2 -\omega^2}{\omega^2 }\bigg| 
\bigg) 
+ i\pi \text{sgn}(\omega) \theta(|v_\star\Lambda_c|-|\omega|)
\ea 
is a complex number for given $\omega$. Thus $\mathcal{E}_{\omega+i0^+,\kk,n}$ is also complex. We separate the $\mathcal{E}_{\omega+i0^+,\kk,n}$ into real and imaginary part such that 
\ba 
\label{eq:real_frequency_pole}
\mathcal{E}_{\omega+i0^+,\kk,n} = \mathcal{R}_{\omega,\kk,n}
- i{\Gamma}_{\omega,\kk,n} 
\ea 

The Green's function can then be expressed as 
\ba 
\mathcal{G}^{\alpha\eta s}(\omega+i0^+,\kk) = 
\sum_{n=1,...,4} \frac{\mathcal{A}_{\omega,\kk}(-1)^{n+1}}{\omega - \mathcal{R}_{\omega,\kk,n}
+ i{\Gamma}_{\omega,\kk,n} }
\ea 
The spectral functions are 
\ba 
\rho^{\alpha \eta s}(\omega,\kk) 
=-\frac{1}{\pi}\mathrm{Im}[\mathcal{G}^{\alpha\eta s}(\omega+i0^+,\kk)]
= -\sum_{n=1,...,4} \frac{(-1)^{n+1}\mathrm{Im}[\mathcal{A}_{\omega,\kk}(\omega - \mathcal{R}_{\omega,\kk,n}
+ i{\Gamma}_{\omega,\kk,n})] }{(\omega - \mathcal{R}_{\omega,\kk,n})^2
+ ({\Gamma}_{\omega,\kk,n})^2 }
\ea 
We can then define the renormalized dispersion and the damping rate from the denominator $(\omega - \mathcal{R}_{\omega,\kk,n})^2 + (\Gamma_{\omega,\kk,n})^2$. 
The renormalized dispersion $\tilde{E}_{\kk,n}$ is defined as the solution of  
\ba 
&\bigg[\omega -\mathcal{R}_{\omega,\kk,n}\bigg]\bigg|_{\omega = \tilde{E}_{\kk,n}} =0
\label{eq:renormalize_M0_def}
\ea 
The corresponding damping rate can be defined as
\ba 
\frac{1}{\tau_{\kk,n}} = 
 {\Gamma}_{\omega,\kk,n}\bigg|_{\omega= \tilde{E}_{\kk,n}}
\ea 

We can write the renormalized dispersion as a series of $\kappa$
\begin{equation}
    \tilde{E}_{\kk,n}=\tilde{E}_{\kk,n}^{(0)}+\tilde{E}_{\kk,n}^{(1)}+\mathcal{O}(\kappa^2)
\end{equation}
where the superscript $(i)$ means the $i$-th order of $\kappa$.
Then up to first order of $\kappa$, \cref{eq:renormalize_M0_def} becomes
\begin{align}
    \tilde{E}^{(0)}_{\kk,n}+\tilde{E}^{(1)}_{\kk,n}=&\mathcal{R}_{\omega,\kk,n}\bigg|_{\omega=\tilde{E}^{(0)}_{\kk,n}}+(\partial_{\omega}\mathcal{R}_{\omega,\kk,n})\bigg|_{\omega=\tilde{E}^{(0)}_{\kk,n}}\times \tilde{E}^{(1)}_{\kk,n}+\mathcal{O}(\kappa^2)
\end{align}
Comparing the zeroth order term we find that
\begin{equation}
    \tilde{E}^{(0)}_{\kk,n}=E_{\kk,n}
\end{equation}
which is nothing but the Hubbard-I dispersion.
On the other hand, we have
\begin{equation}
    \partial_{\omega}\mathcal{R}_{\omega,\kk,n}\propto \kappa \partial_{\omega}\mathrm{Re} f(\omega+i0^+)=\kappa\frac{2|v_{\star}\Lambda_c|^2}{\omega(\omega^2-|v_{\star}\Lambda_c|^2)}
\end{equation}
is already first order of $\kappa$, thus up to first order of $\kappa$ we have
\begin{equation}
    \tilde{E}_{\kk,n}=\tilde{E}^{(0)}_{\kk,n}+\tilde{E}^{(1)}_{\kk,n}+\mathcal{O}(\kappa^2)=\mathcal{R}_{\omega,\kk,n}\bigg|_{\omega=\tilde{E}^{(0)}_{\kk,n}}+\mathcal{O}(\kappa^2)=\mathcal{R}_{\omega,\kk,n}\bigg|_{\omega={E}_{\kk,n}}+\mathcal{O}(\kappa^2)
\end{equation}
To be specific,
\ba 
\tilde{E}_{\kk,n=1} 
\approx &E_{\kk,n=1} 
\bigg[ 
1 -(N_f+1)\pi\kappa [\mathrm{Re}f(E_{\kk,n=1}+i0^+)]  
\frac{U_1^2 }{16E_{\kk,n=1}^2}
\bigg( 1 - \frac{(\gamma^2 +U_1^2/4-|v_\star\kk|^2)}{\sqrt{ (\gamma^2+U_1^2/4+|v_\star\kk|^2)^2 -U_1^2|v_\star\kk|^2  }}
\bigg) 
\bigg] \nonumber\\
\tilde{E}_{\kk,n=2} 
\approx  & E_{\kk,n=2} \bigg[ 
1 -(N_f+1)\pi\kappa [\mathrm{Re}f(E_{\kk,n=2}+i0^+)]   
\frac{U_1^2 }{16E_{\kk,n=2}^2}
\bigg( 1 + \frac{(\gamma^2 +U_1^2/4-|v_\star\kk|^2)}{\sqrt{ (\gamma^2+U_1^2/4+|v_\star\kk|^2)^2 -U_1^2|v_\star\kk|^2  }}
\bigg) 
\bigg] \nonumber\\
\tilde{E}_{\kk,n=3}=&-\tilde{E}_{\kk,n=1} \nonumber\\
\tilde{E}_{\kk,n=4} = &-\tilde{E}_{\kk,n=2}
\ea 
We are particularly interested in two types of features. First, we focus on the Hubbard band located near the Brillouin-zone boundary, described by $\tilde{E}_{\kk,n=1}$. Second, we examine the gapless excitations in the vicinity of the $\Gamma$ point, described by $\tilde{E}_{\kk,n=1}$. 

Near the Brillouin zone edge, we can take $|v_\star\kk| = |v_\star\Lambda_c|$, where $\Lambda_c=\sqrt{\Omega_M/\pi}$. We observe that 
\ba 
\tilde{E}_{\kk,n=1}\bigg|_{|\kk|=\Lambda_c} 
\approx 
\frac{U_1}{2} 
\bigg[ 
1 - \frac{1}{2}\kappa \pi (N_f+1)  \log( \frac{4 \gamma^2}{\pi \kappa U_1^2} ) 
\bigg] + \mathcal{O}(\kappa^2) 
\ea 
which describes the energy renormalization of the Hubbard bands. 
The corresponding scattering rate reads 
\ba 
\frac{1}{\tau_{\kk,n=1}}\bigg|_{|\kk|=\Lambda_c}\approx \frac{U_1\pi^2 (N_f+1) }{4}\kappa +\mathcal{O}(\kappa^2) 
\ea which captures the damping effects of the Hubbard bands.

Near the $\Gamma$ point, we expand $\tilde{E}_{\kk,n},1/\tau_{\kk,n}$ in powers of $|\kk|$. At each order, we obtain its coefficient in powers of $\kappa$. For $\tilde{E}_{\kk,n}$ we find 
\ba 
\tilde{E}_{\kk,n=1}
\approx 
\frac{U_1}{2\sqrt{U_1^2/4+\gamma^2}}
\bigg[ 
1- (N_f+1)\pi \kappa 
\frac{2\gamma^2}{4\gamma^2 +U_1^2} 
\log\bigg( \frac{|v_\star\Lambda_c|^2(U_1^2+4\gamma^2)}{U_1^2|v_\star\kk|^2}
\bigg) 
+\mathcal{O}(\kappa^2)
\bigg] |v_\star \kk| + \mathcal{O}(|v_\star\kk|^2)
\ea 
which described the renormalization of the velocity of the Dirac node. 
In addition, the scattering rate reads
\ba 
\frac{1}{\tau_{\kk,n=1}}\approx 
\frac{(N_f+1)U_1 \gamma^2 \pi^2}{4 (U_1^2/4+\gamma^2)^{3/2}}|v_\star \kk| \kappa 
\label{eq:scattering_time_linear}
\ea

\subsection{Effect of finite $M$}
We next investigate the effect of finite $M$. 
Similarly to the $M=0$ limit, the self-energy of $f$-electron can be written as (\cref{eq:sigma_f_CNP,eq:def_fM})
\ba 
\label{eq:experssion_sigma_f_finite_M}
&\Sigma_f^{CNP}(i\omega) 
=\frac{ U_1^2Z_{M,a}}{4i(i\omega)\omega}   \nonumber\\ 
&{
Z_{M,a}(i\omega) = 
1 -\pi (N_f+1)\kappa f_M(i\omega)=1-\pi(N_f+1)\kappa\bigg[\log\bigg( \frac{|v_\star\Lambda_c|^2+\omega^2}{\omega^2}\bigg)
 - {\color{black}\frac{M}{2i\omega }\log\bigg( \frac{i\omega+M}{i\omega-M}
 \bigg)}
 -\frac{1}{2}\log\bigg(\frac{M^2+\omega^2}{\omega^2}\bigg)\bigg]
}
\ea 
Via the auxiliary field, the effective Green's function is defined as 
\ba 
 & [\tilde{G}^{aux,  \eta s}(\tau,\kk)]_{ij} = -\langle T_\tau [\tilde{\psi}_{\kk,\eta s}]_i(\tau) [\tilde{\psi}_{\kk,\eta s}^\dag ]_j (0) \rangle  \nonumber\\ 
    &\tilde{\psi}_{\kk,\eta s}= \begin{bmatrix}
        f_{\kk,1 \eta s} & f_{\kk,2 \eta s} 
        & c_{\kk,1\eta s}& c_{\kk,2  \eta s}
        & c_{\kk,3\eta s}& c_{\kk,4  \eta s}
        & a_{\kk,1\eta s}   & a_{\kk,2 \eta s}
    \end{bmatrix}
\ea 
which takes the form of 
\ba 
\label{eq:inverse_green_fintie_M}
&[\tilde{G}^{aux,\eta s}(i\omega,\kk)]^{-1}  \nonumber\\ 
= &\begin{bmatrix}
   i\omega & 0  & -\gamma & 0 & 0 & 0 &- \frac{U_1}{2} &0 
    \\ 
    0 &   i\omega  & 0 & -\gamma 
    &0 & 0 & 0 & -\frac{U_1}{2}
    \\ 
    -\gamma & 0 & i\omega & 0 & -v_\star (\eta k_x +ik_y) & 0 & 0 & 0 \\  
    0 & -\gamma & 0 & i\omega & 0 & -v_\star( \eta k_x-ik_y) & 0 & 0 \\
    0 & 0 & -v_\star(\eta k_x-ik_y) & 0 & i\omega & -M & 0 & 0 \\ 
    0 & 0 & 0 & -v_\star(\eta k_x +ik_y) & -M & i\omega & 0 & 0 \\ 
    -\frac{U_1}{2} & 0 & 0 & 0 & 0 & 0 & Z_{M,a}^{-1}(i\omega)i\omega & 0 \\
    0& -\frac{U_1}{2} & 0 & 0 & 0 & 0 & 0 &Z_{M,a}^{-1}(i\omega) i\omega 
\end{bmatrix}
\ea

We now consider the trace of Green's function for the physical degrees of freedom 
\ba 
\mathcal{\tilde{G}}^{\eta s}(i\omega,\kk) 
= \text{Tr}[\tilde{G}^{aux,\alpha \eta s}(i\omega,\kk)\tilde{P}_{phys}] 
\ea 
where the projection matrix only keeps the physical degrees of freedom $\tilde{P}_{phys} = \text{diag}\{1,1,1,1,1,1,0,0\}$. 
The spectral function of the system can be obtained from $\mathcal{\tilde{G}}$ as 
\ba 
\tilde{\rho}^{ \eta s}(\omega,\kk) 
= -\frac{1}{\pi}\mathrm{Im}
\bigg[ 
\mathcal{\tilde{G}}^{\eta s}(i\omega\rightarrow \omega+i0^+,\kk) 
\bigg] 
\ea 

Written explicitly, the trace of Green's function takes the form of 
\ba 
\label{eq:green_fun_finite_M}
\mathcal{\tilde{G}}^{\eta s}(i\omega,\kk)  = \frac{ D_{i\omega,\kk}}{\prod_{n=1}^4[(i\omega)^2 -(\mathcal{E}^{M}_{i\omega,\kk,n})^2]} 
= 
\sum_{n=1}^4 \bigg[ \frac{\mathcal{A}^{M}_{i\omega,\kk,n}}{i\omega - \mathcal{E}^{M}_{i\omega,\kk,n}} - \frac{\mathcal{A}^{M}_{i\omega,\kk,n}}{i\omega + \mathcal{E}^{M}_{i\omega,\kk,n}}\bigg] 
\ea 
with 
\ba 
&D_{i\omega,\kk} \nonumber\\
= &\frac{i\omega}{8 }
\bigg[ 16\gamma^4(i\omega)^2 +
[4(i\omega)^2-U_1^2Z_{M,a}(i\omega)]
[4(3(i\omega)^4+|v_\star\kk|^4-2(i\omega)^2(M^2+2|v_\star\kk|^2)
+U_1^2 (-2(i\omega)^2+M^2+2|v_\star\kk|^2)Z_{M,a}(i\omega)]\nonumber\\
&
+4\gamma^2 
[-16(i\omega)^4 -U_1^2 Z_{M,a}(i\omega)(M^2+|v_\star\kk|^2)+(i\omega)^2[8(M^2+|v_\star\kk|^2)+3U_1^2Z_{M,a}(i\omega)]]\bigg] 
\ea 
and 
\ba 
\mathcal{A}^{M}_{i\omega,\kk,n} 
= \frac{D_{i\omega,\kk} }{
2\mathcal{E}^{M}_{i\omega,\kk,n} }\bigg( 
\sum_{n=1}^4 
\bigg[ 
\prod_{m, m\ne n}[ (i\omega)^2 -(\mathcal{E}^{M}_{i\omega,\kk,m})^2]
\bigg] \bigg)^{-1}
\ea 
We also note that, due to the particle-hole symmetry, we have contributions from both $1/(i\omega -\mathcal{E}^{M}_{i\omega,\kk,n})$ and $1/(i\omega +\mathcal{E}^{M}_{i\omega,\kk,n})
$. It is sufficient to focus on one of them. 
In addition, $\{\mathcal{E}^{M}_{i\omega,\kk,n}) \}_{n=1,...,4}$ corresponds to the solutions of the following equation with unknown variable $x$ 
\ba
\label{eq:eqaution_of_pole_finite_M}
\bigg[ x^4 
-x^2\bigg(\gamma^2+|v_\star\kk|^2 +U_1^2Z_{M,a}(i\omega)/4\bigg) + U_1^2Z_{M,a}(i\omega) |v_\star\kk|^2/4 
\bigg]^2 
-M^2x^2\bigg( x^2 -U_1^2 Z_{M,a}(i\omega)/4 -\gamma^2\bigg)^2=0 
\ea
which is derived from the determinant of the matrix in \cref{eq:inverse_green_fintie_M} (see also the the discussion near \cref{eq:det_decompose_complext_pole}).
Factorization of \cref{eq:eqaution_of_pole_finite_M} gives the following two equations 
\begin{align}
   &x^4 
-x^2\bigg(\gamma^2+|v_\star\kk|^2 +U_1^2Z_{M,a}(i\omega)/4\bigg) + U_1^2Z_{M,a}(i\omega) |v_\star\kk|^2/4 + Mx \bigg( x^2 -U_1^2 Z_{M,a}(i\omega)/4 -\gamma^2\bigg)=0\label{eq:det_M_1}\\
&x^4 
-x^2\bigg(\gamma^2+|v_\star\kk|^2 +U_1^2Z_{M,a}(i\omega)/4\bigg) + U_1^2Z_{M,a}(i\omega) |v_\star\kk|^2/4 - Mx \bigg( x^2 -U_1^2 Z_{M,a}(i\omega)/4 -\gamma^2\bigg)=0\label{eq:det_M_2}
\end{align}
Assume $\{\mathcal{E}^{M}_{i\omega,\kk,n}) \}_{n=1,...,4}$ are the solutions of \cref{eq:det_M_1}.
Then $\{-\mathcal{E}^{M}_{i\omega,\kk,n}) \}_{n=1,...,4}$ will be solutions of \cref{eq:det_M_2}.
Thus we only need to solve \cref{eq:det_M_1}.

We focus on the behaviors near the $\Gamma$ point.
Writing the solution as $x^*=x_0^*+x_1^*$ where $x_0^*$ and $x_1^*$ are zeroth and first order terms of $|v_{\star}\kk|^2$.
Substituting $x=x_0^*+x_1^*$ into \cref{eq:det_M_1} and keep up to the first order of $|v_{\star}\kk|^2$ we have
\begin{align}
    &[(x_0^*)^4+4(x_0^*)^3x_1^*]-[(x_0^*)^2+2x_0^*x_1^*]\bigg(\gamma^2+|v_\star\kk|^2 +U_1^2Z_{M,a}(i\omega)/4\bigg) + U_1^2Z_{M,a}(i\omega) |v_\star\kk|^2/4\nonumber\\
    &+M(x_0^*+x_1^*)\bigg( [(x_0^*)^2+2x_0^*x_1^*] -U_1^2 Z_{M,a}(i\omega)/4 -\gamma^2\bigg)+\mathcal{O}(|v_{\star}\kk|^4)=0
\end{align}
The equation for $x_0^*$ is
\begin{align}
    &(x_0^*)^4 
-(x_0^*)^2\bigg(\gamma^2 +U_1^2Z_{M,a}(i\omega)/4\bigg) + Mx_0^* \bigg( (x_0^*)^2 -U_1^2 Z_{M,a}(i\omega)/4 -\gamma^2\bigg)\nonumber\\
=&x_0^*\bigg[
(x_0^*)^3+M\bigg( (x_0^*)^2 -U_1^2 Z_{M,a}(i\omega)/4 -\gamma^2\bigg)-x_0^*\bigg(\gamma^2 +U_1^2Z_{M,a}(i\omega)/4\bigg)
\bigg]\nonumber\\
=&x_0^*\bigg[
(x_0^*)^2(x_0^*+M)-(M+x_0^*)\bigg(\gamma^2 +U_1^2Z_{M,a}(i\omega)/4\bigg)
\bigg]\nonumber\\
=&x_0^*(x_0^*+M)\bigg(x_0^*-\sqrt{\gamma^2 +U_1^2Z_{M,a}(i\omega)/4}\bigg)\bigg(x_0^*+\sqrt{\gamma^2 +U_1^2Z_{M,a}(i\omega)/4}\bigg)=0
\end{align}
which gives $x_0^*=0,-M,\pm\sqrt{\gamma^2 +U_1^2Z_{M,a}(i\omega)/4}$.
The equation for $x_1^*$ is
\begin{align}
    &4(x_0^*)^3x_1^*-(x_0^*)^2|v_{\star}\kk|^2-2x_0^*x_1^*\bigg(\gamma^2 +U_1^2Z_{M,a}(i\omega)/4\bigg)+U_1^2Z_{M,a}(i\omega) |v_\star\kk|^2/4\nonumber\\
    &+M x_1^*\bigg((x_0^*)^2-U_1^2Z_{M,a}(i\omega)/4-\gamma^2\bigg)+2M(x_0^*)^2x_1^*\nonumber\\
    =&x_1^*\bigg[
    4(x_0^*)^3+3M(x_0^*)^2-(2x_0^*+M)\bigg(\gamma^2 +U_1^2Z_{M,a}(i\omega)/4\bigg)
    \bigg]-(x_0^*)^2|v_{\star}\kk|^2+U_1^2Z_{M,a}(i\omega) |v_\star\kk|^2/4=0
\end{align}
which gives
\begin{equation}
    x_1^*=\frac{(x_0^*)^2|v_{\star}\kk|^2-U_1^2Z_{M,a}(i\omega) |v_\star\kk|^2/4}{4(x_0^*)^3+3M(x_0^*)^2-(2x_0^*+M)\bigg(\gamma^2 +U_1^2Z_{M,a}(i\omega)/4\bigg)}
\end{equation}
Thus, for the four solutions $x_0^*=0,-M,\pm\sqrt{\gamma^2 +U_1^2Z_{M,a}(i\omega)/4}$, the corresponding first order corrections with respect to $|v_{\star}\kk|^2$ are
\begin{align}
    x_1^*\bigg|_{x_0^*=0}=&\frac{U_1^2Z_{M,a}(i\omega) |v_\star\kk|^2/4}{M\bigg(\gamma^2 +U_1^2Z_{M,a}(i\omega)/4\bigg)}\nonumber\\
    x_1^*\bigg|_{x_0^*=-M}=&\frac{M^2|v_{\star}\kk|^2-U_1^2Z_{M,a}(i\omega) |v_\star\kk|^2/4}{-M^3+M\bigg(\gamma^2 +U_1^2Z_{M,a}(i\omega)/4\bigg)}=-\frac{U_1^2Z_{M,a}(i\omega)/4-M^2}{M\bigg(\gamma^2 +U_1^2Z_{M,a}(i\omega)/4-M^2\bigg)}|v_{\star}\kk|^2
    \nonumber\\
    x_1^*\bigg|_{x_0^*=\sqrt{\gamma^2 +U_1^2Z_{M,a}(i\omega)/4}}=&\frac{\gamma^2|v_{\star}\kk|^2}{2\bigg(\gamma^2 +U_1^2Z_{M,a}(i\omega)/4\bigg)\bigg(M+\sqrt{\gamma^2 +U_1^2Z_{M,a}(i\omega)/4}\bigg)}\nonumber\\
    x_1^*\bigg|_{x_0^*=-\sqrt{\gamma^2 +U_1^2Z_{M,a}(i\omega)/4}}=&\frac{\gamma^2|v_{\star}\kk|^2}{2\bigg(\gamma^2 +U_1^2Z_{M,a}(i\omega)/4\bigg)\bigg(M-\sqrt{\gamma^2 +U_1^2Z_{M,a}(i\omega)/4}\bigg)}
    \label{eq:roots_0_1}
\end{align}
In summary, by treating $\kk$ as a small parameter,
we obtain the following roots
\ba
\label{eq:pole_expression_finite_M}
&\mathcal{E}^{M}_{i\omega,\kk,n=1} \approx  
\frac{U_1^2 Z_{M,a}(i\omega)/4}{M[\gamma^2 +U_1^2 Z_{M,a}(i\omega)/4]}|v_\star\kk|^2 
\nonumber\\
&\mathcal{E}^{M}_{i\omega,\kk,n=2} \approx  -M 
- \frac{U_1^2Z_{M,a}(i\omega)/4 -M^2}{M \bigg[
U_1^2Z_{M,a}(i\omega)/4 +\gamma^2  
-M^2 
\bigg]}|v_\star\kk|^2 
\nonumber\\
&\mathcal{E}^{M}_{i\omega,\kk,n=3} \approx \sqrt{ \gamma^2 + U_1^2 Z_{M,a}(i\omega) /4 } 
+ \frac{\gamma^2 |v_\star\kk|^2 }{2(U_1^2Z_{M,a}(i\omega)/4 + \gamma^2) 
\bigg( 
\sqrt{ \gamma^2 +U_1^2Z_{M,a}(i\omega)/4  }+M
\bigg) 
}
\nonumber\\
&\mathcal{E}^{M}_{i\omega,\kk,n=4} 
\approx  -\sqrt{\gamma^2 + U_1^2 Z_{M,a}(i\omega)  /4 } 
- \frac{\gamma^2 |v_\star\kk|^2 }{2(U_1^2Z_{M,a}(i\omega)/4 + \gamma^2) 
\bigg( 
\sqrt{ \gamma^2 +U_1^2Z_{M,a}(i\omega)/4  }-M
\bigg) }
\ea 
In the Hubbard-I limit where $Z_{M,a}(i\omega) =1$, we obtain the following dispersion for each valley and spin sector 
\ba 
E^{M}_{\kk,n=1}=&\mathcal{E}^{M}_{i\omega,\kk,n=1}\bigg|_{Z_{M,a}(i\omega)=1} \approx  
\frac{U_1^2/4}{M[\gamma^2 +U_1^2 /4]}|v_\star\kk|^2 
\nonumber\\
E^{M}_{\kk,n=2}=&\mathcal{E}^{M}_{i\omega,\kk,n=2} \bigg|_{Z_{M,a}(i\omega)=1} \approx  -M  
- \frac{U_1^2/4 -M^2}{M \bigg[
\bigg(U_1^2/4 +\gamma^2 \bigg) 
-M^2 
\bigg]}|v_\star\kk|^2 
\nonumber\\
E^{M}_{\kk,n=3}=&\mathcal{E}^{M}_{i\omega,\kk,n=3}\bigg|_{Z_{M,a}(i\omega)=1}  \approx \sqrt{ \gamma^2 + U_1^2  /4 } 
+ \frac{\gamma^2 |v_\star\kk|^2 }{2(U_1^2/4 + \gamma^2) 
\bigg( 
\sqrt{ \gamma^2 +U_1^2/4 }+M
\bigg) 
}
\nonumber\\
E^{M}_{\kk,n=4}=&\mathcal{E}^{M}_{i\omega,\kk,n=4} \bigg|_{Z_{M,a}(i\omega)=1} 
\approx  -\sqrt{ \gamma^2 + U_1^2 /4 } 
- \frac{\gamma^2 |v_\star\kk|^2 }{2(U_1^2/4 + \gamma^2) 
\bigg( 
\sqrt{ \gamma^2 +U_1^2/4 }-M
\bigg) }
\ea 
Again, due to the particle-hole symmetry, we also have the excitation mode with dispersion $-\mathcal{E}^{M}_{i\omega,\kk,n}|_{Z_{M,a}(i\omega)=1}=-E^{M}_{\kk,n}$. 

We now focus on the low-energy gapless excitation described by $\mathcal{E}^{M}_{i\omega,\kk,n=1}$, which leads to a quadratic gapless excitation in the Hubbard-I limit. We now investigate the effect of self-energy at order of $\kappa$. Similarly to \cref{eq:real_frequency_pole}, we consider its real frequency behaviors and let 
\ba 
\mathcal{E}^{M}_{\omega+i0^+,\kk,1} = \mathcal{R}^{M}_{\omega,\kk,1}
- i{\Gamma}^{M}_{\omega,\kk,1} 
\ea 
such that 
\ba 
\frac{1}{\omega -{\mathcal{E}}^{M}_{\omega+i0^+,\kk,1}} 
= \frac{1}{\omega - {\mathcal{R}}^{M}_{\omega,\kk,1}
+i{{\Gamma}}^{M}_{\omega,\kk,1} }
\ea 
From \cref{eq:pole_expression_finite_M,eq:experssion_sigma_f_finite_M,eq:fM_real_freq}, up to first order of $\kappa$ we find 
\ba 
\mathcal{R}^{M}_{\omega,\kk,1} 
\approx &
 \frac{U_1^2/4}{M(\gamma^2 +U_1^2/4)}
\bigg\{
1-  \frac{\gamma^2 \pi}{\gamma^2+U_1^2/4}(N_f+1)
\kappa 
\bigg[
\log\bigg( \bigg| \frac{|v_\star \Lambda_c|^2 -\omega^2}{\omega^2}\bigg|\bigg)
- \frac{M}{2\omega}\log\bigg(\bigg|\frac{M+\omega}{M-\omega}\bigg|\bigg)\nonumber\\
&
{-\frac{1}{2}\log\bigg(\bigg|\frac{M^2-\omega^2}{\omega^2}\bigg|\bigg)}
\bigg]
+\mathcal{O}(\kappa^2) 
\bigg\}
|v_\star\kk|^2 
\ea 
and 
\ba 
{\Gamma}^{M}_{\omega,\kk,1} = 
\frac{\gamma^2 U_1^2 \pi}{4M(\gamma^2+U_1^2/4)^2}\kappa (N_f+1)   \text{sgn}(\omega) 
\bigg[ \pi\theta(|v_\star\Lambda_c|-|\omega|)
+ \frac{\pi}{2}\frac{{\color{black}M}-|\omega|}{|\omega|}\theta(M-|\omega|) + \mathcal{O}(\kappa^2)\bigg] |v_\star\kk|^2 
\label{eq:scattering_time}
\ea 

Denote the renormalized dispersion as $\tilde{E}^{M}_{\kk,n}$.
Then the renormalized dispersion of the gapless mode $(n=1)$ is determined by 
\ba 
(\omega - {\mathcal{R}}^{M}_{\omega,\kk,1}) \bigg|_{\omega = \tilde{E}^{M}_{\kk,1}} = 0
\label{eq:renormalize_disp_def}
\ea 
Assume $\tilde{E}^{M}_{\kk,1}$ could be written as a series of $\kappa$:
\begin{equation}
    \tilde{E}^{M}_{\kk,1}=\tilde{E}^{M,(0)}_{\kk,1}+\tilde{E}^{M,(1)}_{\kk,1}+\mathcal{O}(\kappa^2)
\end{equation}
where the superscript $(i)$ means the $i$-th order of $\kappa$.
Then, expand \cref{eq:renormalize_disp_def} to first order of $\kappa$ we find
\begin{align}
    \tilde{E}^{M,(0)}_{\kk,1}+\tilde{E}^{M,(1)}_{\kk,1}=&\mathcal{R}^{M}_{\omega,\kk,1}\bigg|_{\omega = \tilde{E}^{M,(0)}_{\kk,1}}+(\partial_{\omega}{\mathcal{R}}^{M}_{\omega,\kk,1})\bigg|_{\omega = \tilde{E}^{M,(0)}_{\kk,1}}\times \tilde{E}^{M,(1)}_{\kk,1}\nonumber\\
    =&
 \frac{U_1^2/4|v_\star\kk|^2}{M(\gamma^2 +U_1^2/4)}
\bigg\{
1-  \frac{\gamma^2 \pi}{\gamma^2+U_1^2/4}(N_f+1)
\kappa 
\bigg[
\log\bigg( \bigg| \frac{|v_\star \Lambda_c|^2 -\omega^2}{\omega^2}\bigg|\bigg)
- \frac{M}{2\omega}\log\bigg(\bigg|\frac{M+\omega}{M-\omega}\bigg|\bigg)\nonumber\\
&
{-\frac{1}{2}\log\bigg(\bigg|\frac{M^2-\omega^2}{\omega^2}\bigg|\bigg)}
\bigg]
\bigg\}
\bigg|_{\omega = \tilde{E}^{M,(0)}_{\kk,1}} \nonumber\\
&-\kappa\tilde{E}^{M,(1)}_{\kk,1}\frac{2\gamma^2\pi(N_f+1)U_1^2|v_{\star}\kk|^2}{M(U_1^2+4\gamma^2)^2\omega^2(-|v_{\star}\Lambda_c|^2+\omega^2)}\bigg[
4|v_{\star}\Lambda_c|^2\omega+M(-|v_{\star}\Lambda_c|^2+\omega^2)\log\bigg(\bigg|\frac{M+\omega}{M-\omega}\bigg|\bigg)
\bigg]\bigg|_{\omega = \tilde{E}^{M,(0)}_{\kk,1}}\nonumber\\
=& \frac{U_1^2/4|v_\star\kk|^2}{M(\gamma^2 +U_1^2/4)}
\bigg\{
1-  \frac{\gamma^2 \pi}{\gamma^2+U_1^2/4}(N_f+1)
\kappa 
\bigg[
\log\bigg( \bigg| \frac{|v_\star \Lambda_c|^2 -\omega^2}{\omega^2}\bigg|\bigg)
- \frac{M}{2\omega}\log\bigg(\bigg|\frac{M+\omega}{M-\omega}\bigg|\bigg)\nonumber\\
&
{-\frac{1}{2}\log\bigg(\bigg|\frac{M^2-\omega^2}{\omega^2}\bigg|\bigg)}
\bigg]
\bigg\}
\bigg|_{\omega = \tilde{E}^{M,(0)}_{\kk,1}}+\mathcal{O}(\kappa^2)
\label{eq:renormalize_up_to_1}
\end{align}
Comparing the zeroth term of $\kappa$ we find
\begin{align}
   \tilde{E}^{M,(0)}_{\kk,1}= &\frac{U_1^2/4|v_\star\kk|^2}{M(\gamma^2 +U_1^2/4)}
   \label{eq:renormalize_0}
\end{align}
Substitute \cref{eq:renormalize_0} to \cref{eq:renormalize_up_to_1} we have
\begin{align}
    \tilde{E}^{M}_{\kk,1}=&\tilde{E}^{M,(0)}_{\kk,1}+\tilde{E}^{M,(1)}_{\kk,1}+\mathcal{O}(\kappa^2)\nonumber\\
    =&\tilde{E}^{M,(0)}_{\kk,1}\bigg\{
1-  \frac{\gamma^2 \pi}{\gamma^2+U_1^2/4}(N_f+1)
\kappa 
\bigg[
\log\bigg( \bigg| \frac{|v_\star \Lambda_c|^2 -(\tilde{E}^{M,(0)}_{\kk,1})^2}{(\tilde{E}^{M,(0)}_{\kk,1})^2}\bigg|\bigg)
- \frac{M}{2\tilde{E}^{M,(0)}_{\kk,1}}\log\bigg(\bigg|\frac{M+\tilde{E}^{M,(0)}_{\kk,1}}{M-\tilde{E}^{M,(0)}_{\kk,1}}\bigg|\bigg)\nonumber\\
&
{-\frac{1}{2}\log\bigg(\bigg|\frac{M^2-(\tilde{E}^{M,(0)}_{\kk,1})^2}{(\tilde{E}^{M,(0)}_{\kk,1})^2}\bigg|\bigg)}
\bigg]
\bigg\}+\mathcal{O}(\kappa^2)
\label{eq:renormalize_middle}
\end{align}
Near $|\kk|\to0$, we have $\tilde{E}^{M,(0)}_{\kk,1}\sim |v_{\star}\kk|^2$. Only keep up to the $|v_{\star}\kk|^2$ term, \cref{eq:renormalize_middle} becomes
\begin{align}
    \tilde{E}^{M}_{\kk,1}=&\tilde{E}^{M,(0)}_{\kk,1}\bigg\{
1-  \frac{\gamma^2 \pi}{\gamma^2+U_1^2/4}(N_f+1)
\kappa 
\bigg[
\log\bigg( \bigg| \frac{|v_\star \Lambda_c|^2 }{(\tilde{E}^{M,(0)}_{\kk,1})^2}\bigg|\bigg)
- 1
{-\frac{1}{2}\log\bigg(\bigg|\frac{M^2}{(\tilde{E}^{M,(0)}_{\kk,1})^2}\bigg|\bigg)}
\bigg]+\mathcal{O}(|v_{\star}\kk|^2)
\bigg\}+\mathcal{O}(\kappa^2)\nonumber\\
=&\frac{U_1^2/4|v_\star\kk|^2}{M(\gamma^2 +U_1^2/4)}\bigg\{
1-  \frac{\gamma^2 \pi}{\gamma^2+U_1^2/4}(N_f+1)
\kappa 
\bigg[2
\log\bigg( \frac{|v_{\star}\Lambda_c|M(\gamma^2 +U_1^2/4)}{U_1^2/4|v_\star\kk|^2}\bigg)
- 1
{-\log\bigg( \frac{M^2(\gamma^2 +U_1^2/4)}{U_1^2/4|v_\star\kk|^2}\bigg)}
\bigg]
\bigg\}\nonumber\\
&+\mathcal{O}(\kappa^2,|v_{\star}\kk|^4)\nonumber\\
=&\frac{U_1^2/4|v_\star\kk|^2}{M(\gamma^2 +U_1^2/4)}\bigg\{
1-  \frac{\gamma^2 \pi}{\gamma^2+U_1^2/4}(N_f+1)
\kappa 
\bigg[
\log\bigg( \frac{|v_{\star}\Lambda_c|^2(4\gamma^2 +U_1^2)}{U_1^2|v_\star\kk|^2}\bigg)
- 1
\bigg]
\bigg\}+\mathcal{O}(\kappa^2,|v_{\star}\kk|^4)
\end{align}
The expression for the scattering time \cref{eq:scattering_time} is already first order term of $\kappa$.
Thus up to first order of $\kappa$, the scattering time in the limit of $|v_{\star}\kk|\to 0$ is
\begin{align}
\frac{1}{{\tau}^{M}_{\kk,n=1}} 
= &
{\Gamma}^{M}_{\omega,\kk,1}
\bigg|_{\omega=\tilde{E}^{M}_{\kk,1}} ={\Gamma}^{M}_{\omega,\kk,1}
\bigg|_{\omega=\tilde{E}^{M,(0)}_{\kk,1}} +\mathcal{O}(\kappa^2)\nonumber\\
=&\frac{\gamma^2 U_1^2\pi |v_{\star}\kk|^2}{4M(\gamma^2+U_1^2/4)^2}\kappa(N_f+1)\mathrm{sgn}(\omega)
\bigg(
\pi+\frac{\pi}{2}\frac{{\color{black}M}-|\omega|}{|\omega|}
\bigg)\bigg|_{\omega=\tilde{E}^{M,(0)}_{\kk,1}}+\mathcal{O}(\kappa^2)\nonumber\\
=&\frac{\gamma^2 U_1^2\pi^2 |v_{\star}\kk|^2}{8M(\gamma^2+U_1^2/4)^2}\kappa(N_f+1)\mathrm{sgn}(\omega)+
{\color{black}\frac{\gamma^2 U_1^2\pi^2 |v_{\star}\kk|^2}{8(\gamma^2+U_1^2/4)^2}\kappa(N_f+1)\frac{1}{\tilde{E}^{M,(0)}_{\kk,1}}}
+\mathcal{O}(\kappa^2)\nonumber\\
=&{\color{black}\frac{M\gamma^2\pi^2}{2(\gamma^2+U_1^2/4)}(N_f+1)\kappa}+\frac{\gamma^2 U_1^2\pi^2 |v_{\star}\kk|^2}{8M(\gamma^2+U_1^2/4)^2}\kappa(N_f+1)\mathrm{sgn}(\omega)+\mathcal{O}(\kappa^2)\nonumber\\
=&{\color{black}\frac{M\gamma^2\pi^2}{2(\gamma^2+U_1^2/4)}(N_f+1)\kappa}+\mathcal{O}(\kappa^2,|v_{\star}\kk|^2)
\label{eq:scattering_time_quad}
\end{align}

Using the relation between $f_{M}(i\omega)$, hybridization function $\Delta_{M}(i\omega)$ and $c$ electron spectral function $\rho_c^{hyb}(\epsilon)$ \cref{eq:def_fM,eq:Delta_rho_relation}, the scattering time could be expressed as (\cref{eq:scattering_time})
\begin{align}
    \frac{1}{{\tau}^{M}_{\kk,n=1}} =&E^{M}_{\kk,1}\frac{\gamma^2(N_f+1)}{\gamma^2+U_1^2/4}{\pi\kappa\mathrm{Im}[f_{M}(\omega+i0^+)]}\bigg|_{\omega=E^{M}_{\kk,1}}\nonumber\\
    =&-E^{M}_{\kk,1}\frac{\gamma^2(N_f+1)}{\gamma^2+U_1^2/4}\frac{\gamma^2\mathrm{Im}[\Delta_{M}(\omega+i0^+)]}{\omega}\bigg|_{\omega=E^{M}_{\kk,1}}\nonumber\\
    =&E^{M}_{\kk,1}\frac{\gamma^2(N_f+1)}{\gamma^2+U_1^2/4}\frac{\pi \rho_c^{hyb}(\omega)}{\omega}\bigg|_{\omega=E^{M}_{\kk,1}}\nonumber\\
    =&\frac{\gamma^2(N_f+1)}{\gamma^2+U_1^2/4}{\pi \rho_c^{hyb}(E^{M}_{\kk,1})}\label{eq:scattering_time_dos_quad}
\end{align}
which is proportional to the density of state of the quadratic gapless mode $E^{M}_{\kk,1}$ (\cref{eq:pole_expression_finite_M}).
As the DOS of a quadratic band is finite, the scattering rate is also finite as given by \cref{eq:scattering_time_quad}.
As a comparison, for the $M=0$ case, from \cref{eq:hyb_poles}, we can show that the scattering time of the gapless mode can also be written in a similar form as
\begin{align}
    \frac{1}{{\tau}_{\kk,n=1}} =&E_{\kk,1}\frac{1}{2}\frac{\gamma^2(N_f+1)}{\gamma^2+U_1^2/4}{\pi\kappa\mathrm{Im}[f(\omega+i0^+)]}\bigg|_{\omega=E_{\kk,1}}\nonumber\\
    =&\frac{1}{2}\frac{\gamma^2(N_f+1)}{\gamma^2+U_1^2/4}\pi \rho_c^{hyb}(E_{\kk,1})\bigg|_{M=0}
    \label{eq:scattering_time_dos_linear}
\end{align}
which is proportional to the density of state linear dispersive gapless mode $E_{\kk,1}$.
Thus the scattering time will be linear in $|\kk|$ as given by \cref{eq:scattering_time_linear}.

Then we investigate the $\pm M$ gapped excitations described by $\pm\mathcal{E}^{M}_{i\omega,\kk,n=2}$.
We write
\begin{equation}
    \mathcal{E}^{M}_{\omega+i0^+,\kk,2} = \mathcal{R}^{M}_{\omega,\kk,2}
- i{\Gamma}^{M}_{\omega,\kk,2} 
\end{equation}
Expanding $\mathcal{E}^{M}_{i\omega,\kk,n=2}$ in \cref{eq:pole_expression_finite_M} to the first order of $\kappa$, we obtain
\begin{align}
    \mathcal{R}^{M}_{\omega,\kk,2}\approx& E^{M}_{\kk,2}+\frac{U_1^2/4 |v_{\star}\kk|^2\gamma^2}{M(U_1^2/4+\gamma^2-M^2)^2}\pi(N_f+1)\kappa \mathrm{Re}[f_{M}(\omega+i0^+)]+\mathcal{O}(\kappa^2)\nonumber\\
    \Gamma^{M}_{\omega,\kk,2}\approx&-\frac{U_1^2/4 |v_{\star}\kk|^2\gamma^2}{M(U_1^2/4+\gamma^2-M^2)^2}\pi(N_f+1)\kappa \mathrm{Im}[f_{M}(\omega+i0^+)]+\mathcal{O}(\kappa^2)
\end{align}
The renormalized dispersion $\tilde{E}^{M}_{\kk,2}$ is given by 
\begin{equation}
    (\omega-\mathcal{R}^{M}_{\omega,\kk,2})\bigg|_{\omega=\tilde{E}^{M}_{\kk,2}}=0
\end{equation}
Write $\tilde{E}^{M}_{\kk,2}$ into series of $\kappa$ as 
\begin{equation}
   \tilde{E}^{M}_{\kk,2}=\tilde{E}^{M,(0)}_{\kk,2}+\tilde{E}^{M,(1)}_{\kk,2} +\mathcal{O}(\kappa^2)
\end{equation}
Then, similar to the case of the gapless excitation, we have
\begin{equation}
    \tilde{E}^{M,(0)}_{\kk,2}=E^{M}_{\kk,2}= -M  
- \frac{(U_1^2/4 -M^2)|v_\star\kk|^2}{M (
U_1^2/4 +\gamma^2 -M^2 
)}
\end{equation}
In addition, the renormalized dispersion and scattering time (up to first order of $\kappa$) are
\begin{align}
\tilde{E}^{M}_{\kk,2}=&\mathcal{R}^{M}_{\omega,\kk,2}\bigg|_{\omega=\tilde{E}^{M,(0)}_{\kk,2}}=E^{M}_{\kk,2}+\frac{U_1^2/4 |v_{\star}\kk|^2\gamma^2}{M(U_1^2/4+\gamma^2-M^2)^2}\pi(N_f+1)\kappa \mathrm{Re}[f_{M}(E^{M}_{\kk,2}+i0^+)]+\mathcal{O}(\kappa^2)\nonumber\\
\frac{1}{{\tau}^{M}_{\kk,n=2}}=&\Gamma^{M}_{\omega,\kk,2}\bigg|_{\omega=\tilde{E}^{M,(0)}_{\kk,2}}=-\frac{U_1^2/4 |v_{\star}\kk|^2\gamma^2}{M(U_1^2/4+\gamma^2-M^2)^2}\pi(N_f+1)\kappa \mathrm{Im}[f_{M}(E^{M}_{\kk,2}+i0^+)]+\mathcal{O}(\kappa^2)
\end{align}
Near $|\kk|\to 0$, $E^{M}_{\kk,2}=-M+\mathcal{O}(|v_{\star}\kk|^2)$.
Besides, as $M<U_1/2$, we always have $E^{M}_{\kk,2}<-M$ and $|E^{M}_{\kk,2}|>M$.
Thus, keep up to the lowest order of $|v_{\star}\kk|^2$ term, $f_{M}(E^{M}_{\kk,2}+i0^+)$ is evaluated as
\begin{align}
    \mathrm{Re}[f_{M}(E^{M}_{\kk,2}+i0^+)]=&\log\bigg(\bigg|\frac{|v_{\star}\Lambda_c|^2-(E^{M}_{\kk,2})^2}{(E^{M}_{\kk,2})^2}\bigg|\bigg)-\frac{M}{2E^{M}_{\kk,2}}\log\bigg(\bigg|\frac{E^{M}_{\kk,2}+M}{E^{M}_{\kk,2}-M}\bigg|\bigg)-\frac{1}{2}\log\bigg(\bigg|\frac{M^2-(E^{M}_{\kk,2})^2}{(E^{M}_{\kk,2})^2}\bigg|\bigg)
    \nonumber\\
    =&\log\bigg[\frac{|v_{\star}\Lambda_c|^2-M^2}{M^2}+\mathcal{O}(|v_{\star}\kk|^2)\bigg]+\frac{1}{2}\log\bigg[\frac{(U_1^2/4-M^2)|v_{\star}\kk|^2}{2M^2(U_1^2/4+\gamma^2-M^2)}+\mathcal{O}(|v_{\star}\kk|^4)\bigg]\nonumber\\
    &-\frac{1}{2}\log\bigg[\frac{2(U_1^2/4-M^2)|v_{\star}\kk|^2}{M^2(U_1^2/4+\gamma^2-M^2)}+\mathcal{O}(|v_{\star}\kk|^4)\bigg]
    \nonumber\\
    =&\log\bigg(\frac{|v_{\star}\Lambda_c|^2-M^2}{2M^2}\bigg)+\mathcal{O}(|v_{\star}\kk|^2)
    \nonumber\\
    \mathrm{Im}[f_{M}(E^{M}_{\kk,2}+i0^+)]=&\pi\mathrm{sgn}(E^{M}_{\kk,2})\theta(|v_{\star}\Lambda_c|-|E^{M}_{\kk,2}|)+\frac{\pi}{2}\frac{M-|E^{M}_{\kk,2}|}{E^{M}_{\kk,2}}\theta(M-|E^{M}_{\kk,2}|)=-\pi
\end{align}
Then the renormalized dispersion and scattering time become
\begin{align}
    \tilde{E}^{M}_{\kk,2}=&-M  
- \frac{(U_1^2/4 -M^2)|v_\star\kk|^2}{M (
U_1^2/4 +\gamma^2 -M^2 
)}
+\frac{U_1^2/4 |v_{\star}\kk|^2\gamma^2}{M(U_1^2/4+\gamma^2-M^2)^2}\pi(N_f+1)\kappa \log\bigg(\frac{|v_{\star}\Lambda_c|^2-M^2}{2M^2}\bigg)+\mathcal{O}(\kappa^2,|v_{\star}\kk|^4)\nonumber\\
\frac{1}{\tilde{\tau}_{\kk,n=2}}=&\frac{U_1^2/4 |v_{\star}\kk|^2\gamma^2}{M(U_1^2/4+\gamma^2-M^2)^2}\pi^2(N_f+1)\kappa +\mathcal{O}(\kappa^2,|v_{\star}\kk|^4)
\end{align}

\section{$c$-electron self-energy from $fc$ interactions at charge neutrality $\nu=0$}\label{app:HJ_effect}
In \cref{app:hyb_exp} to \cref{app:chiral_limit},
we focused on correlation effects arising from the
Hubbard interaction $U_1$, treating the hybridization $\gamma$ perturbatively
and retaining contributions up to order $\gamma^2$.
This correlation effects first generate a dynamical self-energy for the $f$ electrons, which subsequently contribute to the damping rate of the hybridized bands.

In this section, we turn to correlation effects induced by the $fc$ interactions
$H_J$ and $H_W$, which will also generate a dynamical self-energy and finite scattering rate.
We begin in the zero-hybridization limit, where the $fc$
interactions are treated at the Hartree-Fock level previously.
(Although when calculating the hybridization function, we take the approximation $W_1=W_2$ and $J=0$ to get an analytic formula. While in the other parts, the HF contribution of $H_J$ is kept.)
We then incorporate
dynamical correlation effects generated by the $fc$ interactions by expanding
the self-energy to second order in the $fc$ interaction strength, with the first-order term corresponding to the Hartree-Fock contribution.
Throughout this section, we set $\gamma=0$ and focus on charge neutrality to simplify the analysis. Since, in the $\gamma=0$ limit, the low-energy gapless excitations are formed by the $c$ electrons, we focus on the self-energy of the $c$ electrons. 

We consider a generic $fc$ interaction taking the form of 
\ba 
H_{fc}^{int}= 
\frac{1}{N_M}
\sum_{\RR,\kk,\qq}\sum_{iji'j'}\mathcal{J}_{iji'j'}
f_{\RR,i}^\dag f_{\RR,j} 
c_{\kk,i'}^\dag c_{\kk+\qq,j'}  e^{i\qq\cdot\RR}
\ea 
To simplify the notation, we use index $i$ (also $j,i',j'$) to characterize the unified spin, orbital, and valley degrees of freedom. $\mathcal{J}_{iji'j'}$ denotes a generic $fc$ interactions including both $H_{W}$ and $H_J$. 
Written explicitly, via the original spin, valley, and orbital indices, the interaction tensor takes the form
\ba 
\label{eq:def_mathcal_J}
\mathcal{J}_{ \alpha_1\eta_1s_1,\alpha_2\eta_2s_2,a_3\eta_3s_3,a_4\eta_4s_4 } = W \delta_{\alpha_1\eta_1s_1,\alpha_2\eta_2s_2} 
\delta_{a_3\eta_3s_3,a_4\eta_4s_4}
-\frac{J}{2}
\delta_{\alpha_1+2,a_4}
\delta_{\alpha_2+2,a_3}\delta_{\eta_1 s_1,\eta_4s_4}
\delta_{\eta_2s_2,\eta_3s_3}(\eta_1\eta_2 +(-1)^{\alpha_1+\alpha_2})
\ea 
where we also assume $W_1= W_3=W$ for simplicity.

We aim to calculate the effect of $H_{fc}^{int}$ on the $c$ electron self-energy. We separate the system into 
\ba 
&S = S_{decouple} + S_{fc} \nonumber\\ 
&S_{decouple}=\int_{\tau}\sum_{\eta sa,\kk}c^{\dagger}_{\kk,a\eta s}(\tau)\partial_{\tau}c_{\kk,a\eta s}(\tau)+\int_{\tau}\sum_{\RR,\alpha\eta s}f^{\dagger}_{\RR,\alpha\eta s}(\tau)\partial_{\tau}f_{\RR,\alpha\eta s}(\tau)+\int_{\tau}[H'(\tau)|_{H^{(cf,\eta)}=0}]\nonumber\\
&S_{fc} = \int_\tau [H_{fc}^{int}(\tau) - H_{fc}^{HF}(\tau) ]
\ea 
where $H'$ is the meanfield Hamiltonian given by (\cref{eq:generic_Ham,eq:HcHcf})
\ba 
H' =& \sum_{\eta s} \sum_{aa'} \sum_{\kk} ( H^{(c,\eta)}_{a,a'}(\kk+\GG) 
  ) c_{\kk,a\eta s}^\dagger c_{\kk,a'\eta s}
  + \frac1{\sqrt{N_M}} \sum_{\eta s \alpha s} \sum_{\RR} \sum_{\kk}
  \bigg( e^{- i\kk\cdot\RR } H_{a\alpha}^{(cf,\eta)}(\kk) 
    c_{\kk,a\eta s}^\dagger f_{\RR ,\alpha \eta s} + h.c. 
  \bigg) \nonumber\\ 
  &+ \sum_{\RR,\alpha\eta s}(\epsilon_f-\mu) f_{\RR, \alpha \eta s}^\dag f_{\RR,\alpha \eta s}
  + {\frac{U_1}{2} \sum_{\RR,\alpha \eta s\ne \alpha'\eta's'}f_{\RR,\alpha\eta s}^\dag f_{\RR,\alpha\eta s} f_{\RR,\alpha'\eta's'}^\dag f_{\RR,\alpha'\eta's'}}
  \, .
\ea
\begin{align}
 & H^{(c,\eta)}(\kk) = 
  \begin{pmatrix}
    \sigma_0(\epsilon_{c,1}-\mu) & v_\star (\eta k_x \sigma_0 + ik_y \sigma_z) \\
    v_\star (\eta k_x \sigma_0 - ik_y \sigma_z) & \sigma_0(\epsilon_{c,2}-\mu)+ M\sigma_x 
  \end{pmatrix} \nonumber\\
  &
  H^{(cf,\eta)}(\kk) = 
  \begin{pmatrix}
    \gamma \sigma_0 + v_\star' (\eta k_x \sigma_x + k_y \sigma_y) \\
    0_{2\times2}
  \end{pmatrix}
\end{align}
which includes the non-interacting Hamiltonian $H_0$, the Hubbard interactions $H_{U_1}$, Hartree-Fock contributions from the $H_V$ and the Hartree-Fock contributions from $H_{fc}^{int}$ denoted by $H_{fc}^{HF}$. 
By taking the single particle $cf$ hybridization strength to be zero $H^{(cf,\eta)}=0$, the $c$ and $f$ electrons are decoupled and could be separated into 
\begin{align}
    S_{decouple}=&S_c+S_{atom}\nonumber\\
    S_c=&\int_{\tau}\sum_{\eta sa,\kk}c^{\dagger}_{\kk,a\eta s}(\tau)\partial_{\tau}c_{\kk,a\eta s}(\tau)+\int_{\tau}H_c(\tau)\nonumber\\
    S_{atom}=&\int_{\tau}\sum_{\RR,\alpha\eta s}f^{\dagger}_{\RR,\alpha\eta s}(\tau)\partial_{\tau}f_{\RR,\alpha\eta s}(\tau)+\int_{\tau}H_{atom}(\tau)\nonumber\\
    H_c=&\sum_{\eta s} \sum_{aa'} \sum_{\kk} ( H^{(c,\eta)}_{a,a'}(\kk+\GG) 
  ) c_{\kk,a\eta s}^\dagger c_{\kk,a'\eta s}\nonumber\\
  H_{atom}=&\sum_{\RR,\alpha\eta s}(\epsilon_f-\mu) f_{\RR, \alpha \eta s}^\dag f_{\RR,\alpha \eta s}
  + {\frac{U_1}{2} \sum_{\RR,\alpha \eta s\ne \alpha'\eta's'}f_{\RR,\alpha\eta s}^\dag f_{\RR,\alpha\eta s} f_{\RR,\alpha'\eta's'}^\dag f_{\RR,\alpha'\eta's'}}
  \label{eq:H_decouple}
\end{align}
Note that $H_{atom}$ is nothing but the atomic Hamiltonian discussed defined in \cref{eq:atomic_Hamiltonian_rewrite,eq:H_atomic_def}.
In addition, $S_{fc}$ characterizes the interaction effect induced by $H_{fc}^{int}$ that goes beyond the Hartree-Fock approximation. 
Written explicitly, we have 
\ba 
[H_{fc}^{int}- H_{fc}^{HF} ] 
= \frac{1}{N_M}
\sum_{\RR,\kk,\qq} \sum_{iji'j'}
\mathcal{J}_{iji'j'} :f_{\RR,i}^\dag f_{\RR,j} :
: c_{\kk,i'}^\dag c_{\kk+\qq,j'} : 
e^{i\qq\cdot\RR}
\ea 
where we have defined the normal order with respect to the ground state of $S_{decouple}$ as
\ba 
&:f_{\RR,i}^\dag f_{\RR,j} :
=f_{\RR,i}^\dag f_{\RR,j} 
- \langle f_{\RR,i}^\dag f_{\RR,j} \rangle_{S_{decouple}} \nonumber\\ 
&
: c_{\kk,i'}^\dag c_{\kk+\qq,j'} : 
= c_{\kk,i'}^\dag c_{\kk+\qq,j'} 
-\delta_{\qq,0}
\langle  c_{\kk,i'}^\dag c_{\kk,j'} \rangle_{S_{decouple}}
\ea 
where, $\langle O\rangle_{S_{decouple}}$ is defined as
\begin{align}
    \langle O\rangle_{S_{decouple}}=&\frac{1}{Z_{S_{decouple}}}\int D[f,f^{\dagger},c,c^{\dagger}]Oe^{-S_{decouple}}\nonumber\\
    Z_{S_{decouple}}=&\int D[f,f^{\dagger},c,c^{\dagger}]e^{-S_{decouple}}
\end{align}
For operators $O_c$ constructed by $c$ electrons and $O_c$ constructed by $f$, we have
\begin{align}
    \langle O_c O_f\rangle_{S_{decouple}}=&\frac{1}{Z_{S_{decouple}}}\int D[f,f^{\dagger},c,c^{\dagger}]O_cO_fe^{-S_c-S_{atom}}\nonumber\\
    =&\frac{1}{Z_{S_{decouple}}}\bigg[\int D[f,f^{\dagger}]O_f e^{-S_{atom}}\bigg]\bigg[\int D[c,c^{\dagger}]O_c e^{-S_c}\bigg]\nonumber\\
    =&\frac{1}{Z_{S_{decouple}}^2}\bigg[\int D[f,f^{\dagger},c,c^{\dagger}]O_f e^{-S_{decouple}}\bigg]\bigg[\int D[f,f^{\dagger},c,c^{\dagger}]O_c e^{-S_{decouple}}\bigg]\nonumber\\
    =&\langle O_c\rangle_{S_{decouple}}\langle O_f\rangle_{S_{decouple}}
    \label{eq:decouple_expectation}
\end{align}
Besides, by integrating out the $c$ electron, $\langle O_f\rangle_{S_{decouple}}$ can also be written as $\langle O_f\rangle_{\gamma=0}$ defined in \cref{eq:exp_path_integral}.

We now aim to calculate the $c$-electron Green's function by treating $\mathcal{J}_{iji'j'}$ perturbatively. The Green's function of $c$ electron can now be calculated via
\ba 
 &G_{c,ij}(\tau-\tau',\kk)=-\langle T_\tau c_{\kk, i}(\tau) c_{\kk,j }^\dag(\tau')\rangle  \nonumber\\ 
 \approx & 
 -\langle T_\tau c_{\kk,i}(\tau) 
 c_{\kk,j}(\tau') 
 \rangle_{S_{decouple}} \nonumber\\
 &+\frac{1}{N_M}\int_{\tau_1}\sum_{\RR,,\kk,\qq}\sum_{iji'j'}\mathcal{J}_{iji'j'}e^{i\qq\cdot\RR}\langle T_{\tau}:f_{\RR,i}^\dag(\tau_1) f_{\RR,j}(\tau_1) :
: c_{\kk,i'}^\dag(\tau_1) c_{\kk+\qq,j'}(\tau_1) : c_{\kk,i}(\tau)c^{\dagger}_{\kk,j}(\tau')\rangle_{S_{decouple}}
 \nonumber\\
 &- 
 \frac{1}{2N^2}\int_{\tau_1,\tau_2}
 \sum_{\RR_1,\RR_2,\kk_1,\qq_1,\kk_2,\qq_2}
 \sum_{i_1j_1i_1'j_1',i_2j_2i_2'j_2'}
 \mathcal{J}_{i_1j_1i_1'j_1'}
 \mathcal{J}_{i_2j_2i_2'j_2'} 
 e^{i\qq_1\cdot\RR_1 +i\qq_2\cdot\RR_2} \nonumber\\ 
 &
 \langle 
 T_\tau 
 :f_{\RR_1,i_1}^\dag (\tau_1) f_{\RR_1,j_1} (\tau_1):
 : c^\dag_{\kk_1,i_1'}(\tau_1) c_{\kk_1+\qq_1,j_1'}(\tau_1): \nonumber\\
 &
 :f_{\RR_2,i_2}^\dag (\tau_2) f_{\RR_2,j_2} (\tau_2):
 : c^\dag_{\kk_2,i_2'}(\tau_2) c_{\kk_2+\qq_2,j_2'}(\tau_2) :
 c_{\kk,i}(\tau) c_{\kk,j}^\dag(\tau')
 \rangle_{S_{decouple},connect}
\ea 
where the subscript ``connect'' indicates that only the connected contribution is included. 
According to \cref{eq:decouple_expectation}, the first order term in $\mathcal{J}$ is proportional to
\begin{equation}
    \langle :f^{\dagger}_{\RR,i}(\tau_1)f_{\RR,j}(\tau_1):\rangle_{S_{decouple}}=\langle f^{\dagger}_{\RR,i}(\tau_1)f_{\RR,j}(\tau_1)-\langle f^{\dagger}_{\RR,i}(\tau_1)f_{\RR,j}(\tau_1)\rangle_{S_{decouple}}\rangle_{S_{decouple}}=0\label{eq:first_order_vanish}
\end{equation}
Thus, the first order term vanishes.
As for the second order term, using Wick's theorem we find
\begin{align}
   & \langle 
 T_\tau 
 :f_{\RR_1,i_1}^\dag (\tau_1) f_{\RR_1,j_1} (\tau_1):
 : c^\dag_{\kk_1,i_1'}(\tau_1) c_{\kk_1+\qq_1,j_1'}(\tau_1): \nonumber\\
 &
 :f_{\RR_2,i_2}^\dag (\tau_2) f_{\RR_2,j_2} (\tau_2):
 : c^\dag_{\kk_2,i_2'}(\tau_2) c_{\kk_2+\qq_2,j_2'}(\tau_2) :
 c_{\kk,i}(\tau) c_{\kk,j}^\dag(\tau')
 \rangle_{S_{decouple},connect}\nonumber\\
 =&\langle T_{\tau}:f_{\RR_1,i_1}^\dag (\tau_1) f_{\RR_1,j_1} (\tau_1)::f_{\RR_2,i_2}^\dag (\tau_2) f_{\RR_2,j_2} (\tau_2):\rangle_{S_{decouple}}\times\nonumber\\
 &\langle T_{\tau} : c^\dag_{\kk_1,i_1'}(\tau_1) c_{\kk_1+\qq_1,j_1'}(\tau_1):: c^\dag_{\kk_2,i_2'}(\tau_2) c_{\kk_2+\qq_2,j_2'}(\tau_2) :
 c_{\kk,i}(\tau) c_{\kk,j}^\dag(\tau')
 \rangle_{S_{decouple},connect}\nonumber\\
 =&\delta_{\RR_1,\RR_2}\chi_{i_1j_1,i_2j_2}(\tau_1-\tau_2)\langle T_{\tau}c_{\kk,i}(\tau) : c^\dag_{\kk_1,i_1'}(\tau_1) c_{\kk_1+\qq_1,j_1'}(\tau_1):: c^\dag_{\kk_2,i_2'}(\tau_2) c_{\kk_2+\qq_2,j_2'}(\tau_2) :
  c_{\kk,j}^\dag(\tau')
 \rangle_{S_{decouple},connect}\nonumber\\
 =&-\delta_{\RR_1,\RR_2}\chi_{i_1j_1,i_2j_2}(\tau_1-\tau_2)\delta
 _{\kk,\kk_1}G^{S_{decouple}}_{c,ii_1'}(\tau-\tau_1,\kk)\delta_{\kk_1+\qq_1,\kk_2}G_{c,j_1'i_2'}^{S_{decouple}}(\tau_1-\tau_2,\kk_2)\delta_{\kk_2+\qq_2,\kk}G_{c,j_2'j}^{S_{decouple}}(\tau_2-\tau',\kk)
\end{align}
where 
\ba 
\label{eq:def_G_chi}
&G_{c,ij}^{S_{decouple}}(\tau-\tau',\kk) = 
-\langle T_\tau c_{\kk,i}(\tau) c_{\kk,j}^\dag(\tau')\rangle_{S_{decouple}} \nonumber\\
& \chi_{i_1j_1,i_2j_2}(\tau_1-\tau_2) 
= \langle 
 T_\tau 
 :f_{\RR,i_1}^\dag (\tau_1) f_{\RR,j_1} (\tau_1):
 :f_{\RR,i_2}^\dag (\tau_2) f_{\RR,j_2} (\tau_2):
 \rangle_{S_{decouple}}
\ea 
denotes the Hartree-Fock Green's function of $c$ electrons and the susceptibility (four-fermion correlation) function of $f$ electrons at the zero-hybridization limit.
In summary, up to second order of $\mathcal{J}$, the $c$ electron Green's is written as
\ba 
\label{eq:CNP_green_c_pertubration}
& G_{c,ij}(\tau-\tau',\kk)  \nonumber\\
\approx & 
G_{c,ij}^{S_{decouple}}(\tau-\tau',\kk)
{+} 
 \frac{1}{N^2}\int_{\tau_1,\tau_2}
 \sum_{\RR,\qq}
 \sum_{i_1j_1i_1'j_1',i_2j_2i_2'j_2'}
 \mathcal{J}_{i_1j_1i_1'j_1'}
 \mathcal{J}_{i_2j_2i_2'j_2'}   \nonumber\\
 &
 \chi_{i_1j_1,i_2j_2}(\tau_1-\tau_2) 
 G_{c,j_1'i_2'}^{S_{decouple}}(\tau_1-\tau_2,\kk-\qq) 
 G_{c,j_2' j}^{S_{decouple}}(\tau_2-\tau',\kk) 
 G_{c,ii_1'}^{S_{decouple}}(\tau-\tau_1,\kk) 
\ea 
In the Matsubara frequency domain, we find 
\ba 
 \label{eq:gc_w_expand_in_J_with_chi}
G_{c,ij}(i\omega,\kk) = &\frac{1}{\beta}\int_{\tau,\tau'}G_{c,ij}(\tau-\tau',\kk)e^{i\omega(\tau-\tau')} \nonumber\\
\approx &G_{c,ij}^{S_{decouple}}(i\omega,\kk) +\sum_{i_1j_1i_1'j_1',i_2j_2i_2'j_2'}
 \mathcal{J}_{i_1j_1i_1'j_1'}
 \mathcal{J}_{i_2j_2i_2'j_2'}   \nonumber\\
 &
 \frac{1}{\beta}\sum_{i\Omega} \chi_{i_1j_1,i_2j_2}(i\Omega) 
 \frac{1}{N_M}\sum_{\kk'}G_{c,j_1'i_2'}^{S_{decouple}}(\kk',i\omega-i\Omega) 
 G_{c,j_2' j}^{S_{decouple}}(i\omega,\kk) 
 G_{c,ii_1'}^{S_{decouple}}(i\omega,\kk) 
\ea 
where $\chi_{i_1j_1,i_2j_2}(i\Omega) = \int_{\tau} \chi_{i_1j_1,i_2j_2}(\tau)e^{i\Omega\tau}$.
In addition, in the zero hybridization limit and exact integer filling we have $\epsilon_{c,1}=\epsilon_{c,2}=\mu$ with $c$ electron filling $\nu_c=0$ (see discussion in \cref{app:hybridization_func}).
Therefore, the Hartree Fock Green's function of $c$ electron is written as
\begin{align}
    G^{S_{decouple}}_{c,a_1\eta_1s_1,a_2\eta_2s_2}(i\omega,\kk)=&\delta_{\eta_1s_1,\eta_2s_2}\bigg[i\omega\mathbb{I}-H^{(c,\eta_1)}(\kk)\bigg]^{-1}_{a_1,a_2}\nonumber\\
    =&\delta_{s_1,s_2}\delta_{\eta_1,\eta_2}
\bigg[ 
i\omega \mathbb{I} 
- 
  \begin{pmatrix}
    0_{2\times 2}& v_\star (\eta k_x \sigma_0 + ik_y \sigma_z) \\
    v_\star (\eta k_x \sigma_0 - ik_y \sigma_z) &  M\sigma_x 
  \end{pmatrix} 
\bigg]^{-1}_{a_1,a_2}
\label{eq:G_decouple}
\end{align}

We can also use Dyson's equation to obtain the self-energy of $c$ electrons at order $\mathcal{J}^2$. In the matrix form, we define the self-energy as 
\ba 
\label{eq:dyson_of_c}
&\Sigma_c(i\omega,\kk) = [G_{c}^{S_{decouple}}(i\omega,\kk)]^{-1} -[G_{c}(i\omega,\kk) ]^{-1}
\ea 
which indicates 
\ba 
\label{eq:dyson_of_c_2}
G_c(i\omega,\kk) \approx G_{c}^{S_{decouple}}(i\omega,\kk) 
+G_{c}^{S_{decouple}}(i\omega,\kk) \cdot 
\Sigma_c(i\omega,\kk) 
\cdot G_{c}^{S_{decouple}}(i\omega,\kk)  +...
\ea 
 Combining \cref{eq:gc_w_expand_in_J_with_chi,eq:dyson_of_c,eq:dyson_of_c_2}, we find 
\ba 
\label{eq:sig_c_from_chi_G}
 [\Sigma_c(i\omega,\kk) ]_{i_1'j_2'}
\approx & {+}
 \sum_{i_1j_1 j_1',i_2j_2i_2'}
 \mathcal{J}_{i_1j_1i_1'j_1'}
 \mathcal{J}_{i_2j_2i_2'j_2'} 
 \frac{1}{\beta}\sum_{i\Omega}
 \chi_{i_1j_1,i_2j_2}(i\Omega) 
 \frac{1}{N_M}\sum_{\kk'}G_{c,j_1'i_2'}^{S_{decouple}}(i\omega-i\Omega,\kk') 
 + O(\mathcal{J}^3)
\ea

\subsection{Evaluation of $\Sigma_c(i\omega)$}
We now evaluate $\Sigma_c$ explicitly. We focus on charge neutrality with $f$ electron filling $n_f=N_f/2$. 

We first consider the susceptibility of $f$ electron (\cref{eq:def_G_chi}). 
At charge neutrality, we have
\ba 
:f_{\RR,i}^\dag f_{\RR,j}: = f_{\RR,i}^\dag f_{\RR,j} -\langle f_{\RR,i}^\dag f_{\RR,j}\rangle _{S_{decouple}}= 
f_{\RR,i}^\dag f_{\RR,j}
- \delta_{i,j} \frac{n_f}{N_f}
\ea 
Using \cref{eq:exp_equivalence_single_site}, the four Fermion correlation function without normal order is calculated as (in the previous discussion of $\gamma$, the four Fermion correlation function \cref{eq:FT_of_four-point} is evaluated at four different time points, while in this section, the four Fermion correlation function is evaluated at two different time points)
\ba 
\label{eq:f_electron_propagtor}
&  \langle 
 T_\tau 
  f_{\RR,i_1}^\dag (\tau_1) f_{\RR,j_1} (\tau_1)
 f_{\RR,i_2}^\dag (\tau_2) f_{\RR,j_2} (\tau_2)
 \rangle_{S_{decouple}}=\langle 
 T_\tau 
  f_{\RR,i_1}^\dag (\tau_1) f_{\RR,j_1} (\tau_1)
 f_{\RR,i_2}^\dag (\tau_2) f_{\RR,j_2} (\tau_2)
 \rangle_{\gamma=0} \nonumber\\
= &
\begin{cases}
\frac{1}{Z_{atom,\RR}} \sum_{n,m} e^{-(\beta-\tau_1+\tau_2) E_n -(\tau_1-\tau_2) E_m}
\langle n|  f_{\RR,i_1}^\dag  f_{\RR,j_1}| m\rangle 
\langle m| f_{\RR,i_2}^\dag  f_{\RR,j_2} | n\rangle & \tau_1>\tau_2  \\
\frac{1}{Z_{atom,\RR}} \sum_{n,m} e^{-(\beta-\tau_2+\tau_1)E_m -(\tau_2-\tau_1)E_n}
\langle n|  f_{\RR,i_1}^\dag  f_{\RR,j_1}| m\rangle 
\langle m| f_{\RR,i_2}^\dag  f_{\RR,j_2} | n\rangle & \tau_1<\tau_2  \\
\end{cases}
\ea 
where $|n\rangle, |m\rangle$ denote the eigen states of atomic Hamiltonian of $f$ electron with energy $E_n, E_m$ respectively (\cref{eq:H_atomic_def}). We note that if the filling of the state $|n\rangle$ is $n_f$, then the condition $\langle m| f_{\RR,i_2}^\dag  f_{\RR,j_2} | n\rangle \ne 0$ implies that the filling of the state $|m\rangle$ is also $n_f$ and that $E_n = E_m$.
 Thus
\ba 
&  \langle 
 T_\tau 
  f_{\RR,i_1}^\dag (\tau_1) f_{\RR,j_1} (\tau_1)
 f_{\RR,i_2}^\dag (\tau_2) f_{\RR,j_2} (\tau_2)
 \rangle_{S_{decouple}} \nonumber\\
=  &
\frac{1}{Z_{atom,\RR}} \sum_{n,m} e^{- \beta  E_n }
\langle n|  f_{\RR,i_1}^\dag  f_{\RR,j_1}| m\rangle 
\langle m| f_{\RR,i_2}^\dag  f_{\RR,j_2} | n\rangle 
\label{eq:two_point_correlation}
\ea 
We further note that
\begin{itemize}
    \item If $i_1=j_1$, the non-zero contributions are from the term with $\ket{m}=\ket{n}$, thus the form factor becomes
    \begin{align}
        &\delta_{i_1,j_1}\langle n|  f_{\RR,i_1}^\dag  f_{\RR,j_1}| m\rangle 
\langle m| f_{\RR,i_2}^\dag  f_{\RR,j_2} | n\rangle \nonumber\\
=&\delta_{i_1,j_1}\bra{n}f^{\dagger}_{\RR,i_1}f_{\RR,i_1}\ket{n}\bra{n}f^{\dagger}_{\RR,i_2}f_{\RR,j_2}\ket{n}\nonumber\\
=&\delta_{i_1,j_1}\delta_{i_2,j_2}\bra{n}f^{\dagger}_{\RR,i_1}f_{\RR,i_1}\ket{n}\bra{n}f^{\dagger}_{\RR,i_2}f_{\RR,i_2}\ket{n}\nonumber\\
=&\delta_{i_1,j_1}\delta_{i_2,j_2}\delta_{i_1,i_2}\bra{n}f^{\dagger}_{\RR,i_1}f_{\RR,i_1}\ket{n}+\delta_{i_1,j_1}\delta_{i_2,j_2}(1-\delta_{i_1,i_2})\bra{n}f^{\dagger}_{\RR,i_1}f_{\RR,i_1}\ket{n}\bra{n}f^{\dagger}_{\RR,i_2}f_{\RR,i_2}\ket{n}
    \end{align}
    \item If $i_1\ne j_1$, the non-zero contributions come from the term with $\ket{m}=f^{\dagger}_{\RR,j_1}f_{\RR,i_1}\ket{n}$, then the form factors becomes
    \begin{align}
        &(1-\delta_{i_1,j_1})\bra{n}f^{\dagger}_{\RR,i_1}f_{\RR,j_1}\ket{m}\bra{m}f^{\dagger}_{\RR,i_2}f_{\RR,j_2}\ket{n}\nonumber\\
        =&(1-\delta_{i_1,j_1})\bra{n}f^{\dagger}_{\RR,i_1}f_{\RR,j_1}f^{\dagger}_{\RR,j_1}f_{\RR,i_1}\ket{n}\bra{n}f^{\dagger}_{\RR,i_1}f_{\RR,j_1}f^{\dagger}_{\RR,i_2}f_{\RR,j_2}\ket{n}\nonumber\\
        =&(1-\delta_{i_1,j_1})\bra{n}f^{\dagger}_{\RR,i_1}f_{\RR,i_1}\ket{n}\bra{n}f_{\RR,j_1}f^{\dagger}_{\RR,j_1}\ket{n}\bra{n}f^{\dagger}_{\RR,i_1}f_{\RR,j_2}\ket{n}\bra{n}f_{\RR,j_1}f^{\dagger}_{\RR,i_2}\ket{n}\nonumber\\
        =&(1-\delta_{i_1,j_1})\delta
        _{i_1,j_2}\delta_{i_2,j_1}\bra{n}f^{\dagger}_{\RR,i_1}f_{\RR,i_1}\ket{n}\bra{n}f_{\RR,j_1}f^{\dagger}_{\RR,j_1}\ket{n}
    \end{align}
\end{itemize}
In addition, in the low-temperature limit, we have the following relation (\cref{eq:partition_single_site})
\begin{equation}
    \lim_{\beta\to+\infty}\frac{e^{-\beta E_n}}{Z_{atom,\RR}}=\binom{N_f}{n_f}^{-1}\delta_{n,gs}
\end{equation}
Therefore, in the low-temperature limit, \cref{eq:two_point_correlation} becomes
\ba 
&  \langle 
 T_\tau 
  f_{\RR,i_1}^\dag (\tau_1) f_{\RR,j_1} (\tau_1)
 f_{\RR,i_2}^\dag (\tau_2) f_{\RR,j_2} (\tau_2)
 \rangle_{S_{decouple}} \nonumber\\
 =&\binom{N_f}{n_f}^{-1}\sum_{n\in GS}\bigg[
 \delta_{i_1,j_1}\delta_{i_2,j_2}\delta_{i_1,i_2}\bra{n}f^{\dagger}_{\RR,i_1}f_{\RR,i_1}\ket{n}+\delta_{i_1,j_1}\delta_{i_2,j_2}(1-\delta_{i_1,i_2})\bra{n}f^{\dagger}_{\RR,i_1}f_{\RR,i_1}\ket{n}\bra{n}f^{\dagger}_{\RR,i_2}f_{\RR,i_2}\ket{n}\nonumber\\
 &+(1-\delta_{i_1,j_1})\delta
        _{i_1,j_2}\delta_{i_2,j_1}\bra{n}f^{\dagger}_{\RR,i_1}f_{\RR,i_1}\ket{n}\bra{n}f_{\RR,j_1}f^{\dagger}_{\RR,j_1}\ket{n}
 \bigg]\nonumber\\
= &
\delta_{i_1,j_1}\delta_{i_2,j_2 } \delta_{i_1,i_2}
\frac{ 
\binom{N_f-1}{n_f-1}
}{\binom{N_f}{n_f}} 
+ 
\delta_{i_1,j_1}\delta_{i_2,j_2 } (1-\delta_{i_1,i_2})
\frac{ 
\binom{N_f-2}{n_f-2}
}{\binom{N_f}{n_f}} 
+ (1-\delta_{i_1,j_1})\delta_{i_1,j_2}\delta_{i_2,j_1}
\frac{\binom{N_f-2}{n_f-1}}{\binom{N_f}{n_f}}  \nonumber\\ 
=& 
\delta_{i_1,j_1}\delta_{i_2,j_2 } \delta_{i_1,i_2}
\frac{n_f}{N_f}
+ 
\delta_{i_1,j_1}\delta_{i_2,j_2 } (1-\delta_{i_1,i_2})
\frac{n_f(n_f-1)}{N_f(N_f-1)}
+ (1-\delta_{i_1,j_1})\delta_{i_1,j_2}\delta_{i_2,j_1}
\frac{n_f(N_f-n_f)}{N_f(N_f-1)}
\ea 
Then the susceptibility takes the form of 
\ba 
&\chi_{i_1j_1,i_2j_2}(\tau_1-\tau_2) \nonumber\\
= & \langle 
 T_\tau 
 : f_{\RR,i_1}^\dag (\tau_1) f_{\RR,j_1} (\tau_1):
: f_{\RR,i_2}^\dag (\tau_2) f_{\RR,j_2} (\tau_2):
 \rangle_{S_{decouple}} \nonumber\\ 
 =&\langle 
 T_\tau 
  f_{\RR,i_1}^\dag (\tau_1) f_{\RR,j_1} (\tau_1)
 f_{\RR,i_2}^\dag (\tau_2) f_{\RR,j_2} (\tau_2)
 \rangle_{S_{decouple}}\nonumber\\
 &-\delta_{i_1,j_1}\frac{n_f}{N_f}\langle f^{\dagger}_{\RR,i_2}(\tau_2)f_{\RR,j_2}(\tau_2)\rangle_{S_{decouple}}-\delta_{i_2,j_2}\frac{n_f}{N_f}\langle f^{\dagger}_{\RR,i_1}(\tau_1)f_{\RR,j_1}(\tau_1)\rangle_{S_{decouple}}+\delta_{i_1,j_1}\delta_{i_2,j_2}\frac{n_f^2}{N_f^2}
 \nonumber\\
 =&
\delta_{i_1,j_1}\delta_{i_2,j_2 } \delta_{i_1,i_2}
\frac{n_f}{N_f}
+ 
\delta_{i_1,j_1}\delta_{i_2,j_2 } (1-\delta_{i_1,i_2})
\frac{n_f(n_f-1)}{N_f(N_f-1)}
+ (1-\delta_{i_1,j_1})\delta_{i_1,j_2}\delta_{i_2,j_1}
\frac{n_f(N_f-n_f)}{N_f(N_f-1)}-\delta_{i_1,j_1}\delta_{i_2,j_2}\frac{n_f^2}{N_f^2}
\nonumber\\
 = & 
 \delta_{i_1,j_1}\delta_{i_2,j_2 } \delta_{i_1,i_2}
\frac{n_f(N_f-n_f)}{N_f^2}
+ 
\delta_{i_1,j_1}\delta_{i_2,j_2 } (1-\delta_{i_1,i_2})
\frac{n_f(n_f-N_f)}{N_f^2(N_f-1)}
+ (1-\delta_{i_1,j_1})\delta_{i_1,j_2}\delta_{i_2,j_1}
\frac{n_f(N_f-n_f)}{N_f(N_f-1)}
\ea 
In the frequency domain, we find 
\ba 
\label{eq:chi_nonstrain}
&\chi_{i_1j_1,i_2j_2}(i\Omega) \nonumber\\
=& \beta 
\delta_{\Omega,0}
\bigg[  \delta_{i_1,j_1}\delta_{i_2,j_2 } \delta_{i_1,i_2}
\frac{n_f(N_f-n_f)}{N_f^2}
+ 
\delta_{i_1,j_1}\delta_{i_2,j_2 } (1-\delta_{i_1,i_2})
\frac{n_f(n_f-N_f)}{N_f^2(N_f-1)}
+ (1-\delta_{i_1,j_1})\delta_{i_1,j_2}\delta_{i_2,j_1}
\frac{n_f(N_f-n_f)}{N_f(N_f-1)}
\bigg ]
\ea 

We first prove that $H_{W}$ do not contribute to $\Sigma_c(i\omega,\kk)$ in the zero-temperature limit.
From \cref{eq:chi_nonstrain} we have
\begin{align}
    &\sum_{i_1j_1}\delta_{i_1,j_1}\frac{1}{\beta}\chi_{i_1j_1,i_2j_2}(i\Omega=0)\nonumber\\
    =&\sum_{i_1j_1}\delta_{i_1,j_1}\bigg[  \delta_{i_1,j_1}\delta_{i_2,j_2 } \delta_{i_1,i_2}
\frac{n_f(N_f-n_f)}{N_f^2}
+ 
\delta_{i_1,j_1}\delta_{i_2,j_2 } (1-\delta_{i_1,i_2})
\frac{n_f(n_f-N_f)}{N_f^2(N_f-1)}
+ (1-\delta_{i_1,j_1})\delta_{i_1,j_2}\delta_{i_2,j_1}
\frac{n_f(N_f-n_f)}{N_f(N_f-1)}
\bigg ]\nonumber\\
=&\sum_{i_1}\bigg[  \delta_{i_2,j_2 } \delta_{i_1,i_2}
\frac{n_f(N_f-n_f)}{N_f^2}
+ 
\delta_{i_2,j_2 } (1-\delta_{i_1,i_2})
\frac{n_f(n_f-N_f)}{N_f^2(N_f-1)}
\bigg ]\nonumber\\
=&\delta_{i_2,j_2}\frac{n_f(N_f-n_f)}{N_f^2}+\delta_{i_2,j_2 } (N_f-1)
\frac{n_f(n_f-N_f)}{N_f^2(N_f-1)}=0
\end{align}
Similarly, we also have
\begin{equation}
    \sum_{i_2,j_2}\delta_{i_2,j_2}\frac{1}{\beta}\chi_{i_1j_1,i_2j_2}(i\Omega=0)=0
\end{equation}
Denote (\cref{eq:def_mathcal_J})
\begin{align}
    \mathcal{J}_{i_1j_1i_i'j_1'}=&\mathcal{J}^{W}_{i_1j_1i_i'j_1'}+\mathcal{J}^{J}_{i_1j_1i_i'j_1'}\nonumber\\
    \mathcal{J}_{i_1j_1i_i'j_1'}^{W}=&W\delta_{i_1,j_1}\delta_{i_1',j_1'}\nonumber\\
    \mathcal{J}^{J}_{ \alpha_1\eta_1s_1,\alpha_2\eta_2s_2,a_3\eta_3s_3,a_4\eta_4s_4 } = &
-\frac{J}{2}
\delta_{\alpha_1+2,a_4}
\delta_{\alpha_2+2,a_3}\delta_{\eta_1 s_1,\eta_4s_4}
\delta_{\eta_2s_2,\eta_3s_3}(\eta_1\eta_2 +(-1)^{\alpha_1+\alpha_2})
\label{eq:mathcal_J_J}
\end{align}
The self-energy of $c$ electron now reads (from \cref{eq:sig_c_from_chi_G})
\begin{align}
    [\Sigma_{c}(i\omega,\kk)]_{i_1',j_2'}\approx &\sum_{i_1j_1 j_1',i_2j_2i_2'}
 [W\delta_{i_1,j_1}\delta_{i_1',j_1'}+\mathcal{J}^{J}_{i_1j_1i_1'j_1'}]
 \mathcal{J}_{i_2j_2i_2'j_2'}\frac{1}{\beta}
 \chi_{i_1j_1,i_2j_2}(i\Omega=0) 
 \frac{1}{N_M}\sum_{\kk'}G_{c,j_1'i_2'}^{S_{decouple}}(i\omega,\kk') \nonumber\\
 =&W\bigg[\sum_{i_1j_1}\delta_{i_1,j_1}\frac{1}{\beta}\chi_{i_1j_1,i_2j_2}(i\Omega=0)\bigg]
 \sum_{j_1',i_2j_2i_2'}\delta_{i_1',j_1'}\mathcal{J}_{i_2j_2i_2'j_2'}
 \frac{1}{N_M}\sum_{\kk'}G_{c,j_1'i_2'}^{S_{decouple}}(i\omega,\kk') \nonumber\\
 &+\sum_{i_1j_1 j_1',i_2j_2i_2'}
 \mathcal{J}^{J}_{i_1j_1i_1'j_1'}
 [W\delta_{i_2,j_2}\delta_{i_2',j_2'}+\mathcal{J}^{J}_{i_2j_2i_2'j_2'}]\frac{1}{\beta}
 \chi_{i_1j_1,i_2j_2}(i\Omega=0) 
 \frac{1}{N_M}\sum_{\kk'}G_{c,j_1'i_2'}^{S_{decouple}}(i\omega,\kk') \nonumber\\
 =&\sum_{i_1j_1 j_1',i_2j_2i_2'}
 \mathcal{J}^{J}_{i_1j_1i_1'j_1'}
 \mathcal{J}^{J}_{i_2j_2i_2'j_2'}\frac{1}{\beta}
 \chi_{i_1j_1,i_2j_2}(i\Omega=0) 
 \frac{1}{N_M}\sum_{\kk'}G_{c,j_1'i_2'}^{S_{decouple}}(i\omega,\kk')
 \label{eq:Sigma_c_no_W}
\end{align}
We can see that only $H_J$ term contributes to the dynamical self-energy at low temperature through the $f$ electron local-moment fluctuations.
On the other hand, as the charge fluctuations of $f$ electron induced by $H_{W}$ are frozen at low temperature, $H_W$ term does not contribute any self-energy at second order. 
The self-energy contributed by $H_J$ is
\ba 
&[\Sigma_c(i\omega,\kk)]_{a_1\eta_1s_1, a_2\eta_2s_2} \nonumber\\ 
=& {+}J^2 [\delta_{a_1,3}+\delta_{a_1,4}]
[\delta_{a_2,3}+\delta_{a_2,4}]
\frac{1}{N_M}\sum_{\kk'}\sum_{\eta_3s_3,\eta_4s_4}
\sum_{a_3,a_4\in \{3,4\} }
G_{c,a_3\eta_3s_3,a_4\eta_4s_4}^{S_{decouple}}(i\omega,\kk') \nonumber\\
&
\bigg[ \frac{n_f(N_f-n_f)}{N_f^2}\delta_{a_1\eta_1s_1,a_2\eta_2s_2}\delta_{a_2\eta_2s_2,a_3\eta_3s_3}\delta_{a_3\eta_3s_3,a_4\eta_4s_4} 
+ \frac{n_f(n_f-N_f)}{N_f^2(N_f-1)}
\delta_{a_1\eta_1s_1,a_3\eta_3s_3}
\delta_{a_2\eta_2s_2,a_4\eta_4s_4} (1-\delta_{a_1\eta_1s_1,a_2\eta_2s_2}) \nonumber\\
&
+ \frac{n_f(N_f-n_f)}{N_f(N_f-1)}[\delta_{\eta_1,\eta_3}\delta_{a_1,a_3}
+\delta_{\eta_1,-\eta_3}(1-\delta_{a_1,a_3})]
\delta_{a_1\eta_1s_1,a_2\eta_2s_2}
\delta_{a_3\eta_3s_3,a_4\eta_4s_4}(1-\delta_{a_1\eta_1s_1,a_3\eta_3s_3})
\bigg] 
\label{eq:sigma_c_with_suscep}
\ea 
We can observe that only $\Gamma_{1}\oplus \Gamma_2$ electrons acquire a self-energy contribution coming from the $H_J$ term. 

\subsection{Evaluation of the Green's function}
We now evaluate the Green's function of $c$ electron at charge neutrality. 
Using the eigenvalues and eigenvectors of the $c$ electron Hamiltonian \cref{eq:Hc_eigs_evcs,eq:Hc_evcs_normalize}, the $c$ electron Green's function \cref{eq:G_decouple} is expressed as
\begin{align}
    G_{c,a_1\eta_1s_1,a_2\eta_2s_2}^{S_{decouple}}(i\omega,\kk)=&\delta_{s_1,s_2}\delta_{\eta_1,\eta_2}\sum_{n=1}^{4}\frac{U^{\eta_1,c}_{\kk,a_1n}[U^{\eta_1,c}_{\kk,a_2n}]^*}{i\omega-E^{\eta_1,c}_{\kk,n}}
    \label{eq:Gc}
\end{align}
From \cref{eq:sigma_c_with_suscep}, the self-energy only depends on the Green's function of $\Gamma_1\oplus \Gamma_2$ electrons with $a=3,4$.
We focus on the $\Gamma_1\oplus \Gamma_2$ electrons and introduce the orbital diagonal and off-diagonal local Green's functions as
\begin{align}
    g_c(i\omega)=&\frac{1}{N_M}\sum_\kk G_{c,3\eta_1s_1,3\eta_1 s_1}^{S_{decouple}}(i\omega,\kk)=\frac{1}{N_M}\sum_\kk G_{c,4\eta_1s_1,4\eta_1 s_1}^{S_{decouple}}(i\omega,\kk)\nonumber\\
    =&\frac{1}{N_M}\sum_{\kk}\bigg[(A_{1,\kk})^2|v_{\star}\kk|^4\bigg(\frac{1}{i\omega - E^{\eta_1,c}_{\kk,n=1}}+\frac{1}{i\omega - E^{\eta_1,c}_{\kk,n=4}}\bigg)
+(A_{2,\kk})^2|v_{\star}\kk|^4\bigg(\frac{1}{i\omega - E^{\eta_1,c}_{\kk,n=2}}+\frac{1}{i\omega - E^{\eta_1,c}_{\kk,n=3}}\bigg)
\bigg]
\nonumber\\
    =&\frac{1}{N_M}\sum_{\kk}\bigg[
\frac{|v_{\star}\kk|^2}{\sqrt{M^2+4|v_{\star}\kk|^2}(\sqrt{M^2+4|v_{\star}\kk|^2}-M)}\bigg(\frac{1}{i\omega - E^{\eta_1,c}_{\kk,n=1}}+\frac{1}{i\omega - E^{\eta_1,c}_{\kk,n=4}}\bigg)\nonumber\\
&+\frac{|v_{\star}\kk|^2}{\sqrt{M^2+4|v_{\star}\kk|^2}(\sqrt{M^2+4|v_{\star}\kk|^2}+M)}\bigg(\frac{1}{i\omega - E^{\eta_1,c}_{\kk,n=2}}+\frac{1}{i\omega - E^{\eta_1,c}_{\kk,n=3}}\bigg)\
\bigg]
\nonumber\\
=&\frac{1}{N_M}\sum_{\kk}\frac{1}{4}\bigg[
\frac{\sqrt{M^2+4|v_{\star}\kk|^2}+M}{\sqrt{M^2+4|v_{\star}\kk|^2}}\bigg(\frac{1}{i\omega - E^{\eta_1,c}_{\kk,n=1}}+\frac{1}{i\omega - E^{\eta_1,c}_{\kk,n=4}}\bigg)\nonumber\\
&+\frac{\sqrt{M^2+4|v_{\star}\kk|^2}-M}{\sqrt{M^2+4|v_{\star}\kk|^2}}\bigg(\frac{1}{i\omega - E^{\eta_1,c}_{\kk,n=2}}+\frac{1}{i\omega - E^{\eta_1,c}_{\kk,n=3}}\bigg)\
\bigg]
\nonumber\\
=&\frac{1}{N_M}\sum_{\kk}\frac{1}{4}\bigg[\bigg(1+\frac{M}{\sqrt{M^2+4|v_{\star}\kk|^2}}\bigg)
\bigg(\frac{1}{i\omega - E^{\eta_1,c}_{\kk,n=1}}+\frac{1}{i\omega - E^{\eta_1,c}_{\kk,n=4}}\bigg)\nonumber\\
&+\bigg(1-\frac{M}{\sqrt{M^2+4|v_{\star}\kk|^2}}\bigg)\bigg(\frac{1}{i\omega - E^{\eta_1,c}_{\kk,n=2}}+\frac{1}{i\omega - E^{\eta_1,c}_{\kk,n=3}}\bigg)\
\bigg]
\nonumber\\
g_{c,off}(i\omega)=&\frac{1}{N_M}\sum_\kk G_{c,3\eta_1s_1,4\eta_1 s_1}^{S_{decouple}}(i\omega,\kk)=\frac{1}{N_M}\sum_\kk G_{c,4\eta_1s_1,3\eta_1 s_1}^{S_{decouple}}(i\omega,\kk)\nonumber\\
=&\frac{1}{N_M}\sum_{\kk}\bigg[(A_{1,\kk})^2|v_{\star}\kk|^4\bigg(-\frac{1}{i\omega - E^{\eta_1,c}_{\kk,n=1}}+\frac{1}{i\omega - E^{\eta_1,c}_{\kk,n=4}}\bigg)
+(A_{2,\kk})^2|v_{\star}\kk|^4\bigg(\frac{1}{i\omega - E^{\eta_1,c}_{\kk,n=2}}-\frac{1}{i\omega - E^{\eta_1,c}_{\kk,n=3}}\bigg)
\bigg]
\nonumber\\
=&\frac{1}{N_M}\sum_{\kk}\frac{1}{4}\bigg[\bigg(1+\frac{M}{\sqrt{M^2+4|v_{\star}\kk|^2}}\bigg)
\bigg(-\frac{1}{i\omega - E^{\eta_1,c}_{\kk,n=1}}+\frac{1}{i\omega - E^{\eta_1,c}_{\kk,n=4}}\bigg)\nonumber\\
&+\bigg(1-\frac{M}{\sqrt{M^2+4|v_{\star}\kk|^2}}\bigg)\bigg(\frac{1}{i\omega - E^{\eta_1,c}_{\kk,n=2}}-\frac{1}{i\omega - E^{\eta_1,c}_{\kk,n=3}}\bigg)\
\bigg]
\label{eq:local_gc_G_12}
\end{align}
Combining \cref{eq:sigma_c_with_suscep,eq:Gc,eq:local_gc_G_12}, we find 
\begin{align}
    &[\Sigma_c(i\omega,\kk)]_{a_1\eta_1s_1,a_2\eta_2s_2}\nonumber\\
    =&J^2(\delta_{a_1,3}+\delta_{a_1,4})(\delta_{a_2,3}+\delta_{a_2,4})\sum_{\eta_3s_3,\eta_4s_4}\delta_{\eta_3,\eta_4}\delta_{s_3,s_4}\sum_{a_3,a_4\in\{3.4\}}[\delta_{a_3,a_4}g_c(i\omega)+(1-\delta_{a_3,a_4})g_{c,off}(i\omega)]\nonumber\\
    &
\bigg[ \frac{n_f(N_f-n_f)}{N_f^2}\delta_{a_1\eta_1s_1,a_2\eta_2s_2}\delta_{a_2\eta_2s_2,a_3\eta_3s_3}\delta_{a_3\eta_3s_3,a_4\eta_4s_4} 
+ \frac{n_f(n_f-N_f)}{N_f^2(N_f-1)}
\delta_{a_1\eta_1s_1,a_3\eta_3s_3}
\delta_{a_2\eta_2s_2,a_4\eta_4s_4} (1-\delta_{a_1\eta_1s_1,a_2\eta_2s_2}) \nonumber\\
&
+ \frac{n_f(N_f-n_f)}{N_f(N_f-1)}[\delta_{\eta_1,\eta_3}\delta_{a_1,a_3}
+\delta_{\eta_1,-\eta_3}(1-\delta_{a_1,a_3})]
\delta_{a_1\eta_1s_1,a_2\eta_2s_2}
\delta_{a_3\eta_3s_3,a_4\eta_4s_4}(1-\delta_{a_1\eta_1s_1,a_3\eta_3s_3})
\bigg] \nonumber\\
=&J^2(\delta_{a_1,3}+\delta_{a_1,4})(\delta_{a_2,3}+\delta_{a_2,4})\sum_{a_3,a_4\in\{3.4\}}[\delta_{a_3,a_4}g_c(i\omega)+(1-\delta_{a_3,a_4})g_{c,off}(i\omega)]\nonumber\\
    &
\bigg[ \frac{n_f(N_f-n_f)}{N_f^2}\delta_{a_1\eta_1s_1,a_2\eta_2s_2}\delta_{a_2,a_3}\delta_{a_3,a_4}
+ \frac{n_f(n_f-N_f)}{N_f^2(N_f-1)}
\delta_{a_1\eta_1s_1,a_3\eta_2s_2}
\delta_{a_2,a_4} (1-\delta_{a_1\eta_1s_1,a_2\eta_2s_2}) \nonumber\\
&+\frac{n_f(N_f-n_f)}{N_f(N_f-1)}[\delta_{a_1,a_3}
\delta_{a_1\eta_1s_1,a_2\eta_2s_2}
\delta_{a_3,a_4}(2-\delta_{a_1,a_3})
+2(1-\delta_{a_1,a_3})
\delta_{a_1\eta_1s_1,a_2\eta_2s_2}
\delta_{a_3,a_4}]
\bigg] \nonumber\\
=&J^2\delta_{\eta_1,\eta_2}\delta_{s_1,s_2}(\delta_{a_1,3}+\delta_{a_1,4})(\delta_{a_2,3}+\delta_{a_2,4})\sum_{a_3,a_4\in\{3.4\}}[\delta_{a_3,a_4}g_c(i\omega)+(1-\delta_{a_3,a_4})g_{c,off}(i\omega)]\nonumber\\
    &
\bigg[ \frac{n_f(N_f-n_f)}{N_f^2}\delta_{a_1,a_2}\delta_{a_2,a_3}\delta_{a_3,a_4}
+ \frac{n_f(n_f-N_f)}{N_f^2(N_f-1)}
\delta_{a_1,a_3}
\delta_{a_2,a_4} (1-\delta_{a_1,a_2})+\frac{n_f(N_f-n_f)}{N_f(N_f-1)}
(2-\delta_{a_1,a_3})
\delta_{a_1,a_2}
\delta_{a_3,a_4}
\bigg] \nonumber\\
=&J^2\delta_{\eta_1,\eta_2}\delta_{s_1,s_2}(\delta_{a_1,3}+\delta_{a_1,4})(\delta_{a_2,3}+\delta_{a_2,4})\nonumber\\
&\bigg[
g_c(i\omega)\delta_{a_1,a_2}\bigg(\frac{n_f(N_f-n_f)}{N_f^2}+3\frac{n_f(N_f-n_f)}{N_f(N_f-1)}\bigg)
+g_{c,off}(i\omega)(1-\delta_{a_1,a_2})\frac{n_f(n_f-N_f)}{N_f^2(N_f-1)}
\bigg]\nonumber\\
=&J^2\delta_{\eta_1,\eta_2}\delta_{s_1,s_2}(\delta_{a_1,3}+\delta_{a_1,4})(\delta_{a_2,3}+\delta_{a_2,4})\bigg[
g_c(i\omega)\delta_{a_1,a_2}\frac{n_f(N_f-n_f)(4N_f-1)}{N_f^2(N_f-1)}
+g_{c,off}(i\omega)(1-\delta_{a_1,a_2})\frac{n_f(n_f-N_f)}{N_f^2(N_f-1)}
\bigg]
\end{align}
To be specific, we have
\ba 
\label{eq:sigma_c_from_gc}
&[\Sigma_c(i\omega,\kk)]_{3\eta_1s_1,3\eta_2s_2}=[\Sigma_c(i\omega,\kk)]_{4\eta_1s_1,4\eta_2s_2}=\delta_{\eta_1s_1,\eta_2s_2}\frac{n_f(N_f-n_f)(4N_f-1)}{N_f^2(N_f-1)}J^2g_c(i\omega)=\delta_{\eta_1s_1,\eta_2s_2} \frac{31J^2}{28} g_{c}(i\omega) \nonumber\\ 
&[\Sigma_c(i\omega,\kk)]_{3\eta_1s_1,4\eta_2s_2}=[\Sigma_c(i\omega,\kk)]_{4\eta_1s_1,3\eta_2s_2}=\delta_{\eta_1s_1,\eta_2s_2}\frac{n_f(n_f-N_f)}{N_f^2(N_f-1)}J^2g_{c,off}(i\omega)=-\delta_{\eta_1s_1,\eta_2s_2} \frac{J^2}{28} g_{c,off}(i\omega) 
\ea 
where we take $N_f=8$ and $n_f=N_f/2=4$ in the final expression.
Denote the diagonal and off-diagonal $c$ electron self-energy as
\begin{align}
    \Sigma_{c,dia}(i\omega)=&[\Sigma_c(i\omega,\kk)]_{3\eta_1s_1,3\eta_1s_1}=[\Sigma_c(i\omega,\kk)]_{4\eta_1s_1,4\eta_1s_1}=\frac{n_f(N_f-n_f)(4N_f-1)}{N_f^2(N_f-1)}J^2g_c(i\omega)\nonumber\\
    \Sigma_{c,off}(i\omega)=&[\Sigma_c(i\omega,\kk)]_{3\eta_1s_1,4\eta_1s_1}=[\Sigma_c(i\omega,\kk)]_{4\eta_1s_1,3\eta_1s_1}=\frac{n_f(n_f-N_f)}{N_f^2(N_f-1)}J^2g_{c,off}(i\omega)
    \label{eq:Sigma_c_dia_off}
\end{align}
Then the $c$ electron self-energy could be written as
\begin{equation}
    [\Sigma_{c}(i\omega,\kk)]_{a_1\eta_1,s_1,a_2\eta_2 s_2}=\delta_{\eta_1s_1,\eta_2s_2}\begin{pmatrix}
        0_{2\times 2} & 0_{2\times2}\\
        0_{2\times 2} & \Sigma_{c,dia}(i\omega)\sigma_0+\Sigma_{c,off}(i\omega)\sigma_x
    \end{pmatrix}_{a_1,a_2}
    \label{eq:Sigma_c_full}
\end{equation}

To calculate $g_c(i\omega)$ and $g_{c,off}(i\omega)$, we introduce the spectral function
\ba 
&\rho_c(\epsilon) = \frac{1}{\pi}\mathrm{Im}[g_c(\epsilon-i0^+)]\nonumber\\
&\rho_{c,off}(\epsilon) = \frac{1}{\pi}\mathrm{Im}[g_{c,off}(\epsilon-i0^+)]
\ea 
which are expressed as
\begin{align}
    \rho_c(\epsilon)=&\frac{1}{\Omega_M} \int_{|\kk|<\Lambda_c}
\bigg[ 
\frac{1}{4}\bigg(1+\frac{M}{\sqrt{M^2+4|v_\star\kk|^2}}\bigg) [\delta(\epsilon- E^{\eta,c}_{\kk,n=1})+\delta(\epsilon -E^{\eta,c}_{\kk,n=4})] \nonumber\\
&+ 
\frac{1}{4}\bigg(1-\frac{M}{\sqrt{M^2+4|v_\star\kk|^2}}\bigg) [\delta(\epsilon- E^{\eta,c}_{\kk,n=2})+\delta(\epsilon -E^{\eta,c}_{\kk,n=3})]
\bigg]d^2k \nonumber\\
=&\frac{2\pi}{\Omega_M}
\int_0^{\Lambda_c}
\frac{1}{4} \bigg[
\bigg(1+ \frac{M}{\sqrt{M^2+4|v_\star k|^2}}\bigg)
\delta( |\epsilon| - \frac{M+\sqrt{4|v_\star k|^2+M^2}}{2})\nonumber\\
&+
\bigg(1- \frac{M}{\sqrt{M^2+4|v_\star k|^2}}\bigg)
\delta( |\epsilon| - \frac{-M+\sqrt{4|v_\star k|^2+M^2}}{2})
\bigg] kdk \nonumber\\
\rho_{c,off}(\epsilon)=&\frac{1}{\Omega_M} \int_{|\kk|<\Lambda_c}
\bigg[ 
\frac{1}{4}\bigg(1+\frac{M}{\sqrt{M^2+4|v_\star\kk|^2}}\bigg) [-\delta(\epsilon- E^{\eta,c}_{\kk,n=1})+\delta(\epsilon -E^{\eta,c}_{\kk,n=4})] \nonumber\\
&+ 
\frac{1}{4}\bigg(1-\frac{M}{\sqrt{M^2+4|v_\star\kk|^2}}\bigg) [\delta(\epsilon- E^{\eta,c}_{\kk,n=2})-\delta(\epsilon -E^{\eta,c}_{\kk,n=3})]
\bigg]d^2k \nonumber\\
=&\frac{2\pi}{\Omega_M}
\int_0^{\Lambda_c}
\frac{1}{4} \bigg[
\bigg(1+ \frac{M}{\sqrt{M^2+4|v_\star k|^2}}\bigg)
\delta( |\epsilon| - \frac{M+\sqrt{4|v_\star k|^2+M^2}}{2})\mathrm{sgn}(\epsilon)\nonumber\\
&-
\bigg(1- \frac{M}{\sqrt{M^2+4|v_\star k|^2}}\bigg)
\delta( |\epsilon| - \frac{-M+\sqrt{4|v_\star k|^2+M^2}}{2})\mathrm{sgn}(\epsilon)
\bigg] kdk 
\end{align}
Using the integration strategy in \cref{eq:delta_func_integration}, the spectrum functions are further calculated as
\begin{align}
    \rho_c(\epsilon)=&\frac{\pi}{2|v_{\star}|^2\Omega_M}\bigg[|\epsilon|\theta(|\epsilon|-M)\theta(\frac{M+\sqrt{4|v_{\star}\Lambda_c|^2+M^2}}{2}-|\epsilon|) +|\epsilon|\theta(|\epsilon|+M)\theta(\frac{-M+\sqrt{4|v_{\star}\Lambda_c|^2+M^2}}{2}-|\epsilon|) \bigg]\nonumber\\
    =&\frac{\pi}{2|v_{\star}|^2\Omega_M}|\epsilon|[\theta(|\epsilon|-M)\theta(|v_{\star}\Lambda_c|-|\epsilon|)+\theta(|v_{\star}\Lambda_c|-|\epsilon|)]\nonumber\\
    =&\frac{\pi}{|v_{\star}|^2\Omega_{M}}|\epsilon|\theta(|v_{\star}\Lambda_c|-|\epsilon|)-\frac{\pi}{2|v_{\star}|^2\Omega_{M}}|\epsilon|\theta(M-|\epsilon|)
    \nonumber\\
    \rho_{c,off}(\epsilon)=&\frac{\pi}{2|v_{\star}|^2\Omega_M}\mathrm{sgn}(\epsilon)\bigg[|\epsilon|\theta(|\epsilon|-M)\theta(\frac{M+\sqrt{4|v_{\star}\Lambda_c|^2+M^2}}{2}-|\epsilon|) -|\epsilon|\theta(|\epsilon|+M)\theta(\frac{-M+\sqrt{4|v_{\star}\Lambda_c|^2+M^2}}{2}-|\epsilon|) \bigg]\nonumber\\
    =&\frac{\pi}{2|v_{\star}|^2\Omega_M}\epsilon[\theta(|\epsilon|-M)\theta(|v_{\star}\Lambda_c|-|\epsilon|)-\theta(|v_{\star}\Lambda_c|-|\epsilon|)]\nonumber\\
    =&-\frac{\pi}{2|v_{\star}|^2\Omega_M}\epsilon\theta(M-|\epsilon|)
\end{align}
where we have used $|v_{\star}\Lambda_c|\gg M$ so that $\theta(M-|\epsilon|)\theta(|v_{\star}\Lambda_c|-|\epsilon|)=\theta(M-|\epsilon|)$.
Then similar to \cref{eq:self_energy_in_f_M} we obtain
\begin{align}
    g_{c}(i\omega)=&\int_{\epsilon}\rho_c(\epsilon)\frac{1}{i\omega-\epsilon}=-i\omega \frac{\pi}{|v_\star^2|\Omega_M}
\log\bigg( \frac{|v_\star\Lambda_c|^2 +\omega^2}{\omega^2} 
\bigg) +{i\omega \frac{\pi}{2|v_\star^2|\Omega_M}
\log\bigg( \frac{M^2 +\omega^2}{\omega^2} 
\bigg)}\nonumber\\
g_{c,off}(i\omega)=&\int_{\epsilon}\rho_{c,off}(\epsilon)\frac{1}{i\omega-\epsilon}=\frac{\pi M}{|v_{\star}|^2\Omega_M}-i\omega\frac{\pi}{2|v_{\star}|^2\Omega_M}\log\bigg(\frac{i\omega+M}{i\omega-M}\bigg)
\end{align}

Using the same analytic continuation method as in \cref{eq:fM_real_freq}, the real-frequency Green's function reads
\ba 
\label{eq:real_freuqnecy_g_c}
&g_c(\omega+i0^+) 
= -\frac{\pi }{|v_\star|^2\Omega_M }
\omega \bigg[ 
\log\bigg( 
\bigg| 
\frac{|v_\star\Lambda_c|^2 -\omega^2 }
{\omega^2}
\bigg| 
\bigg) 
-\frac{1}{2}\log\bigg( 
\bigg| 
\frac{M^2 -\omega^2 }
{\omega^2}
\bigg| 
\bigg) 
\bigg] -i \pi \rho_c(\omega) \nonumber\\ 
&g_{c,off}(\omega+i0^+)  = \frac{\pi }{|v_\star|^2\Omega_M }\bigg[M+\frac{1}{2}
 \omega \log\bigg(\bigg|
\frac{\omega-M}{\omega+M}\bigg|\bigg) \bigg]
 -i \pi \rho_{c,off}(\omega)
\ea 
Note that in the hybridization function calculation \cref{eq:self_energy_in_f_M}, we are evaluating the $c$-electron Greens's function in $\Gamma_3$ sector. Here we are computing in the $\Gamma_1\oplus\Gamma_2$ sector.

\subsubsection{$M=0$}
We first discuss the self-energy of $c$ electron in the $M=0$ limit. 
For $M=0$, we find 
\ba 
\label{eq:local_gc_M=0}
g_c(i\omega) = -\frac{\pi}{|v_\star|^2\Omega_M}
i\omega 
\log\bigg(
\frac{(i\omega)^2 -|v_\star \Lambda_c|^2}{(i\omega)^2}
\bigg) ,\quad g_{c,off}(i\omega)=0
\ea 
The $c$ electron self-energy then reads (from \cref{eq:Sigma_c_dia_off})
\begin{align}
    \Sigma_{c,dia}(i\omega)=&-j\frac{n_f(N_f-n_f)(4N_f-1)}{N_f^2(N_f-1)}\pi i\omega \log\bigg(
\frac{(i\omega)^2 -|v_\star \Lambda_c|^2}{(i\omega)^2}
\bigg) \nonumber\\
\Sigma_{c,off}(i\omega)=&0
\end{align}
where we have introduced the dimensionless parameter (\cref{eq:define_small_para})
\ba 
j = \frac{J^2}{|v_\star|^2\Omega_M }
\ea 
In real frequency we have
\begin{align}
    \Sigma_{c,dia}(\omega+i0^+)=&-j\frac{n_f(N_f-n_f)(4N_f-1)}{N_f^2(N_f-1)}\pi \bigg[\omega \log\bigg(\bigg|
\frac{\omega^2 -|v_\star \Lambda_c|^2}{\omega^2}\bigg|
\bigg)+i\pi|\omega|\theta(|v_{\star}\Lambda_c|-|\omega|)\bigg]\nonumber\\
\Sigma_{c,off}(\omega+i0^+)=&0
\end{align}

At $M=0$, the non-interacting Green's function of $c$ electron reads (\cref{eq:G_decouple})
\ba 
[G_{c}^{S_{decouple}}(i\omega,\kk)]^{-1}_{a_1\eta_1s_1,a_2\eta_2s_2}
= \delta_{s_1,s_2}\delta_{\eta_1,\eta_2}
\bigg[ 
i\omega \mathbb{I} 
- 
  \begin{pmatrix}
    0_{2\times 2}& v_\star (\eta k_x \sigma_0 + ik_y \sigma_z) \\
    v_\star (\eta k_x \sigma_0 - ik_y \sigma_z) &  0_{2\times 2}
  \end{pmatrix} 
\bigg]_{a_1,a_2}
\ea 
The interacting Green's function follows (\cref{eq:dyson_of_c,eq:Sigma_c_full}) 
\ba 
[G_{c}(i\omega,\kk)]^{-1}_{a_1\eta_1s_1,a_2\eta_2s_2}
=\delta_{s_1,s_2}\delta_{\eta_1,\eta_2}
\bigg[ 
i\omega \mathbb{I} 
- 
  \begin{pmatrix}
    0_{2\times 2}& v_\star (\eta k_x \sigma_0 + ik_y \sigma_z) \\
    v_\star (\eta k_x \sigma_0 - ik_y \sigma_z) &  \Sigma_{c,dia}(i\omega)\sigma_0
  \end{pmatrix} 
\bigg]_{a_1,a_2}
\ea 
The trace of the Green's function now reads 
\begin{align}
    \mathrm{Tr}[G_c(i\omega,\kk)]=&\frac{8[2i\omega-\Sigma_{c,dia}(i\omega)]}{(i\omega)^2-i\omega\Sigma_{c,dia}(i\omega)-|v_{\star}\kk|^2}\nonumber\\
    =&\frac{8[2i\omega-\Sigma_{c,dia}(i\omega)]}{(i\omega-\mathcal{E}^{j}_{i\omega,\kk,1})(i\omega-\mathcal{E}^{j}_{i\omega,\kk,2})}\nonumber\\
    =&\frac{8[2i\omega-\Sigma_{c,dia}(i\omega)]}{\mathcal{E}^j_{i\omega,\kk,1}-\mathcal{E}^j_{i\omega,\kk,2}}\bigg(\frac{1}{i\omega-\mathcal{E}^j_{i\omega,\kk,1}}-\frac{1}{i\omega-\mathcal{E}^j_{i\omega,\kk,2}}\bigg)
\end{align}
where 
\begin{align}
    \mathcal{E}^j_{i\omega,\kk,1}=&\Sigma_{c,dia}(i\omega)/2-\sqrt{[\Sigma_{c,dia}(i\omega)]^2/4+|v_{\star}\kk|^2}=-|v_{\star}\kk|+\Sigma_{c,dia}(i\omega)/2+\mathcal{O}(j^2)\nonumber\\
    \mathcal{E}^j_{i\omega,\kk,2}=&\Sigma_{c,dia}(i\omega)/2+\sqrt{[\Sigma_{c,dia}(i\omega)]^2/4+|v_{\star}\kk|^2}=|v_{\star}\kk|+\Sigma_{c,dia}(i\omega)/2+\mathcal{O}(j^2)
\end{align}
At $j=0$ and $\Sigma_c(i\omega)=0$, the bare dispersions are
\begin{equation}
    E^{j}_{\kk,1}=-|v_{\star}\kk|,\quad E^{j}_{\kk,2}=|v_{\star}\kk|
\end{equation}

In real frequency, we write
\begin{equation}
    \mathcal{E}^{j}_{\omega+i0^+,\kk,n}=\mathcal{R}^{j}_{\omega,\kk,n}-i\Gamma^{j}_{\omega,\kk,n}
\end{equation}
Then the renormalized dispersions $\tilde{E}^{j}_{\kk,n}$ and scattering times $\tau^j_{\kk,n}$ are determined by
\begin{align}
    &(\omega-\mathcal{R}^{j}_{\omega,\kk,n})\bigg|_{\omega=\tilde{E}^j_{\kk,n}}=0,\quad\frac{1}{\tau^j_{\kk,n}}=\Gamma^{j}_{\omega,\kk,n}\bigg|_{\omega=\tilde{E}^j_{\kk,n}}
\end{align}
Then up to first order of $j$ we have
\begin{align}
    \tilde{E}^j_{\kk,n}=\mathcal{R}^j_{\omega,\kk,n}\bigg|_{\omega=E^j_{\kk,n}},\quad
    \frac{1}{\tau^j_{\kk,n}}=\Gamma^j_{\omega,\kk,n}\bigg|_{\omega=E^j_{\kk,n}}
\end{align}
To be specific,
\begin{align}
    \tilde{E}^{j}_{\kk,1}=-\tilde{E}^{j}_{\kk,2}=&-|v_{\star}\kk|+\mathrm{Re}[\Sigma_{c,dia}(-|v_{\star}\kk|+i0^+)]/2\nonumber\\
    =&-|v_{\star}\kk|+|v_{\star}\kk|\frac{n_f(N_f-n_f)(4N_f-1)\pi j}{2N_f^2(N_f-1)}\log\bigg(\bigg|
\frac{|v_{\star}\kk|^2 -|v_\star \Lambda_c|^2}{|v_{\star}\kk|^2}\bigg|
\bigg)+\mathcal{O}(j^2)\nonumber\\
\approx&-|v_{\star}\kk|+|v_{\star}\kk|\frac{n_f(N_f-n_f)(4N_f-1)\pi j}{2N_f^2(N_f-1)}\log\bigg(
\frac{|v_\star \Lambda_c|^2}{|v_{\star}\kk|^2}
\bigg)+\mathcal{O}(j^2,|v_{\star}\kk|^2)\nonumber\\
\frac{1}{\tau^{j}_{\kk,1}}=\frac{1}{\tau^{j}_{\kk,2}}=&-\mathrm{Im}[\Sigma_{c,dia}(-|v_{\star}\kk|+i0^+)]/2\nonumber\\
=&|v_{\star}\kk|\frac{n_f(N_f-n_f)(4N_f-1)\pi^2 j}{2N_f^2(N_f-1)}+\mathcal{O}(j^2)
\end{align}

At charge neutrality with $N_f=8$ and $n_f=4$, we have
\begin{align}
    \tilde{E}^{j}_{\kk,1}=-\tilde{E}^{j}_{\kk,2}\approx&-|v_{\star}\kk|+|v_{\star}\kk|\frac{31\pi j}{56}\log\bigg(
\frac{|v_\star \Lambda_c|^2}{|v_{\star}\kk|^2}
\bigg)+\mathcal{O}(j^2,|v_{\star}\kk|^2)\nonumber\\
\frac{1}{\tau^{j}_{\kk,1}}=\frac{1}{\tau^{j}_{\kk,2}}
=&|v_{\star}\kk|\frac{31\pi^2 j}{56}+\mathcal{O}(j^2)
\end{align}

Taking the parameters of the topological heavy-fermion model in Ref.~\cite{PhysRevLett.129.047601}, we find $j \approx 0.006$, corresponding to $\frac{31\pi^2 j}{56} \approx 0.03$, which leads to an almost vanishing damping rate compared to the bare dispersion $|v_\star\kk|$.
As a comparison, the damping rate of the gapless modes induced by the single-particle Hybridization $\gamma$ is approximately $0.18|v_{\star}\kk|$.

\subsubsection{Finite $M$}
We next consider the effect of a finite $M$.
The $c$ electron self-energy is evaluated as
\begin{align}
    \Sigma_{c,dia}(i\omega)=&-j\frac{n_f(N_f-n_f)(4N_f-1)}{N_f^2(N_f-1)}\pi i\omega \bigg[\log\bigg(
\frac{(i\omega)^2 -|v_\star \Lambda_c|^2}{(i\omega)^2}
\bigg)-\frac{1}{2}\log\bigg(\frac{(i\omega)^2-M^2}{(i\omega)^2}\bigg)
\bigg] \nonumber\\
\Sigma_{c,off}(i\omega)=&j\frac{n_f(n_f-N_f)}{N_f^2(N_f-1)}\pi\bigg[
M+\frac{i\omega}{2}\log\bigg(\frac{i\omega-M}{i\omega+M}\bigg)
\bigg]
\end{align}
which in real frequency becomes
\begin{align}
    \Sigma_{c,dia}(\omega+i0^+)=&-j\frac{n_f(N_f-n_f)(4N_f-1)}{N_f^2(N_f-1)}\pi\omega \bigg[ \log\bigg(\bigg|
\frac{\omega^2 -|v_\star \Lambda_c|^2}{\omega^2}\bigg|
\bigg)-\frac{1}{2}\log\bigg(\bigg|\frac{M^2-\omega^2}{\omega^2}\bigg|\bigg)
\bigg]\nonumber\\
&-j\frac{n_f(N_f-n_f)(4N_f-1)}{N_f^2(N_f-1)}i\pi^2|\omega|\bigg[\theta(|v_{\star}\Lambda_c|-|\omega|)-\frac{1}{2}\theta(M-|\omega|) \bigg]
\nonumber\\
\Sigma_{c,off}(\omega+i0^+)=&j\frac{n_f(n_f-N_f)}{N_f^2(N_f-1)}\pi\bigg[
M+\frac{\omega}{2}\log\bigg(\bigg|\frac{\omega-M}{\omega+M}\bigg|\bigg)
\bigg]+j\frac{n_f(n_f-N_f)}{N_f^2(N_f-1)}i\pi^2\omega\theta(M-|\omega|)
\label{eq:Sigma_c_real_freq}
\end{align}
Besides, the interacting Green's function is written as
\ba 
[G_{c}^{-1}(i\omega,\kk)]_{a_1\eta_1s_1,a_2\eta_2s_2}
=\delta_{s_1,s_2}\delta_{\eta_1,\eta_2}
\bigg[ 
i\omega \mathbb{I} 
- 
  \begin{pmatrix}
    0_{2\times 2}& v_\star (\eta k_x \sigma_0 + ik_y \sigma_z) \\
    v_\star (\eta k_x \sigma_0 - ik_y \sigma_z) & \Sigma_{c,dia}(i\omega)\sigma_0 + \Sigma_{c,off}(i\omega)\sigma_x +   M\sigma_x  
  \end{pmatrix} 
\bigg]_{a_1,a_2}
\ea 
The trace of the Green's function reads
\begin{align}
    \mathrm{Tr}[G_c(i\omega,\kk)]=&\frac{4[2i\omega-M-\Sigma_{c,off}(i\omega)-\Sigma_{c,dia}(i\omega)]}{(i\omega)^2-i\omega[M+\Sigma_{c,off}(i\omega)+\Sigma_{c,dia}(i\omega)]-|v_{\star}\kk|^2}+\frac{4[2i\omega+M+\Sigma_{c,off}(i\omega)-\Sigma_{c,dia}(i\omega)]}{(i\omega)^2+i\omega[M+\Sigma_{c,off}(i\omega)-\Sigma_{c,dia}(i\omega)]-|v_{\star}\kk|^2}\nonumber\\
    =&\frac{4[2i\omega-M-\Sigma_{c,off}(i\omega)-\Sigma_{c,dia}(i\omega)]}{\mathcal{E}^{j,M}_{i\omega,\kk,1}-\mathcal{E}^{j,M}_{i\omega,\kk,2}}\bigg(\frac{1}{i\omega-\mathcal{E}^{j,M}_{i\omega,\kk,1}}-\frac{1}{i\omega-\mathcal{E}^{j,M}_{i\omega,\kk,2}}\bigg)\nonumber\\
    &+\frac{4[2i\omega+M+\Sigma_{c,off}(i\omega)-\Sigma_{c,dia}(i\omega)]}{\mathcal{E}^{j,M}_{i\omega,\kk,3}-\mathcal{E}^{j,M}_{i\omega,\kk,4}}\bigg(\frac{1}{i\omega-\mathcal{E}^{j,M}_{i\omega,\kk,3}}-\frac{1}{i\omega-\mathcal{E}^{j,M}_{i\omega,\kk,4}}\bigg)
\end{align}
where 
\begin{align}
    \mathcal{E}^{j,M}_{i\omega,\kk,1}=&\frac{M+\Sigma_{c,off}(i\omega)+\Sigma_{c,dia}(i\omega)}{2}-\sqrt{|v_{\star}\kk|^2+[M+\Sigma_{c,off}(i\omega)+\Sigma_{c,dia}(i\omega)]^2/4}\nonumber\\
    \mathcal{E}^{j,M}_{i\omega,\kk,2}=&\frac{M+\Sigma_{c,off}(i\omega)+\Sigma_{c,dia}(i\omega)}{2}+\sqrt{|v_{\star}\kk|^2+[M+\Sigma_{c,off}(i\omega)+\Sigma_{c,dia}(i\omega)]^2/4}\nonumber\\
    \mathcal{E}^{j,M}_{i\omega,\kk,3}=&-\frac{M+\Sigma_{c,off}(i\omega)-\Sigma_{c,dia}(i\omega)}{2}-\sqrt{|v_{\star}\kk|^2+[M+\Sigma_{c,off}(i\omega)-\Sigma_{c,dia}(i\omega)]^2/4}\nonumber\\
    \mathcal{E}^{j,M}_{i\omega,\kk,4}=&-\frac{M+\Sigma_{c,off}(i\omega)-\Sigma_{c,dia}(i\omega)}{2}+\sqrt{|v_{\star}\kk|^2+[M+\Sigma_{c,off}(i\omega)-\Sigma_{c,dia}(i\omega)]^2/4}
\end{align}
Without the dynamical self-energy $j=0$, we recover the bare dispersions $E^{j,M}_{\kk,n}=\mathcal{E}^{j,M}_{i\omega,\kk,n}\bigg|_{j=0}$ with
\begin{align}
    E^{j,M}_{\kk,1}=&-E^{j,M}_{\kk,4}=M/2-\sqrt{|v_{\star}\kk|^2+M^2/4}\nonumber\\
    E^{j,M}_{\kk,2}=&-E^{j,M}_{\kk,3}=M/2+\sqrt{|v_{\star}\kk|^2+M^2/4}
\end{align}
Then expand up to first order of $j$ we have
\begin{align}
    \mathcal{E}^{j,M}_{i\omega,\kk,1}=&E^{j,M}_{\kk,1}\bigg[1-\frac{\Sigma_{c,dia}(i\omega)+\Sigma_{c,off}(i\omega)}{\sqrt{M^2+4|v_{\star}\kk|^2}}\bigg]+\mathcal{O}(j^2)\nonumber\\
    \mathcal{E}^{j,M}_{i\omega,\kk,2}=&E^{j,M}_{\kk,2}\bigg[1+\frac{\Sigma_{c,dia}(i\omega)+\Sigma_{c,off}(i\omega)}{\sqrt{M^2+4|v_{\star}\kk|^2}}\bigg]+\mathcal{O}(j^2)\nonumber\\
    \mathcal{E}^{j,M}_{i\omega,\kk,3}=&E^{j,M}_{\kk,3}\bigg[1-\frac{\Sigma_{c,dia}(i\omega)-\Sigma_{c,off}(i\omega)}{\sqrt{M^2+4|v_{\star}\kk|^2}}\bigg]+\mathcal{O}(j^2)\nonumber\\
    \mathcal{E}^{j,M}_{i\omega,\kk,4}=&E^{j,M}_{\kk,4}\bigg[1+\frac{\Sigma_{c,dia}(i\omega)-\Sigma_{c,off}(i\omega)}{\sqrt{M^2+4|v_{\star}\kk|^2}}\bigg]+\mathcal{O}(j^2)\nonumber\\
\end{align}
In real frequency, denote
\begin{equation}
    \mathcal{E}^{j,M}_{\omega+i0^+,\kk,n}=\mathcal{R}^{j,M}_{\omega,\kk,n}-i\Gamma^{j,M}_{\omega,\kk,n}
\end{equation}
Then the renormalized dispersions $\tilde{E}^{j,M}_{\kk,n}$ and scattering times $\tau^{j,M}_{\kk,n}$ are determined by
\begin{align}
    &(\omega-\mathcal{R}^{j}_{\omega,\kk,n})\bigg|_{\omega=\tilde{E}^j_{\kk,n}}=0,\quad\frac{1}{\tau^j_{\kk,n}}=\Gamma^{j}_{\omega,\kk,n}\bigg|_{\omega=\tilde{E}^j_{\kk,n}}
\end{align}
which, up to first order of $j$ are 
\begin{equation}
    \tilde{E}^{j,M}_{\kk,n}=\mathcal{R}^{j,M}_{\omega,\kk,n}\bigg|_{\omega=E^{j,M}_{\kk,n}}+\mathcal{O}(j^2),\quad \frac{1}{\tau^{j,M}_{\kk,n}}=\Gamma^{j,M}_{\omega,\kk,n}\bigg|_{\omega=E^{j,M}_{\kk,n}}+\mathcal{O}(j^2)
\end{equation}

To get the final expression, we need to evaluate $\Sigma_{c}(E^{j,M}_{\kk,n}+i0^+)$.
Here we focus on small $|\kk|$.
Up to the lowest order of $|v_{\star}\kk|$, the bare dispersions are
\begin{align}
    E^{j,M}_{\kk,1}=-E^{j,M}_{\kk,4}\approx& -\frac{|v_{\star}\kk|^2}{M}+\mathcal{O}(|v_{\star}\kk|^4)\nonumber\\
    E^{j,M}_{\kk,2}=-E^{j,M}_{\kk,3}\approx& M +\frac{|v_{\star}\kk|^2}{M}+\mathcal{O}(|v_{\star}\kk|^4)
\end{align}
Then the self-energies are (\cref{eq:Sigma_c_real_freq})
\begin{align}
    \Sigma_{c,dia}(E^{j,M}_{\kk,1}+i0^+)=&-j\frac{n_f(N_f-n_f)(4N_f-1)}{N_f^2(N_f-1)}\pi\bigg(-\frac{|v_{\star}\kk|^2}{M}\bigg) \bigg[ \log\bigg(\bigg|
\frac{|v_{\star}\kk|^4 -M^2|v_\star \Lambda_c|^2}{|v_{\star}\kk|^4}\bigg|
\bigg)-\frac{1}{2}\log\bigg(\bigg|\frac{M^4-|v_{\star}\kk|^4}{|v_{\star}\kk|^4}\bigg|\bigg)
\bigg]\nonumber\\
&-j\frac{n_f(N_f-n_f)(4N_f-1)}{2N_f^2(N_f-1)}i\pi^2\frac{|v_{\star}\kk|^2}{M}+\mathcal{O}(|v_{\star}\kk|^4)
\nonumber\\
\approx&j\frac{n_f(N_f-n_f)(4N_f-1)}{N_f^2(N_f-1)}\pi\frac{|v_{\star}\kk|^2}{M} \log\bigg(
\frac{|v_\star \Lambda_c|^2}{|v_{\star}\kk|^2}
\bigg)-j\frac{n_f(N_f-n_f)(4N_f-1)}{2N_f^2(N_f-1)}i\pi^2\frac{|v_{\star}\kk|^2}{M}+\mathcal{O}(|v_{\star}\kk|^4)
\nonumber\\
\Sigma_{c,off}(E^{j,M}_{\kk,1}+i0^+)=&j\frac{n_f(n_f-N_f)}{N_f^2(N_f-1)}\pi\bigg[
M-\frac{|v_{\star}\kk|^2}{2M}\log\bigg(\bigg|\frac{M^2+|v_{\star}\kk|^2}{M^2-|v_{\star}\kk|^2}\bigg|\bigg)
\bigg]-j\frac{n_f(n_f-N_f)}{N_f^2(N_f-1)}i\pi^2\frac{|v_{\star}\kk|^2}{M}+\mathcal{O}(|v_{\star}\kk|^4)\nonumber\\
=&j\frac{n_f(n_f-N_f)}{N_f^2(N_f-1)}\pi
M-j\frac{n_f(n_f-N_f)}{N_f^2(N_f-1)}i\pi^2\frac{|v_{\star}\kk|^2}{M}+\mathcal{O}(|v_{\star}\kk|^4)\nonumber\\
\Sigma_{c,dia}(E^{j,M}_{\kk,2}+i0^+)=&-j\frac{n_f(N_f-n_f)(4N_f-1)}{N_f^2(N_f-1)}\pi M \bigg[ \log\bigg(\frac{|v_{\star}\Lambda_c|^2-M^2}{M^2}
\bigg)-\frac{1}{2}\log\bigg(\frac{2|v_{\star}\kk|^2}{M^2}\bigg)
\bigg]\nonumber\\
&-j\frac{n_f(N_f-n_f)(4N_f-1)}{N_f^2(N_f-1)}i\pi^2 M+\mathcal{O}(|v_{\star}\kk|^2)
\nonumber\\
=&-j\frac{n_f(N_f-n_f)(4N_f-1)}{N_f^2(N_f-1)}\pi M \log\bigg(\frac{|v_{\star}\Lambda_c|^2-M^2}{\sqrt{2}|v_{\star}\kk|M}
\bigg)
-j\frac{n_f(N_f-n_f)(4N_f-1)}{N_f^2(N_f-1)}i\pi^2 M+\mathcal{O}(|v_{\star}\kk|^2)
\nonumber\\
\Sigma_{c,off}(E^{j,M}_{\kk,2}+i0^+)=&j\frac{n_f(n_f-N_f)}{N_f^2(N_f-1)}\pi\bigg[
M+\frac{M}{2}\log\bigg(\frac{|v_{\star}\kk|^2}{2M^2}\bigg)
\bigg]+\mathcal{O}(|v_{\star}\kk|^2)\nonumber\\
\Sigma_{c,dia}(E^{j,M}_{\kk,3}+i0^+)=&\Sigma_{c,dia}(-E^{j,M}_{\kk,2}+i0^+)=-[\Sigma_{c,dia}(E^{j,M}_{\kk,2}+i0^+)]^*\nonumber\\
=&j\frac{n_f(N_f-n_f)(4N_f-1)}{N_f^2(N_f-1)}\pi M \log\bigg(\frac{|v_{\star}\Lambda_c|^2-M^2}{\sqrt{2}|v_{\star}\kk|M}
\bigg)
-j\frac{n_f(N_f-n_f)(4N_f-1)}{N_f^2(N_f-1)}i\pi^2 M+\mathcal{O}(|v_{\star}\kk|^2)
\nonumber\\
\Sigma_{c,off}(E^{j,M}_{\kk,3}+i0^+)=&\Sigma_{c,off}(-E^{j,M}_{\kk,2}+i0^+)=[\Sigma_{c,off}(E^{j,M}_{\kk,2}+i0^+)]^*\nonumber\\
=&j\frac{n_f(n_f-N_f)}{N_f^2(N_f-1)}\pi\bigg[
M+\frac{M}{2}\log\bigg(\frac{|v_{\star}\kk|^2}{2M^2}\bigg)
\bigg]+\mathcal{O}(|v_{\star}\kk|^2)\nonumber\\
\Sigma_{c,dia}(E^{j,M}_{\kk,4}+i0^+)=&\Sigma_{c,dia}(-E^{j,M}_{\kk,1}+i0^+)=-[\Sigma_{c,dia}(E^{j,M}_{\kk,1}+i0^+)]^*\nonumber\\
=&-j\frac{n_f(N_f-n_f)(4N_f-1)}{N_f^2(N_f-1)}\pi\frac{|v_{\star}\kk|^2}{M} \log\bigg(
\frac{|v_\star \Lambda_c|^2}{|v_{\star}\kk|^2}
\bigg)-j\frac{n_f(N_f-n_f)(4N_f-1)}{2N_f^2(N_f-1)}i\pi^2\frac{|v_{\star}\kk|^2}{M}+\mathcal{O}(|v_{\star}\kk|^4)
\nonumber\\
\Sigma_{c,off}(E^{j,M}_{\kk,4}+i0^+)=&\Sigma_{c,off}(-E^{j,M}_{\kk,1}+i0^+)=[\Sigma_{c,off}(E^{j,M}_{\kk,1}+i0^+)]^*\nonumber\\
=&j\frac{n_f(n_f-N_f)}{N_f^2(N_f-1)}\pi
M+j\frac{n_f(n_f-N_f)}{N_f^2(N_f-1)}i\pi^2\frac{|v_{\star}\kk|^2}{M}+\mathcal{O}(|v_{\star}\kk|^4)\nonumber\\
\end{align}
Finally, the renormalized dispersions and scattering times are obtained as
\begin{align}
    \tilde{E}^{j,M}_{\kk,1}=&-\tilde{E}^{j,M}_{\kk,4}=E^{j,M}_{\kk,1}\bigg[1-\frac{\mathrm{Re}[\Sigma_{c,dia}(E^{j,M}_{\kk,1}+i0^+)+\Sigma_{c,off}(E^{j,M}_{\kk,1}+i0^+)]}{\sqrt{M^2+4|v_{\star}\kk|^2}}\bigg]+\mathcal{O}(j^2)\nonumber\\
    =&E^{j,M}_{\kk,1}\bigg[1-\frac{j\frac{n_f(n_f-N_f)}{N_f^2(N_f-1)}\pi M}{\sqrt{M^2+4|v_{\star}\kk|^2}}+\mathcal{O}(|v_{\star}\kk|^2)\bigg]+\mathcal{O}(j^2)\nonumber\\
    =&-\frac{|v_{\star}\kk|^2}{M}\bigg[1-j\frac{n_f(n_f-N_f)}{N_f^2(N_f-1)}\pi\bigg]+\mathcal{O}(j^2,|v_{\star}\kk|^4)\nonumber\\
    \frac{1}{\tau^{j,M}_{\kk,1}}=&\frac{1}{\tau^{j,M}_{\kk,4}}=E^{j,M}_{\kk,1}\frac{\mathrm{Im}[\Sigma_{c,dia}(E^{j,M}_{\kk,1}+i0^+)+\Sigma_{c,off}(E^{j,M}_{\kk,1}+i0^+)]}{\sqrt{M^2+4|v_{\star}\kk|^2}}+\mathcal{O}(j^2)\nonumber\\
    =&\frac{|v_{\star}\kk|^4}{M^2}\frac{\pi^2 j}{M}\bigg[\frac{n_f(N_f-n_f)(4N_f-1)}{2N_f^2(N_f-1)}+\frac{n_f(n_f-N_f)}{N_f^2(N_f-1)}\bigg]+\mathcal{O}(j^2,|v_{\star}\kk|^6)\nonumber\\
    =&\frac{|v_{\star}\kk|^4}{M^2}\frac{\pi^2 j}{M}\frac{n_f(N_f-n_f)(4N_f-3)}{2N_f^2(N_f-1)}+\mathcal{O}(j^2,|v_{\star}\kk|^6)\nonumber\\
    \tilde{E}^{j,M}_{\kk,2}=&-\tilde{E}^{j,M}_{\kk,3}=E^{j,M}_{\kk,2}\bigg[1+\frac{\mathrm{Re}[\Sigma_{c,dia}(E^{j,M}_{\kk,2}+i0^+)+\Sigma_{c,off}(E^{j,M}_{\kk,2}+i0^+)]}{\sqrt{M^2+4|v_{\star}\kk|^2}}\bigg]+\mathcal{O}(j^2)\nonumber\\
    =&\bigg(M+\frac{|v_{\star}\kk|^2}{M}+\mathcal{O}(|v_{\star}\kk|^4)\bigg)\bigg[1-j\frac{n_f(N_f-n_f)(4N_f-1)}{N_f^2(N_f-1)}\pi \log\bigg(\frac{|v_{\star}\Lambda_c|^2-M^2}{\sqrt{2}|v_{\star}\kk|M}
\bigg)\nonumber\\
&+j\frac{n_f(n_f-N_f)}{N_f^2(N_f-1)}\pi\bigg[
1+\frac{1}{2}\log\bigg(\frac{|v_{\star}\kk|^2}{2M^2}\bigg)
\bigg]+\mathcal{O}(|v_{\star}\kk|^2)
\bigg]+\mathcal{O}(j^2)\nonumber\\
=&M\bigg[
1-j\frac{n_f(N_f-n_f)(4N_f-1)}{N_f^2(N_f-1)}\pi \log\bigg(\frac{|v_{\star}\Lambda_c|^2-M^2}{\sqrt{2}|v_{\star}\kk|M}
\bigg)+j\frac{n_f(n_f-N_f)}{N_f^2(N_f-1)}\pi\bigg[
1+\log\bigg(\frac{|v_{\star}\kk|}{\sqrt{2}M}\bigg)
\bigg]
\bigg]+\mathcal{O}(j^2,|v_{\star}\kk|^2)\nonumber\\
\frac{1}{\tau^{j,M}_{\kk,2}}=&\frac{1}{\tau^{j,M}_{\kk,3}}=-E^{j,M}_{\kk,2}\frac{\mathrm{Im}[\Sigma_{c,dia}(E^{j,M}_{\kk,2}+i0^+)+\Sigma_{c,off}(E^{j,M}_{\kk,2}+i0^+)]}{\sqrt{M^2+4|v_{\star}\kk|^2}}+\mathcal{O}(j^2)\nonumber\\
=&j\frac{n_f(N_f-n_f)(4N_f-1)}{N_f^2(N_f-1)}\pi^2 M+\mathcal{O}(j^2,|v_{\star}\kk|^2)
\end{align}

For the gapless modes $n=1,4$, we observe that at first order of $j$, the self-energy introduces a weak damping proportional to $|v_{\star}\kk|^4$. 
Physically, the self-energy correction arising from the $J$ term acts on the $\Gamma_1 \oplus \Gamma_2$ electrons, which are already gapped by the $M$ term. As a result, the gapless mode formed by the $\Gamma_3$ electrons does not experience a strong damping effect near $\Gamma$ point. 
As a comparison, the gapped modes $n=2,3$ formed by $\Gamma_1\oplus\Gamma_2$ electrons do have a finite damping rate at $\Gamma$ point.
At charge neutrality, the damping rate reads
\begin{align}
    \frac{1}{\tau_{\kk,n=1}^{j,M}}=&\frac{1}{\tau_{\kk,n=4}^{j,M}}=\bigg[  0 +\mathcal{O}(j^2)\bigg] |v_\star \kk|^2 + \mathcal{O}(|v_\star\kk|^3)\nonumber\\
    \frac{1}{\tau_{\kk,n=2}^{j,M}}=&\frac{1}{\tau_{\kk,n=3}^{j,M}}=M\frac{31\pi^2 j}{56}
\end{align}
Taking the parameters of the topological heavy-fermion model in Ref.~\cite{PhysRevLett.129.047601}, we find the finite scattering rate of the gapped modes is $0.03M\approx 0.012 $meV.

\subsection{Effect of strain}
In \cref{eq:Sigma_c_no_W,eq:sigma_c_with_suscep}, we proved that at low temperature, only the $\Gamma_1\oplus\Gamma_2$ $c$ electrons acquire a dynamical self-energy at order $j\sim J^2$ due to the local-moment fluctuations of $f$ electrons.
Thus, we expect that if the local-momentum fluctuations are quenched, the damping rate will also be suppressed at order $j$.
To verify this, we investigate the effect of finite strain, which will frozen the local-momentum fluctuations at charge neutrality in the low-temperature limit. 
For simplicity, we focus on the case of $M=0$. 

The strained Hamiltonian of the $f$ sector takes the form of
\ba 
   H_{strain,f}= \sum_{\RR,\alpha\alpha',\eta s}
    [M_f(\varepsilon_{xy} \sigma_x + \varepsilon_- \eta \sigma_y)_{\alpha\alpha'} -\mu]
     f_{\RR, \alpha \eta s}^\dag f_{\RR,\alpha'\eta s}
\ea 
We can perform a rotation characterized by 
\ba 
\label{eq:eigenbasis_strain}
U_{\alpha \tilde{\alpha}'}^\eta  =\frac{1}{\sqrt{2}}
\begin{bmatrix}
    \frac{\varepsilon_{xy} -i \eta \varepsilon_-}{\sqrt{\varepsilon_{xy}^2 + \varepsilon_-^2 }} & - \frac{\varepsilon_{xy} -i \eta \varepsilon_-}{\sqrt{\varepsilon_{xy}^2 + \varepsilon_-^2 }} \\ 
    1 & 1 
\end{bmatrix}_{\alpha \tilde{\alpha}'}
\ea 
and let 
\ba 
\label{eq:eigenbasis_strain_f}
f_{\RR,\alpha \eta s } = \sum_{\tilde{\alpha}'}U^\eta_{\alpha\tilde{\alpha}'} \tilde{f}_{\RR,\tilde{\alpha}'\eta s}
\ea 
where we use $\tilde{\alpha}=1,2$ to label the new orbital basis and use $\tilde{f}$ to denote the corresponding annihilation operators.
In the new electron basis $\tilde{f}$, the Hamiltonian reads 
\begin{align}
\label{H_atomic}
     H_{strain,f}=& \sum_{\RR,\tilde{\alpha}\eta s} [\varepsilon(-1)^{\tilde{\alpha}+1}-\mu]\tilde{f}_{\RR, \tilde{\alpha} \eta s}^\dag \tilde{f}_{\RR,\tilde{\alpha}'\eta s}
\end{align} 
where 
\ba 
&\varepsilon = M_f\sqrt{ \varepsilon_{xy}^2 + \varepsilon_-^2}
\ea 
The atomic Hamiltonian of $f$ electrons at charge neutrality ($\mu=0$) now becomes 
\ba 
H_{atom,strain} =&\sum_{\RR} H_{\RR,strain}\nonumber\\
H_{\RR,strain} =&\sum_{\tilde{\alpha}\eta s} \varepsilon(-1)^{\tilde{\alpha}+1}\tilde{f}_{ \RR,\tilde{\alpha} \eta s}^\dag \tilde{f}_{ \RR,\tilde{\alpha}'\eta s}+ \frac{U_1}{2} 
\bigg( \sum_{\tilde{\alpha} \eta s} (\tilde{f}_{ \RR,\tilde{\alpha} \eta s}^\dag \tilde{f}_{ \RR,\tilde{\alpha}'\eta s}-1/2)\bigg)^2
\ea 
For a given site, the eigenstate of $H_{\RR,strain}$ can be characterized by the filling of $\tilde{f}$ electrons of each orbital $\tilde{\alpha}$. We let 
\ba 
|n_1,n_2,i \rangle
\ea 
denote the state with 
\ba 
\sum_{\eta s}\tilde{f}_{ \RR,\tilde{\alpha}\eta s}^\dag \tilde{f}_{ \RR,\tilde{\alpha}\eta s}|n_1,n_2,i\rangle = n_{\tilde{\alpha}} |n_1,n_2,i \rangle 
\ea 
and $i=1,...,d_{n_1,n_2}$ labels the degeneracy of the states with the same filling factor $n_1,n_2$ and 
\ba d_{n_1,n_2}
 = \binom{4}{n_1}
\binom{4}{n_2}
\ea 
The energy of $|n_1,n_2,i\rangle$ state is 
\ba 
&H_{\RR,strain} |n_1,n_2,i\rangle =E_{n_1,n_2} |n_1,n_2,i\rangle \nonumber\\ 
&E_{n_1,n_2} 
=  \varepsilon(n_1-n_2) + \frac{U_1}{2}(n_1+n_2-4)^2
\ea 
Then the partition function is
\begin{equation}
    Z_{\RR,strain}=\sum_{n_1,n_2}d_{n_1,n_2}e^{-\beta E_{n_1,n_2}}
\end{equation}
At charge neutrality, the ground state $|n_1=0,n_2=4,i=1\rangle $ is singly degenerate $d_{0,4}=1$, with ground state energy
\begin{equation}
    E_{gs,strain}=E_{n_1=0,n_2=4}=-4\varepsilon
\end{equation}
Thus in the low-temperature limit, we have
\begin{equation}
    \lim_{\beta\to +\infty}Z_{\RR,strain}=e^{-\beta E_{gs,strain}}=e^{-\beta E_{0,4}}
    \label{eq:ZRR_strain_zero_T}
\end{equation}

We now investigate the $\tilde{f}$ electron correlation function 
\ba 
&  \langle 
 T_\tau 
  \tilde{f}_{\RR,\tilde{\alpha}_1\eta_1 s_1}^\dag (\tau_1) \tilde{f}_{\RR,\tilde{\alpha}_2\eta_2 s_2} (\tau_1)
\tilde{f}_{\RR,\tilde{\alpha}_3\eta_3 s_3}^\dag (\tau_2)\tilde{f}_{\RR,\tilde{\alpha}_4\eta_4 s_4}(\tau_2)
 \rangle_{H_{atom,strain}} \nonumber\\
= &
\frac{1}{Z_{\RR,strain}}\sum_{n_1,n_2,m_1,m_2,i,j} \langle n_1,n_2,i|  \tilde{f}_{\RR,\tilde{\alpha}_1\eta_1 s_1}^\dag  \tilde{f}_{\RR,\tilde{\alpha}_2\eta_2 s_2}| m_1,m_2,j\rangle 
\langle m_1,m_2,j|\tilde{f}_{\RR,\tilde{\alpha}_3\eta_3 s_3}^\dag \tilde{f}_{\RR,\tilde{\alpha}_4\eta_4 s_4}| n_1,n_2,i\rangle
\nonumber\\
&
\begin{cases}
 e^{-(\beta-\tau_1+\tau_2) E_{n_1,n_2} -(\tau_1-\tau_2) E_{m_1,m_2}} & \tau_1>\tau_2  \\
e^{-(\beta-\tau_2+\tau_1)E_{m_1,m_2} -(\tau_2-\tau_1)E_{n_1,n_2}}& \tau_1<\tau_2  \\
\end{cases}
\ea 
In the Matsubara frequency domain, we find 
\ba 
& \frac{1}{\beta}\int_{\tau_1,\tau_2} \langle 
 T_\tau 
  \tilde{f}_{\RR,\tilde{\alpha}_1\eta_1 s_1}^\dag (\tau_1) \tilde{f}_{\RR,\tilde{\alpha}_2\eta_2 s_2} (\tau_1)
\tilde{f}_{\RR,\tilde{\alpha}_3\eta_3 s_3}^\dag (\tau_2)\tilde{f}_{\RR,\tilde{\alpha}_4\eta_4 s_4}(\tau_2)
 \rangle_{H_{atom,strain}} e^{i\Omega(\tau_1-\tau_2)} \nonumber\\ 
 =& 
 \frac{1}{Z_{\RR,strain}}\sum_{n_1,n_2,m_1,m_2,i,j} \langle n_1,n_2,i|  \tilde{f}_{\RR,\tilde{\alpha}_1\eta_1 s_1}^\dag  \tilde{f}_{\RR,\tilde{\alpha}_2\eta_2 s_2}| m_1,m_2,j\rangle 
\langle m_1,m_2,j|\tilde{f}_{\RR,\tilde{\alpha}_3\eta_3 s_3}^\dag \tilde{f}_{\RR,\tilde{\alpha}_4\eta_4 s_4}| n_1,n_2,i\rangle
\chi(i\Omega, E_{m_1,m_2},E_{n_1,n_2}) 
\ea 
where 
\ba 
\chi(i\Omega,E_{m_1,m_2},E_{n_1,n_2} ) 
= (1-\delta_{E_{m_1,m_2},E_{n_1,n_2}}) \frac{ 
e^{-\beta E_{m_1,m_2}}-e^{-\beta E_{n_1,n_2}}}{i\Omega +E_{n_1,n_2} - E_{m_1,m_2}}
+ \delta_{E_{m_1,m_2},E_{n_1,n_2}}\beta e^{-\beta E_{n_1,n_2}}
\ea 
Using \cref{eq:ZRR_strain_zero_T}, we obtain
\begin{align}
    &\lim_{\beta\to +\infty}\frac{1}{Z_{\RR,strain}}\chi(i\Omega,E_{m_1,m_2},E_{n_1,n_2})\nonumber\\
    =&(1-\delta_{(m_1,m_2),(n_1n_2)})\frac{\delta_{(m_1,m_2),(0,4)}-\delta_{(n_1,n_2),(0,4)}}{i\Omega+E_{n_1,n_2}-E_{m_1,m_2}}+\beta\delta_{(m_1,m_2),(n_1,n_2)}\delta_{(m_1,m_2),(0,4)}
\end{align}
Thus in the low-temperature limit, we have
\begin{align}
   & \lim_{\beta\to +\infty} \frac{1}{\beta}\int_{\tau_1,\tau_2} \langle 
 T_\tau 
  \tilde{f}_{\RR,\tilde{\alpha}_1\eta_1 s_1}^\dag (\tau_1) \tilde{f}_{\RR,\tilde{\alpha}_2\eta_2 s_2} (\tau_1)
\tilde{f}_{\RR,\tilde{\alpha}_3\eta_3 s_3}^\dag (\tau_2)\tilde{f}_{\RR,\tilde{\alpha}_4\eta_4 s_4}(\tau_2)
 \rangle_{H_{atom,strain}} e^{i\Omega(\tau_1-\tau_2)} \nonumber\\ 
 =&\sum_{(n_1,n_2)\ne(0,4),i} \langle n_1,n_2,i|  \tilde{f}_{\RR,\tilde{\alpha}_1\eta_1 s_1}^\dag  \tilde{f}_{\RR,\tilde{\alpha}_2\eta_2 s_2}| 0,4,1\rangle 
\langle 0,4,1|\tilde{f}_{\RR,\tilde{\alpha}_3\eta_3 s_3}^\dag \tilde{f}_{\RR,\tilde{\alpha}_4\eta_4 s_4}| n_1,n_2,i\rangle \frac{1}{i\Omega+E_{n_1,n_2}-E_{0,4}}\nonumber\\
&-\sum_{(m_1,m_2)\ne(0,4),j} \langle 0,4,1|  \tilde{f}_{\RR,\tilde{\alpha}_1\eta_1 s_1}^\dag  \tilde{f}_{\RR,\tilde{\alpha}_2\eta_2 s_2}| m_1,m_2,j\rangle 
\langle m_1,m_2,j|\tilde{f}_{\RR,\tilde{\alpha}_3\eta_3 s_3}^\dag \tilde{f}_{\RR,\tilde{\alpha}_4\eta_4 s_4}| 0,4,1\rangle \frac{1}{i\Omega+E_{0,4}-E_{m_1,m_2}}\nonumber\\
&+\beta\langle 0,4,1|  \tilde{f}_{\RR,\tilde{\alpha}_1\eta_1 s_1}^\dag  \tilde{f}_{\RR,\tilde{\alpha}_2\eta_2 s_2}| 0,4,1\rangle 
\langle 0,4,1|\tilde{f}_{\RR,\tilde{\alpha}_3\eta_3 s_3}^\dag \tilde{f}_{\RR,\tilde{\alpha}_4\eta_4 s_4}| 0,4,1\rangle
\end{align}
The matrix element is evaluated as
\begin{align}
    \langle n_1,n_2,i|  \tilde{f}_{\RR,\tilde{\alpha}_1\eta_1 s_1}^\dag  \tilde{f}_{\RR,\tilde{\alpha}_2\eta_2 s_2}| 0,4,1\rangle =\delta_{\tilde{\alpha}_2,2}\delta_{(\tilde{\alpha}_1,\eta_1,s_1),(\tilde{\alpha}_2,\eta_2,s_2)}\delta_{(n_1,n_2),(0,4)}+\delta_{\tilde{\alpha}_1,1}\delta_{\tilde{\alpha}_2,2}\delta_{\ket{n_1,n_2,i},\tilde{f}_{\RR,\tilde{\alpha}_1\eta_1 s_1}^\dag  \tilde{f}_{\RR,\tilde{\alpha}_2\eta_2 s_2}\ket{0,4,1}}
\end{align}
Thus, for $(n_1,n_2)\ne(0,4)$, we have
\begin{align}
    &\langle n_1,n_2,i|  \tilde{f}_{\RR,\tilde{\alpha}_1\eta_1 s_1}^\dag  \tilde{f}_{\RR,\tilde{\alpha}_2\eta_2 s_2}| 0,4,1\rangle 
\langle 0,4,1|\tilde{f}_{\RR,\tilde{\alpha}_3\eta_3 s_3}^\dag \tilde{f}_{\RR,\tilde{\alpha}_4\eta_4 s_4}| n_1,n_2,i\rangle\nonumber\\
=&\delta_{\tilde{\alpha}_1,1}\delta_{\tilde{\alpha}_2,2}\delta_{\ket{n_1,n_2,i},\tilde{f}_{\RR,\tilde{\alpha}_1\eta_1 s_1}^\dag  \tilde{f}_{\RR,\tilde{\alpha}_2\eta_2 s_2}\ket{0,4,1}}
\delta_{\tilde{\alpha}_4,1}\delta_{\tilde{\alpha}_3,2}\delta_{\ket{n_1,n_2,i},\tilde{f}_{\RR,\tilde{\alpha}_4\eta_4 s_4}^\dag  \tilde{f}_{\RR,\tilde{\alpha}_3\eta_3 s_3}\ket{0,4,1}}\nonumber\\
=&\delta_{(\tilde{\alpha}_1,\eta_1,s_1),(\tilde{\alpha}_4,\eta_4,s_4)}\delta_{(\tilde{\alpha}_2,\eta_2,s_2),(\tilde{\alpha}_3,\eta_3,s_3)}\delta_{\tilde{\alpha}_1,1}\delta_{\tilde{\alpha}_2,2}\delta
_{(n_1,n_2),(1,3)}\delta_{\ket{n_1,n_2,i},\tilde{f}_{\RR,\tilde{\alpha}_1\eta_1 s_1}^\dag  \tilde{f}_{\RR,\tilde{\alpha}_2\eta_2 s_2}\ket{0,4,1}}
\end{align}
Similarly, for $(m_1,m_2)\ne(0,4)$ we have
\begin{align}
    &\langle 0,4,1|  \tilde{f}_{\RR,\tilde{\alpha}_1\eta_1 s_1}^\dag  \tilde{f}_{\RR,\tilde{\alpha}_2\eta_2 s_2}| m_1,m_2,j\rangle 
\langle m_1,m_2,j|\tilde{f}_{\RR,\tilde{\alpha}_3\eta_3 s_3}^\dag \tilde{f}_{\RR,\tilde{\alpha}_4\eta_4 s_4}| 0,4,1\rangle \nonumber\\
=&\delta_{(\tilde{\alpha}_1,\eta_1,s_1),(\tilde{\alpha}_4,\eta_4,s_4)}\delta_{(\tilde{\alpha}_2,\eta_2,s_2),(\tilde{\alpha}_3,\eta_3,s_3)}\delta_{\tilde{\alpha}_1,2}\delta_{\tilde{\alpha}_2,1}\delta
_{(n_1,n_2),(1,3)}\delta_{\ket{m_1,m_2,j},\tilde{f}_{\RR,\tilde{\alpha}_2\eta_2 s_2}^\dag  \tilde{f}_{\RR,\tilde{\alpha}_1\eta_1 s_1}\ket{0,4,1}}
\end{align}
In addition, 
\begin{align}
    &\langle 0,4,1|  \tilde{f}_{\RR,\tilde{\alpha}_1\eta_1 s_1}^\dag  \tilde{f}_{\RR,\tilde{\alpha}_2\eta_2 s_2}| 0,4,1\rangle 
\langle 0,4,1|\tilde{f}_{\RR,\tilde{\alpha}_3\eta_3 s_3}^\dag \tilde{f}_{\RR,\tilde{\alpha}_4\eta_4 s_4}| 0,4,1\rangle\nonumber\\
=&\delta_{\tilde{\alpha}_2,2}\delta_{(\tilde{\alpha}_1,\eta_1,s_1),(\tilde{\alpha}_2,\eta_2,s_2)}\delta_{\tilde{\alpha}_3,2}\delta_{(\tilde{\alpha}_3,\eta_3,s_3),(\tilde{\alpha}_4,\eta_4,s_4)}
\end{align}
Accumulating all contributions, we find 
\ba 
 &\lim_{\beta\to+\infty}\frac{1}{\beta}\int_{\tau_1,\tau_2} \langle 
 T_\tau 
  \tilde{f}_{\RR,\tilde{\alpha}_1\eta_1 s_1}^\dag (\tau_1) \tilde{f}_{\RR,\tilde{\alpha}_2\eta_2 s_2} (\tau_1)
\tilde{f}_{\RR,\tilde{\alpha}_3\eta_3 s_3}^\dag (\tau_2)\tilde{f}_{\RR,\tilde{\alpha}_4\eta_4 s_4}(\tau_2)
 \rangle_{H_{atom,strain}} e^{i\Omega(\tau_1-\tau_2)} \nonumber\\ 
 \approx &
  \delta_{(\tilde{\alpha}_1,\eta_1,s_1),(\tilde{\alpha}_2,\eta_2,s_2)}\delta_{(\tilde{\alpha}_3,\eta_3,s_3),(\tilde{\alpha}_4,\eta_4,s_4)}\delta_{\tilde{\alpha}_2,2}\delta_{\tilde{\alpha}_3,2} \beta \nonumber\\
  & + 
  \delta_{(\tilde{\alpha}_1,\eta_1,s_1),(\tilde{\alpha}_4,\eta_4,s_4)}\delta_{(\tilde{\alpha}_2,\eta_2,s_2),(\tilde{\alpha}_3,\eta_3,s_3)} (\delta_{\tilde{\alpha}_1,2}\delta_{\tilde{\alpha}_2,1}\frac{-1}{i\Omega -2\varepsilon}
  +\delta_{\tilde{\alpha}_1,1}\delta_{\tilde{\alpha}_2,2}
  \frac{1}{i\Omega + 2\varepsilon})
\ea 
where we have used $E_{1,3}-E_{0,4}=2\varepsilon$.
Define the normal order $:O: = O - \langle O \rangle_{H_{atom,strain}} $.
The expectation value of two Fermion operator is
\begin{align}
    \langle \tilde{f}_{\RR,\tilde{\alpha}_1\eta_1 s_1}^\dag  \tilde{f}_{\RR,\tilde{\alpha}_2\eta_2 s_2} \rangle_{H_{atom,strain}}=\delta_{(\tilde{\alpha}_1,\eta_1,s_1),(\tilde{\alpha}_2,\eta_2,s_2)}\delta_{\tilde{\alpha}_1,2}
\end{align}
Therefore
\begin{align}
    &\langle 
 T_\tau 
 : \tilde{f}_{\RR,\tilde{\alpha}_1\eta_1 s_1}^\dag (\tau_1) \tilde{f}_{\RR,\tilde{\alpha}_2\eta_2 s_2} (\tau_1)::
\tilde{f}_{\RR,\tilde{\alpha}_3\eta_3 s_3}^\dag (\tau_2)\tilde{f}_{\RR,\tilde{\alpha}_4\eta_4 s_4}(\tau_2):
 \rangle_{H_{atom,strain}}\nonumber\\
 =&\langle 
 T_\tau 
  \tilde{f}_{\RR,\tilde{\alpha}_1\eta_1 s_1}^\dag (\tau_1) \tilde{f}_{\RR,\tilde{\alpha}_2\eta_2 s_2} (\tau_1)
\tilde{f}_{\RR,\tilde{\alpha}_3\eta_3 s_3}^\dag (\tau_2)\tilde{f}_{\RR,\tilde{\alpha}_4\eta_4 s_4}(\tau_2)
 \rangle_{H_{atom,strain}}\nonumber\\
 &-\langle \tilde{f}_{\RR,\tilde{\alpha}_1\eta_1 s_1}^\dag(\tau_1) \tilde{f}_{\RR,\tilde{\alpha}_2\eta_2 s_2}(\tau_1) \rangle_{H_{atom,strain}}\langle \tilde{f}_{\RR,\tilde{\alpha}_3\eta_3 s_3}^\dag  (\tau_2)\tilde{f}_{\RR,\tilde{\alpha}_4\eta_4 s_4} (\tau_2)\rangle_{H_{atom,strain}}\nonumber\\
 =&\langle 
 T_\tau 
  \tilde{f}_{\RR,\tilde{\alpha}_1\eta_1 s_1}^\dag (\tau_1) \tilde{f}_{\RR,\tilde{\alpha}_2\eta_2 s_2} (\tau_1)
\tilde{f}_{\RR,\tilde{\alpha}_3\eta_3 s_3}^\dag (\tau_2)\tilde{f}_{\RR,\tilde{\alpha}_4\eta_4 s_4}(\tau_2)
 \rangle_{H_{atom,strain}}-\delta_{(\tilde{\alpha}_1,\eta_1,s_1),(\tilde{\alpha}_2,\eta_2,s_2)}\delta_{(\tilde{\alpha}_3,\eta_3,s_3),(\tilde{\alpha}_4,\eta_4,s_4)}\delta_{\tilde{\alpha}_2,2}\delta_{\tilde{\alpha}_3,2}
\end{align}
and
\ba 
& \frac{1}{\beta}\int_{\tau_1,\tau_2} \langle 
 T_\tau 
 : \tilde{f}_{\RR,\tilde{\alpha}_1\eta_1 s_1}^\dag (\tau_1) \tilde{f}_{\RR,\tilde{\alpha}_2\eta_2 s_2} (\tau_1)::
\tilde{f}_{\RR,\tilde{\alpha}_3\eta_3 s_3}^\dag (\tau_2)\tilde{f}_{\RR,\tilde{\alpha}_4\eta_4 s_4}(\tau_2):
 \rangle_{H_{atom,strain}} e^{i\Omega(\tau_1-\tau_2)}\nonumber\\ 
 = &  \delta_{(\tilde{\alpha}_1,\eta_1,s_1),(\tilde{\alpha}_4,\eta_4,s_4)}\delta_{(\tilde{\alpha}_2,\eta_2,s_2),(\tilde{\alpha}_3,\eta_3,s_3)} (\delta_{\tilde{\alpha}_1,2}\delta_{\tilde{\alpha}_2,1}\frac{-1}{i\Omega -2\varepsilon}
  +\delta_{\tilde{\alpha}_1,1}\delta_{\tilde{\alpha}_2,2}
  \frac{1}{i\Omega + 2\varepsilon})
\ea 
In the original $f$-electron basis (using \cref{eq:eigenbasis_strain}), we find 
\ba 
\label{eq:sus_strain_f_basis}
\chi_{\alpha_1\eta_1s_1,\alpha_2\eta_2s_2, \alpha_3\eta_3s_3,\alpha_4\eta_4s_4}(i\Omega) = \sum_{\tilde{\alpha}}
 \delta_{(\eta_1,s_1),(\eta_4,s_4)}\delta_{(\eta_2,s_2),(\eta_3,s_3)} U^{\eta_1,*}_{\alpha_1, \tilde{\alpha}}
 U^{\eta_2}_{\alpha_2,3-\tilde{\alpha} }U^{\eta_2,*}_{\alpha_3,3-\tilde{\alpha}}U^{\eta_1}_{\alpha_4,\tilde{\alpha}}\frac{1}{2\varepsilon 
 + (-1)^{\tilde{\alpha}+1}i\Omega 
 }
\ea 
We then utilize ~\cref{eq:sig_c_from_chi_G,eq:local_gc_M=0,eq:sus_strain_f_basis} and obtain 
\ba 
\label{eq:sigma_c_strain_freq}
[\Sigma_c(i\omega,\kk)]_{a \eta s, a'\eta's'}
\approx - \delta_{a \eta s,a'\eta's'}\frac{1}{\beta}\sum_{i\Omega}
g_c(i\omega-i\Omega) J^2 \frac{4\varepsilon}{(2\varepsilon)^2 -(i\Omega)^2}
\ea 
where, for simplicity, we have dropped the effect of strain on the $c$-electron. 
We utilize the following spectral representation of Green's function 
\ba 
\label{eq:gc_spectral_rep}
&g_c(i\omega) = \int_E \frac{\rho_c(E)}{i\omega-E},\quad 
\rho_c(E) = \frac{1}{2 \Omega_M}\int_{|\kk|<\Lambda_c}\bigg[ \delta(E - |v_\star \kk|) + \delta(E+|v_\star \kk|)\bigg] 
= \frac{\pi|E|}{\Omega_M|v_\star|^2} \theta(|v_\star\Lambda_c|-|E|)
\ea 
We find 
\ba 
\label{eq:sigc_strain_spec}
[\Sigma_c(i\omega,\kk)]_{a \eta s, a'\eta's'}
\approx - \delta_{a \eta s,a'\eta's'}\frac{1}{\beta}\sum_{i\Omega}J^2
\int_E \frac{\rho_c(E)}{i\omega-i\Omega-E} 
\bigg[ 
\frac{1}{i\Omega+2\varepsilon} + \frac{1}{-i\Omega +2\varepsilon}
\bigg] 
\ea 
We then use (in the $\beta\rightarrow \infty$ limit)
\ba 
\label{eq:fermi_boson_freq}
&
\frac{1}{\beta}\sum_{i\Omega} \frac{1}{i\omega-i\Omega-E}\bigg[ 
\frac{1}{i\Omega+2\varepsilon} + \frac{1}{-i\Omega + 2 \varepsilon}
\bigg] \approx 
\frac{\theta(E)}{i\omega -2\varepsilon -E}
+ \frac{\theta(-E)}{i\omega + 2\varepsilon -E}
\ea 
Combining \cref{eq:fermi_boson_freq,eq:sigc_strain_spec}, we find 
\ba 
[\Sigma_c(i\omega,\kk)]_{a \eta s, a'\eta's'}
\approx &- \delta_{a \eta s,a'\eta's'}J^2
\int_E \rho_c(E) 
\bigg[ \frac{\theta(E)}{i\omega -2\varepsilon -E}
+ \frac{\theta(-E)}{i\omega + 2\varepsilon -E} \bigg] \nonumber\\
=& - \delta_{a \eta s,a'\eta's'}
\frac{\pi J^2}{\Omega_M|v_\star|^2}
\bigg[
-(i\omega-2\varepsilon) \log\bigg( 
\frac{i\omega-2\varepsilon-|v_\star\Lambda_c|}{i\omega-2\varepsilon}
\bigg)
-(i\omega+2\varepsilon) \log\bigg( 
\frac{i\omega+2\varepsilon+|v_\star\Lambda_c|}{i\omega+2\varepsilon}
\bigg)
\bigg]
\ea 
In the real-frequency, we find 
\ba 
[\Sigma_c(\kk,\omega-i0^+)]_{a \eta s, a'\eta's'}
\approx &\delta_{a \eta s,a'\eta's'}
\frac{\pi J^2}{\Omega_M|v_\star|^2}
\bigg[
(\omega-2\varepsilon) \log\bigg( \bigg|
\frac{\omega-2\varepsilon-|v_\star\Lambda_c|}{\omega-2\varepsilon}\bigg|
\bigg)
+(\omega+2\varepsilon) \log\bigg( \bigg|
\frac{\omega+2\varepsilon+|v_\star\Lambda_c|}{\omega+2\varepsilon}
\bigg)\bigg|
\bigg] \nonumber\\
& + i\delta_{a \eta s,a'\eta's'}\pi \frac{\pi J^2}{\Omega_M|v_\star|^2}\bigg(|\omega|-2\varepsilon\bigg) 
\theta( |\omega|-2\varepsilon) 
\theta(|v_\star\Lambda_c+2\varepsilon|-|\omega|) 
\ea 
We can observe that, the scattering rate of the $c$ electrons ($\mathrm{Im}[\Sigma_c(\kk,\omega-i0^+)]$) vanishes at low-energy limit with $|\omega|<2\varepsilon$. 
When a finite strain is applied, the fluctuations of the
$f$ electrons are quenched at low temperatures, because the $f$-electron ground state becomes non-degenerate in the atomic limit. 
In the absence of low-energy $f$-electron
fluctuations, the scattering rate is correspondingly suppressed and
approaches zero.

\newpage 
\newpage

%% file: tbg.bib
@article{ANG19,
  title = {Valley {{Jahn-Teller Effect}} in {{Twisted Bilayer Graphene}}},
  author = {Angeli, M. and Tosatti, E. and Fabrizio, M.},
  year = {2019},
  month = oct,
  journal = {Phys. Rev. X},
  volume = {9},
  number = {4},
  pages = {041010},
  publisher = {American Physical Society},
  doi = {10.1103/PhysRevX.9.041010},
  urldate = {2023-12-22}
}

@article{BEN21,
  title = {Measuring Local Moir{\'e} Lattice Heterogeneity of Twisted Bilayer Graphene},
  author = {Benschop, Tjerk and {de Jong}, Tobias A. and Stepanov, Petr and Lu, Xiaobo and Stalman, Vincent and {van der Molen}, Sense Jan and Efetov, Dmitri K. and Allan, Milan P.},
  year = {2021},
  month = feb,
  journal = {Phys. Rev. Res.},
  volume = {3},
  number = {1},
  pages = {013153},
  publisher = {American Physical Society},
  doi = {10.1103/PhysRevResearch.3.013153},
  urldate = {2023-12-22}
}

@article{BER21,
  title = {Twisted Bilayer Graphene. {{I}}. {{Matrix}} Elements, Approximations, Perturbation Theory, and a $\vec{k}\cdot\vec{p}$ Two-Band Model},
  author = {Bernevig, B. Andrei and Song, Zhi-Da and Regnault, Nicolas and Lian, Biao},
  year = {2021},
  month = may,
  journal = {Phys. Rev. B},
  volume = {103},
  number = {20},
  pages = {205411},
  publisher = {American Physical Society},
  doi = {10.1103/PhysRevB.103.205411},
  urldate = {2021-09-13}
}

@article{BER21a,
  title = {Twisted Bilayer Graphene. {{III}}. {{Interacting Hamiltonian}} and Exact Symmetries},
  author = {Bernevig, B. Andrei and Song, Zhi-Da and Regnault, Nicolas and Lian, Biao},
  year = {2021},
  month = may,
  journal = {Phys. Rev. B},
  volume = {103},
  number = {20},
  pages = {205413},
  publisher = {American Physical Society},
  doi = {10.1103/PhysRevB.103.205413},
  urldate = {2021-09-13}
}

@article{bistritzer_moire_2011,
	title = {Moiré bands in twisted double-layer graphene},
	volume = {108},
	issn = {0027-8424, 1091-6490},
	url = {https://www.pnas.org/content/108/30/12233},
	doi = {10.1073/pnas.1108174108},
	number = {30},
	urldate = {2019-07-21},
	journal = {Proceedings of the National Academy of Sciences},
	author = {Bistritzer, Rafi and MacDonald, Allan H.},
	month = jul,
	year = {2011},
	pmid = {21730173},
	pages = {12233--12237}
}

@article{BLA22,
  title = {Local {{Kekul{\'e}}} Distortion Turns Twisted Bilayer Graphene into Topological {{Mott}} Insulators and Superconductors},
  author = {Blason, Andrea and Fabrizio, Michele},
  year = {2022},
  month = dec,
  journal = {Phys. Rev. B},
  volume = {106},
  number = {23},
  pages = {235112},
  publisher = {American Physical Society},
  doi = {10.1103/PhysRevB.106.235112},
  urldate = {2023-12-22}
}

@article{BRI22,
  title = {Analytical Renormalization Group Approach to Competing Orders at Charge Neutrality in Twisted Bilayer Graphene},
  author = {Brillaux, Eric and Carpentier, David and Fedorenko, Andrei A. and Savary, Lucile},
  year = {2022},
  month = aug,
  journal = {Phys. Rev. Res.},
  volume = {4},
  number = {3},
  pages = {033168},
  publisher = {American Physical Society},
  doi = {10.1103/PhysRevResearch.4.033168},
  urldate = {2023-12-22}
}

@article{BUL20a,
  title = {Ground {{State}} and {{Hidden Symmetry}} of {{Magic-Angle Graphene}} at {{Even Integer Filling}}},
  author = {Bultinck, Nick and Khalaf, Eslam and Liu, Shang and Chatterjee, Shubhayu and Vishwanath, Ashvin and Zaletel, Michael P.},
  year = {2020},
  month = aug,
  journal = {Phys. Rev. X},
  volume = {10},
  number = {3},
  pages = {031034},
  publisher = {American Physical Society},
  doi = {10.1103/PhysRevX.10.031034},
  urldate = {2020-09-29}
}

@article{BUL20b,
  title = {Mechanism for {{Anomalous Hall Ferromagnetism}} in {{Twisted Bilayer Graphene}}},
  author = {Bultinck, Nick and Chatterjee, Shubhayu and Zaletel, Michael P.},
  year = {2020},
  month = apr,
  journal = {Phys. Rev. Lett.},
  volume = {124},
  number = {16},
  pages = {166601},
  publisher = {American Physical Society},
  doi = {10.1103/PhysRevLett.124.166601},
  urldate = {2021-01-23}
}

@article{CAL20,
  title = {Interactions in the 8-Orbital Model for Twisted Bilayer Graphene},
  author = {Calder{\'o}n, M. J. and Bascones, E.},
  year = {2020},
  month = oct,
  journal = {Phys. Rev. B},
  volume = {102},
  number = {15},
  pages = {155149},
  publisher = {American Physical Society},
  doi = {10.1103/PhysRevB.102.155149},
  urldate = {2023-12-22}
}

@article{CAL21,
  title = {Twisted Symmetric Trilayer Graphene: {{Single-particle}} and Many-Body {{Hamiltonians}} and Hidden Nonlocal Symmetries of Trilayer Moir{\'e} Systems with and without Displacement Field},
  shorttitle = {Twisted Symmetric Trilayer Graphene},
  author = {C{\u a}lug{\u a}ru, Dumitru and Xie, Fang and Song, Zhi-Da and Lian, Biao and Regnault, Nicolas and Bernevig, B. Andrei},
  year = {2021},
  month = may,
  journal = {Phys. Rev. B},
  volume = {103},
  number = {19},
  pages = {195411},
  publisher = {American Physical Society},
  doi = {10.1103/PhysRevB.103.195411},
  urldate = {2021-06-15}
}

@article{CAL22d,
  title = {Spectroscopy of {{Twisted Bilayer Graphene Correlated Insulators}}},
  author = {C{\u a}lug{\u a}ru, Dumitru and Regnault, Nicolas and Oh, Myungchul and Nuckolls, Kevin P. and Wong, Dillon and Lee, Ryan L. and Yazdani, Ali and Vafek, Oskar and Bernevig, B. Andrei},
  year = {2022},
  month = sep,
  journal = {Phys. Rev. Lett.},
  volume = {129},
  number = {11},
  pages = {117602},
  publisher = {American Physical Society},
  doi = {10.1103/PhysRevLett.129.117602},
  urldate = {2023-12-22}
}

@article{CAO18,
  title = {Correlated Insulator Behaviour at Half-Filling in Magic-Angle Graphene Superlattices},
  author = {Cao, Yuan and Fatemi, Valla and Demir, Ahmet and Fang, Shiang and Tomarken, Spencer L. and Luo, Jason Y. and {Sanchez-Yamagishi}, Javier D. and Watanabe, Kenji and Taniguchi, Takashi and Kaxiras, Efthimios and Ashoori, Ray C. and {Jarillo-Herrero}, Pablo},
  year = {2018},
  month = apr,
  journal = {Nature},
  volume = {556},
  number = {7699},
  pages = {80--84},
  publisher = {Nature Publishing Group},
  issn = {1476-4687},
  doi = {10.1038/nature26154},
  urldate = {2020-09-20},
  copyright = {2018 Macmillan Publishers Limited, part of Springer Nature. All rights reserved.},
  langid = {english}
}

@article{CAO18a,
  title = {Unconventional Superconductivity in Magic-Angle Graphene Superlattices},
  author = {Cao, Yuan and Fatemi, Valla and Fang, Shiang and Watanabe, Kenji and Taniguchi, Takashi and Kaxiras, Efthimios and {Jarillo-Herrero}, Pablo},
  year = {2018},
  month = apr,
  journal = {Nature},
  volume = {556},
  number = {7699},
  pages = {43--50},
  publisher = {Nature Publishing Group},
  issn = {1476-4687},
  doi = {10.1038/nature26160},
  urldate = {2020-09-20},
  copyright = {2018 Macmillan Publishers Limited, part of Springer Nature. All rights reserved.},
  langid = {english}
}

@article{CAO20,
  title = {Strange Metal in Magic-Angle Graphene with near Planckian Dissipation},
  author = {Cao, Yuan and Chowdhury, Debanjan and {Rodan-Legrain}, Daniel and {Rubies-Bigorda}, Oriol and Watanabe, Kenji and Taniguchi, Takashi and Senthil, T. and {Jarillo-Herrero}, Pablo},
  year = {2020},
  month = feb,
  journal = {Phys. Rev. Lett.},
  volume = {124},
  number = {7},
  pages = {076801},
  publisher = {American Physical Society},
  doi = {10.1103/PhysRevLett.124.076801},
  urldate = {2021-01-23}
}

@article{CAO21,
  title = {Nematicity and Competing Orders in Superconducting Magic-Angle Graphene},
  author = {Cao, Yuan and {Rodan-Legrain}, Daniel and Park, Jeong Min and Yuan, Noah F. Q. and Watanabe, Kenji and Taniguchi, Takashi and Fernandes, Rafael M. and Fu, Liang and {Jarillo-Herrero}, Pablo},
  year = {2021},
  month = apr,
  journal = {Science},
  volume = {372},
  number = {6539},
  pages = {264--271},
  publisher = {American Association for the Advancement of Science},
  doi = {10.1126/science.abc2836},
  urldate = {2021-10-25}
}

@article{CAO21a,
  title = {Ab Initio Four-Band {{Wannier}} Tight-Binding Model for Generic Twisted Graphene Systems},
  author = {Cao, Jin and Wang, Maoyuan and Qian, Shi-Feng and Liu, Cheng-Cheng and Yao, Yugui},
  year = {2021},
  month = aug,
  journal = {Phys. Rev. B},
  volume = {104},
  number = {8},
  pages = {L081403},
  publisher = {American Physical Society},
  doi = {10.1103/PhysRevB.104.L081403},
  urldate = {2021-11-29}
}

@article{CAR19,
  title = {Derivation of {{Wannier}} Orbitals and Minimal-Basis Tight-Binding {{Hamiltonians}} for Twisted Bilayer Graphene: {{First-principles}} Approach},
  shorttitle = {Derivation of {{Wannier}} Orbitals and Minimal-Basis Tight-Binding {{Hamiltonians}} for Twisted Bilayer Graphene},
  author = {Carr, Stephen and Fang, Shiang and Po, Hoi Chun and Vishwanath, Ashvin and Kaxiras, Efthimios},
  year = {2019},
  month = nov,
  journal = {Phys. Rev. Res.},
  volume = {1},
  number = {3},
  pages = {033072},
  publisher = {American Physical Society},
  doi = {10.1103/PhysRevResearch.1.033072},
  urldate = {2023-12-22}
}

@article{CAR19a,
  title = {Exact Continuum Model for Low-Energy Electronic States of Twisted Bilayer Graphene},
  author = {Carr, Stephen and Fang, Shiang and Zhu, Ziyan and Kaxiras, Efthimios},
  year = {2019},
  month = aug,
  journal = {Phys. Rev. Res.},
  volume = {1},
  number = {1},
  pages = {013001},
  publisher = {American Physical Society},
  doi = {10.1103/PhysRevResearch.1.013001},
  urldate = {2023-12-22}
}

@article{CAR20,
  title = {Ultraheavy and {{Ultrarelativistic Dirac Quasiparticles}} in {{Sandwiched Graphenes}}},
  author = {Carr, Stephen and Li, Chenyuan and Zhu, Ziyan and Kaxiras, Efthimios and Sachdev, Subir and Kruchkov, Alexander},
  year = {2020},
  month = may,
  journal = {Nano Lett.},
  volume = {20},
  number = {5},
  pages = {3030--3038},
  publisher = {American Chemical Society},
  issn = {1530-6984},
  doi = {10.1021/acs.nanolett.9b04979},
  urldate = {2021-01-24}
}

@article{CAR20a,
  title = {Electronic-Structure Methods for Twisted Moir{\'e} Layers},
  author = {Carr, Stephen and Fang, Shiang and Kaxiras, Efthimios},
  year = {2020},
  month = oct,
  journal = {Nat. Rev. Mater.},
  volume = {5},
  number = {10},
  pages = {748--763},
  publisher = {Nature Publishing Group},
  issn = {2058-8437},
  doi = {10.1038/s41578-020-0214-0},
  urldate = {2021-01-24},
  copyright = {2020 Springer Nature Limited},
  langid = {english}
}

@article{CEA20,
  title = {Band Structure and Insulating States Driven by {{Coulomb}} Interaction in Twisted Bilayer Graphene},
  author = {Cea, Tommaso and Guinea, Francisco},
  year = {2020},
  month = jul,
  journal = {Phys. Rev. B},
  volume = {102},
  number = {4},
  pages = {045107},
  publisher = {American Physical Society},
  doi = {10.1103/PhysRevB.102.045107},
  urldate = {2021-01-23}
}

@article{CHA20,
  title = {Symmetry Breaking and Skyrmionic Transport in Twisted Bilayer Graphene},
  author = {Chatterjee, Shubhayu and Bultinck, Nick and Zaletel, Michael P.},
  year = {2020},
  month = apr,
  journal = {Phys. Rev. B},
  volume = {101},
  number = {16},
  pages = {165141},
  publisher = {American Physical Society},
  doi = {10.1103/PhysRevB.101.165141},
  urldate = {2021-10-25}
}

@article{CHA21,
  title = {Strange {{Metals}} from {{Melting Correlated Insulators}} in {{Twisted Bilayer Graphene}}},
  author = {Cha, Peter and Patel, Aavishkar A. and Kim, Eun-Ah},
  year = {2021},
  month = dec,
  journal = {Phys. Rev. Lett.},
  volume = {127},
  number = {26},
  pages = {266601},
  publisher = {American Physical Society},
  doi = {10.1103/PhysRevLett.127.266601},
  urldate = {2023-01-31}
}

@article{CHA22,
  title = {Skyrmion Superconductivity: {{DMRG}} Evidence for a Topological Route to Superconductivity},
  shorttitle = {Skyrmion Superconductivity},
  author = {Chatterjee, Shubhayu and Ippoliti, Matteo and Zaletel, Michael P.},
  year = {2022},
  month = jul,
  journal = {Phys. Rev. B},
  volume = {106},
  number = {3},
  pages = {035421},
  publisher = {American Physical Society},
  doi = {10.1103/PhysRevB.106.035421},
  urldate = {2023-12-22}
}

@article{CHE20b,
  title = {Tunable Correlated {{Chern}} Insulator and Ferromagnetism in a Moir{\'e} Superlattice},
  author = {Chen, Guorui and Sharpe, Aaron L. and Fox, Eli J. and Zhang, Ya-Hui and Wang, Shaoxin and Jiang, Lili and Lyu, Bosai and Li, Hongyuan and Watanabe, Kenji and Taniguchi, Takashi and Shi, Zhiwen and Senthil, T. and {Goldhaber-Gordon}, David and Zhang, Yuanbo and Wang, Feng},
  year = {2020},
  month = mar,
  journal = {Nature},
  volume = {579},
  number = {7797},
  pages = {56--61},
  publisher = {Nature Publishing Group},
  issn = {1476-4687},
  doi = {10.1038/s41586-020-2049-7},
  urldate = {2021-01-23},
  copyright = {2020 The Author(s), under exclusive licence to Springer Nature Limited},
  langid = {english}
}

@article{CHE21,
  title = {Realization of Topological {{Mott}} Insulator in a Twisted Bilayer Graphene Lattice Model},
  author = {Chen, Bin-Bin and Liao, Yuan Da and Chen, Ziyu and Vafek, Oskar and Kang, Jian and Li, Wei and Meng, Zi Yang},
  year = {2021},
  month = sep,
  journal = {Nat. Commun.},
  volume = {12},
  number = {1},
  pages = {5480},
  publisher = {Nature Publishing Group},
  issn = {2041-1723},
  doi = {10.1038/s41467-021-25438-1},
  urldate = {2021-10-25},
  copyright = {2021 The Author(s)},
  langid = {english}
}

@article{CHI20a,
  title = {Nematic Superconductivity in Twisted Bilayer Graphene},
  author = {Chichinadze, Dmitry V. and Classen, Laura and Chubukov, Andrey V.},
  year = {2020},
  month = jun,
  journal = {Phys. Rev. B},
  volume = {101},
  number = {22},
  pages = {224513},
  publisher = {American Physical Society},
  doi = {10.1103/PhysRevB.101.224513},
  urldate = {2021-10-25}
}

@article{CHI20b,
  title = {Valley Magnetism, Nematicity, and Density Wave Orders in Twisted Bilayer Graphene},
  author = {Chichinadze, Dmitry V. and Classen, Laura and Chubukov, Andrey V.},
  year = {2020},
  month = sep,
  journal = {Phys. Rev. B},
  volume = {102},
  number = {12},
  pages = {125120},
  publisher = {American Physical Society},
  doi = {10.1103/PhysRevB.102.125120},
  urldate = {2021-10-28}
}

@article{CHO19,
  title = {Electronic Correlations in Twisted Bilayer Graphene near the Magic Angle},
  author = {Choi, Youngjoon and Kemmer, Jeannette and Peng, Yang and Thomson, Alex and Arora, Harpreet and Polski, Robert and Zhang, Yiran and Ren, Hechen and Alicea, Jason and Refael, Gil and {von Oppen}, Felix and Watanabe, Kenji and Taniguchi, Takashi and {Nadj-Perge}, Stevan},
  year = {2019},
  month = nov,
  journal = {Nat. Phys.},
  volume = {15},
  number = {11},
  pages = {1174--1180},
  publisher = {Nature Publishing Group},
  issn = {1745-2481},
  doi = {10.1038/s41567-019-0606-5},
  urldate = {2021-01-23},
  copyright = {2019 The Author(s), under exclusive licence to Springer Nature Limited},
  langid = {english}
}

@article{CHO20,
  title = {Tracing out {{Correlated Chern Insulators}} in {{Magic Angle Twisted Bilayer Graphene}}},
  author = {Choi, Youngjoon and Kim, Hyunjin and Peng, Yang and Thomson, Alex and Lewandowski, Cyprian and Polski, Robert and Zhang, Yiran and Arora, Harpreet Singh and Watanabe, Kenji and Taniguchi, Takashi and Alicea, Jason and {Nadj-Perge}, Stevan},
  year = {2020},
  month = aug,
  journal = {arXiv:2008.11746 [cond-mat]},
  eprint = {2008.11746},
  primaryclass = {cond-mat},
  urldate = {2021-01-23},
  archiveprefix = {arXiv}
}

@article{CHO21,
  title = {Correlation-Driven Topological Phases in Magic-Angle Twisted Bilayer Graphene},
  author = {Choi, Youngjoon and Kim, Hyunjin and Peng, Yang and Thomson, Alex and Lewandowski, Cyprian and Polski, Robert and Zhang, Yiran and Arora, Harpreet Singh and Watanabe, Kenji and Taniguchi, Takashi and Alicea, Jason and {Nadj-Perge}, Stevan},
  year = {2021},
  month = jan,
  journal = {Nature},
  volume = {589},
  number = {7843},
  pages = {536--541},
  publisher = {Nature Publishing Group},
  issn = {1476-4687},
  doi = {10.1038/s41586-020-03159-7},
  urldate = {2021-10-25},
  copyright = {2021 The Author(s), under exclusive licence to Springer Nature Limited},
  langid = {english}
}

@article{CHO21a,
  title = {Interaction-Driven Band Flattening and Correlated Phases in Twisted Bilayer Graphene},
  author = {Choi, Youngjoon and Kim, Hyunjin and Lewandowski, Cyprian and Peng, Yang and Thomson, Alex and Polski, Robert and Zhang, Yiran and Watanabe, Kenji and Taniguchi, Takashi and Alicea, Jason and {Nadj-Perge}, Stevan},
  year = {2021},
  month = dec,
  journal = {Nat. Phys.},
  volume = {17},
  number = {12},
  pages = {1375--1381},
  publisher = {Nature Publishing Group},
  issn = {1745-2481},
  doi = {10.1038/s41567-021-01359-0},
  urldate = {2022-07-21},
  copyright = {2021 The Author(s), under exclusive licence to Springer Nature Limited},
  langid = {english}
}

@article{CHO21c,
  title = {Correlation-{{Induced Triplet Pairing Superconductivity}} in {{Graphene-Based Moir{\'e} Systems}}},
  author = {Chou, Yang-Zhi and Wu, Fengcheng and Sau, Jay D. and Das Sarma, Sankar},
  year = {2021},
  month = nov,
  journal = {Phys. Rev. Lett.},
  volume = {127},
  number = {21},
  pages = {217001},
  publisher = {American Physical Society},
  doi = {10.1103/PhysRevLett.127.217001},
  urldate = {2023-12-22}
}

@article{CHO21d,
  title = {Dichotomy of {{Electron-Phonon Coupling}} in {{Graphene Moir{\'e} Flat Bands}}},
  author = {Choi, Young Woo and Choi, Hyoung Joon},
  year = {2021},
  month = oct,
  journal = {Phys. Rev. Lett.},
  volume = {127},
  number = {16},
  pages = {167001},
  publisher = {American Physical Society},
  doi = {10.1103/PhysRevLett.127.167001},
  urldate = {2023-12-22}
}

@article{CHR20,
  title = {Superconductivity, Correlated Insulators, and {{Wess}}--{{Zumino}}--{{Witten}} Terms in Twisted Bilayer Graphene},
  author = {Christos, Maine and Sachdev, Subir and Scheurer, Mathias S.},
  year = {2020},
  month = nov,
  journal = {PNAS},
  volume = {117},
  number = {47},
  pages = {29543--29554},
  publisher = {National Academy of Sciences},
  issn = {0027-8424, 1091-6490},
  doi = {10.1073/pnas.2014691117},
  urldate = {2021-01-23},
  chapter = {Physical Sciences},
  copyright = {{\copyright} 2020 . https://www.pnas.org/site/aboutpnas/licenses.xhtmlPublished under the PNAS license.},
  langid = {english},
  pmid = {33168719}
}

@article{CHR22,
  title = {Correlated {{Insulators}}, {{Semimetals}}, and {{Superconductivity}} in {{Twisted Trilayer Graphene}}},
  author = {Christos, Maine and Sachdev, Subir and Scheurer, Mathias S.},
  year = {2022},
  month = apr,
  journal = {Phys. Rev. X},
  volume = {12},
  number = {2},
  pages = {021018},
  publisher = {American Physical Society},
  doi = {10.1103/PhysRevX.12.021018},
  urldate = {2023-08-16}
}

@article{CLA19,
  title = {Competing Phases of Interacting Electrons on Triangular Lattices in Moir{\'e} Heterostructures},
  author = {Classen, Laura and Honerkamp, Carsten and Scherer, Michael M.},
  year = {2019},
  month = may,
  journal = {Phys. Rev. B},
  volume = {99},
  number = {19},
  pages = {195120},
  publisher = {American Physical Society},
  doi = {10.1103/PhysRevB.99.195120},
  urldate = {2021-01-23}
}

@article{CLA22,
  title = {Interaction-Induced Velocity Renormalization in Magic-Angle Twisted Multilayer Graphene},
  author = {Classen, Laura and Pixley, J. H. and K{\"o}nig, Elio J.},
  year = {2022},
  month = jun,
  journal = {2D Mater.},
  volume = {9},
  number = {3},
  pages = {031001},
  publisher = {IOP Publishing},
  issn = {2053-1583},
  doi = {10.1088/2053-1583/ac6e71},
  urldate = {2023-12-22},
  langid = {english}
}

@article{DA19,
  title = {Valence {{Bond Orders}} at {{Charge Neutrality}} in a {{Possible Two-Orbital Extended Hubbard Model}} for {{Twisted Bilayer Graphene}}},
  author = {Da Liao, Yuan and Meng, Zi Yang and Xu, Xiao Yan},
  year = {2019},
  month = oct,
  journal = {Phys. Rev. Lett.},
  volume = {123},
  number = {15},
  pages = {157601},
  publisher = {American Physical Society},
  doi = {10.1103/PhysRevLett.123.157601},
  urldate = {2021-01-23}
}

@article{DA21,
  title = {Correlation-{{Induced Insulating Topological Phases}} at {{Charge Neutrality}} in {{Twisted Bilayer Graphene}}},
  author = {Da Liao, Yuan and Kang, Jian and Brei{\o}, Clara N. and Xu, Xiao Yan and Wu, Han-Qing and Andersen, Brian M. and Fernandes, Rafael M. and Meng, Zi Yang},
  year = {2021},
  month = jan,
  journal = {Phys. Rev. X},
  volume = {11},
  number = {1},
  pages = {011014},
  publisher = {American Physical Society},
  doi = {10.1103/PhysRevX.11.011014},
  urldate = {2021-10-08}
}

@article{DAI16,
  title = {Twisted {{Bilayer Graphene}}: {{Moir{\'e}}} with a {{Twist}}},
  shorttitle = {Twisted {{Bilayer Graphene}}},
  author = {Dai, Shuyang and Xiang, Yang and Srolovitz, David J.},
  year = {2016},
  month = sep,
  journal = {Nano Lett.},
  volume = {16},
  number = {9},
  pages = {5923--5927},
  publisher = {American Chemical Society},
  issn = {1530-6984},
  doi = {10.1021/acs.nanolett.6b02870},
  urldate = {2020-12-07}
}

@article{DAS21,
  title = {Symmetry-Broken {{Chern}} Insulators and {{Rashba-like Landau-level}} Crossings in Magic-Angle Bilayer Graphene},
  author = {Das, Ipsita and Lu, Xiaobo and {Herzog-Arbeitman}, Jonah and Song, Zhi-Da and Watanabe, Kenji and Taniguchi, Takashi and Bernevig, B. Andrei and Efetov, Dmitri K.},
  year = {2021},
  month = jun,
  journal = {Nat. Phys.},
  volume = {17},
  number = {6},
  pages = {710--714},
  publisher = {Nature Publishing Group},
  issn = {1745-2481},
  doi = {10.1038/s41567-021-01186-3},
  urldate = {2021-11-29},
  copyright = {2021 The Author(s), under exclusive licence to Springer Nature Limited},
  langid = {english}
}

@article{DAS22,
  title = {Observation of {{Reentrant Correlated Insulators}} and {{Interaction-Driven Fermi-Surface Reconstructions}} at {{One Magnetic Flux Quantum}} per {{Moir{\'e} Unit Cell}} in {{Magic-Angle Twisted Bilayer Graphene}}},
  author = {Das, Ipsita and Shen, Cheng and Jaoui, Alexandre and {Herzog-Arbeitman}, Jonah and Chew, Aaron and Cho, Chang-Woo and Watanabe, Kenji and Taniguchi, Takashi and Piot, Benjamin A. and Bernevig, B. Andrei and Efetov, Dmitri K.},
  year = {2022},
  month = may,
  journal = {Phys. Rev. Lett.},
  volume = {128},
  number = {21},
  pages = {217701},
  publisher = {American Physical Society},
  doi = {10.1103/PhysRevLett.128.217701},
  urldate = {2023-01-31}
}

@article{DAT23,
  title = {Heavy Quasiparticles and Cascades without Symmetry Breaking in Twisted Bilayer Graphene},
  author = {Datta, Anushree and Calder{\'o}n, M. J. and Camjayi, A. and Bascones, E.},
  year = {2023},
  month = aug,
  journal = {Nat. Commun.},
  volume = {14},
  number = {1},
  pages = {5036},
  publisher = {Nature Publishing Group},
  issn = {2041-1723},
  doi = {10.1038/s41467-023-40754-4},
  urldate = {2023-12-22},
  copyright = {2023 The Author(s)},
  langid = {english}
}

@article{DAV22,
  title = {Construction of Low-Energy Symmetric {{Hamiltonians}} and {{Hubbard}} Parameters for Twisted Multilayer Systems Using Ab Initio Input},
  author = {Davydov, Arkadiy and Choo, Kenny and Fischer, Mark H. and Neupert, Titus},
  year = {2022},
  month = apr,
  journal = {Phys. Rev. B},
  volume = {105},
  number = {16},
  pages = {165153},
  publisher = {American Physical Society},
  doi = {10.1103/PhysRevB.105.165153},
  urldate = {2022-07-21}
}

@article{DE21a,
  title = {Gate-Defined {{Josephson}} Junctions in Magic-Angle Twisted Bilayer Graphene},
  author = {{de Vries}, Folkert K. and Portol{\'e}s, El{\'i}as and Zheng, Giulia and Taniguchi, Takashi and Watanabe, Kenji and Ihn, Thomas and Ensslin, Klaus and Rickhaus, Peter},
  year = {2021},
  month = jul,
  journal = {Nat. Nanotechnol.},
  volume = {16},
  number = {7},
  pages = {760--763},
  publisher = {Nature Publishing Group},
  issn = {1748-3395},
  doi = {10.1038/s41565-021-00896-2},
  urldate = {2023-12-22},
  copyright = {2021 The Author(s), under exclusive licence to Springer Nature Limited},
  langid = {english}
}

@article{DI22a,
  title = {Revealing the {{Thermal Properties}} of {{Superconducting Magic-Angle Twisted Bilayer Graphene}}},
  author = {Di Battista, Giorgio and Seifert, Paul and Watanabe, Kenji and Taniguchi, Takashi and Fong, Kin Chung and Principi, Alessandro and Efetov, Dmitri K.},
  year = {2022},
  month = aug,
  journal = {Nano Lett.},
  volume = {22},
  number = {16},
  pages = {6465--6470},
  publisher = {American Chemical Society},
  issn = {1530-6984},
  doi = {10.1021/acs.nanolett.1c04512},
  urldate = {2023-12-22}
}

@article{DIE23,
  title = {Symmetry-Broken {{Josephson}} Junctions and Superconducting Diodes in Magic-Angle Twisted Bilayer Graphene},
  author = {{D{\'i}ez-M{\'e}rida}, J. and {D{\'i}ez-Carl{\'o}n}, A. and Yang, S. Y. and Xie, Y.-M. and Gao, X.-J. and Senior, J. and Watanabe, K. and Taniguchi, T. and Lu, X. and Higginbotham, A. P. and Law, K. T. and Efetov, Dmitri K.},
  year = {2023},
  month = apr,
  journal = {Nat. Commun.},
  volume = {14},
  number = {1},
  pages = {2396},
  publisher = {Nature Publishing Group},
  issn = {2041-1723},
  doi = {10.1038/s41467-023-38005-7},
  urldate = {2024-02-14},
  copyright = {2023 The Author(s)},
  langid = {english}
}

@article{DOD18,
  title = {Phases of a Phenomenological Model of Twisted Bilayer Graphene},
  author = {Dodaro, J. F. and Kivelson, S. A. and Schattner, Y. and Sun, X. Q. and Wang, C.},
  year = {2018},
  month = aug,
  journal = {Phys. Rev. B},
  volume = {98},
  number = {7},
  pages = {075154},
  publisher = {American Physical Society},
  doi = {10.1103/PhysRevB.98.075154},
  urldate = {2021-01-23}
}

@article{EFI18,
  title = {Helical Network Model for Twisted Bilayer Graphene},
  author = {Efimkin, Dmitry K. and MacDonald, Allan H.},
  year = {2018},
  month = jul,
  journal = {Phys. Rev. B},
  volume = {98},
  number = {3},
  pages = {035404},
  publisher = {American Physical Society},
  doi = {10.1103/PhysRevB.98.035404},
  urldate = {2021-01-23}
}

@article{EUG20,
  title = {{{DMRG}} Study of Strongly Interacting $\mathbb{Z}_2$ Flatbands: A Toy Model Inspired by Twisted Bilayer Graphene},
  shorttitle = {{{DMRG}} Study of Strongly Interacting \${\textbackslash}mathbb\{\vphantom\}{{Z}}\vphantom\{\}\_2\$ Flatbands},
  author = {Eugenio, Paul and Dag, Ceren},
  year = {2020},
  month = dec,
  journal = {SciPost Phys.},
  volume = {3},
  number = {2},
  pages = {015},
  issn = {2666-9366},
  doi = {10.21468/SciPostPhysCore.3.2.015},
  urldate = {2021-01-23},
  langid = {english}
}

@article{FAN19,
  title = {Angle-{{Dependent}} {{{\emph{Ab}}}}{\emph{ Initio}} {{Low-Energy Hamiltonians}} for a {{Relaxed Twisted Bilayer Graphene Heterostructure}}},
  author = {Fang, Shiang and Carr, Stephen and Zhu, Ziyan and Massatt, Daniel and Kaxiras, Efthimios},
  year = {2019},
  month = jul,
  journal = {arXiv:1908.00058 [cond-mat]},
  eprint = {1908.00058},
  primaryclass = {cond-mat},
  doi = {10.48550/arXiv.1908.00058},
  urldate = {2023-12-22},
  archiveprefix = {arXiv}
}

@article{FER20,
  title = {Nematicity with a Twist: {{Rotational}} Symmetry Breaking in a Moir{\'e} Superlattice},
  shorttitle = {Nematicity with a Twist},
  author = {Fernandes, Rafael M. and Venderbos, J{\"o}rn W. F.},
  year = {2020},
  month = aug,
  journal = {Sci. Adv.},
  volume = {6},
  number = {32},
  pages = {eaba8834},
  publisher = {American Association for the Advancement of Science},
  issn = {2375-2548},
  doi = {10.1126/sciadv.aba8834},
  urldate = {2021-01-23},
  chapter = {Research Article},
  copyright = {Copyright {\copyright} 2020 The Authors, some rights reserved; exclusive licensee American Association for the Advancement of Science. No claim to original U.S. Government Works. Distributed under a Creative Commons Attribution License 4.0 (CC BY).. https://creativecommons.org/licenses/by/4.0/ This is an open-access article distributed under the terms of the Creative Commons Attribution license, which permits unrestricted use, distribution, and reproduction in any medium, provided the original work is properly cited.},
  langid = {english}
}

@article{FER21,
  title = {Charge-$4e$ {{Superconductivity}} from {{Multicomponent Nematic Pairing}}: {{Application}} to {{Twisted Bilayer Graphene}}},
  shorttitle = {Charge-\$4e\$ {{Superconductivity}} from {{Multicomponent Nematic Pairing}}},
  author = {Fernandes, Rafael M. and Fu, Liang},
  year = {2021},
  month = jul,
  journal = {Phys. Rev. Lett.},
  volume = {127},
  number = {4},
  pages = {047001},
  publisher = {American Physical Society},
  doi = {10.1103/PhysRevLett.127.047001},
  urldate = {2021-11-29}
}

@article{FIS22,
  title = {Unconventional Superconductivity in Magic-Angle Twisted Trilayer Graphene},
  author = {Fischer, Ammon and Goodwin, Zachary A. H. and Mostofi, Arash A. and Lischner, Johannes and Kennes, Dante M. and Klebl, Lennart},
  year = {2022},
  month = jan,
  journal = {npj Quantum Mater.},
  volume = {7},
  number = {1},
  pages = {1--10},
  publisher = {Nature Publishing Group},
  issn = {2397-4648},
  doi = {10.1038/s41535-021-00410-w},
  urldate = {2023-12-22},
  copyright = {2022 The Author(s)},
  langid = {english}
}

@article{FU20,
  title = {Magic-Angle Semimetals},
  author = {Fu, Yixing and K{\"o}nig, Elio J. and Wilson, Justin H. and Chou, Yang-Zhi and Pixley, Jedediah H.},
  year = {2020},
  month = oct,
  journal = {npj Quantum Mater.},
  volume = {5},
  number = {1},
  pages = {1--8},
  publisher = {Nature Publishing Group},
  issn = {2397-4648},
  doi = {10.1038/s41535-020-00271-9},
  urldate = {2021-01-23},
  copyright = {2020 The Author(s)},
  langid = {english}
}

@article{GHA22,
  title = {Breakdown of Semiclassical Description of Thermoelectricity in Near-Magic Angle Twisted Bilayer Graphene},
  author = {Ghawri, Bhaskar and Mahapatra, Phanibhusan S. and Garg, Manjari and Mandal, Shinjan and Bhowmik, Saisab and Jayaraman, Aditya and Soni, Radhika and Watanabe, Kenji and Taniguchi, Takashi and Krishnamurthy, H. R. and Jain, Manish and Banerjee, Sumilan and Chandni, U. and Ghosh, Arindam},
  year = {2022},
  month = mar,
  journal = {Nat. Commun.},
  volume = {13},
  number = {1},
  pages = {1522},
  publisher = {Nature Publishing Group},
  issn = {2041-1723},
  doi = {10.1038/s41467-022-29198-4},
  urldate = {2022-11-13},
  copyright = {2022 The Author(s)},
  langid = {english}
}

@article{GON19,
  title = {Kohn-{{Luttinger Superconductivity}} in {{Twisted Bilayer Graphene}}},
  author = {Gonz{\'a}lez, J. and Stauber, T.},
  year = {2019},
  month = jan,
  journal = {Phys. Rev. Lett.},
  volume = {122},
  number = {2},
  pages = {026801},
  publisher = {American Physical Society},
  doi = {10.1103/PhysRevLett.122.026801},
  urldate = {2021-01-23}
}

@article{GON23,
  title = {Ising Superconductivity Induced from Spin-Selective Valley Symmetry Breaking in Twisted Trilayer Graphene},
  author = {Gonz{\'a}lez, J. and Stauber, T.},
  year = {2023},
  month = may,
  journal = {Nat. Commun.},
  volume = {14},
  number = {1},
  pages = {2746},
  publisher = {Nature Publishing Group},
  issn = {2041-1723},
  doi = {10.1038/s41467-023-38250-w},
  urldate = {2023-12-22},
  copyright = {2023 The Author(s)},
  langid = {english}
}

@article{GRO22,
  title = {Chern Mosaic and {{Berry-curvature}} Magnetism in Magic-Angle Graphene},
  author = {Grover, Sameer and Bocarsly, Matan and Uri, Aviram and Stepanov, Petr and Di Battista, Giorgio and Roy, Indranil and Xiao, Jiewen and Meltzer, Alexander Y. and Myasoedov, Yuri and Pareek, Keshav and Watanabe, Kenji and Taniguchi, Takashi and Yan, Binghai and Stern, Ady and Berg, Erez and Efetov, Dmitri K. and Zeldov, Eli},
  year = {2022},
  month = aug,
  journal = {Nat. Phys.},
  volume = {18},
  number = {8},
  pages = {885--892},
  publisher = {Nature Publishing Group},
  issn = {1745-2481},
  doi = {10.1038/s41567-022-01635-7},
  urldate = {2023-01-13},
  copyright = {2022 The Author(s), under exclusive licence to Springer Nature Limited},
  langid = {english}
}

@article{GUE22,
  title = {Higher-Order {{Van Hove}} Singularity in Magic-Angle Twisted Trilayer Graphene},
  author = {Guerci, Daniele and Simon, Pascal and Mora, Christophe},
  year = {2022},
  month = feb,
  journal = {Phys. Rev. Res.},
  volume = {4},
  number = {1},
  pages = {L012013},
  publisher = {American Physical Society},
  doi = {10.1103/PhysRevResearch.4.L012013},
  urldate = {2023-12-22}
}

@article{GUI18,
  title = {Electrostatic Effects, Band Distortions, and Superconductivity in Twisted Graphene Bilayers},
  author = {Guinea, Francisco and Walet, Niels R.},
  year = {2018},
  month = dec,
  journal = {PNAS},
  volume = {115},
  number = {52},
  pages = {13174--13179},
  publisher = {National Academy of Sciences},
  issn = {0027-8424, 1091-6490},
  doi = {10.1073/pnas.1810947115},
  urldate = {2021-01-23},
  chapter = {Physical Sciences},
  copyright = {Copyright {\copyright} 2018 the Author(s). Published by PNAS.. https://creativecommons.org/licenses/by-nc-nd/4.0/This open access article is distributed under Creative Commons Attribution-NonCommercial-NoDerivatives License 4.0 (CC BY-NC-ND).},
  langid = {english},
  pmid = {30538203}
}

@article{GUO18,
  title = {Pairing Symmetry of Interacting Fermions on a Twisted Bilayer Graphene Superlattice},
  author = {Guo, Huaiming and Zhu, Xingchuan and Feng, Shiping and Scalettar, Richard T.},
  year = {2018},
  month = jun,
  journal = {Phys. Rev. B},
  volume = {97},
  number = {23},
  pages = {235453},
  publisher = {American Physical Society},
  doi = {10.1103/PhysRevB.97.235453},
  urldate = {2021-01-23}
}

@article{HEJ21,
  title = {Hybrid {{Wannier Chern}} Bands in Magic Angle Twisted Bilayer Graphene and the Quantized Anomalous {{Hall}} Effect},
  author = {Hejazi, Kasra and Chen, Xiao and Balents, Leon},
  year = {2021},
  month = mar,
  journal = {Phys. Rev. Research},
  volume = {3},
  number = {1},
  pages = {013242},
  publisher = {American Physical Society},
  doi = {10.1103/PhysRevResearch.3.013242},
  urldate = {2021-11-29}
}

@article{HES21,
  title = {Observation of Interband Collective Excitations in Twisted Bilayer Graphene},
  author = {Hesp, Niels C. H. and Torre, Iacopo and {Rodan-Legrain}, Daniel and Novelli, Pietro and Cao, Yuan and Carr, Stephen and Fang, Shiang and Stepanov, Petr and {Barcons-Ruiz}, David and Herzig Sheinfux, Hanan and Watanabe, Kenji and Taniguchi, Takashi and Efetov, Dmitri K. and Kaxiras, Efthimios and {Jarillo-Herrero}, Pablo and Polini, Marco and Koppens, Frank H. L.},
  year = {2021},
  month = oct,
  journal = {Nat. Phys.},
  volume = {17},
  number = {10},
  pages = {1162--1168},
  publisher = {Nature Publishing Group},
  issn = {1745-2481},
  doi = {10.1038/s41567-021-01327-8},
  urldate = {2023-01-13},
  copyright = {2021 The Author(s), under exclusive licence to Springer Nature Limited},
  langid = {english}
}

@article{HOF22,
  title = {Fermionic {{Monte Carlo Study}} of a {{Realistic Model}} of {{Twisted Bilayer Graphene}}},
  author = {Hofmann, Johannes S. and Khalaf, Eslam and Vishwanath, Ashvin and Berg, Erez and Lee, Jong Yeon},
  year = {2022},
  month = mar,
  journal = {Phys. Rev. X},
  volume = {12},
  number = {1},
  pages = {011061},
  publisher = {American Physical Society},
  doi = {10.1103/PhysRevX.12.011061},
  urldate = {2023-01-16}
}

@article{HON22,
  title = {Detecting {{Symmetry Breaking}} in {{Magic Angle Graphene Using Scanning Tunneling Microscopy}}},
  author = {Hong, Jung Pyo and Soejima, Tomohiro and Zaletel, Michael P.},
  year = {2022},
  month = sep,
  journal = {Phys. Rev. Lett.},
  volume = {129},
  number = {14},
  pages = {147001},
  publisher = {American Physical Society},
  doi = {10.1103/PhysRevLett.129.147001},
  urldate = {2024-02-08}
}

@article{HU19a,
  title = {Geometric and {{Conventional Contribution}} to the {{Superfluid Weight}} in {{Twisted Bilayer Graphene}}},
  author = {Hu, Xiang and Hyart, Timo and Pikulin, Dmitry I. and Rossi, Enrico},
  year = {2019},
  month = dec,
  journal = {Phys. Rev. Lett.},
  volume = {123},
  number = {23},
  pages = {237002},
  publisher = {American Physical Society},
  doi = {10.1103/PhysRevLett.123.237002},
  urldate = {2023-12-22}
}

@article{HUA19,
  title = {Antiferromagnetically Ordered {{Mott}} Insulator and $d+i d$ Superconductivity in Twisted Bilayer Graphene: A Quantum {{Monte Carlo}} Study},
  shorttitle = {Antiferromagnetically Ordered {{Mott}} Insulator and D+id Superconductivity in Twisted Bilayer Graphene},
  author = {Huang, Tongyun and Zhang, Lufeng and Ma, Tianxing},
  year = {2019},
  month = mar,
  journal = {Sci. Bull.},
  volume = {64},
  number = {5},
  pages = {310--314},
  issn = {2095-9273},
  doi = {10.1016/j.scib.2019.01.026},
  urldate = {2021-01-23},
  langid = {english}
}

@article{HUA20a,
  title = {Quasi-Flat-Band Physics in a Two-Leg Ladder Model and Its Relation to Magic-Angle Twisted Bilayer Graphene},
  author = {Huang, Yixuan and Hosur, Pavan and Pal, Hridis K.},
  year = {2020},
  month = oct,
  journal = {Phys. Rev. B},
  volume = {102},
  number = {15},
  pages = {155429},
  publisher = {American Physical Society},
  doi = {10.1103/PhysRevB.102.155429},
  urldate = {2021-01-23}
}

@article{HUB22,
  title = {Nonlinear Intensity Dependence of Photogalvanics and Photoconductance Induced by Terahertz Laser Radiation in Twisted Bilayer Graphene Close to Magic Angle},
  author = {Hubmann, S. and Soul, P. and Di Battista, G. and Hild, M. and Watanabe, K. and Taniguchi, T. and Efetov, D. K. and Ganichev, S. D.},
  year = {2022},
  month = feb,
  journal = {Phys. Rev. Mater.},
  volume = {6},
  number = {2},
  pages = {024003},
  publisher = {American Physical Society},
  doi = {10.1103/PhysRevMaterials.6.024003},
  urldate = {2023-12-22}
}

@article{ISO18,
  title = {Unconventional {{Superconductivity}} and {{Density Waves}} in {{Twisted Bilayer Graphene}}},
  author = {Isobe, Hiroki and Yuan, Noah F. Q. and Fu, Liang},
  year = {2018},
  month = dec,
  journal = {Phys. Rev. X},
  volume = {8},
  number = {4},
  pages = {041041},
  publisher = {American Physical Society},
  doi = {10.1103/PhysRevX.8.041041},
  urldate = {2020-09-20}
}

@article{JAI16,
  title = {Structure of Twisted and Buckled Bilayer Graphene},
  author = {Jain, Sandeep K. and Juri{\v c}i{\'c}, Vladimir and Barkema, Gerard T.},
  year = {2016},
  month = nov,
  journal = {2D Mater.},
  volume = {4},
  number = {1},
  pages = {015018},
  publisher = {IOP Publishing},
  issn = {2053-1583},
  doi = {10.1088/2053-1583/4/1/015018},
  urldate = {2020-12-07},
  langid = {english}
}

@article{JAO22,
  title = {Quantum Critical Behaviour in Magic-Angle Twisted Bilayer Graphene},
  author = {Jaoui, Alexandre and Das, Ipsita and Di Battista, Giorgio and {D{\'i}ez-M{\'e}rida}, Jaime and Lu, Xiaobo and Watanabe, Kenji and Taniguchi, Takashi and Ishizuka, Hiroaki and Levitov, Leonid and Efetov, Dmitri K.},
  year = {2022},
  month = jun,
  journal = {Nat. Phys.},
  volume = {18},
  number = {6},
  pages = {633--638},
  publisher = {Nature Publishing Group},
  issn = {1745-2481},
  doi = {10.1038/s41567-022-01556-5},
  urldate = {2023-01-13},
  copyright = {2022 The Author(s), under exclusive licence to Springer Nature Limited},
  langid = {english}
}

@article{JIA19,
  title = {Charge Order and Broken Rotational Symmetry in Magic-Angle Twisted Bilayer Graphene},
  author = {Jiang, Yuhang and Lai, Xinyuan and Watanabe, Kenji and Taniguchi, Takashi and Haule, Kristjan and Mao, Jinhai and Andrei, Eva Y.},
  year = {2019},
  month = sep,
  journal = {Nature},
  volume = {573},
  number = {7772},
  pages = {91--95},
  publisher = {Nature Publishing Group},
  issn = {1476-4687},
  doi = {10.1038/s41586-019-1460-4},
  urldate = {2021-01-23},
  copyright = {2019 The Author(s), under exclusive licence to Springer Nature Limited},
  langid = {english}
}

@article{JUL20,
  title = {Superfluid Weight and {{Berezinskii-Kosterlitz-Thouless}} Transition Temperature of Twisted Bilayer Graphene},
  author = {Julku, A. and Peltonen, T. J. and Liang, L. and Heikkil{\"a}, T. T. and T{\"o}rm{\"a}, P.},
  year = {2020},
  month = feb,
  journal = {Phys. Rev. B},
  volume = {101},
  number = {6},
  pages = {060505},
  publisher = {American Physical Society},
  doi = {10.1103/PhysRevB.101.060505},
  urldate = {2021-01-23}
}

@article{KAN18,
  title = {Symmetry, {{Maximally Localized Wannier States}}, and a {{Low-Energy Model}} for {{Twisted Bilayer Graphene Narrow Bands}}},
  author = {Kang, Jian and Vafek, Oskar},
  year = {2018},
  month = sep,
  journal = {Phys. Rev. X},
  volume = {8},
  number = {3},
  pages = {031088},
  publisher = {American Physical Society},
  doi = {10.1103/PhysRevX.8.031088},
  urldate = {2021-01-23}
}

@article{KAN19,
  title = {Strong {{Coupling Phases}} of {{Partially Filled Twisted Bilayer Graphene Narrow Bands}}},
  author = {Kang, Jian and Vafek, Oskar},
  year = {2019},
  month = jun,
  journal = {Phys. Rev. Lett.},
  volume = {122},
  number = {24},
  pages = {246401},
  publisher = {American Physical Society},
  doi = {10.1103/PhysRevLett.122.246401},
  urldate = {2020-10-09}
}

@article{KAN20a,
  title = {Non-{{Abelian Dirac}} Node Braiding and near-Degeneracy of Correlated Phases at Odd Integer Filling in Magic-Angle Twisted Bilayer Graphene},
  author = {Kang, Jian and Vafek, Oskar},
  year = {2020},
  month = jul,
  journal = {Phys. Rev. B},
  volume = {102},
  number = {3},
  pages = {035161},
  publisher = {American Physical Society},
  doi = {10.1103/PhysRevB.102.035161},
  urldate = {2021-01-23}
}

@article{KAN21,
  title = {Cascades between {{Light}} and {{Heavy Fermions}} in the {{Normal State}} of {{Magic-Angle Twisted Bilayer Graphene}}},
  author = {Kang, Jian and Bernevig, B. Andrei and Vafek, Oskar},
  year = {2021},
  month = dec,
  journal = {Phys. Rev. Lett.},
  volume = {127},
  number = {26},
  pages = {266402},
  publisher = {American Physical Society},
  doi = {10.1103/PhysRevLett.127.266402},
  urldate = {2022-07-21}
}

@article{KAN23b,
  title = {Pseudomagnetic Fields, Particle-Hole Asymmetry, and Microscopic Effective Continuum {{Hamiltonians}} of Twisted Bilayer Graphene},
  author = {Kang, Jian and Vafek, Oskar},
  year = {2023},
  month = feb,
  journal = {Phys. Rev. B},
  volume = {107},
  number = {7},
  pages = {075408},
  publisher = {American Physical Society},
  doi = {10.1103/PhysRevB.107.075408},
  urldate = {2023-12-22}
}

@article{KEN18,
  title = {Strong Correlations and  $d+i d$  Superconductivity in Twisted Bilayer Graphene},
  author = {Kennes, Dante M. and Lischner, Johannes and Karrasch, Christoph},
  year = {2018},
  month = dec,
  journal = {Phys. Rev. B},
  volume = {98},
  number = {24},
  pages = {241407},
  publisher = {American Physical Society},
  doi = {10.1103/PhysRevB.98.241407},
  urldate = {2021-01-23}
}

@article{KER19,
  title = {Maximized Electron Interactions at the Magic Angle in Twisted Bilayer Graphene},
  author = {Kerelsky, Alexander and McGilly, Leo J. and Kennes, Dante M. and Xian, Lede and Yankowitz, Matthew and Chen, Shaowen and Watanabe, K. and Taniguchi, T. and Hone, James and Dean, Cory and Rubio, Angel and Pasupathy, Abhay N.},
  year = {2019},
  month = aug,
  journal = {Nature},
  volume = {572},
  number = {7767},
  pages = {95--100},
  publisher = {Nature Publishing Group},
  issn = {1476-4687},
  doi = {10.1038/s41586-019-1431-9},
  urldate = {2021-01-23},
  copyright = {2019 The Author(s), under exclusive licence to Springer Nature Limited},
  langid = {english}
}

@article{KHA19,
  title = {Magic Angle Hierarchy in Twisted Graphene Multilayers},
  author = {Khalaf, Eslam and Kruchkov, Alex J. and Tarnopolsky, Grigory and Vishwanath, Ashvin},
  year = {2019},
  month = aug,
  journal = {Phys. Rev. B},
  volume = {100},
  number = {8},
  pages = {085109},
  publisher = {American Physical Society},
  doi = {10.1103/PhysRevB.100.085109},
  urldate = {2020-12-07}
}

@article{KHA21,
  title = {Charged Skyrmions and Topological Origin of Superconductivity in Magic-Angle Graphene},
  author = {Khalaf, Eslam and Chatterjee, Shubhayu and Bultinck, Nick and Zaletel, Michael P. and Vishwanath, Ashvin},
  year = {2021},
  month = may,
  journal = {Sci. Adv.},
  volume = {7},
  number = {19},
  pages = {eabf5299},
  publisher = {American Association for the Advancement of Science},
  doi = {10.1126/sciadv.abf5299},
  urldate = {2021-10-08}
}

@article{KON20,
  title = {Spin Magnetometry as a Probe of Stripe Superconductivity in Twisted Bilayer Graphene},
  author = {K{\"o}nig, E. J. and Coleman, Piers and Tsvelik, A. M.},
  year = {2020},
  month = sep,
  journal = {Phys. Rev. B},
  volume = {102},
  number = {10},
  pages = {104514},
  publisher = {American Physical Society},
  doi = {10.1103/PhysRevB.102.104514},
  urldate = {2021-01-23}
}

@article{KOS18,
  title = {Maximally {{Localized Wannier Orbitals}} and the {{Extended Hubbard Model}} for {{Twisted Bilayer Graphene}}},
  author = {Koshino, Mikito and Yuan, Noah F. Q. and Koretsune, Takashi and Ochi, Masayuki and Kuroki, Kazuhiko and Fu, Liang},
  year = {2018},
  month = sep,
  journal = {Phys. Rev. X},
  volume = {8},
  number = {3},
  pages = {031087},
  publisher = {American Physical Society},
  doi = {10.1103/PhysRevX.8.031087},
  urldate = {2021-01-23}
}

@article{KWA20,
  title = {Twisted Bilayer Graphene in a Parallel Magnetic Field},
  author = {Kwan, Yves H. and Parameswaran, S. A. and Sondhi, S. L.},
  year = {2020},
  month = may,
  journal = {Phys. Rev. B},
  volume = {101},
  number = {20},
  pages = {205116},
  publisher = {American Physical Society},
  doi = {10.1103/PhysRevB.101.205116},
  urldate = {2023-12-22}
}

@article{KWA21,
  title = {Kekul{\'e} {{Spiral Order}} at {{All Nonzero Integer Fillings}} in {{Twisted Bilayer Graphene}}},
  author = {Kwan, Y. H. and Wagner, G. and Soejima, T. and Zaletel, M. P. and Simon, S. H. and Parameswaran, S. A. and Bultinck, N.},
  year = {2021},
  month = dec,
  journal = {Phys. Rev. X},
  volume = {11},
  number = {4},
  pages = {041063},
  publisher = {American Physical Society},
  doi = {10.1103/PhysRevX.11.041063},
  urldate = {2022-07-21}
}

@article{KWA21a,
  title = {Exciton {{Band Topology}} in {{Spontaneous Quantum Anomalous Hall Insulators}}: {{Applications}} to {{Twisted Bilayer Graphene}}},
  shorttitle = {Exciton {{Band Topology}} in {{Spontaneous Quantum Anomalous Hall Insulators}}},
  author = {Kwan, Yves H. and Hu, Yichen and Simon, Steven H. and Parameswaran, S. A.},
  year = {2021},
  month = mar,
  journal = {Phys. Rev. Lett.},
  volume = {126},
  number = {13},
  pages = {137601},
  publisher = {American Physical Society},
  doi = {10.1103/PhysRevLett.126.137601},
  urldate = {2023-12-22}
}

@article{KWA21b,
  title = {Domain Wall Competition in the {{Chern}} Insulating Regime of Twisted Bilayer Graphene},
  author = {Kwan, Yves H. and Wagner, Glenn and Chakraborty, Nilotpal and Simon, Steven H. and Parameswaran, S. A.},
  year = {2021},
  month = sep,
  journal = {Phys. Rev. B},
  volume = {104},
  number = {11},
  pages = {115404},
  publisher = {American Physical Society},
  doi = {10.1103/PhysRevB.104.115404},
  urldate = {2023-12-22}
}

@article{KWA22,
  title = {Skyrmions in {{Twisted Bilayer Graphene}}: {{Stability}}, {{Pairing}}, and {{Crystallization}}},
  shorttitle = {Skyrmions in {{Twisted Bilayer Graphene}}},
  author = {Kwan, Yves H. and Wagner, Glenn and Bultinck, Nick and Simon, Steven H. and Parameswaran, S. A.},
  year = {2022},
  month = jul,
  journal = {Phys. Rev. X},
  volume = {12},
  number = {3},
  pages = {031020},
  publisher = {American Physical Society},
  doi = {10.1103/PhysRevX.12.031020},
  urldate = {2023-12-22}
}

@article{KWA23,
  title = {Electron-Phonon Coupling and Competing {{Kekul{\'e}}} Orders in Twisted Bilayer Graphene},
  author = {Kwan, Yves H. and Wagner, Glenn and Bultinck, Nick and Simon, Steven H. and Berg, Erez and Parameswaran, S. A.},
  year = {2023},
  month = mar,
  journal = {arXiv:2303.13602 [cond-mat]},
  eprint = {2303.13602},
  primaryclass = {cond-mat},
  doi = {10.48550/arXiv.2303.13602},
  urldate = {2023-12-22},
  archiveprefix = {arXiv}
}

@article{LAK21,
  title = {Reentrant Superconductivity through a Quantum {{Lifshitz}} Transition in Twisted Trilayer Graphene},
  author = {Lake, Ethan and Senthil, T.},
  year = {2021},
  month = nov,
  journal = {Phys. Rev. B},
  volume = {104},
  number = {17},
  pages = {174505},
  publisher = {American Physical Society},
  doi = {10.1103/PhysRevB.104.174505},
  urldate = {2023-12-22}
}

@article{LED21,
  title = {{{TB}} or Not {{TB}}? {{Contrasting}} Properties of Twisted Bilayer Graphene and the Alternating Twist $n$-Layer Structures ($n=3, 4, 5, \dots$)},
  shorttitle = {{{TB}} or Not {{TB}}?},
  author = {Ledwith, Patrick J. and Khalaf, Eslam and Zhu, Ziyan and Carr, Stephen and Kaxiras, Efthimios and Vishwanath, Ashvin},
  year = {2021},
  month = nov,
  journal = {arXiv:2111.11060 [cond-mat]},
  eprint = {2111.11060},
  primaryclass = {cond-mat},
  doi = {10.48550/arXiv.2111.11060},
  urldate = {2023-12-08},
  archiveprefix = {arXiv}
}

@article{LEI21,
  title = {Mirror Symmetry Breaking and Lateral Stacking Shifts in Twisted Trilayer Graphene},
  author = {Lei, Chao and Linhart, Lukas and Qin, Wei and Libisch, Florian and MacDonald, Allan H.},
  year = {2021},
  month = jul,
  journal = {Phys. Rev. B},
  volume = {104},
  number = {3},
  pages = {035139},
  publisher = {American Physical Society},
  doi = {10.1103/PhysRevB.104.035139},
  urldate = {2021-11-29}
}

@article{LEW21,
  title = {Pairing in Magic-Angle Twisted Bilayer Graphene: {{Role}} of Phonon and Plasmon Umklapp},
  shorttitle = {Pairing in Magic-Angle Twisted Bilayer Graphene},
  author = {Lewandowski, Cyprian and Chowdhury, Debanjan and Ruhman, Jonathan},
  year = {2021},
  month = jun,
  journal = {Phys. Rev. B},
  volume = {103},
  number = {23},
  pages = {235401},
  publisher = {American Physical Society},
  doi = {10.1103/PhysRevB.103.235401},
  urldate = {2021-10-08}
}

@article{LI19,
  title = {Electronic {{Structure}} of {{Single-Twist Trilayer Graphene}}},
  author = {Li, Xiao and Wu, Fengcheng and MacDonald, Allan H.},
  year = {2019},
  month = jul,
  journal = {arXiv:1907.12338 [cond-mat]},
  eprint = {1907.12338},
  primaryclass = {cond-mat},
  urldate = {2021-01-24},
  archiveprefix = {arXiv}
}

@article{LI22c,
  title = {Induced Superconductivity in Magic-Angle Twisted Trilayer Graphene through Graphene-Metal Contacts},
  author = {Li, Shujin and Zheng, Guanyuan and Huang, Junlin},
  year = {2022},
  month = jan,
  journal = {arXiv:2111.04451 [cond-mat]},
  eprint = {2111.04451},
  primaryclass = {cond-mat},
  doi = {10.48550/arXiv.2111.04451},
  urldate = {2023-12-22},
  archiveprefix = {arXiv}
}

@article{LIA19,
  title = {Twisted {{Bilayer Graphene}}: {{A Phonon-Driven Superconductor}}},
  shorttitle = {Twisted {{Bilayer Graphene}}},
  author = {Lian, Biao and Wang, Zhijun and Bernevig, B. Andrei},
  year = {2019},
  month = jun,
  journal = {Phys. Rev. Lett.},
  volume = {122},
  number = {25},
  pages = {257002},
  publisher = {American Physical Society},
  doi = {10.1103/PhysRevLett.122.257002},
  urldate = {2020-09-20}
}

@article{LIA21,
  title = {Twisted Bilayer Graphene. {{IV}}. {{Exact}} Insulator Ground States and Phase Diagram},
  author = {Lian, Biao and Song, Zhi-Da and Regnault, Nicolas and Efetov, Dmitri K. and Yazdani, Ali and Bernevig, B. Andrei},
  year = {2021},
  month = may,
  journal = {Phys. Rev. B},
  volume = {103},
  number = {20},
  pages = {205414},
  publisher = {American Physical Society},
  doi = {10.1103/PhysRevB.103.205414},
  urldate = {2021-09-13}
}

@article{LIA21c,
  title = {Heating Freezes Electrons in Twisted Bilayer Graphene},
  author = {Lian, Biao},
  year = {2021},
  month = apr,
  journal = {Nature},
  volume = {592},
  number = {7853},
  pages = {191--193},
  publisher = {Nature Publishing Group},
  doi = {10.1038/d41586-021-00843-0},
  urldate = {2023-12-22},
  copyright = {2021 Nature},
  langid = {english}
}

@article{LIN22,
  title = {Energetic Stability and Spatial Inhomogeneity in the Local Electronic Structure of Relaxed Twisted Trilayer Graphene},
  author = {Lin, Xianqing and Li, Cheng and Su, Kelu and Ni, Jun},
  year = {2022},
  month = aug,
  journal = {Phys. Rev. B},
  volume = {106},
  number = {7},
  pages = {075423},
  publisher = {American Physical Society},
  doi = {10.1103/PhysRevB.106.075423},
  urldate = {2023-12-22}
}

@article{LIS21,
  title = {Observation of Flat Bands in Twisted Bilayer Graphene},
  author = {Lisi, Simone and Lu, Xiaobo and Benschop, Tjerk and {de Jong}, Tobias A. and Stepanov, Petr and Duran, Jose R. and Margot, Florian and Cucchi, Ir{\`e}ne and Cappelli, Edoardo and Hunter, Andrew and Tamai, Anna and Kandyba, Viktor and Giampietri, Alessio and Barinov, Alexei and Jobst, Johannes and Stalman, Vincent and Leeuwenhoek, Maarten and Watanabe, Kenji and Taniguchi, Takashi and Rademaker, Louk and {van der Molen}, Sense Jan and Allan, Milan P. and Efetov, Dmitri K. and Baumberger, Felix},
  year = {2021},
  month = feb,
  journal = {Nat. Phys.},
  volume = {17},
  number = {2},
  pages = {189--193},
  publisher = {Nature Publishing Group},
  issn = {1745-2481},
  doi = {10.1038/s41567-020-01041-x},
  urldate = {2023-01-13},
  copyright = {2020 The Author(s), under exclusive licence to Springer Nature Limited},
  langid = {english}
}

@article{LIU18a,
  title = {Chiral {{Spin Density Wave}} and $d+id$  {{Superconductivity}} in the {{Magic-Angle-Twisted Bilayer Graphene}}},
  author = {Liu, Cheng-Cheng and Zhang, Li-Da and Chen, Wei-Qiang and Yang, Fan},
  year = {2018},
  month = nov,
  journal = {Phys. Rev. Lett.},
  volume = {121},
  number = {21},
  pages = {217001},
  publisher = {American Physical Society},
  doi = {10.1103/PhysRevLett.121.217001},
  urldate = {2020-09-20}
}

@article{LIU19,
  title = {Pseudo {{Landau}} Level Representation of Twisted Bilayer Graphene: {{Band}} Topology and Implications on the Correlated Insulating Phase},
  shorttitle = {Pseudo {{Landau}} Level Representation of Twisted Bilayer Graphene},
  author = {Liu, Jianpeng and Liu, Junwei and Dai, Xi},
  year = {2019},
  month = apr,
  journal = {Phys. Rev. B},
  volume = {99},
  number = {15},
  pages = {155415},
  publisher = {American Physical Society},
  doi = {10.1103/PhysRevB.99.155415},
  urldate = {2021-01-23}
}

@article{LIU19a,
  title = {Quantum {{Valley Hall Effect}}, {{Orbital Magnetism}}, and {{Anomalous Hall Effect}} in {{Twisted Multilayer Graphene Systems}}},
  author = {Liu, Jianpeng and Ma, Zhen and Gao, Jinhua and Dai, Xi},
  year = {2019},
  month = aug,
  journal = {Phys. Rev. X},
  volume = {9},
  number = {3},
  pages = {031021},
  publisher = {American Physical Society},
  doi = {10.1103/PhysRevX.9.031021},
  urldate = {2021-01-23}
}

@article{LIU21,
  title = {Nematic Topological Semimetal and Insulator in Magic-Angle Bilayer Graphene at Charge Neutrality},
  author = {Liu, Shang and Khalaf, Eslam and Lee, Jong Yeon and Vishwanath, Ashvin},
  year = {2021},
  month = jan,
  journal = {Phys. Rev. Research},
  volume = {3},
  number = {1},
  pages = {013033},
  publisher = {American Physical Society},
  doi = {10.1103/PhysRevResearch.3.013033},
  urldate = {2021-01-23}
}

@article{LIU21a,
  title = {Theories for the Correlated Insulating States and Quantum Anomalous {{Hall}} Effect Phenomena in Twisted Bilayer Graphene},
  author = {Liu, Jianpeng and Dai, Xi},
  year = {2021},
  month = jan,
  journal = {Phys. Rev. B},
  volume = {103},
  number = {3},
  pages = {035427},
  publisher = {American Physical Society},
  doi = {10.1103/PhysRevB.103.035427},
  urldate = {2021-10-08}
}

@article{LIU21c,
  title = {Tuning Electron Correlation in Magic-Angle Twisted Bilayer Graphene Using {{Coulomb}} Screening},
  author = {Liu, Xiaoxue and Wang, Zhi and Watanabe, K. and Taniguchi, T. and Vafek, Oskar and Li, J. I. A.},
  year = {2021},
  month = mar,
  journal = {Science},
  volume = {371},
  number = {6535},
  pages = {1261--1265},
  publisher = {American Association for the Advancement of Science},
  doi = {10.1126/science.abb8754},
  urldate = {2021-11-29}
}

@article{LOP07,
  title = {Graphene {{Bilayer}} with a {{Twist}}: {{Electronic Structure}}},
  shorttitle = {Graphene {{Bilayer}} with a {{Twist}}},
  author = {{Lopes~dos~Santos}, J. M. B. and Peres, N. M. R. and Castro~Neto, A. H.},
  year = {2007},
  month = dec,
  journal = {Phys. Rev. Lett.},
  volume = {99},
  number = {25},
  pages = {256802},
  publisher = {American Physical Society},
  doi = {10.1103/PhysRevLett.99.256802},
  urldate = {2020-09-20}
}

@article{LOP20,
  title = {Electrical Band Flattening, Valley Flux, and Superconductivity in Twisted Trilayer Graphene},
  author = {{Lopez-Bezanilla}, Alejandro and Lado, J. L.},
  year = {2020},
  month = sep,
  journal = {Phys. Rev. Research},
  volume = {2},
  number = {3},
  pages = {033357},
  publisher = {American Physical Society},
  doi = {10.1103/PhysRevResearch.2.033357},
  urldate = {2021-01-24}
}

@article{LU19,
  title = {Superconductors, Orbital Magnets and Correlated States in Magic-Angle Bilayer Graphene},
  author = {Lu, Xiaobo and Stepanov, Petr and Yang, Wei and Xie, Ming and Aamir, Mohammed Ali and Das, Ipsita and Urgell, Carles and Watanabe, Kenji and Taniguchi, Takashi and Zhang, Guangyu and Bachtold, Adrian and MacDonald, Allan H. and Efetov, Dmitri K.},
  year = {2019},
  month = oct,
  journal = {Nature},
  volume = {574},
  number = {7780},
  pages = {653--657},
  publisher = {Nature Publishing Group},
  issn = {1476-4687},
  doi = {10.1038/s41586-019-1695-0},
  urldate = {2020-09-20},
  copyright = {2019 The Author(s), under exclusive licence to Springer Nature Limited},
  langid = {english}
}

@article{LU21,
  title = {Multiple Flat Bands and Topological {{Hofstadter}} Butterfly in Twisted Bilayer Graphene Close to the Second Magic Angle},
  author = {Lu, Xiaobo and Lian, Biao and Chaudhary, Gaurav and Piot, Benjamin A. and Romagnoli, Giulio and Watanabe, Kenji and Taniguchi, Takashi and Poggio, Martino and MacDonald, Allan H. and Bernevig, B. Andrei and Efetov, Dmitri K.},
  year = {2021},
  month = jul,
  journal = {PNAS},
  volume = {118},
  number = {30},
  pages = {e2100006118},
  publisher = {National Academy of Sciences},
  issn = {0027-8424, 1091-6490},
  doi = {10.1073/pnas.2100006118},
  urldate = {2021-11-29},
  chapter = {Physical Sciences},
  copyright = {{\copyright} 2021 . https://www.pnas.org/site/aboutpnas/licenses.xhtmlPublished under the PNAS license.},
  isbn = {9782100006113},
  langid = {english},
  pmid = {34301893}
}

@article{MOR19,
  title = {Flatbands and {{Perfect Metal}} in {{Trilayer Moir{\'e} Graphene}}},
  author = {Mora, Christophe and Regnault, Nicolas and Bernevig, B. Andrei},
  year = {2019},
  month = jul,
  journal = {Phys. Rev. Lett.},
  volume = {123},
  number = {2},
  pages = {026402},
  publisher = {American Physical Society},
  doi = {10.1103/PhysRevLett.123.026402},
  urldate = {2020-11-10}
}

@article{NAM17,
  title = {Lattice Relaxation and Energy Band Modulation in Twisted Bilayer Graphene},
  author = {Nam, Nguyen N. T. and Koshino, Mikito},
  year = {2017},
  month = aug,
  journal = {Phys. Rev. B},
  volume = {96},
  number = {7},
  pages = {075311},
  publisher = {American Physical Society},
  doi = {10.1103/PhysRevB.96.075311},
  urldate = {2023-12-22}
}

@article{NUC20,
  title = {Strongly Correlated {{Chern}} Insulators in Magic-Angle Twisted Bilayer Graphene},
  author = {Nuckolls, Kevin P. and Oh, Myungchul and Wong, Dillon and Lian, Biao and Watanabe, Kenji and Taniguchi, Takashi and Bernevig, B. Andrei and Yazdani, Ali},
  year = {2020},
  month = dec,
  journal = {Nature},
  volume = {588},
  number = {7839},
  pages = {610--615},
  publisher = {Nature Publishing Group},
  issn = {1476-4687},
  doi = {10.1038/s41586-020-3028-8},
  urldate = {2021-01-18},
  copyright = {2020 The Author(s), under exclusive licence to Springer Nature Limited},
  langid = {english}
}

@article{NUC23,
  title = {Quantum Textures of the Many-Body Wavefunctions in Magic-Angle Graphene},
  author = {Nuckolls, Kevin P. and Lee, Ryan L. and Oh, Myungchul and Wong, Dillon and Soejima, Tomohiro and Hong, Jung Pyo and C{\u a}lug{\u a}ru, Dumitru and {Herzog-Arbeitman}, Jonah and Bernevig, B. Andrei and Watanabe, Kenji and Taniguchi, Takashi and Regnault, Nicolas and Zaletel, Michael P. and Yazdani, Ali},
  year = {2023},
  month = aug,
  journal = {Nature},
  volume = {620},
  number = {7974},
  pages = {525--532},
  publisher = {Nature Publishing Group},
  issn = {1476-4687},
  doi = {10.1038/s41586-023-06226-x},
  urldate = {2023-08-18},
  copyright = {2023 The Author(s), under exclusive licence to Springer Nature Limited},
  langid = {english}
}

@article{OCH18,
  title = {Possible Correlated Insulating States in Magic-Angle Twisted Bilayer Graphene under Strongly Competing Interactions},
  author = {Ochi, Masayuki and Koshino, Mikito and Kuroki, Kazuhiko},
  year = {2018},
  month = aug,
  journal = {Phys. Rev. B},
  volume = {98},
  number = {8},
  pages = {081102},
  publisher = {American Physical Society},
  doi = {10.1103/PhysRevB.98.081102},
  urldate = {2021-01-23}
}

@article{OH21,
  title = {Evidence for Unconventional Superconductivity in Twisted Bilayer Graphene},
  author = {Oh, Myungchul and Nuckolls, Kevin P. and Wong, Dillon and Lee, Ryan L. and Liu, Xiaomeng and Watanabe, Kenji and Taniguchi, Takashi and Yazdani, Ali},
  year = {2021},
  month = dec,
  journal = {Nature},
  volume = {600},
  number = {7888},
  pages = {240--245},
  publisher = {Nature Publishing Group},
  issn = {1476-4687},
  doi = {10.1038/s41586-021-04121-x},
  urldate = {2023-12-22},
  copyright = {2021 The Author(s), under exclusive licence to Springer Nature Limited},
  langid = {english}
}

@article{PAD18,
  title = {Doped {{Twisted Bilayer Graphene}} near {{Magic Angles}}: {{Proximity}} to {{Wigner Crystallization}}, {{Not Mott Insulation}}},
  shorttitle = {Doped {{Twisted Bilayer Graphene}} near {{Magic Angles}}},
  author = {Padhi, Bikash and Setty, Chandan and Phillips, Philip W.},
  year = {2018},
  month = oct,
  journal = {Nano Lett.},
  volume = {18},
  number = {10},
  pages = {6175--6180},
  publisher = {American Chemical Society},
  issn = {1530-6984},
  doi = {10.1021/acs.nanolett.8b02033},
  urldate = {2023-01-16}
}

@article{PAR20,
  title = {Gate-Tunable Topological Flat Bands in Twisted Monolayer-Bilayer Graphene},
  author = {Park, Youngju and Chittari, Bheema Lingam and Jung, Jeil},
  year = {2020},
  month = jul,
  journal = {Phys. Rev. B},
  volume = {102},
  number = {3},
  pages = {035411},
  publisher = {American Physical Society},
  doi = {10.1103/PhysRevB.102.035411},
  urldate = {2021-01-24}
}

@article{PAR21a,
  title = {Strain-{{Induced Quantum Phase Transitions}} in {{Magic-Angle Graphene}}},
  author = {Parker, Daniel E. and Soejima, Tomohiro and Hauschild, Johannes and Zaletel, Michael P. and Bultinck, Nick},
  year = {2021},
  month = jul,
  journal = {Phys. Rev. Lett.},
  volume = {127},
  number = {2},
  pages = {027601},
  publisher = {American Physical Society},
  doi = {10.1103/PhysRevLett.127.027601},
  urldate = {2021-10-25}
}

@article{PAR21c,
  title = {Flavour {{Hund}}'s Coupling, {{Chern}} Gaps and Charge Diffusivity in Moir{\'e} Graphene},
  author = {Park, Jeong Min and Cao, Yuan and Watanabe, Kenji and Taniguchi, Takashi and {Jarillo-Herrero}, Pablo},
  year = {2021},
  month = apr,
  journal = {Nature},
  volume = {592},
  number = {7852},
  pages = {43--48},
  publisher = {Nature Publishing Group},
  issn = {1476-4687},
  doi = {10.1038/s41586-021-03366-w},
  urldate = {2021-11-29},
  copyright = {2021 The Author(s), under exclusive licence to Springer Nature Limited},
  langid = {english}
}

@article{PAU22,
  title = {Interaction-Driven Giant Thermopower in Magic-Angle Twisted Bilayer Graphene},
  author = {Paul, Arup Kumar and Ghosh, Ayan and Chakraborty, Souvik and Roy, Ujjal and Dutta, Ranit and Watanabe, K. and Taniguchi, T. and Panda, Animesh and Agarwala, Adhip and Mukerjee, Subroto and Banerjee, Sumilan and Das, Anindya},
  year = {2022},
  month = jun,
  journal = {Nat. Phys.},
  volume = {18},
  number = {6},
  pages = {691--698},
  publisher = {Nature Publishing Group},
  issn = {1745-2481},
  doi = {10.1038/s41567-022-01574-3},
  urldate = {2022-11-13},
  copyright = {2022 The Author(s), under exclusive licence to Springer Nature Limited},
  langid = {english}
}

@article{PEL18,
  title = {Mean-Field Theory for Superconductivity in Twisted Bilayer Graphene},
  author = {Peltonen, Teemu J. and Ojaj{\"a}rvi, Risto and Heikkil{\"a}, Tero T.},
  year = {2018},
  month = dec,
  journal = {Phys. Rev. B},
  volume = {98},
  number = {22},
  pages = {220504},
  publisher = {American Physical Society},
  doi = {10.1103/PhysRevB.98.220504},
  urldate = {2020-09-20}
}

@article{PHO21,
  title = {Band Structure and Superconductivity in Twisted Trilayer Graphene},
  author = {Phong, V{\~o} Ti\'{\^e}n and Pantale{\'o}n, Pierre A. and Cea, Tommaso and Guinea, Francisco},
  year = {2021},
  month = sep,
  journal = {Phys. Rev. B},
  volume = {104},
  number = {12},
  pages = {L121116},
  publisher = {American Physical Society},
  doi = {10.1103/PhysRevB.104.L121116},
  urldate = {2023-12-22}
}

@article{PIE21,
  title = {Unconventional Sequence of Correlated {{Chern}} Insulators in Magic-Angle Twisted Bilayer Graphene},
  author = {Pierce, Andrew T. and Xie, Yonglong and Park, Jeong Min and Khalaf, Eslam and Lee, Seung Hwan and Cao, Yuan and Parker, Daniel E. and Forrester, Patrick R. and Chen, Shaowen and Watanabe, Kenji and Taniguchi, Takashi and Vishwanath, Ashvin and {Jarillo-Herrero}, Pablo and Yacoby, Amir},
  year = {2021},
  month = nov,
  journal = {Nat. Phys.},
  volume = {17},
  number = {11},
  pages = {1210--1215},
  publisher = {Nature Publishing Group},
  issn = {1745-2481},
  doi = {10.1038/s41567-021-01347-4},
  urldate = {2021-11-29},
  copyright = {2021 The Author(s), under exclusive licence to Springer Nature Limited},
  langid = {english}
}

@article{PO18a,
  title = {Origin of {{Mott Insulating Behavior}} and {{Superconductivity}} in {{Twisted Bilayer Graphene}}},
  author = {Po, Hoi Chun and Zou, Liujun and Vishwanath, Ashvin and Senthil, T.},
  year = {2018},
  month = sep,
  journal = {Phys. Rev. X},
  volume = {8},
  number = {3},
  pages = {031089},
  publisher = {American Physical Society},
  doi = {10.1103/PhysRevX.8.031089},
  urldate = {2020-09-20}
}

@article{PO19,
  title = {Faithful Tight-Binding Models and Fragile Topology of Magic-Angle Bilayer Graphene},
  author = {Po, Hoi Chun and Zou, Liujun and Senthil, T. and Vishwanath, Ashvin},
  year = {2019},
  month = may,
  journal = {Phys. Rev. B},
  volume = {99},
  number = {19},
  pages = {195455},
  publisher = {American Physical Society},
  doi = {10.1103/PhysRevB.99.195455},
  urldate = {2020-09-17}
}

@article{POL19,
  title = {Large Linear-in-Temperature Resistivity in Twisted Bilayer Graphene},
  author = {Polshyn, Hryhoriy and Yankowitz, Matthew and Chen, Shaowen and Zhang, Yuxuan and Watanabe, K. and Taniguchi, T. and Dean, Cory R. and Young, Andrea F.},
  year = {2019},
  month = oct,
  journal = {Nat. Phys.},
  volume = {15},
  number = {10},
  pages = {1011--1016},
  publisher = {Nature Publishing Group},
  issn = {1745-2481},
  doi = {10.1038/s41567-019-0596-3},
  urldate = {2021-01-23},
  copyright = {2019 The Author(s), under exclusive licence to Springer Nature Limited},
  langid = {english}
}

@article{POT21,
  title = {Exact {{Diagonalization}} for {{Magic-Angle Twisted Bilayer Graphene}}},
  author = {Potasz, Pawel and Xie, Ming and MacDonald, A. H.},
  year = {2021},
  month = sep,
  journal = {Phys. Rev. Lett.},
  volume = {127},
  number = {14},
  pages = {147203},
  publisher = {American Physical Society},
  doi = {10.1103/PhysRevLett.127.147203},
  urldate = {2021-11-29}
}

@article{QIN21,
  title = {In-{{Plane Critical Magnetic Fields}} in {{Magic-Angle Twisted Trilayer Graphene}}},
  author = {Qin, Wei and MacDonald, Allan H.},
  year = {2021},
  month = aug,
  journal = {Phys. Rev. Lett.},
  volume = {127},
  number = {9},
  pages = {097001},
  publisher = {American Physical Society},
  doi = {10.1103/PhysRevLett.127.097001},
  urldate = {2023-12-22}
}

@article{RAD18,
  title = {Charge-Transfer Insulation in Twisted Bilayer Graphene},
  author = {Rademaker, Louk and Mellado, Paula},
  year = {2018},
  month = dec,
  journal = {Phys. Rev. B},
  volume = {98},
  number = {23},
  pages = {235158},
  publisher = {American Physical Society},
  doi = {10.1103/PhysRevB.98.235158},
  urldate = {2023-03-14}
}

@article{RAD19,
  title = {Charge Smoothening and Band Flattening Due to {{Hartree}} Corrections in Twisted Bilayer Graphene},
  author = {Rademaker, Louk and Abanin, Dmitry A. and Mellado, Paula},
  year = {2019},
  month = nov,
  journal = {Phys. Rev. B},
  volume = {100},
  number = {20},
  pages = {205114},
  publisher = {American Physical Society},
  doi = {10.1103/PhysRevB.100.205114},
  urldate = {2023-03-14}
}

@article{RAI23a,
  title = {Dynamical {{Correlations}} and {{Order}} in {{Magic-Angle Twisted Bilayer Graphene}}},
  author = {Rai, Gautam and Crippa, Lorenzo and C{\u a}lug{\u a}ru, Dumitru and Hu, Haoyu and Paoletti, Francesca and {de' Medici}, Luca and Georges, Antoine and Bernevig, B. Andrei and Valent{\'i}, Roser and Sangiovanni, Giorgio and Wehling, Tim},
  year = {2024},
  month = sep,
  journal = {Phys. Rev. X},
  volume = {14},
  number = {3},
  pages = {031045},
  publisher = {American Physical Society},
  doi = {10.1103/PhysRevX.14.031045},
  urldate = {2024-10-24}
}

@article{RAM21,
  title = {Emulating {{Heavy Fermions}} in {{Twisted Trilayer Graphene}}},
  author = {Ramires, Aline and Lado, Jose L.},
  year = {2021},
  month = jul,
  journal = {Phys. Rev. Lett.},
  volume = {127},
  number = {2},
  pages = {026401},
  publisher = {American Physical Society},
  doi = {10.1103/PhysRevLett.127.026401},
  urldate = {2023-12-22}
}

@article{REN21,
  title = {{{WKB Estimate}} of {{Bilayer Graphene}}'s {{Magic Twist Angles}}},
  author = {Ren, Yafei and Gao, Qiang and MacDonald, A. H. and Niu, Qian},
  year = {2021},
  month = jan,
  journal = {Phys. Rev. Lett.},
  volume = {126},
  number = {1},
  pages = {016404},
  publisher = {American Physical Society},
  doi = {10.1103/PhysRevLett.126.016404},
  urldate = {2021-10-25}
}

@article{REP20,
  title = {Ferromagnetism in {{Narrow Bands}} of {{Moir{\'e} Superlattices}}},
  author = {Repellin, C{\'e}cile and Dong, Zhihuan and Zhang, Ya-Hui and Senthil, T.},
  year = {2020},
  month = may,
  journal = {Phys. Rev. Lett.},
  volume = {124},
  number = {18},
  pages = {187601},
  publisher = {American Physical Society},
  doi = {10.1103/PhysRevLett.124.187601},
  urldate = {2021-01-23}
}

@article{ROY19,
  title = {Unconventional Superconductivity in Nearly Flat Bands in Twisted Bilayer Graphene},
  author = {Roy, Bitan and Juri{\v c}i{\'c}, Vladimir},
  year = {2019},
  month = mar,
  journal = {Phys. Rev. B},
  volume = {99},
  number = {12},
  pages = {121407},
  publisher = {American Physical Society},
  doi = {10.1103/PhysRevB.99.121407},
  urldate = {2021-01-27}
}

@article{ROZ21,
  title = {Entropic Evidence for a {{Pomeranchuk}} Effect in Magic-Angle Graphene},
  author = {Rozen, Asaf and Park, Jeong Min and Zondiner, Uri and Cao, Yuan and {Rodan-Legrain}, Daniel and Taniguchi, Takashi and Watanabe, Kenji and Oreg, Yuval and Stern, Ady and Berg, Erez and {Jarillo-Herrero}, Pablo and Ilani, Shahal},
  year = {2021},
  month = apr,
  journal = {Nature},
  volume = {592},
  number = {7853},
  pages = {214--219},
  publisher = {Nature Publishing Group},
  issn = {1476-4687},
  doi = {10.1038/s41586-021-03319-3},
  urldate = {2021-11-29},
  copyright = {2021 The Author(s), under exclusive licence to Springer Nature Limited},
  langid = {english}
}

@article{SAI20,
  title = {Independent Superconductors and Correlated Insulators in Twisted Bilayer Graphene},
  author = {Saito, Yu and Ge, Jingyuan and Watanabe, Kenji and Taniguchi, Takashi and Young, Andrea F.},
  year = {2020},
  month = sep,
  journal = {Nat. Phys.},
  volume = {16},
  number = {9},
  pages = {926--930},
  publisher = {Nature Publishing Group},
  issn = {1745-2481},
  doi = {10.1038/s41567-020-0928-3},
  urldate = {2021-01-23},
  copyright = {2020 The Author(s), under exclusive licence to Springer Nature Limited},
  langid = {english}
}

@article{SAI21,
  title = {Hofstadter Subband Ferromagnetism and Symmetry-Broken {{Chern}} Insulators in Twisted Bilayer Graphene},
  author = {Saito, Yu and Ge, Jingyuan and Rademaker, Louk and Watanabe, Kenji and Taniguchi, Takashi and Abanin, Dmitry A. and Young, Andrea F.},
  year = {2021},
  month = jan,
  journal = {Nat. Phys.},
  volume = {17},
  number = {4},
  pages = {478--481},
  publisher = {Nature Publishing Group},
  issn = {1745-2481},
  doi = {10.1038/s41567-020-01129-4},
  urldate = {2021-01-23},
  copyright = {2020 The Author(s), under exclusive licence to Springer Nature Limited},
  langid = {english}
}

@article{SAI21a,
  title = {Isospin {{Pomeranchuk}} Effect in Twisted Bilayer Graphene},
  author = {Saito, Yu and Yang, Fangyuan and Ge, Jingyuan and Liu, Xiaoxue and Taniguchi, Takashi and Watanabe, Kenji and Li, J. I. A. and Berg, Erez and Young, Andrea F.},
  year = {2021},
  month = apr,
  journal = {Nature},
  volume = {592},
  number = {7853},
  pages = {220--224},
  publisher = {Nature Publishing Group},
  issn = {1476-4687},
  doi = {10.1038/s41586-021-03409-2},
  urldate = {2021-10-25},
  copyright = {2021 The Author(s), under exclusive licence to Springer Nature Limited},
  langid = {english}
}

@article{SAM22,
  title = {Moir{\'e} Phonons and Impact of Electronic Symmetry Breaking in Twisted Trilayer Graphene},
  author = {Samajdar, Rhine and Teng, Yanting and Scheurer, Mathias S.},
  year = {2022},
  month = nov,
  journal = {Phys. Rev. B},
  volume = {106},
  number = {20},
  pages = {L201403},
  publisher = {American Physical Society},
  doi = {10.1103/PhysRevB.106.L201403},
  urldate = {2023-12-22}
}

@article{SCA22,
  title = {Theory of Zero-Field Superconducting Diode Effect in Twisted Trilayer Graphene},
  author = {Scammell, Harley D. and Li, J. I. A. and Scheurer, Mathias S.},
  year = {2022},
  month = mar,
  journal = {2D Mater.},
  volume = {9},
  number = {2},
  pages = {025027},
  publisher = {IOP Publishing},
  issn = {2053-1583},
  doi = {10.1088/2053-1583/ac5b16},
  urldate = {2023-12-22},
  langid = {english}
}

@article{SEO19,
  title = {Ferromagnetic {{Mott}} State in {{Twisted Graphene Bilayers}} at the {{Magic Angle}}},
  author = {Seo, Kangjun and Kotov, Valeri N. and Uchoa, Bruno},
  year = {2019},
  month = jun,
  journal = {Phys. Rev. Lett.},
  volume = {122},
  number = {24},
  pages = {246402},
  publisher = {American Physical Society},
  doi = {10.1103/PhysRevLett.122.246402},
  urldate = {2020-09-20}
}

@article{SER20,
  title = {Intrinsic Quantized Anomalous {{Hall}} Effect in a Moir{\'e} Heterostructure},
  author = {Serlin, M. and Tschirhart, C. L. and Polshyn, H. and Zhang, Y. and Zhu, J. and Watanabe, K. and Taniguchi, T. and Balents, L. and Young, A. F.},
  year = {2020},
  month = feb,
  journal = {Science},
  volume = {367},
  number = {6480},
  pages = {900--903},
  publisher = {American Association for the Advancement of Science},
  issn = {0036-8075, 1095-9203},
  doi = {10.1126/science.aay5533},
  urldate = {2021-01-23},
  chapter = {Report},
  copyright = {Copyright {\copyright} 2020 The Authors, some rights reserved; exclusive licensee American Association for the Advancement of Science. No claim to original U.S. Government Works. http://www.sciencemag.org/about/science-licenses-journal-article-reuseThis is an article distributed under the terms of the Science Journals Default License.},
  langid = {english},
  pmid = {31857492}
}

@article{SHA19,
  title = {Emergent Ferromagnetism near Three-Quarters Filling in Twisted Bilayer Graphene},
  author = {Sharpe, Aaron L. and Fox, Eli J. and Barnard, Arthur W. and Finney, Joe and Watanabe, Kenji and Taniguchi, Takashi and Kastner, M. A. and {Goldhaber-Gordon}, David},
  year = {2019},
  month = aug,
  journal = {Science},
  volume = {365},
  number = {6453},
  pages = {605--608},
  publisher = {American Association for the Advancement of Science},
  issn = {0036-8075, 1095-9203},
  doi = {10.1126/science.aaw3780},
  urldate = {2020-09-20},
  chapter = {Report},
  copyright = {Copyright {\copyright} 2019 The Authors, some rights reserved; exclusive licensee American Association for the Advancement of Science. No claim to original U.S. Government Works. http://www.sciencemag.org/about/science-licenses-journal-article-reuseThis is an article distributed under the terms of the Science Journals Default License.},
  langid = {english},
  pmid = {31346139}
}

@article{SHE21,
  title = {Chiral Magic-Angle Twisted Bilayer Graphene in a Magnetic Field: {{Landau}} Level Correspondence, Exact Wave Functions, and Fractional {{Chern}} Insulators},
  shorttitle = {Chiral Magic-Angle Twisted Bilayer Graphene in a Magnetic Field},
  author = {Sheffer, Yarden and Stern, Ady},
  year = {2021},
  month = sep,
  journal = {Phys. Rev. B},
  volume = {104},
  number = {12},
  pages = {L121405},
  publisher = {American Physical Society},
  doi = {10.1103/PhysRevB.104.L121405},
  urldate = {2023-01-16}
}

@article{SHI21,
  title = {Stacking and Gate-Tunable Topological Flat Bands, Gaps, and Anisotropic Strip Patterns in Twisted Trilayer Graphene},
  author = {Shin, Jiseon and Chittari, Bheema Lingam and Jung, Jeil},
  year = {2021},
  month = jul,
  journal = {Phys. Rev. B},
  volume = {104},
  number = {4},
  pages = {045413},
  publisher = {American Physical Society},
  doi = {10.1103/PhysRevB.104.045413},
  urldate = {2023-12-22}
}

@article{shi_heavy-fermion_2022,
	title = {Heavy-fermion representation for twisted bilayer graphene systems},
	volume = {106},
	url = {https://link.aps.org/doi/10.1103/PhysRevB.106.245129},
	doi = {10.1103/PhysRevB.106.245129},
	number = {24},
	urldate = {2022-12-28},
	journal = {Physical Review B},
	author = {Shi, Hao and Dai, Xi},
	month = dec,
	year = {2022},
	note = {Publisher: American Physical Society},
	pages = {245129}
}

@article{SHI23,
  title = {Electronic Structure of Biased Alternating-Twist Multilayer Graphene},
  author = {Shin, Kyungjin and Jang, Yunsu and Shin, Jiseon and Jung, Jeil and Min, Hongki},
  year = {2023},
  month = jun,
  journal = {Phys. Rev. B},
  volume = {107},
  number = {24},
  pages = {245139},
  publisher = {American Physical Society},
  doi = {10.1103/PhysRevB.107.245139},
  urldate = {2023-12-22}
}

@article{SOE20,
  title = {Efficient Simulation of Moir{\'e} Materials Using the Density Matrix Renormalization Group},
  author = {Soejima, Tomohiro and Parker, Daniel E. and Bultinck, Nick and Hauschild, Johannes and Zaletel, Michael P.},
  year = {2020},
  month = nov,
  journal = {Phys. Rev. B},
  volume = {102},
  number = {20},
  pages = {205111},
  publisher = {American Physical Society},
  doi = {10.1103/PhysRevB.102.205111},
  urldate = {2021-01-23}
}

@article{STE20,
  title = {Untying the Insulating and Superconducting Orders in Magic-Angle Graphene},
  author = {Stepanov, Petr and Das, Ipsita and Lu, Xiaobo and Fahimniya, Ali and Watanabe, Kenji and Taniguchi, Takashi and Koppens, Frank H. L. and Lischner, Johannes and Levitov, Leonid and Efetov, Dmitri K.},
  year = {2020},
  month = jul,
  journal = {Nature},
  volume = {583},
  number = {7816},
  pages = {375--378},
  publisher = {Nature Publishing Group},
  issn = {1476-4687},
  doi = {10.1038/s41586-020-2459-6},
  urldate = {2021-01-23},
  copyright = {2020 The Author(s), under exclusive licence to Springer Nature Limited},
  langid = {english}
}

@article{STE21,
  title = {Competing {{Zero-Field Chern Insulators}} in {{Superconducting Twisted Bilayer Graphene}}},
  author = {Stepanov, Petr and Xie, Ming and Taniguchi, Takashi and Watanabe, Kenji and Lu, Xiaobo and MacDonald, Allan H. and Bernevig, B. Andrei and Efetov, Dmitri K.},
  year = {2021},
  month = nov,
  journal = {Phys. Rev. Lett.},
  volume = {127},
  number = {19},
  pages = {197701},
  publisher = {American Physical Society},
  doi = {10.1103/PhysRevLett.127.197701},
  urldate = {2023-01-16}
}

@article{SUA10,
  title = {Flat Bands in Slightly Twisted Bilayer Graphene: {{Tight-binding}} Calculations},
  shorttitle = {Flat Bands in Slightly Twisted Bilayer Graphene},
  author = {Su{\'a}rez Morell, E. and Correa, J. D. and Vargas, P. and Pacheco, M. and Barticevic, Z.},
  year = {2010},
  month = sep,
  journal = {Phys. Rev. B},
  volume = {82},
  number = {12},
  pages = {121407},
  publisher = {American Physical Society},
  doi = {10.1103/PhysRevB.82.121407},
  urldate = {2020-09-20}
}

@article{TAR19,
  title = {Origin of {{Magic Angles}} in {{Twisted Bilayer Graphene}}},
  author = {Tarnopolsky, Grigory and Kruchkov, Alex Jura and Vishwanath, Ashvin},
  year = {2019},
  month = mar,
  journal = {Phys. Rev. Lett.},
  volume = {122},
  number = {10},
  pages = {106405},
  publisher = {American Physical Society},
  doi = {10.1103/PhysRevLett.122.106405},
  urldate = {2020-09-17}
}

@article{THO18,
  title = {Triangular Antiferromagnetism on the Honeycomb Lattice of Twisted Bilayer Graphene},
  author = {Thomson, Alex and Chatterjee, Shubhayu and Sachdev, Subir and Scheurer, Mathias S.},
  year = {2018},
  month = aug,
  journal = {Phys. Rev. B},
  volume = {98},
  number = {7},
  pages = {075109},
  publisher = {American Physical Society},
  doi = {10.1103/PhysRevB.98.075109},
  urldate = {2021-01-23}
}

@article{THO21,
  title = {Recovery of Massless {{Dirac}} Fermions at Charge Neutrality in Strongly Interacting Twisted Bilayer Graphene with Disorder},
  author = {Thomson, Alex and Alicea, Jason},
  year = {2021},
  month = mar,
  journal = {Phys. Rev. B},
  volume = {103},
  number = {12},
  pages = {125138},
  publisher = {American Physical Society},
  doi = {10.1103/PhysRevB.103.125138},
  urldate = {2023-12-22}
}

@article{TIA23,
  title = {Evidence for {{Dirac}} Flat Band Superconductivity Enabled by Quantum Geometry},
  author = {Tian, Haidong and Gao, Xueshi and Zhang, Yuxin and Che, Shi and Xu, Tianyi and Cheung, Patrick and Watanabe, Kenji and Taniguchi, Takashi and Randeria, Mohit and Zhang, Fan and Lau, Chun Ning and Bockrath, Marc W.},
  year = {2023},
  month = feb,
  journal = {Nature},
  volume = {614},
  number = {7948},
  pages = {440--444},
  publisher = {Nature Publishing Group},
  issn = {1476-4687},
  doi = {10.1038/s41586-022-05576-2},
  urldate = {2024-02-14},
  copyright = {2023 The Author(s), under exclusive licence to Springer Nature Limited},
  langid = {english}
}

@article{TOM19,
  title = {Electronic {{Compressibility}} of {{Magic-Angle Graphene Superlattices}}},
  author = {Tomarken, S. L. and Cao, Y. and Demir, A. and Watanabe, K. and Taniguchi, T. and {Jarillo-Herrero}, P. and Ashoori, R. C.},
  year = {2019},
  month = jul,
  journal = {Phys. Rev. Lett.},
  volume = {123},
  number = {4},
  pages = {046601},
  publisher = {American Physical Society},
  doi = {10.1103/PhysRevLett.123.046601},
  urldate = {2021-10-25}
}

@article{TRI20,
  title = {Electronic Structure Calculations of Twisted Multi-Layer Graphene Superlattices},
  author = {Tritsaris, Georgios A. and Carr, Stephen and Zhu, Ziyan and Xie, Yiqi and Torrisi, Steven B. and Tang, Jing and Mattheakis, Marios and Larson, Daniel T. and Kaxiras, Efthimios},
  year = {2020},
  month = jun,
  journal = {2D Mater.},
  volume = {7},
  number = {3},
  pages = {035028},
  publisher = {IOP Publishing},
  issn = {2053-1583},
  doi = {10.1088/2053-1583/ab8f62},
  urldate = {2023-12-22},
  langid = {english}
}

@article{TSC21,
  title = {Imaging Orbital Ferromagnetism in a Moir{\'e} {{Chern}} Insulator},
  author = {Tschirhart, C. L. and Serlin, M. and Polshyn, H. and Shragai, A. and Xia, Z. and Zhu, J. and Zhang, Y. and Watanabe, K. and Taniguchi, T. and Huber, M. E. and Young, A. F.},
  year = {2021},
  month = jun,
  journal = {Science},
  volume = {372},
  number = {6548},
  pages = {1323--1327},
  publisher = {American Association for the Advancement of Science},
  doi = {10.1126/science.abd3190},
  urldate = {2021-10-25}
}

@article{UCH14,
  title = {Atomic Corrugation and Electron Localization Due to {{Moir{\'e}}} Patterns in Twisted Bilayer Graphenes},
  author = {Uchida, Kazuyuki and Furuya, Shinnosuke and Iwata, Jun-Ichi and Oshiyama, Atsushi},
  year = {2014},
  month = oct,
  journal = {Phys. Rev. B},
  volume = {90},
  number = {15},
  pages = {155451},
  publisher = {American Physical Society},
  doi = {10.1103/PhysRevB.90.155451},
  urldate = {2020-12-07}
}

@article{VAF20,
  title = {Renormalization {{Group Study}} of {{Hidden Symmetry}} in {{Twisted Bilayer Graphene}} with {{Coulomb Interactions}}},
  author = {Vafek, Oskar and Kang, Jian},
  year = {2020},
  month = dec,
  journal = {Phys. Rev. Lett.},
  volume = {125},
  number = {25},
  pages = {257602},
  publisher = {American Physical Society},
  doi = {10.1103/PhysRevLett.125.257602},
  urldate = {2021-01-22}
}

@article{VAF21,
  title = {Lattice Model for the {{Coulomb}} Interacting Chiral Limit of Magic-Angle Twisted Bilayer Graphene: {{Symmetries}}, Obstructions, and Excitations},
  shorttitle = {Lattice Model for the {{Coulomb}} Interacting Chiral Limit of Magic-Angle Twisted Bilayer Graphene},
  author = {Vafek, Oskar and Kang, Jian},
  year = {2021},
  month = aug,
  journal = {Phys. Rev. B},
  volume = {104},
  number = {7},
  pages = {075143},
  publisher = {American Physical Society},
  doi = {10.1103/PhysRevB.104.075143},
  urldate = {2023-01-16}
}

@article{VAF23,
  title = {Continuum Effective {{Hamiltonian}} for Graphene Bilayers for an Arbitrary Smooth Lattice Deformation from Microscopic Theories},
  author = {Vafek, Oskar and Kang, Jian},
  year = {2023},
  month = feb,
  journal = {Phys. Rev. B},
  volume = {107},
  number = {7},
  pages = {075123},
  publisher = {American Physical Society},
  doi = {10.1103/PhysRevB.107.075123},
  urldate = {2023-12-22}
}

@article{VEN18,
  title = {Correlations and Electronic Order in a Two-Orbital Honeycomb Lattice Model for Twisted Bilayer Graphene},
  author = {Venderbos, J{\"o}rn W. F. and Fernandes, Rafael M.},
  year = {2018},
  month = dec,
  journal = {Phys. Rev. B},
  volume = {98},
  number = {24},
  pages = {245103},
  publisher = {American Physical Society},
  doi = {10.1103/PhysRevB.98.245103},
  urldate = {2021-01-23}
}

@article{WAG22,
  title = {Global {{Phase Diagram}} of the {{Normal State}} of {{Twisted Bilayer Graphene}}},
  author = {Wagner, Glenn and Kwan, Yves H. and Bultinck, Nick and Simon, Steven H. and Parameswaran, S. A.},
  year = {2022},
  month = apr,
  journal = {Phys. Rev. Lett.},
  volume = {128},
  number = {15},
  pages = {156401},
  publisher = {American Physical Society},
  doi = {10.1103/PhysRevLett.128.156401},
  urldate = {2022-07-21}
}

@article{WAG23,
  title = {Coulomb-Driven Band Unflattening Suppresses $K$-Phonon Pairing in Moir{\'e} Graphene},
  author = {Wagner, Glenn and Kwan, Yves H. and Bultinck, Nick and Simon, Steven H. and Parameswaran, S. A.},
  year = {2024},
  month = mar,
  journal = {Phys. Rev. B},
  volume = {109},
  number = {10},
  pages = {104504},
  publisher = {American Physical Society},
  doi = {10.1103/PhysRevB.109.104504},
  urldate = {2024-08-19}
}

@article{WAG23a,
  title = {Superconductivity from repulsive interactions in Bernal-stacked bilayer graphene},
  author = {Wagner, Glenn and Kwan, Yves H. and Bultinck, Nick and Simon, Steven H. and Parameswaran, S. A.},
  journal = {Phys. Rev. B},
  volume = {110},
  issue = {21},
  pages = {214517},
  numpages = {7},
  year = {2024},
  month = {Dec},
  publisher = {American Physical Society},
  doi = {10.1103/PhysRevB.110.214517},
  url = {https://link.aps.org/doi/10.1103/PhysRevB.110.214517}
}

@article{WAN21,
  title = {Topological and Nematic Superconductivity Mediated by Ferro-{{SU}}(4) Fluctuations in Twisted Bilayer Graphene},
  author = {Wang, Yuxuan and Kang, Jian and Fernandes, Rafael M.},
  year = {2021},
  month = jan,
  journal = {Phys. Rev. B},
  volume = {103},
  number = {2},
  pages = {024506},
  publisher = {American Physical Society},
  doi = {10.1103/PhysRevB.103.024506},
  urldate = {2021-02-05}
}

@article{WAN21a,
  title = {Chiral Approximation to Twisted Bilayer Graphene: {{Exact}} Intravalley Inversion Symmetry, Nodal Structure, and Implications for Higher Magic Angles},
  shorttitle = {Chiral Approximation to Twisted Bilayer Graphene},
  author = {Wang, Jie and Zheng, Yunqin and Millis, Andrew J. and Cano, Jennifer},
  year = {2021},
  month = may,
  journal = {Phys. Rev. Research},
  volume = {3},
  number = {2},
  pages = {023155},
  publisher = {American Physical Society},
  doi = {10.1103/PhysRevResearch.3.023155},
  urldate = {2021-11-29}
}

@article{WAN23c,
  title = {Kekul{\'e} Spirals and Charge Transfer Cascades in Twisted Symmetric Trilayer Graphene},
  author = {Wang, Ziwei and Kwan, Yves H. and Wagner, Glenn and Bultinck, Nick and Simon, Steven H. and Parameswaran, S. A.},
  year = {2024},
  month = may,
  journal = {Phys. Rev. B},
  volume = {109},
  number = {20},
  pages = {L201119},
  publisher = {American Physical Society},
  doi = {10.1103/PhysRevB.109.L201119},
  urldate = {2024-08-19}
}

@article{WAN24,
  title = {Molecular {{Pairing}} in {{Twisted Bilayer Graphene Superconductivity}}},
  author = {Wang, Yi-Jie and Zhou, Geng-Dong and Peng, Shi-Yu and Lian, Biao and Song, Zhi-Da},
  year = {2024},
  month = feb,
  journal = {arXiv:2402.00869 [cond-mat]},
  eprint = {2402.00869},
  primaryclass = {cond-mat},
  doi = {10.48550/arXiv.2402.00869},
  urldate = {2024-02-21},
  archiveprefix = {arXiv}
}

@article{WIJ15,
  title = {Relaxation of Moir{\'e} Patterns for Slightly Misaligned Identical Lattices: Graphene on Graphite},
  shorttitle = {Relaxation of Moir{\'e} Patterns for Slightly Misaligned Identical Lattices},
  author = {van Wijk, M. M. and Schuring, A. and Katsnelson, M. I. and Fasolino, A.},
  year = {2015},
  month = jul,
  journal = {2D Mater.},
  volume = {2},
  number = {3},
  pages = {034010},
  publisher = {IOP Publishing},
  issn = {2053-1583},
  doi = {10.1088/2053-1583/2/3/034010},
  urldate = {2020-12-07},
  langid = {english}
}

@article{WIL20,
  title = {Disorder in Twisted Bilayer Graphene},
  author = {Wilson, Justin H. and Fu, Yixing and Das Sarma, S. and Pixley, J. H.},
  year = {2020},
  month = jun,
  journal = {Phys. Rev. Research},
  volume = {2},
  number = {2},
  pages = {023325},
  publisher = {American Physical Society},
  doi = {10.1103/PhysRevResearch.2.023325},
  urldate = {2021-01-23}
}

@article{WON20,
  title = {Cascade of Electronic Transitions in Magic-Angle Twisted Bilayer Graphene},
  author = {Wong, Dillon and Nuckolls, Kevin P. and Oh, Myungchul and Lian, Biao and Xie, Yonglong and Jeon, Sangjun and Watanabe, Kenji and Taniguchi, Takashi and Bernevig, B. Andrei and Yazdani, Ali},
  year = {2020},
  month = jun,
  journal = {Nature},
  volume = {582},
  number = {7811},
  pages = {198--202},
  publisher = {Nature Publishing Group},
  issn = {1476-4687},
  doi = {10.1038/s41586-020-2339-0},
  urldate = {2021-01-23},
  copyright = {2020 The Author(s), under exclusive licence to Springer Nature Limited},
  langid = {english}
}

@article{WU18,
  title = {Theory of {{Phonon-Mediated Superconductivity}} in {{Twisted Bilayer Graphene}}},
  author = {Wu, Fengcheng and MacDonald, A. H. and Martin, Ivar},
  year = {2018},
  month = dec,
  journal = {Phys. Rev. Lett.},
  volume = {121},
  number = {25},
  pages = {257001},
  publisher = {American Physical Society},
  doi = {10.1103/PhysRevLett.121.257001},
  urldate = {2020-09-20}
}

@article{WU19,
  title = {Coupled-Wire Description of the Correlated Physics in Twisted Bilayer Graphene},
  author = {Wu, Xiao-Chuan and Jian, Chao-Ming and Xu, Cenke},
  year = {2019},
  month = apr,
  journal = {Phys. Rev. B},
  volume = {99},
  number = {16},
  pages = {161405},
  publisher = {American Physical Society},
  doi = {10.1103/PhysRevB.99.161405},
  urldate = {2021-01-23}
}

@article{WU19a,
  title = {Phonon-Induced Giant Linear-in-$T$ Resistivity in Magic Angle Twisted Bilayer Graphene: {{Ordinary}} Strangeness and Exotic Superconductivity},
  shorttitle = {Phonon-Induced Giant Linear-in-\${{T}}\$ Resistivity in Magic Angle Twisted Bilayer Graphene},
  author = {Wu, Fengcheng and Hwang, Euyheon and Das Sarma, Sankar},
  year = {2019},
  month = apr,
  journal = {Phys. Rev. B},
  volume = {99},
  number = {16},
  pages = {165112},
  publisher = {American Physical Society},
  doi = {10.1103/PhysRevB.99.165112},
  urldate = {2021-01-23}
}

@article{WU20,
  title = {Collective {{Excitations}} of {{Quantum Anomalous Hall Ferromagnets}} in {{Twisted Bilayer Graphene}}},
  author = {Wu, Fengcheng and Das Sarma, Sankar},
  year = {2020},
  month = jan,
  journal = {Phys. Rev. Lett.},
  volume = {124},
  number = {4},
  pages = {046403},
  publisher = {American Physical Society},
  doi = {10.1103/PhysRevLett.124.046403},
  urldate = {2021-01-23}
}

@article{WU21a,
  title = {Chern Insulators, van {{Hove}} Singularities and Topological Flat Bands in Magic-Angle Twisted Bilayer Graphene},
  author = {Wu, Shuang and Zhang, Zhenyuan and Watanabe, K. and Taniguchi, T. and Andrei, Eva Y.},
  year = {2021},
  month = apr,
  journal = {Nat. Mater.},
  volume = {20},
  number = {4},
  pages = {488--494},
  publisher = {Nature Publishing Group},
  issn = {1476-4660},
  doi = {10.1038/s41563-020-00911-2},
  urldate = {2021-11-29},
  copyright = {2021 The Author(s), under exclusive licence to Springer Nature Limited},
  langid = {english}
}

@article{WU21c,
  title = {Magic Angle and Plasmon Mode Engineering in Twisted Trilayer Graphene with Pressure},
  author = {Wu, Zewen and Kuang, Xueheng and Zhan, Zhen and Yuan, Shengjun},
  year = {2021},
  month = nov,
  journal = {Phys. Rev. B},
  volume = {104},
  number = {20},
  pages = {205104},
  publisher = {American Physical Society},
  doi = {10.1103/PhysRevB.104.205104},
  urldate = {2023-12-22}
}

@article{WU21d,
  title = {Lattice Relaxation, Mirror Symmetry and Magnetic Field Effects on Ultraflat Bands in Twisted Trilayer Graphene},
  author = {Wu, Zewen and Zhan, Zhen and Yuan, Shengjun},
  year = {2021},
  month = apr,
  journal = {Sci. China Phys. Mech. Astron.},
  volume = {64},
  number = {6},
  pages = {267811},
  issn = {1869-1927},
  doi = {10.1007/s11433-020-1690-4},
  urldate = {2024-02-14},
  langid = {english}
}

@article{XIE19,
  title = {Spectroscopic Signatures of Many-Body Correlations in Magic-Angle Twisted Bilayer Graphene},
  author = {Xie, Yonglong and Lian, Biao and J{\"a}ck, Berthold and Liu, Xiaomeng and Chiu, Cheng-Li and Watanabe, Kenji and Taniguchi, Takashi and Bernevig, B. Andrei and Yazdani, Ali},
  year = {2019},
  month = aug,
  journal = {Nature},
  volume = {572},
  number = {7767},
  pages = {101--105},
  publisher = {Nature Publishing Group},
  issn = {1476-4687},
  doi = {10.1038/s41586-019-1422-x},
  urldate = {2020-09-20},
  copyright = {2019 The Author(s), under exclusive licence to Springer Nature Limited},
  langid = {english}
}

@article{XIE20,
  title = {Topology-{{Bounded Superfluid Weight}} in {{Twisted Bilayer Graphene}}},
  author = {Xie, Fang and Song, Zhida and Lian, Biao and Bernevig, B. Andrei},
  year = {2020},
  month = apr,
  journal = {Phys. Rev. Lett.},
  volume = {124},
  number = {16},
  pages = {167002},
  publisher = {American Physical Society},
  doi = {10.1103/PhysRevLett.124.167002},
  urldate = {2020-09-20}
}

@article{XIE20b,
  title = {Nature of the {{Correlated Insulator States}} in {{Twisted Bilayer Graphene}}},
  author = {Xie, Ming and MacDonald, A. H.},
  year = {2020},
  month = mar,
  journal = {Phys. Rev. Lett.},
  volume = {124},
  number = {9},
  pages = {097601},
  publisher = {American Physical Society},
  doi = {10.1103/PhysRevLett.124.097601},
  urldate = {2021-01-23}
}

@article{XIE21,
  title = {Twisted Bilayer Graphene. {{VI}}. {{An}} Exact Diagonalization Study at Nonzero Integer Filling},
  author = {Xie, Fang and Cowsik, Aditya and Song, Zhi-Da and Lian, Biao and Bernevig, B. Andrei and Regnault, Nicolas},
  year = {2021},
  month = may,
  journal = {Phys. Rev. B},
  volume = {103},
  number = {20},
  pages = {205416},
  publisher = {American Physical Society},
  doi = {10.1103/PhysRevB.103.205416},
  urldate = {2021-09-13}
}

@article{XIE21a,
  title = {Weak-{{Field Hall Resistivity}} and {{Spin-Valley Flavor Symmetry Breaking}} in {{Magic-Angle Twisted Bilayer Graphene}}},
  author = {Xie, Ming and MacDonald, A. H.},
  year = {2021},
  month = nov,
  journal = {Phys. Rev. Lett.},
  volume = {127},
  number = {19},
  pages = {196401},
  publisher = {American Physical Society},
  doi = {10.1103/PhysRevLett.127.196401},
  urldate = {2021-11-29}
}

@article{XIE21b,
  title = {Twisted Symmetric Trilayer Graphene. {{II}}. {{Projected Hartree-Fock}} Study},
  author = {Xie, Fang and Regnault, Nicolas and C{\u a}lug{\u a}ru, Dumitru and Bernevig, B. Andrei and Lian, Biao},
  year = {2021},
  month = sep,
  journal = {Phys. Rev. B},
  volume = {104},
  number = {11},
  pages = {115167},
  publisher = {American Physical Society},
  doi = {10.1103/PhysRevB.104.115167},
  urldate = {2023-07-29}
}

@article{XIE21d,
  title = {Fractional {{Chern}} Insulators in Magic-Angle Twisted Bilayer Graphene},
  author = {Xie, Yonglong and Pierce, Andrew T. and Park, Jeong Min and Parker, Daniel E. and Khalaf, Eslam and Ledwith, Patrick and Cao, Yuan and Lee, Seung Hwan and Chen, Shaowen and Forrester, Patrick R. and Watanabe, Kenji and Taniguchi, Takashi and Vishwanath, Ashvin and {Jarillo-Herrero}, Pablo and Yacoby, Amir},
  year = {2021},
  month = dec,
  journal = {Nature},
  volume = {600},
  number = {7889},
  pages = {439--443},
  publisher = {Nature Publishing Group},
  issn = {1476-4687},
  doi = {10.1038/s41586-021-04002-3},
  urldate = {2023-12-22},
  copyright = {2021 The Author(s)},
  langid = {english}
}

@article{XIE23a,
  title = {Phase Diagram of Twisted Bilayer Graphene at Filling Factor $\ensuremath{\nu}=\ifmmode\pm\else\textpm\fi{}3$},
  author = {Xie, Fang and Kang, Jian and Bernevig, B. Andrei and Vafek, Oskar and Regnault, Nicolas},
  year = {2023},
  month = feb,
  journal = {Phys. Rev. B},
  volume = {107},
  number = {7},
  pages = {075156},
  publisher = {American Physical Society},
  doi = {10.1103/PhysRevB.107.075156},
  urldate = {2023-12-22}
}

@article{XU18,
  title = {Topological {{Superconductivity}} in {{Twisted Multilayer Graphene}}},
  author = {Xu, Cenke and Balents, Leon},
  year = {2018},
  month = aug,
  journal = {Phys. Rev. Lett.},
  volume = {121},
  number = {8},
  pages = {087001},
  publisher = {American Physical Society},
  doi = {10.1103/PhysRevLett.121.087001},
  urldate = {2020-09-20}
}

@article{XU18b,
  title = {Kekul{\'e} Valence Bond Order in an Extended {{Hubbard}} Model on the Honeycomb Lattice with Possible Applications to Twisted Bilayer Graphene},
  author = {Xu, Xiao Yan and Law, K. T. and Lee, Patrick A.},
  year = {2018},
  month = sep,
  journal = {Phys. Rev. B},
  volume = {98},
  number = {12},
  pages = {121406},
  publisher = {American Physical Society},
  doi = {10.1103/PhysRevB.98.121406},
  urldate = {2021-01-23}
}

@article{YAN19,
  title = {Tuning Superconductivity in Twisted Bilayer Graphene},
  author = {Yankowitz, Matthew and Chen, Shaowen and Polshyn, Hryhoriy and Zhang, Yuxuan and Watanabe, K. and Taniguchi, T. and Graf, David and Young, Andrea F. and Dean, Cory R.},
  year = {2019},
  month = mar,
  journal = {Science},
  volume = {363},
  number = {6431},
  pages = {1059--1064},
  publisher = {American Association for the Advancement of Science},
  issn = {0036-8075, 1095-9203},
  doi = {10.1126/science.aav1910},
  urldate = {2021-01-21},
  chapter = {Research Article},
  copyright = {Copyright {\copyright} 2019 The Authors, some rights reserved; exclusive licensee American Association for the Advancement of Science. No claim to original U.S. Government Works. http://www.sciencemag.org/about/science-licenses-journal-article-reuseThis is an article distributed under the terms of the Science Journals Default License.},
  langid = {english},
  pmid = {30679385}
}

@article{YOU19,
  title = {Superconductivity from Valley Fluctuations and Approximate {{SO}}(4) Symmetry in a Weak Coupling Theory of Twisted Bilayer Graphene},
  author = {You, Yi-Zhuang and Vishwanath, Ashvin},
  year = {2019},
  month = apr,
  journal = {npj Quantum Mater.},
  volume = {4},
  number = {1},
  pages = {1--12},
  publisher = {Nature Publishing Group},
  issn = {2397-4648},
  doi = {10.1038/s41535-019-0153-4},
  urldate = {2021-01-23},
  copyright = {2019 The Author(s)},
  langid = {english}
}

@article{YU22,
  title = {Euler-Obstructed {{Cooper}} Pairing: {{Nodal}} Superconductivity and Hinge {{Majorana}} Zero Modes},
  shorttitle = {Euler-Obstructed {{Cooper}} Pairing},
  author = {Yu, Jiabin and Chen, Yu-An and Das Sarma, Sankar},
  year = {2022},
  month = mar,
  journal = {Phys. Rev. B},
  volume = {105},
  number = {10},
  pages = {104515},
  publisher = {American Physical Society},
  doi = {10.1103/PhysRevB.105.104515},
  urldate = {2024-02-14}
}

@article{YU23a,
  title = {Magic-Angle Twisted Symmetric Trilayer Graphene as a Topological Heavy-Fermion Problem},
  author = {Yu, Jiabin and Xie, Ming and Bernevig, B. Andrei and Das Sarma, Sankar},
  year = {2023},
  month = jul,
  journal = {Phys. Rev. B},
  volume = {108},
  number = {3},
  pages = {035129},
  publisher = {American Physical Society},
  doi = {10.1103/PhysRevB.108.035129},
  urldate = {2023-08-10}
}

@article{YU23c,
  title = {Spin Skyrmion Gaps as Signatures of Strong-Coupling Insulators in Magic-Angle Twisted Bilayer Graphene},
  author = {Yu, Jiachen and Foutty, Benjamin A. and Kwan, Yves H. and Barber, Mark E. and Watanabe, Kenji and Taniguchi, Takashi and Shen, Zhi-Xun and Parameswaran, Siddharth A. and Feldman, Benjamin E.},
  year = {2023},
  month = oct,
  journal = {Nat. Commun.},
  volume = {14},
  number = {1},
  eprint = {2206.11304},
  primaryclass = {cond-mat},
  pages = {6679},
  issn = {2041-1723},
  doi = {10.1038/s41467-023-42275-6},
  urldate = {2023-12-22},
  archiveprefix = {arXiv}
}

@article{YUA18,
  title = {Model for the Metal-Insulator Transition in Graphene Superlattices and Beyond},
  author = {Yuan, Noah F. Q. and Fu, Liang},
  year = {2018},
  month = jul,
  journal = {Phys. Rev. B},
  volume = {98},
  number = {4},
  pages = {045103},
  publisher = {American Physical Society},
  doi = {10.1103/PhysRevB.98.045103},
  urldate = {2021-01-23}
}

@article{ZHA20,
  title = {Correlated Insulating Phases of Twisted Bilayer Graphene at Commensurate Filling Fractions: {{A Hartree-Fock}} Study},
  shorttitle = {Correlated Insulating Phases of Twisted Bilayer Graphene at Commensurate Filling Fractions},
  author = {Zhang, Yi and Jiang, Kun and Wang, Ziqiang and Zhang, Fuchun},
  year = {2020},
  month = jul,
  journal = {Phys. Rev. B},
  volume = {102},
  number = {3},
  pages = {035136},
  publisher = {American Physical Society},
  doi = {10.1103/PhysRevB.102.035136},
  urldate = {2021-01-23}
}

@article{ZHA21,
  title = {Momentum {{Space Quantum Monte Carlo}} on {{Twisted Bilayer Graphene}}},
  author = {Zhang, Xu and Pan, Gaopei and Zhang, Yi and Kang, Jian and Meng, Zi Yang},
  year = {2021},
  month = jul,
  journal = {Chinese Phys. Lett.},
  volume = {38},
  number = {7},
  pages = {077305},
  publisher = {{Chinese Physical Society and IOP Publishing Ltd}},
  issn = {0256-307X},
  doi = {10.1088/0256-307X/38/7/077305},
  urldate = {2023-01-16},
  langid = {english}
}

@article{ZHA23a,
  title = {Polynomial Sign Problem and Topological {{Mott}} Insulator in Twisted Bilayer Graphene},
  author = {Zhang, Xu and Pan, Gaopei and Chen, Bin-Bin and Li, Heqiu and Sun, Kai and Meng, Zi Yang},
  year = {2023},
  month = jun,
  journal = {Phys. Rev. B},
  volume = {107},
  number = {24},
  pages = {L241105},
  publisher = {American Physical Society},
  doi = {10.1103/PhysRevB.107.L241105},
  urldate = {2024-02-14}
}

@article{ZHO23a,
  title = {Coexistence of Reconstructed and Unreconstructed Structures in the Structural Transition Regime of Twisted Bilayer Graphene},
  author = {Zhou, Xiao-Feng and Liu, Yi-Wen and Hao, Chen-Yue and Yan, Chao and Zheng, Qi and Ren, Ya-Ning and Zhao, Ya-Xin and Watanabe, Kenji and Taniguchi, Takashi and He, Lin},
  year = {2023},
  month = mar,
  journal = {Phys. Rev. B},
  volume = {107},
  number = {12},
  pages = {125410},
  publisher = {American Physical Society},
  doi = {10.1103/PhysRevB.107.125410},
  urldate = {2023-12-22}
}

@article{ZON20,
  title = {Cascade of Phase Transitions and {{Dirac}} Revivals in Magic-Angle Graphene},
  author = {Zondiner, U. and Rozen, A. and {Rodan-Legrain}, D. and Cao, Y. and Queiroz, R. and Taniguchi, T. and Watanabe, K. and Oreg, Y. and {von Oppen}, F. and Stern, Ady and Berg, E. and {Jarillo-Herrero}, P. and Ilani, S.},
  year = {2020},
  month = jun,
  journal = {Nature},
  volume = {582},
  number = {7811},
  pages = {203--208},
  publisher = {Nature Publishing Group},
  issn = {1476-4687},
  doi = {10.1038/s41586-020-2373-y},
  urldate = {2021-01-23},
  copyright = {2020 The Author(s), under exclusive licence to Springer Nature Limited},
  langid = {english}
}

@article{ZOU18,
  title = {Band Structure of Twisted Bilayer Graphene: {{Emergent}} Symmetries, Commensurate Approximants, and {{Wannier}} Obstructions},
  shorttitle = {Band Structure of Twisted Bilayer Graphene},
  author = {Zou, Liujun and Po, Hoi Chun and Vishwanath, Ashvin and Senthil, T.},
  year = {2018},
  month = aug,
  journal = {Phys. Rev. B},
  volume = {98},
  number = {8},
  pages = {085435},
  publisher = {American Physical Society},
  doi = {10.1103/PhysRevB.98.085435},
  urldate = {2020-09-20}
}

@article{song_magic-angle_2022,
	title = {Magic-{Angle} {Twisted} {Bilayer} {Graphene} as a {Topological} {Heavy} {Fermion} {Problem}},
	volume = {129},
	url = {https://link.aps.org/doi/10.1103/PhysRevLett.129.047601},
	doi = {10.1103/PhysRevLett.129.047601},
	number = {4},
	urldate = {2022-09-14},
	journal = {Physical Review Letters},
	author = {Song, Zhi-Da and Bernevig, B. Andrei},
	month = jul,
	year = {2022},
	note = {Publisher: American Physical Society},
	pages = {047601}
}

@article{calugaru_tbg_2023,
	title = {{TBG} as {Topological} {Heavy} {Fermion}: {II}. {Analytical} approximations of the model parameters},
	volume = {49},
	issn = {1063-777X, 1090-6517},
	shorttitle = {{TBG} as {Topological} {Heavy} {Fermion}},
	url = {http://arxiv.org/abs/2303.03429},
	doi = {10.1063/10.0019421},
	number = {6},
	urldate = {2023-09-20},
	journal = {Low Temperature Physics},
	author = {Călugăru, Dumitru and Borovkov, Maksim and Lau, Liam L. H. and Coleman, Piers and Song, Zhi-Da and Bernevig, B. Andrei},
	month = jun,
	year = {2023},
	note = {arXiv:2303.03429 [cond-mat]},
	pages = {640--654}
}

@misc{ledwith_nonlocal_2024,
	title = {Nonlocal {Moments} in the {Chern} {Bands} of {Twisted} {Bilayer} {Graphene}},
	url = {https://arxiv.org/abs/2408.16761v1},
	urldate = {2024-09-12},
	publisher = {arXiv},
    note={arXiv:2408.16761 [cond-mat.str-el]},
	author = {Ledwith, Patrick J. and Dong, Junkai and Vishwanath, Ashvin and Khalaf, Eslam},
	month = aug,
	year = {2024}
}

@article{chou_kondo_2023,
	title = {Kondo {Lattice} {Model} in {Magic}-{Angle} {Twisted} {Bilayer} {Graphene}},
	volume = {131},
	url = {https://link.aps.org/doi/10.1103/PhysRevLett.131.026501},
	doi = {10.1103/PhysRevLett.131.026501},
	number = {2},
	urldate = {2023-08-18},
	journal = {Physical Review Letters},
	author = {Chou, Yang-Zhi and Das Sarma, Sankar},
	month = jul,
	year = {2023},
	note = {Publisher: American Physical Society},
	pages = {026501}
}

@misc{lau_topological_2023,
	title = {Topological {Mixed} {Valence} {Model} for {Twisted} {Bilayer} {Graphene}},
	url = {http://arxiv.org/abs/2303.02670},
	doi = {10.48550/arXiv.2303.02670},
	urldate = {2023-11-13},
	publisher = {arXiv},
	author = {Lau, Liam L. H. and Coleman, Piers},
	month = mar,
	year = {2023},
	note = {arXiv:2303.02670 [cond-mat]}
}

@article{hu_symmetric_2023,
	title = {Symmetric {Kondo} {Lattice} {States} in {Doped} {Strained} {Twisted} {Bilayer} {Graphene}},
	volume = {131},
	url = {https://link.aps.org/doi/10.1103/PhysRevLett.131.166501},
	doi = {10.1103/PhysRevLett.131.166501},
	number = {16},
	urldate = {2023-12-24},
	journal = {Physical Review Letters},
	author = {Hu, Haoyu and Rai, Gautam and Crippa, Lorenzo and Herzog-Arbeitman, Jonah and Călugăru, Dumitru and Wehling, Tim and Sangiovanni, Giorgio and Valentí, Roser and Tsvelik, Alexei M. and Bernevig, B. Andrei},
	month = oct,
	year = {2023},
	note = {Publisher: American Physical Society},
	pages = {166501}
}

@article{zhou_kondo_2024,
	title = {Kondo phase in twisted bilayer graphene},
	volume = {109},
	url = {https://link.aps.org/doi/10.1103/PhysRevB.109.045419},
	doi = {10.1103/PhysRevB.109.045419},
	number = {4},
	urldate = {2024-01-16},
	journal = {Physical Review B},
	author = {Zhou, Geng-Dong and Wang, Yi-Jie and Tong, Ninghua and Song, Zhi-Da},
	month = jan,
	year = {2024},
	note = {Publisher: American Physical Society},
	pages = {045419}
}

@article{hu_kondo_2023,
	title = {Kondo {Lattice} {Model} of {Magic}-{Angle} {Twisted}-{Bilayer} {Graphene}: {Hund}'s {Rule}, {Local}-{Moment} {Fluctuations}, and {Low}-{Energy} {Effective} {Theory}},
	volume = {131},
	shorttitle = {Kondo {Lattice} {Model} of {Magic}-{Angle} {Twisted}-{Bilayer} {Graphene}},
	url = {https://link.aps.org/doi/10.1103/PhysRevLett.131.026502},
	doi = {10.1103/PhysRevLett.131.026502},
	number = {2},
	urldate = {2023-12-24},
	journal = {Physical Review Letters},
	author = {Hu, Haoyu and Bernevig, B. Andrei and Tsvelik, Alexei M.},
	month = jul,
	year = {2023},
	note = {Publisher: American Physical Society},
	pages = {026502}
}

@misc{rai_dynamical_2023,
	title = {Dynamical correlations and order in magic-angle twisted bilayer graphene},
	url = {http://arxiv.org/abs/2309.08529},
	doi = {10.48550/arXiv.2309.08529},
	urldate = {2023-11-07},
	publisher = {arXiv},
	author = {Rai, Gautam and Crippa, Lorenzo and Călugăru, Dumitru and Hu, Haoyu and Medici, Luca de' and Georges, Antoine and Bernevig, B. Andrei and Valentí, Roser and Sangiovanni, Giorgio and Wehling, Tim},
	month = sep,
	year = {2023},
	note = {arXiv:2309.08529 [cond-mat, physics:quant-ph]}
}

@article{PhysRevB.110.045133,
  title = {Electron--$K$-phonon interaction in twisted bilayer graphene},
  author = {Liu, Chao-Xing and Chen, Yulin and Yazdani, Ali and Bernevig, B. Andrei},
  journal = {Phys. Rev. B},
  volume = {110},
  issue = {4},
  pages = {045133},
  numpages = {7},
  year = {2024},
  month = {Jul},
  publisher = {American Physical Society},
  doi = {10.1103/PhysRevB.110.045133},
  url = {https://link.aps.org/doi/10.1103/PhysRevB.110.045133}
}

@article{datta_heavy_2023,
	title = {Heavy quasiparticles and cascades without symmetry breaking in twisted bilayer graphene},
	volume = {14},
	issn = {2041-1723},
	url = {http://arxiv.org/abs/2301.13024},
	doi = {10.1038/s41467-023-40754-4},
	number = {1},
	urldate = {2023-12-24},
	journal = {Nature Communications},
	author = {Datta, Anushree and Calderón, M. J. and Camjayi, A. and Bascones, E.},
	month = aug,
	year = {2023},
	note = {{T}his work used a lattice model in a similar spirit with the THF model. Correlation comes from the localized f-orbital, while c-electron - after breaking the particle-hole symmetry - is realized on a lattice.},
	pages = {5036}
}

@article{wang_molecular_2024,
	title = {Molecular {Pairing} in {Twisted} {Bilayer} {Graphene} {Superconductivity}},
	volume = {133},
	copyright = {All rights reserved},
	url = {https://link.aps.org/doi/10.1103/PhysRevLett.133.146001},
	doi = {10.1103/PhysRevLett.133.146001},
	number = {14},
	urldate = {2024-10-01},
	journal = {Physical Review Letters},
	author = {Wang, Yi-Jie and Zhou, Geng-Dong and Peng, Shi-Yu and Lian, Biao and Song, Zhi-Da},
	month = sep,
	year = {2024},
	note = {Publisher: American Physical Society},
	pages = {146001}
}

@unpublished{scattering1,
  author = {Vituri, Yaar and Berg, Erez},
  note = {manuscript to appear simultaneously on arXiv},
}

@unpublished{scattering2,
  author = {Nemin Wei and Felix von Oppen and Leonid I. Glazman},
  note = {manuscript to appear simultaneously on arXiv},
}

@misc{youn_hundness_2024,
	title = {Hundness in twisted bilayer graphene: correlated gaps and pairing},
	shorttitle = {Hundness in twisted bilayer graphene},
	url = {http://arxiv.org/abs/2412.03108},
	doi = {10.48550/arXiv.2412.03108},
	urldate = {2024-12-05},
	publisher = {arXiv},
	author = {Youn, Seongyeon and Goh, Beomjoon and Zhou, Geng-Dong and Song, Zhi-Da and Lee, Seung-Sup B.},
	month = dec,
	year = {2024},
	note = {arXiv:2412.03108 [cond-mat]}
}

@misc{crippa2025dynamicalcorrelationeffectstwisted,
      title={Dynamical correlation effects in twisted bilayer graphene under strain and lattice relaxation}, 
      author={Lorenzo Crippa and Gautam Rai and Dumitru Călugăru and Haoyu Hu and Jonah Herzog-Arbeitman and B. Andrei Bernevig and Roser Valentí and Giorgio Sangiovanni and Tim Wehling},
      year={2025},
      eprint={2509.19436},
      archivePrefix={arXiv},
      primaryClass={cond-mat.str-el},
      url={https://arxiv.org/abs/2509.19436}, 
}

@Article{Birkbeck2025,
author={Birkbeck, J.
and Xiao, J.
and Inbar, A.
and Taniguchi, T.
and Watanabe, K.
and Berg, E.
and Glazman, L.
and Guinea, F.
and von Oppen, F.
and Ilani, S.},
title={Quantum twisting microscopy of phonons in twisted bilayer graphene},
journal={Nature},
year={2025},
month={May},
day={01},
volume={641},
number={8062},
pages={345-351},
issn={1476-4687},
doi={10.1038/s41586-025-08881-8},
url={https://doi.org/10.1038/s41586-025-08881-8}
}

@misc{xiao2025interactingenergybandsmagic,
      title={The Interacting Energy Bands of Magic Angle Twisted Bilayer Graphene Revealed by the Quantum Twisting Microscope}, 
      author={J. Xiao and A. Inbar and J. Birkbeck and N. Gershon and Y. Zamir and T. Taniguchi and K. Watanabe and E. Berg and S. Ilani},
      year={2025},
      eprint={2506.20738},
      archivePrefix={arXiv},
      primaryClass={cond-mat.mes-hall},
      url={https://arxiv.org/abs/2506.20738}, 
}

@misc{hu2025projectedsolvabletopologicalheavy,
      title={Projected and Solvable Topological Heavy Fermion Model of Twisted Bilayer Graphene}, 
      author={Haoyu Hu and Zhi-Da Song and B. Andrei Bernevig},
      year={2025},
      eprint={2502.14039},
      archivePrefix={arXiv},
      primaryClass={cond-mat.str-el},
      url={https://arxiv.org/abs/2502.14039}, 
}

@article{PhysRevLett.129.047601,
  title = {Magic-Angle Twisted Bilayer Graphene as a Topological Heavy Fermion Problem},
  author = {Song, Zhi-Da and Bernevig, B. Andrei},
  journal = {Phys. Rev. Lett.},
  volume = {129},
  issue = {4},
  pages = {047601},
  numpages = {10},
  year = {2022},
  month = {Jul},
  publisher = {American Physical Society},
  doi = {10.1103/PhysRevLett.129.047601},
  url = {https://link.aps.org/doi/10.1103/PhysRevLett.129.047601}
}

@article{PhysRevX.14.031045,
  title = {Dynamical Correlations and Order in Magic-Angle Twisted Bilayer Graphene},
  author = {Rai, Gautam and Crippa, Lorenzo and C\ifmmode \u{a}\else \u{a}\fi{}lug\ifmmode \u{a}\else \u{a}\fi{}ru, Dumitru and Hu, Haoyu and Paoletti, Francesca and de' Medici, Luca and Georges, Antoine and Bernevig, B. Andrei and Valent\'{\i}, Roser and Sangiovanni, Giorgio and Wehling, Tim},
  journal = {Phys. Rev. X},
  volume = {14},
  issue = {3},
  pages = {031045},
  numpages = {22},
  year = {2024},
  month = {Sep},
  publisher = {American Physical Society},
  doi = {10.1103/PhysRevX.14.031045},
  url = {https://link.aps.org/doi/10.1103/PhysRevX.14.031045}
}

@article{PhysRevLett.131.026502,
  title = {Kondo Lattice Model of Magic-Angle Twisted-Bilayer Graphene: Hund's Rule, Local-Moment Fluctuations, and Low-Energy Effective Theory},
  author = {Hu, Haoyu and Bernevig, B. Andrei and Tsvelik, Alexei M.},
  journal = {Phys. Rev. Lett.},
  volume = {131},
  issue = {2},
  pages = {026502},
  numpages = {8},
  year = {2023},
  month = {Jul},
  publisher = {American Physical Society},
  doi = {10.1103/PhysRevLett.131.026502},
  url = {https://link.aps.org/doi/10.1103/PhysRevLett.131.026502}
}

@misc{calugaruObtainingSpectralFunction2025,
  title = {Obtaining the {{Spectral Function}} of {{Moir\'e Graphene Heavy-Fermions Using Iterative Perturbation Theory}}},
  author = {C{\u a}lug{\u a}ru, Dumitru and Hu, Haoyu and Crippa, Lorenzo and Rai, Gautam and Regnault, Nicolas and Wehling, Tim O. and Valent{\'i}, Roser and Sangiovanni, Giorgio and Bernevig, B. Andrei},
  year = 2025,
  month = sep,
  number = {arXiv:2509.18256},
  eprint = {2509.18256},
  primaryclass = {cond-mat},
  publisher = {arXiv},
  doi = {10.48550/arXiv.2509.18256},
  urldate = {2025-09-30},
  archiveprefix = {arXiv}
}

@misc{SM,
  title        = {Supplemental Material},
  note         = {See Supplemental Material for additional derivations of the self-energy and scattering rate}
}

@article{PhysRevLett.131.166501,
  title = {Symmetric Kondo Lattice States in Doped Strained Twisted Bilayer Graphene},
  author = {Hu, Haoyu and Rai, Gautam and Crippa, Lorenzo and Herzog-Arbeitman, Jonah and C\ifmmode \u{a}\else \u{a}\fi{}lug\ifmmode \u{a}\else \u{a}\fi{}ru, Dumitru and Wehling, Tim and Sangiovanni, Giorgio and Valent\'{\i}, Roser and Tsvelik, Alexei M. and Bernevig, B. Andrei},
  journal = {Phys. Rev. Lett.},
  volume = {131},
  issue = {16},
  pages = {166501},
  numpages = {8},
  year = {2023},
  month = {Oct},
  publisher = {American Physical Society},
  doi = {10.1103/PhysRevLett.131.166501},
  url = {https://link.aps.org/doi/10.1103/PhysRevLett.131.166501}
}

@misc{MaxFragility,
      title={Fragility of local moments against hybridization with flat bands}, 
      author={Max Fischer and Arianna Poli and Lorenzo Crippa and Dumitru Călugăru and Sergio Ciuchi and Matthias Vojta and Alessandro Toschi and Giorgio Sangiovanni},
      year={2025},
      eprint={2503.14326},
      archivePrefix={arXiv},
      primaryClass={cond-mat.str-el},
      url={https://arxiv.org/abs/2503.14326}, 
}

@misc{CrippaStrain,
      title={Dynamical correlation effects in twisted bilayer graphene under strain and lattice relaxation}, 
      author={Lorenzo Crippa and Gautam Rai and Dumitru Călugăru and Haoyu Hu and Jonah Herzog-Arbeitman and B. Andrei Bernevig and Roser Valentí and Giorgio Sangiovanni and Tim Wehling},
      year={2025},
      eprint={2509.19436},
      archivePrefix={arXiv},
      primaryClass={cond-mat.str-el},
      url={https://arxiv.org/abs/2509.19436}, 
}
